\documentclass[lettersize,journal]{IEEEtran}
\usepackage{amsmath,amsfonts}
\usepackage{algorithmic}
\usepackage{algorithm}
\usepackage{array}
\usepackage[caption=false,font=normalsize,labelfont=sf,textfont=sf]{subfig}
\usepackage{cite}
\usepackage{amsmath,amssymb,amsfonts}
\usepackage{textcomp}
\usepackage{stfloats}
\usepackage{url}
\usepackage{booktabs}
\usepackage{verbatim}
\usepackage{graphicx}
\usepackage{multicol}
\usepackage{array}
\usepackage{textcomp}
\usepackage{stfloats}
\usepackage{lipsum}
\usepackage{tikz,hyperref}
\hypersetup{colorlinks,linkcolor={blue},citecolor={blue},urlcolor={blue}} 
\usepackage{makecell}
\usepackage{pdflscape}
\usepackage{amssymb}
\usepackage{multirow}
\usepackage{etoolbox}
\usepackage{enumitem}
\usepackage{balance}
\usepackage{bbding}
\usepackage{pifont}

\UseRawInputEncoding

\definecolor{lime}{HTML}{A6CE39}
\DeclareRobustCommand{\orcidicon}{%
	\begin{tikzpicture}
	\draw[lime, fill=lime] (0,0) 
	circle [radius=0.16] 
	node[white] {{\fontfamily{qag}\selectfont \tiny ID}};
	\draw[white, fill=white] (-0.0625,0.095) 
	circle [radius=0.007];
	\end{tikzpicture}
	\hspace{-2mm}
}

\foreach \x in {A, ..., Z}{%
	\expandafter\xdef\csname orcid\x\endcsname{\noexpand\href{https://orcid.org/\csname orcidauthor\x\endcsname}{\noexpand\orcidicon}}
}

\begin{document}

\title{Zero-Shot Visual Deepfake Detection: Can AI Predict and Prevent Fake Content Before It’s Created?}

\author{Ayan Sar\orcidA,~\IEEEmembership{Student Member,~IEEE}, Sampurna Roy\orcidD,~\IEEEmembership{Student Member,~IEEE}, \\ Tanupriya Choudhury\orcidB,~\IEEEmembership{Senior Member,~IEEE}, Ajith Abhraham\orcidC,~\IEEEmembership{Fellow,~IEEE}
\thanks{Corresponding Author: Tanupriya Choudhury (email: tanupriya@ddn.upes.ac.in)}
\thanks{Ayan Sar, Sampurna Roy, and Tanupriya Choudhury are with the School of Computer Sciences, University of Petroleum and Energy Studies (UPES), Dehradun, 248007, Uttarakhand, India. (email: 500096946@stu.upes.ac.in, sampurna200430@gmail.com, tanupriya@ddn.upes.ac.in)}
\thanks{Ajith Abhraham is with the School of Artificial Intelligence, Sai University, Chennai 603104, India. (e-mail: abraham.ajith@gmail.com)}}

\markboth{Foundations and Trends in Signal Processing}%
{Ayan Sar \MakeLowercase{\textit{et al.}}: Zero-Shot Visual Deepfake Detection: Can AI Predict and Prevent Fake Content Before It’s Created?}


\maketitle

\begin{abstract}
Generative adversarial networks (GANs) and diffusion models have dramatically advanced deepfake technology, and its threats to digital security, media integrity, and public trust have increased rapidly. This research explored zero-shot deepfake detection – an emerging method even when the models have never seen a particular deepfake variation. In this work, we studied self-supervised learning, transformer-based zero-shot classifier, generative model fingerprinting, and meta-learning techniques that better adapt to the ever-evolving deepfake threat. In addition, we suggested AI-driven prevention strategies that mitigated the underlying generation pipeline of the deepfakes before they occurred. They consisted of adversarial perturbations for creating deepfake generators, digital watermarking for content authenticity verification, real-time AI monitoring for content creation pipelines, and blockchain-based content verification frameworks. Despite these advancements, zero-shot detection and prevention faced critical challenges such as adversarial attacks, scalability constraints, ethical dilemmas, and the absence of standardized evaluation benchmarks. These limitations were addressed by discussing future research directions on explainable AI for deepfake detection, multimodal fusion based on image, audio, and text analysis, quantum AI for enhanced security, and federated learning for privacy-preserving deepfake detection. This further highlighted the need for an integrated defense framework for digital authenticity that utilized zero-shot learning in combination with preventive deepfake mechanisms. Finally, we highlighted the important role of interdisciplinary collaboration between AI researchers, cybersecurity experts, and policymakers to create resilient defenses against the rising tide of deepfake attacks. 
\end{abstract}


\section{Introduction}
\begin{figure*}[t]
    \centering
    \includegraphics[width=\linewidth]{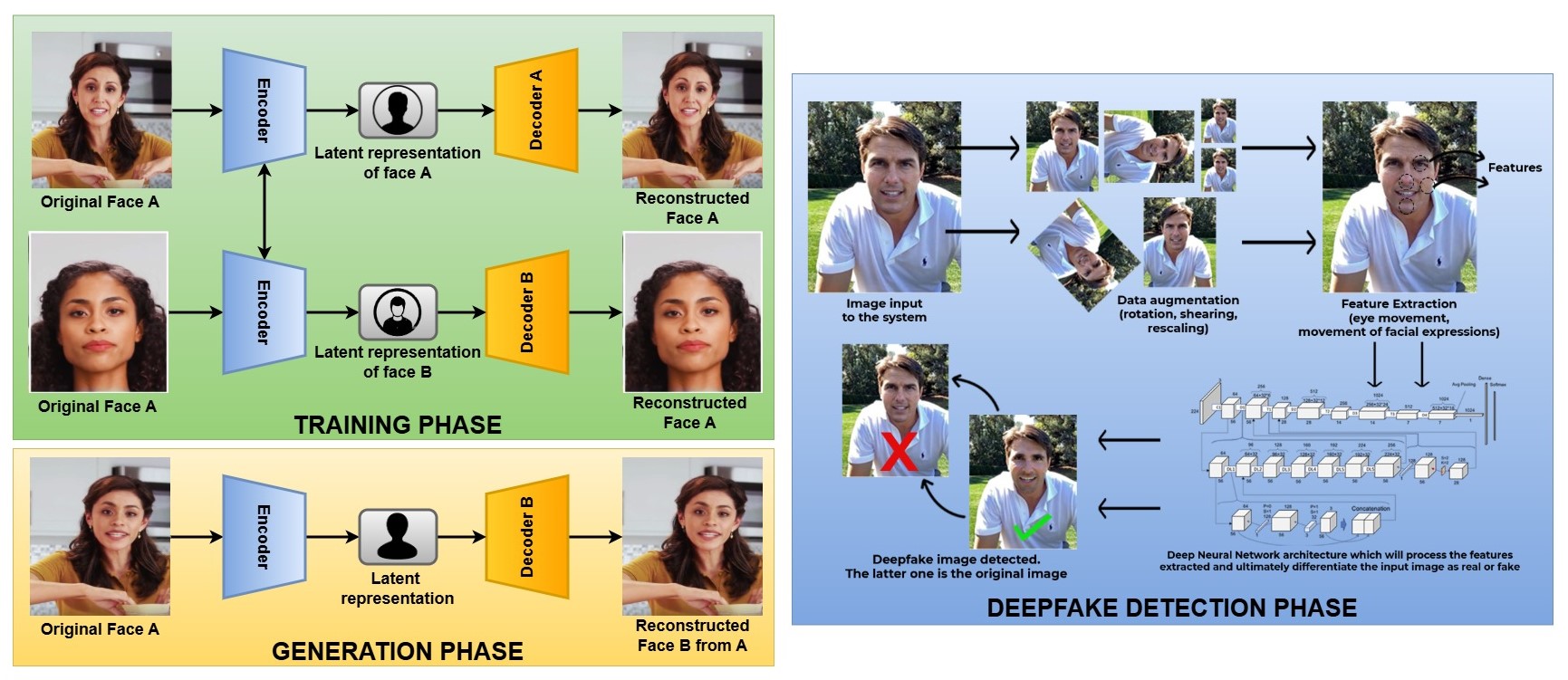}
    \caption{Methodological diagram showing the process of deepfake creating in training phase and generation phase, as well as the deepfake detection phase}
    \label{fig:dvsg}
\end{figure*}

\begin{figure}[t]
    \centering
    \includegraphics[width=\linewidth]{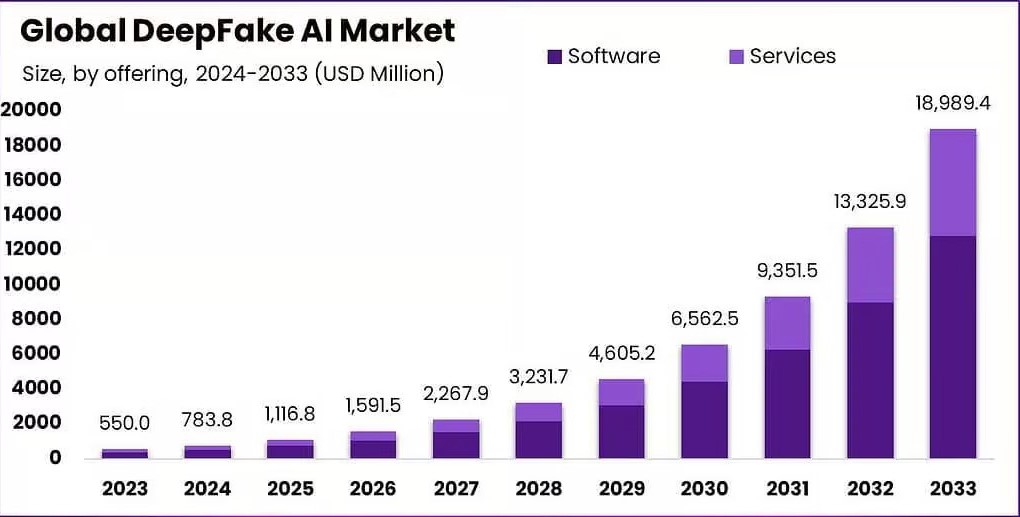}
    \caption{Year-by-year growth of Deepfake AI Market from 2024 to 2033 in USD Million}
    \label{fig:daitrend}
\end{figure}

\begin{figure}[t]
    \centering
    \includegraphics[width=\linewidth]{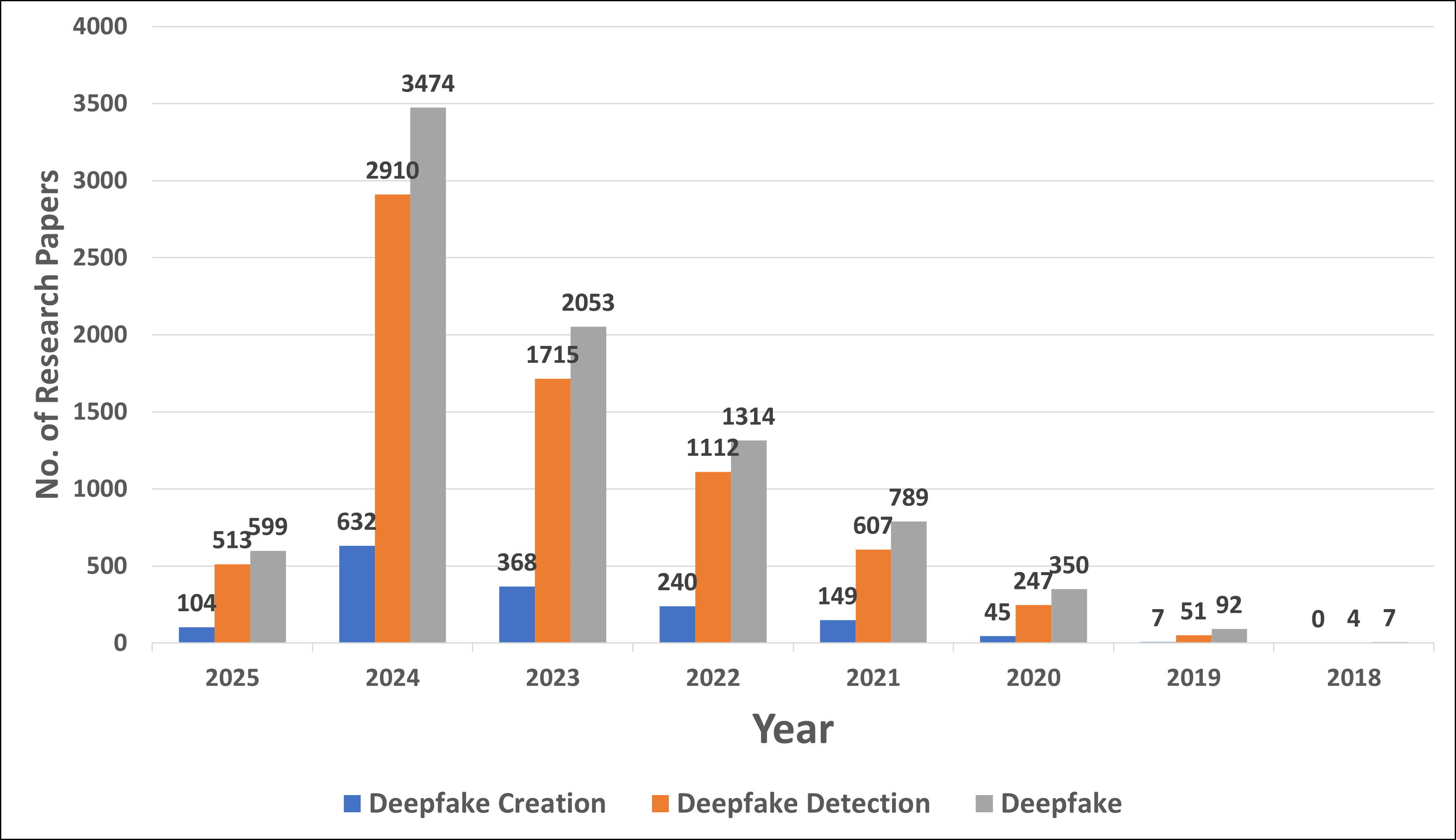}
    \caption{No. of Research studies published from 2018 to 2025 till date (12-03-2025) on Deepfake Creation, and Deepfake Detection. The data is taken from the Scopus database.}
    \label{fig:vs}
\end{figure}

\IEEEPARstart{W}{ith} the rapid advancement in artificial intelligence these days, deepfake technology has come into existence with the help of deep learning models, especially Generative Adversarial Networks (GANs) \cite{Mishra_2024} \cite{Sharma_2024} and diffusion models \cite{Dagar_2022} to create visual and audio content by synthetic manipulation with extreme reality. Deepfakes were first built for fun entertainment \cite{Chawki_2024}, filmmaking \cite{Rathi_2024}, and virtual reality \cite{Tchaptchet_2025}. Still, with the advancement of deep learning, the tools have become a digital threat and have been used to spread misinformation \cite{AL_KHAZRAJI_2023}, identity theft \cite{Maniyal_2024}, political manipulation \cite{Samoilenko_2023}, and financial fraud \cite{Gamb_n_2024}, among other things. With deepfake techniques becoming more advanced, there is a serious security risk in blurring the lines between real and synthesized content \cite{Barrientos_B_ez_2024}. Deepfake models are becoming more sophisticated than traditional detection methods, and it has become harder to separate authentic media from manipulated content \cite{Nagarhalli_2024} \cite{Babaei_2025}. Modern deepfake models have exceeded many limitations using adversarial training and the ability to synthesize high resolution. More recently, the emergence of real-time deepfake generators \cite{Ramadhani_2020} \cite{Rana_2024}, able to modify videos and voices in milliseconds, complicates the efforts in enabling deepfake detection and mitigation, as shown in Fig. \ref{fig:dvsg}. According to market.us report \cite{key}, the global DeepFake AI market is expected to be worth around USD 18,989.4 Million By 2033, from USD 550 Million in 2023, growing at a CAGR of 42.5\% during the forecast period from 2024 to 2033., with year by year growth shown in Fig. \ref{fig:daitrend}. Deepfake technology has developed rapidly; intelligent, adaptive, and proactive defense mechanisms have become increasingly important. Deepfake detection models of this kind struggle to generalize to other types of deepfakes \cite{Le_2023}, need large labeled datasets \cite{Zhalgasbayev_2024} \cite{Kaur_2024}, and are frequently retrained \cite{Chamot_2022} \cite{Pang_2023}. The number of studies published in deepfake detection and creation in recent years from 2018 till date in Fig. \ref{fig:vs} also shows the increasing importance of accurate deepfake detection frameworks. As a result, researchers have started to explore zero-shot deepfake detection, an emerging trend where AIs can tell deepfakes without any particular fake samples in their database.

Despite notable advancements in deepfake detection research, most current approaches are based on supervised deep-learning models that rely heavily on large-scale labeled datasets \cite{10924158}. These models faced several key limitations, as outlined below.

\begin{enumerate}[wide, labelwidth=!, labelindent=0pt]
    \item Challenges in generalization where deepfake detection models, which were trained on a specific dataset, often failed to detect deepfakes generated by unseen and advanced architectures \cite{Khormali_2024} \cite{Atamna_2023}.
    \item The dependency on data increases the need for extensive labeled deepfake datasets, making it difficult to keep up with rapidly evolving deepfake techniques \cite{Agarwal_2024} \cite{Stanciu_2024} \cite{Rana_2024} \cite{Li_2023}.
    \item Many detection models require high processing power, limiting their real-time deployment on edge devices \cite{Karathanasis_2025} \cite{Mohzary_2023} \cite{Sridevi_2022}.
    \item Attackers can train deepfake models to bypass detection by introducing adversarial perturbations \cite{Gandhi_2020} \cite{Ma_2019} \cite{Le_2023}.
    \item There is no confirmed way of preventing the making of deepfakes from multimedia content \cite{Bagaria_2024}.
\end{enumerate}

Given these limitations, zero-shot learning (ZSL) presented an innovative alternative, unlike traditional supervised learning, which enabled deepfake detection without the requirement of prior exposure to specific content \cite{Guo_2023}. Zero-shot learning is an emerging AI paradigm that identifies patterns in previously unseen data with generalized feature representations rather than relying on labeled samples \cite{Chen_2021} \cite{Verma_2024}. In this context, ZSL enabled the models to detect anomalies in facial movements, texture inconsistencies, and frequency distortions without requiring labeled fake data \cite{guo2018zero}. Secondly, it enabled to use of the potential of multimodal learning by analyzing images, videos, and audio cues simultaneously \cite{Cao_2023} \cite{Parida_2020} \cite{Hong_2023}. Adapting new deepfake generation methods without retraining the model from scratch was also one advantage of ZSL \cite{Verma_2024} \cite{Khare_2020}. Several key approaches drove the process of zero-shot deepfake detection, including

\begin{enumerate}[wide, labelwidth=!, labelindent=0pt]
    \item Self-supervised learning where the deepfake characteristics are learned from unaltered real-world data with identification of deviations. 
    \item Using pretrained models like CLIP (Contrastice Language-Image Pretraining) and Vision Transformers (ViTs) could detect synthetic content using textual prompts.
    \item Generative model fingerprinting identified latent space artifacts that were unique to synthetic media.
    \item Few shot and meta-learning methods trained models on minimal deepfake samples and made them adapt for generalized detection.
\end{enumerate}

Integrating these techniques with zero-shot learning provided a robust and future-proof alternative to the traditional deepfake detection models.

\subsection{Contributions}
This research manuscript made several key contributions toward advancing zero-shot deepfake detection and prevention by addressing the critical gaps in existing methodologies. The key contributions here are as follows:

\begin{enumerate}[wide, labelwidth=!, labelindent=0pt]
    \item We provided an in-depth review of how deepfake technology had evolved from early GAN-based synthesis to modern transformer-driven and real-time deepfake models. This examined the widening threat landscape, including misinformation campaigns, fraud identification, cybersecurity risks, and ethical concerns with synthetic media. We mapped the trajectory of deepfake evolution and highlighted the urgent need for adaptable and proactive defense mechanisms.
    \item We explored zero-shot deepfake detection techniques that enabled AI models to identify fake content without prior exposure to specific synthetic variations.
    \item We also discussed AI-driven prevention mechanisms that focused on stopping deepfakes at their source point. We explored several techniques, such as adversarial perturbations, digital watermarking, blockchain authentication, and real-time AI monitoring, mainly focusing on prevention at the earliest possible point.
    \item We also examined key challenges in the limitations of zero-shot deepfake detection, spanning across different angles from adversarial robustness to scalability to standardized benchmarks and ethical concerns.
    \item We proposed several future research directions that could enhance deepfake detection and prevention. We suggested several explainable AI techniques, multimodal systems, quantum AI, and federated learning, which would enable decentralized, privacy-preserving, and, at the same time, interpretable frameworks.
\end{enumerate}

\subsection{Organization of the manuscript}
The remainder of the paper is structured systematically to provide a clear, step-by-step exploration of zero-shot deepfake detection and prevention strategies. Section \ref{sec2} provides an in-depth review of the latest advancements in deepfake generation techniques, which included GAN-based deepfake models, diffusion-based deepfake generation, transformer-driven deepfake synthesis, and real-time deepfake generators with their implications for security and media integrity. Section \ref{sec3} delved into the state-of-the-art zero-shot deepfake detection methodologies, which covered self-supervised anomaly detection, transformer-based zero-shot classifiers, and latent space analysis. Next, Section \ref{sec4} introduced proactive AI strategies for limiting deepfake generation, which included different methods from adversarial perturbations to blockchain-based verification framework to counter the generation of deepfake from multimedia contents. As everything has its limitations, Section \ref{sec5} shows the hurdles and its limitations, along with Section \ref{sec6} outlining promising future directions for the existing challenges. At last, Section \ref{sec7} concluded the paper by summarizing the key insights emphasizing the importance of a multi-layered defense strategy that combines the detection and prevention strategies to combat deepfake threats.

\section{Advances in Deepfake Generation and Their Implications} \label{sec2}
Deepfake generation has witnessed rapid advancements, which were driven by breakthroughs in deep learning, generative adversarial networks (GANs), transformers, and diffusion models. These advancements created highly realistic and indistinguishable synthetic media, which posed severe threats to digital security, misinformation control, and public trust. This section explored the evolution of deepfake generation, key technical improvements, and its far-reaching implications across various formats.

\subsection{Evolution of Deepfake generation technologies}

\subsubsection{Generative Adversarial Networks (GANs)}
Generative Adversarial Networks (GANs) have revolutionized artificial intelligence and deep learning, specifically in generating synthetic content. GANs were first conceived by Ian Goodfellow et al. in 2014 \cite{NIPS2014_f033ed80}. They revolutionized the applications of machine learning that involve media synthesis without direct human intervention, allowing them to produce pictures, videos, and audio that look just as real as human-made. Adversarial architecture underlying GANs involves a propped-up pair of (G, D) neural networks that play games with each other in a zero-sum game to generate better-sounding generated content \cite{Trevisan_de_Souza_2023} \cite{bhat2025review}. However, by leveraging this adversarial training framework, GANs were put into this position of supporting modern deepfake technology by being able to create lifelike facial animations, voice clones, and full-body motion synthesis \cite{J_2024} \cite{Cao_2019} \cite{Aarti_2021} \cite{Purwins_2019}. Over the years, the deepfakes became far better, more realistic, and so powerful that it was difficult to tell them apart using these architectures.

\begin{figure*}[t]
    \centering
    \includegraphics[width=\textwidth]{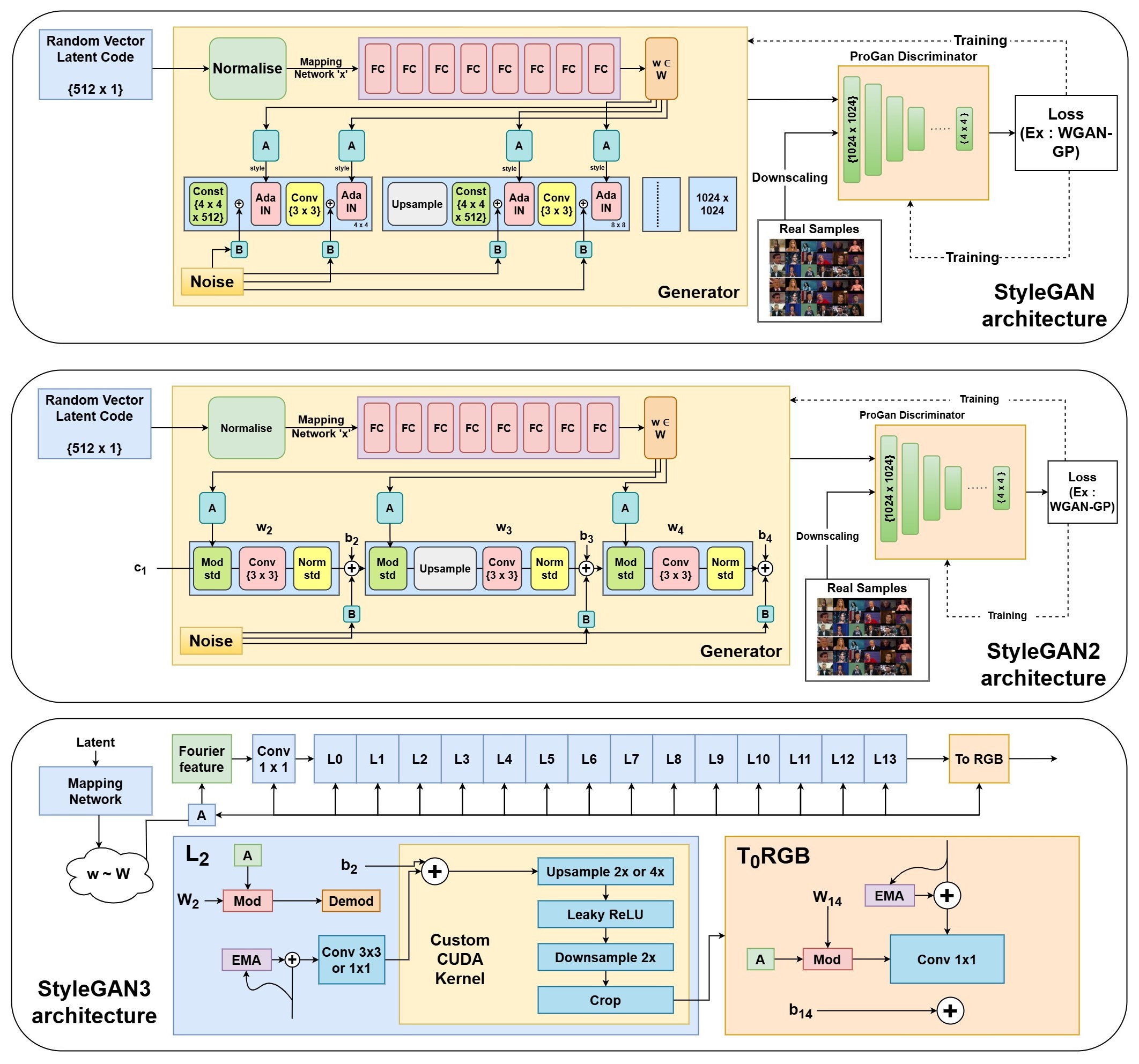}
    \caption{The architecture of three different models developed by NVIDIA named StyleGAN, StyleGAN2, and StyleGAN3 is shown here.}
    \label{fig:sgan}
\end{figure*}

At its core, GANs consisted of two deep neural networks that operated in opposition. First is the generator, which creates synthetic data closely resembling real-world data. This started with random noise and, through different iterative learning, generated output that was increasingly indistinguishable from genuine content. The other was a discriminator, which acted as a binary classifier, distinguishing between real data (from the dataset) and fake data (from the generator). This gave the generator feedback, which helped improve over multiple iterations. The training process is adversarial, where both networks compete continuously. The generator learned to fool the discriminator by creating increasingly realistic content, and the discriminator refined its ability to differentiate between real and fake data. This iterative back-and-forth optimization pushed both networks to enhance their respective performance, which led to high-quality synthetic outputs \cite{Kang_2020} \cite{Prasanthi_2024} \cite{Trevisan_de_Souza_2023}. The objective function of GANs is mathematically expressed in Eq. \ref{eq1}.

\begin{equation}
\begin{aligned}
    \min_G \max_D V(D,G) &= \mathbb{E}_{x \sim p_{data}(x)}[\log D(x)] \\
    &\quad + \mathbb{E}_{z \sim p_{z}(z)}[\log(1 - D(G(z)))]
\end{aligned}
\label{eq1}
\end{equation}

Here, in Eq. \ref{eq1}, $G(z)$ generates the synthetic data from noise vector $z$ and $D(x)$ classified whether the data $x$ is real or fake. The generator ($G$) aimed to minimize the function (i.e. produced outputs that fooled the discriminator), and the discriminator ($D$) aimed to maximize the function (i.e. correctly distinguishing real from fake).

Since the original GAN model was proposed, various architectural advancements have significantly improved image quality, training stability, and realism in deepfake generation. Some notable architectures include:

\begin{enumerate}[wide, labelwidth=!, labelindent=0pt]
    \item Deep Convolutional GANs(DCGANs): Introduced in 2015, DCGANs \cite{radford2016unsupervisedrepresentationlearningdeep} incorporated convolutional layers instead of traditional fully connected layers, which enabled better feature extraction and image synthesis. DCGANs were the first major improvement in GAN-based deepfake generation, which allowed for more stable training and higher-resolution images.
    \item Proposed by Karras (2017) \cite{karras2018progressivegrowinggansimproved} PGGANs introduced a progressive training approach, where image resolution was gradually increased over multiple training phases. This prevented the mode collapse ( which was a common problem in early GANs where the generator produced limited variations) and significantly improved texture details and realism.
    \item Developed by Nvidia, StyleGAN(2018) \cite{karras2019stylebasedgeneratorarchitecturegenerative}, and StyleGAN2 (2019) \cite{karras2020analyzingimprovingimagequality} brought a major leap forward in deepfake realism. The key improvements included style mixing, which enabled finer control over features like hair, skin tone, and facial attributes. Next, Adaptive Instance Normalization (AdaIN) allowed for highly detailed and customizable face synthesis. The improved disentanglement reduced visual artifacts and improved the overall fidelity of generated images. Next, StyleGAN3 \cite{Karras2021} further enhanced realism by eliminating texture inconsistencies and achieving seamless facial blending in high-resolution images. The architecture of the three frameworks is shown in Fig. \ref{fig:sgan}.
    \item CycleGAN (2017) \cite{zhu2020unpairedimagetoimagetranslationusing} introduced the concept of image-to-image translation, which allowed for deepfake transformations, such as changing a person's facial expression or converting a day-time scene into night-time. StarGAN \cite{choi2018starganunifiedgenerativeadversarial} extended this concept by enabling multi-domain deepfake manipulations, such as age progression, gender transformation, and facial attribute modification in a single model.
\end{enumerate}

While GANs empowered the deepfake generation, they also played a key role in deepfake detection through zero-shot learning techniques. Some AI-driven detection models use adversarially trained GANs to identify hidden patterns in deepfakes using residual inconsistencies, generate synthetic training samples for detecting previously unseen deepfake manipulations, and deploy anti-GAN countermeasures that disrupt the process of deepfake generation. Further advancements in GAN-based deepfake prevention could include self-supervised learning, meta-learning, and differential privacy approaches to mitigate deepfake threats before they are created.

\subsubsection{Transformers}
Deepfake technology has evolved beyond traditional GANs to use the potential of transformer-based architectures, significantly enhancing the realism and complexity of AI-generated synthetic media. Initially developed for natural language processing (NLP), transformers have demonstrated remarkable performance in image, video, and speech synthesis \cite{Zhang_2025} \cite{Arshed_2024} \cite{Fnu_2024}. Their ability to model long-range dependencies, understand contextual relationships, and handle high-dimensional data has made them tools for deepfake generation. Transformers, introduced in 2017 \cite{vaswani2023attentionneed}, revolutionized AI by introducing a self-attention mechanism that enabled models to weigh the importance of different input features dynamically. Unlike convolutional or recurrent neural networks, which process data sequentially or in local patches, transformers handle entire sequences simultaneously, which allows them to generate more coherent and contextually aware outputs. In the process of deepfake generation, transformers were applied in the face and video manipulation to generate highly realistic and expressive synthetic faces \cite{Adriana_Mercioni_2024}. In speech and lip-syncing, AI-generated voices are synchronized through transformers with facial movements. The cross-modal synthesis included integrating text, images, and audio to create highly convincing multimedia deepfakes \cite{Khan_2022} \cite{Naik_2024}.

DALL-E, which was developed by OpenAI, and Imagen, developed by Google, represented state-of-the-art models capable of generating hyper-realistic images from textual descriptions. These models utilized large-scale transformer-based architectures to produce synthetic media often indistinguishable from real content. DALL-E (2021) generated photo-realistic human faces, objects, and environments using simple text prompts \cite{Sudha_2024}. Imagen (2022) outperformed previous models' fidelity and detail, producing AI-generated visuals with near-perfect realism \cite{Xu_2024} \cite{Li_2023}. These models posed significant deepfake risks, as they allowed users to create entirely synthetic people who do not exist, which could be easily used for identifying fraud, misinformation , and synthetic propaganda \cite{Saha_2024} \cite{Gonzales_2023}. Transformer-based models \cite{Qi_2024} \cite{Liu_2022} had significantly enhanced facial reenactment techniques , enabling real-time deepfake applications \cite{Wu_2018} \cite{Chapagain_2024} such as DeepFaceLab, one of the most widely used tools that leverage transformers for facial transformation and replacement \cite{Gonzales_2023}. FaceShifter utilized multi-layer self-attention networks to accurately capture facial features and expressions in videos, producing seamless face swaps. These tools surpassed traditional GAN-based methods with better texture consistency \cite{Xu_2023}, improved facial expression alignment \cite{Xu_2023}, and real-time processing capability \cite{Wu_2018} \cite{Wang_2023}. The combination of transformers and facial reenactment algorithms had made video manipulation more accessible than ever, which raised serious ethical and security concerns \cite{Brey_2004} \cite{Hu_2021}. One of the most challenging aspects of deepfake videos is synchronizing AI-generated speech with realistic lip movements. Transformer models like Wav2Lip \cite{Rafiei_Oskooei_2024}, and advanced Text-to-Speech (TTS) engines have solved this issue, making AI-powered voice cloning \cite{Chen_2024} and lip-syncing nearly undetectable. Wav2Lip (2020) \cite{Rafiei_Oskooei_2024} synchronized lip movements in real-time using transformers, which allowed deepfakes to speak with perfect lip coordination. Neural TTS \cite{Ujoodha_2024} (Tacotron 2, FastSpeech  \cite{Sang_2021}, VITS) enabled deepfake speech \cite{Galyashina_2022} synthesis with human-like cadence, emotion, and tonality, making AI-generated voices indistinguishable from real ones. These advancements heightened the risks of fake news, AI-generated propaganda, and misinformation as individuals could be made to say things they never actually said with the most accurate realism possible \cite{Borrelli_2021}.

\subsubsection{Diffusion Models}
Deepfake generation has evolved significantly from early machine learning techniques to sophisticated GANs and, more recently, diffusion models. Diffusion models presented a paradigm shift in deepfake synthesis, which offered high fidelity, improved realism, and greater robustness against different detection mechanisms \cite{Li_2023} \cite{Wang_2024} \cite{Li_2023}. These models had already begun to outperform GANs in tasks such as image and video synthesis, facial animation, and deepfake text-to-image generation, raising new challenges in combatting AI-generated misinformation \cite{Kumar_2024} \cite{Solano_Carrillo_2023}. Diffusion models are mainly probabilistic generative models based on the principle of iterative noise reduction. Unlike GANs, which heavily rely on adversarial processes, diffusion models learned to reverse a noise-adding process, which generated high-quality images and videos through a stepwise denoising approach \cite {Preeti_2023} \cite{Maniyal_2024}. In the forward diffusion process, the model gradually adds Gaussian noise to an image, which destroys its structure over multiple steps until it resembles pure noise. In the reverse diffusion process \cite{Hu_2023} \cite{Mirza_2023}, the model learned for progressive noise removal at each step while reconstructing the original image or generating a new synthetic image from noise \cite{Hu_2023}. The key advantage of this approach is that diffusion models generated more diverse and stable outputs than GANs, which avoided problems like mode collapse (a situation where the model produced limited variations of outputs) \cite{Duan_2023}.

\begin{figure}[]
    \centering
    \includegraphics[width=\linewidth]{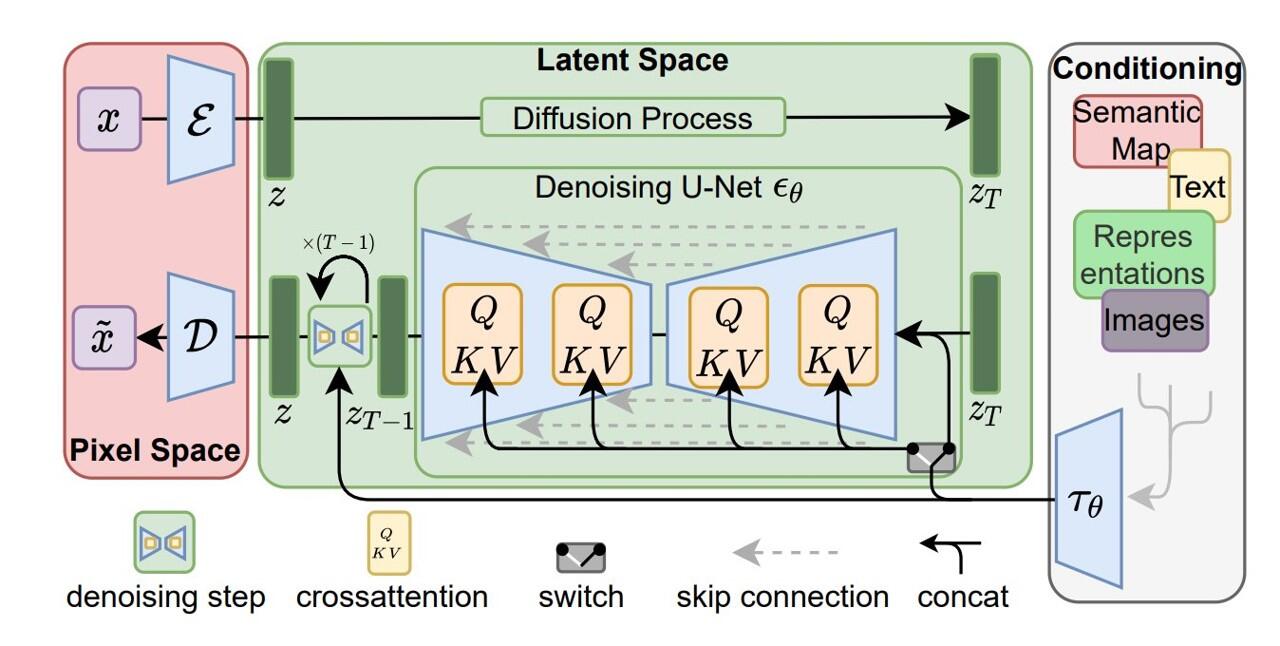}
    \caption{The architecture of Stable Diffusion}
    \label{fig:diff}
\end{figure}

Diffusion models had significantly advanced deepfake creation with improved image resolution, enabling cross-modal generation (text-to-image/video) and enhanced realism in synthetic content. Stable Diffusion (Stability AI) (see Fig. \ref{fig:diff}) emerged as an open-source diffusion model that allowed users to generate high-quality images on consumer hardware. Unlike closed systems such as DALL-E 2, Stable diffusion offered unrestricted access, meaning anyone with a decent GPU could create realistic deepfakes without advanced experience or expertise. This sparked ethical concerns, as bad actors could manipulate public figures' images, create any sort of misleading news visuals, or produce hyper-realistic forgeries with minimal effort \cite{AL_KHAZRAJI_2023} \cite{Bagaria_2024}. On the other hand, while early deepfake models focused primarily on static images, diffusion-based video synthesis emerged as a game changer. Imagen video (Google Deepmind) and Make-A-Video (Meta AI) used diffusion-based denoising techniques for generating full-motion deepfake videos. Unlike traditional frame-based GANs, these models generated coherent, fluid animations with consistent facial expressions, background details, and lighting effects. The implications were concerning where such models could be used to fabricate realistic videos of political leaders, celebrities, or even ordinary individuals, making them indistinguishable from genuine recordings \cite{Shin_2022} \cite{Plazas_Olmedo_2022} \cite{Lalancette_2017}.

Diffusion models offered several advantages over GANs, which made them more effective yet more dangerous in the hands of malicious users.

\begin{enumerate}[wide, labelwidth=!, labelindent=0pt]
    \item The higher image quality produced higher-resolution deepfake images and videos with fewer artifacts, better texture details, and improved color accuracy \cite{Shen_2024} \cite{Ali_Raza_2024} \cite{Tan_2024}. GAN-generated deepfakes often suffer from blurring, pixelation, and unnatural facial features, which are less prevalent in diffusion-based outputs \cite{Liu_2024} \cite{Phung_2023}.
    \item The better diversity and mode coverage allowed diffusion models to generate a broader range of faces, expressions, and movements, which made it harder to identify deepfakes using traditional pattern-based detection methods \cite{Li_2023}.
    \item The greater control over image manipulation and text-to-image and text-to-video capabilities allowed for precise control over deepfake attributes \cite{Arshad_2024} \cite{Singh_2023} . This level of control enhanced their potential for deception \cite{Arshad_2024}, which made it easier to fabricate highly personalized and convincing fake content\cite{Hasan_Fuad_2023}.
    \item Many deepfake detection techniques rely on artifacts left by GAN-generated faces. The high resilience against deepfake detection methods produced more naturally blended images and videos, making them significantly harder to detect existing forensic analysis tools \cite{Kang_2022} \cite{Huang_2020} \cite{Liu_2023}.
\end{enumerate}

As diffusion-based models continue to improve deepfake generation, there is an urgent need to develop zero-shot detection mechanisms and AI-based countermeasures \cite{Khan_2024} \cite{Zhang_2024}. Diffusion models have revolutionized deepfake synthesis, surpassing GANs in quality, diversity, and realism. While these advancements unlock new creative possibilities, they pose serious security risks due to their resilience against detection \cite{Lu_2024} \cite{Liu_2024}. The next phase of AI research must develop robust countermeasures to detect and prevent misuse before deepfake content becomes indistinguishable from reality \cite{Yoon_2024}.

\subsubsection{Real-Time Deepfake Generation Applications}

\renewcommand{\arraystretch}{1.3}
\begin{table*}[t!]
    \centering
    \caption{Comprehensive Comparison of Deepfake Tools: Face-Swapping, Lip-Syncing, and Voice Cloning Technologies}
    {\fontsize{7pt}{7pt}\selectfont \begin{tabular}{>{\centering\arraybackslash}m{2cm}>{\centering\arraybackslash}m{4cm}>{\centering\arraybackslash}m{3cm}>{\centering\arraybackslash}m{2cm}>{\centering\arraybackslash}m{2cm}>{\centering\arraybackslash}m{2cm}} \hline
    \multicolumn{6}{c}{\textbf{Face-Swapping and Video Manipulation Tools}} \\ \hline
         \textbf{Application} & \textbf{Description} & \textbf{Key Features} & \textbf{Platform} & \textbf{Processing Time} & \textbf{Privacy Concerns}  \\ \hline \hline
         Reface &	AI-based face-swapping app for GIFs and videos & 	Real-time face swap, animation & 	iOS, Android & 	~5 sec	 & Stores user face data on servers \\ \hline
         Zao &		Chinese deepfake app with pre-trained celebrity faces &		One-tap video face swap &		iOS, Android &		~8 sec &		High data privacy risks\\ \hline
        DeepFaceLive &		Live deepfake streaming tool for real-time video calls &		Works with Zoom, Skype, Discord &		Windows, Linux &		Instant	 &	Real-time manipulation risk\\ \hline
        FaceMagic	 &	User-friendly deepfake app for video manipulation &		Multiple face swap, high-quality rendering &		iOS, Android &		~6 sec	 &	Stores user-generated content\\ \hline
        Xpression Camera &		AI-driven live face-swap tool for streaming &		Works with Twitch, YouTube &		Windows, Mac &		< 1 sec	 &	No clear data policy\\ \hline
        Avatarify &		AI-based real-time avatar animation &		Works on Zoom, Skype	 &	iOS, Windows &		< 3 sec &		Captures real-time face movement\\ \hline
        DeepFakeWeb &		Online deepfake face swap tool &		No installation needed, cloud-based &		Web-based &		~15 sec &		Cloud storage risk\\ \hline
        Morphin	 &	AI-generated 3D GIF-based deepfake tool &		Cartoon-like avatars with face swap &		iOS, Android	 &	~10 sec &		No clear encryption policy\\ \hline
    \end{tabular}}
    \vspace{0.1cm}
    {\fontsize{7pt}{7pt}\selectfont \begin{tabular}{>{\centering\arraybackslash}m{2cm}>{\centering\arraybackslash}m{4cm}>{\centering\arraybackslash}m{3cm}>{\centering\arraybackslash}m{2cm}>{\centering\arraybackslash}m{2cm}>{\centering\arraybackslash}m{2cm}}
    \hline
     \multicolumn{6}{c}{\textbf{Comparison of Lip-Syncing \& Voice Cloning Deepfake Tools}} \\ \hline
    \textbf{Tool} & \textbf{Description} & \textbf{Key Features} & \textbf{Platform} & \textbf{Processing Speed} & \textbf{Misuse Risk} \\
    \hline \hline
    \textbf{Wav2Lip} & AI-powered lip-syncing tool & Real-time audio-to-video synchronization & Windows, Linux & < 1 sec & Misinformation \& impersonation \\
    \hline
    \textbf{Synthesia} & AI-generated talking avatars & Text-to-speech with facial animations & Web-based & < 2 sec & Deepfake marketing risk \\
    \hline
    \textbf{iFake} & Voice-cloning AI & Celebrity voice synthesis & Web-based & ~5 sec & Identity theft risk \\
    \hline
    \textbf{Resemble AI} & High-quality voice cloning & Real-time text-to-speech & Web, API-based & < 3 sec & Phishing scams \\
    \hline
    \textbf{Descript Overdub} & AI-powered voice cloning & Ultra-realistic text-to-speech synthesis & Windows, Mac & < 2 sec & Fake news risk \\
    \hline
    \textbf{ElevenLabs AI} & High-fidelity AI-generated voice synthesis & Multilingual deepfake voices & Web-based & < 1 sec & Election misinformation risk \\
    \hline
    \textbf{VALL-E (Microsoft)} & Text-to-voice AI that mimics speech & Can replicate voice tones and accents & Web-based & ~5 sec & Political deepfakes \\
    \hline
    \textbf{Voicemy.ai} & Custom AI voice models & Celebrity voice cloning & Web, Mobile & < 3 sec & Prank calls \& scams \\
    \hline
    \textbf{Uberduck AI} & Text-to-speech with deepfake voices & AI-powered song generation & Web-based & < 2 sec & Audio manipulation threats \\
    \hline
    \end{tabular}}
    \label{tab:faceswap}
\end{table*}

The advent of real-time deepfake generation had significantly lowered the wall to entry, which enabled individuals without technical knowledge to create highly realistic and convincing synthetic media \cite{Morris_2024} \cite{Singh_2021}. Unlike early deepfake technologies that required hours to process high-quality videos, modern applications leveraged advanced deep learning architectures, cloud computing \cite{Choquette_2021}, and GPU acceleration \cite{Gschwind_2017} for generating deepfakes in real-time. Real-time deepfake synthesis relied on the following technological advancements:

\begin{enumerate}[wide, labelwidth=!, labelindent=0pt]
    \item Efficient model architectures like autoencoders and GANs where optimized encoder-decoder architecture allowed for real-time processing with learning compressed latent representation of faces \cite{Kwak_2020} \cite{Chen_2021}. The transformer-based models \cite{Sun_2022} used self-attention mechanisms, improving contextual accuracy for facial expressions \cite{Yang_2023} and lip-syncing. Though primarily used for image generation, diffusion models \cite{Han_2024} synthesized ultra-realistic face swaps in real-time.
    \item Realtime deepfake generation required high computational power. Nvidia RTX series GPUs and Google TPUs significantly reduced latency, enabling deepfake applications to process videos at over 30 FPS (frames per second).
    \item The implementation of Edge AI (e.g. TensorFlow Lite, ONNX) allowed mobile devices to process deepfakes with reduced dependency on cloud servers locally \cite{Hao_2021} \cite{Gill_2024}. Moreover, Cloud-based AI inferences (AWS, Google Cloud) improved deepfake processing speeds by offloading computation to high-performance servers \cite{Wu_2021} \cite{Shrestha_2023}.
\end{enumerate}

The rise of AI-powered deepfake applications has transformed how users interact with synthetic media. These applications were widely used for entertainment, social media, gaming, and cyber threats. Real-time face-swapping apps \cite{den_Uyl_2015} are widely used in social media, video editing, and virtual communication. Real-time deepfake voice cloning and lip-syncing tools \cite{Priya_2023} \cite{Datta_2024} have transformed content creation, customer service, and cybercrime \cite{Blancaflor_2024}. Table \ref{tab:faceswap} shows the expanded comparison of these tools. While deepfake technology had serious risks, it also positively impacted various domains. The film industry and special effects segment have reduced CGI costs by enabling digital face-swapping for actors \cite{Mittal_2024} \cite{Singh_2023}. This was also used in de-aging effects (e.g. Marvel movies). This enhanced social media content with AI-generated avatars and enabled real-time video dubbing and translation \cite{Sardana_2024}. On the other hand, this also helps generate AI-powered educational videos in multiple languages \cite{Nasar_2020}. AI avatars also help in interactive learning for younger children \cite{Teresa_2023}. Speech-impaired individuals could communicate through AI-generated voice synthesis \cite{Regondi_2025}  and create virtual customer service agents with lifelike animations\cite{Fink_2024}. Real-time deepfake technology has revolutionized digital media and introduced unprecedented ethical, security, and privacy challenges \cite{de_Ruiter_2021} \cite{Sharma_2024}. Developing AI-powered detection tools, legal frameworks, and public awareness strategies is imperative to counteract deepfake misinformation and fraud \cite{Matli_2024}.

\subsection{Implications of Advanced Deepfake Generation with Case Studies}
Deepfake technology \cite{Maniyal_2024} has rapidly evolved from experimental AI capability to sophisticated tools with important societal, economic, political, and security implications. With deepfake technology having legitimate applications in entertainment, education, and historic preservation \cite{Mittal_2024} \cite{Lundberg_2024}, its misuse has led to severe consequences like identity theft, political disinformation, financial fraud, and psychological manipulation \cite{Malik_2024} \cite{Kopecky_2024}. The widespread accessibility of deepfake generation tools has amplified these risks, as even non-experts could create deceptive content using publicly available AI models \cite{Shakil_2024} \cite{Chapagain_2024}. 

One of the most alarming risks deepfake technology poses is identity theft and digital security breaches. With the cloning of faces and voices, deepfake technology allowed cybercriminals to bypass biometric authentication \cite{Volkova_2023} \cite{Do_2021} (e.g. facial recognition and voice recognition systems), impersonate individuals \cite{Gilbert_2024} \cite{Agarwal_2023} in video calls or social media interactions, and engage in financial fraud and phishing attacks by imitating executives, government officials or family members \cite{de_Rancourt_Raymond_2022} \cite{Khan_2024}.

While this study focuses on visual deepfake detection, we briefly included a seminal example of audio deepfake threats to provide a complete threat landscape. This inclusion underscores the growing need for zero-shot learning approaches that can be extended or adapted across modalities, especially as attackers increasingly employ cross-modal deception techniques.

\subsubsection{Case Study 1: CEO Voice Deepfake Scam (2019, UK-Germany)}
According to the new report in The Wall Street Journal \cite{Stupp_2019}, the CEO of an unnamed UK-based energy company thought he was speaking with his boss (the chief executive) at the German parent company's firm when he followed the orders of €220,000 (approx. \$243,000) to be transferred to the bank account of a Hungarian supplier. The voice was traced to a man illegally using AI technology to mimic the German chief executive. WSJ was provided with the information by Rüdiger Kirsch of the firm's insurance company, Euler Hermes Group SA. He noted that the CEO understood the slight German accent in his boss’s voice but, even more importantly, knew it conveyed the man's ‘melody.’ The yet unidentified fraudster called the company three times: once to start the transfer, twice for a false reimbursement, and once more for another payment. At this stage, the victim became suspicious; he saw that the fabricated reimbursement had not gone through and that the call had been made from an Austrian phone number.

\subsubsection{Case Study 2: Deepfake Bypass of Biometric Authentication (2023, China)}
In 2023, Chinese cybercriminals employed advanced deepfake technology to bypass facial recognition systems used in mobile banking applications. These attackers successfully tricked biometric authentication mechanisms by creating highly realistic AI-generated videos, leading to unauthorized access and significant financial losses \cite{TheHackerNews}. 

Deepfake-generated fake news, speeches, and altered media have been increasingly used for manipulating public opinion and interfering in elections. Political deepfakes could fabricate speeches and public statements by world leaders, alter any historical records for distortion of reality, and create hyper-realistic propaganda videos to sway voter perception. 

\subsubsection{Case Study 3: Deepfake Video of Ukrainian President Volodymyr Zelenskyy (2022)}
Hackers put a fake, heavily manipulated video of Ukrainian President Volodymyr Zelenskyy out on the web on a Ukrainian news site \cite{Allyn_2022}, thwarted Facebook's attempts to remove it, and it spread widely through social media all the same. However, the video was a so-called deepfake, and it lasted around a minute, in which a rendering of the Ukrainian president was apparently filmed telling his soldiers to lay down their arms and surrender the fight against Russia. 

\subsubsection{Case Study 4: Deepfake of Indian Politician in Elections (2020, India)}
During the 2020 Delhi elections, political advertising was carried out with deepfake tech.
A video of BJP MP Manoj Tiwari was tampered with to make it look like the politician speaks multiple languages fluently \cite{Alavi_Achom_2020}. AI shone a light on its power to create new political messaging, while the deepfake was not malicious in itself, merely indicative of possible manipulation.

The increasing use of deepfake technology raised significant ethical concerns regarding consent, privacy, and psychological distress. Over 96\% of deepfakes generated and posted online involved non-consensual sensitive content, which targeted women disproportionately. Here, victims often face severe emotional trauma, career damage, and social stigma because of this. Numerous celebrities have been targeted by AI-generated explicit videos, including Scarlett Johansson publicly criticizing deepfake technology as "The internet is a vast wormhole of darkness that eats itself" \cite{Harwell_2018}. While these platforms attempted to ban such content, it spread through different dark web forums.

\section{Zero-Shot Deepfake Detection: Methods and Approaches} \label{sec3}
With the continuing progress of deepfake generation methods, traditional deepfake detection models cannot keep up with the newly emerging ones. Supervised learning-based approaches \cite{K_t_k__2024} have been the predominant ones, but they depend on labeled datasets for training and hence have limited power to generalize to unknown deepfakes variation \cite{Lin_2022}. In its attempt to overcome this limitation, zero-shot deepfake detection (ZSDD) permits AI to identify deepfakes without prior exposure to specific fake samples. It is achieved through self-supervised learning, anomaly detection \cite{Wang_2024}, transformer-based classification\cite{Patil_2024}, and generative model fingerprinting \cite{Pontorno_2024}. In this section, we gave attention to many zero-shot deepfake detection methodologies and classified them mostly by several approaches that make the system more adaptable, robust, and efficient.

\subsection{Contrastive Learning-Based Deepfake Detection}
Generative models are constantly maturing, and detecting deepfakes is very difficult. Traditional supervised learning is not amenable to successfully learning new deepfake techniques when only a large amount of labeled real material is available for deepfake supervised learning. A self-supervised learning approach that learns interesting feature representation \cite{Fung_2021} from unlabeled data is called contrastive learning, which suits zero-shot deepfake detection very well \cite{Liu_2022} \cite{Dai_2023}. In contrastive learning, the model has to learn to understand the similarities and differences between pairs of images \cite{Moskalev_2022} \cite{Wang_2024}. Similar images (real vs. real) should be closer to the feature space's core than the representation of different images (real vs. deepfake)\cite{John_2022} \cite{Arshad_2024}. It allows the model to learn that these don't exist without it having been exposed to them previously. The key concepts associated is shown in Table \ref{tab:con}.

\begin{table}[t!]
    \centering
    \caption{Key Concepts in Contrastive Learning for Deepfake Detection}
    {\fontsize{7pt}{7pt}\selectfont \begin{tabular}{>{\centering\arraybackslash}m{2cm}>{\centering\arraybackslash}m{5.6cm}} \hline
        \textbf{Concept} & \textbf{Description} \\ \hline \hline
        Anchor Image & A reference image, usually a real face \\ \hline
        Positive Pair & An augmented version of the anchor image (real image with minor transformations)\\ \hline
        Negative Pair & A deepfake image that should be distinguished from the anchor \\ \hline
    Embedding Space & 	A high-dimensional space where similar images have closer vector representations\\ \hline
    Distance Metric	 & A mathematical function (e.g., cosine similarity, Euclidean distance) to measure similarity\\ \hline
    Contrastive Loss & 	A loss function that minimizes the distance between positive pairs and maximizes it between negative pairs\\ \hline
    \end{tabular}}
    \label{tab:con}
\end{table}

Several contrastive learning frameworks have been developed to improve deepfake detection without requiring explicit labels. The most widely used are described below:

\subsubsection{SimCLR (Simple Contrastive Learning of Representations)}
Generative models are evolving very fast, so the replacement of deepfake detection lacks relevance. With the help of a self-supervised learning framework, SimCLR \cite{Pan_2023} (Simple Contrastive Learning of Representations), we can perform deepfake detection without the requirement of label data. SimCLR learns feature representations such that even in a zero-shot setting \cite{Yu_Diong_2024}, it can distinguish real from fake content based on data augmentation and contrastive loss \cite{Yu_Diong_2024} \cite{Pan_2023}. Google Brain introduced SimCLR as a simple but powerful contrastive learning approach. Compared to traditional supervised learning that relies on huge amounts of prelabeled datasets, SimCLR leverages the invariance properties of data by taking contrasting pairs from similar and dissimilar data.

\begin{figure}[t!]
    \centering
    \includegraphics[width=\linewidth]{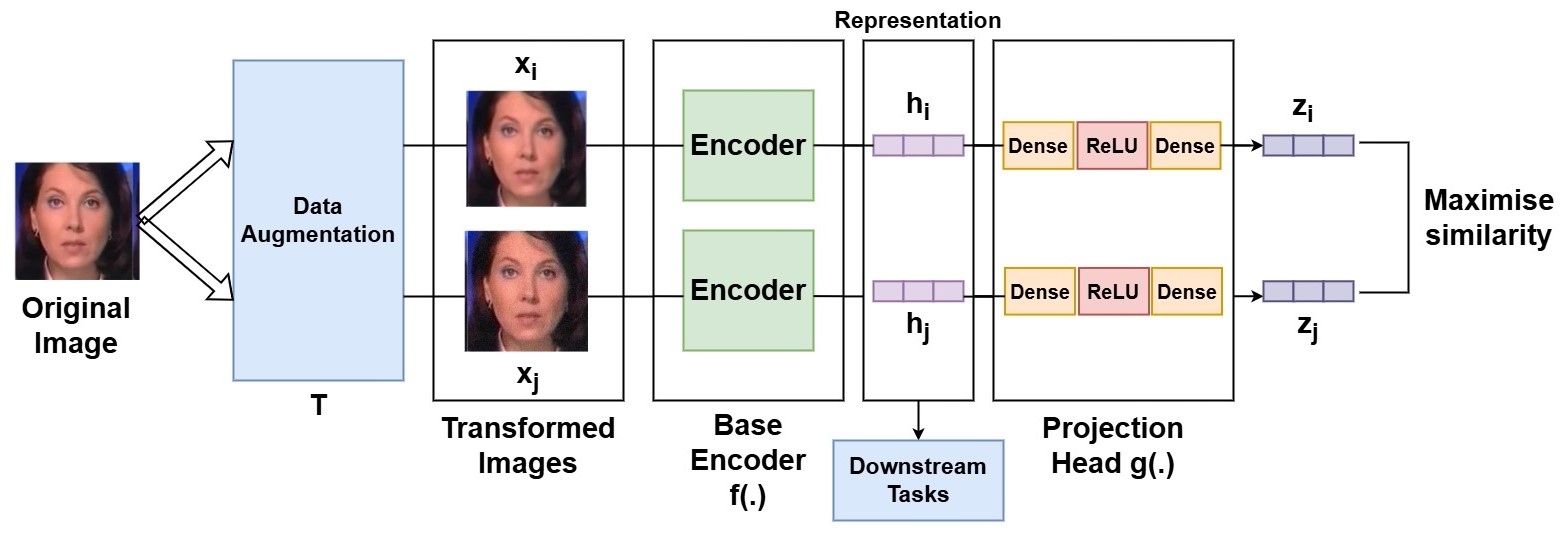}
    \caption{The four-stage pipeline architecture of SimCLR (Simple Contrastive Learning of Representations)}
    \label{fig:simclr}
\end{figure}

SimCLR follows a four-stage pipeline for learning feature representations effectively, as shown in Fig. \ref{fig:simclr}. This starts with data augmentation, where SimCLR creates multiple augmented versions of the same image. This allowed the model to learn invariant features, which made it robust to variations in deepfake synthesis techniques. Table \ref{tab:aug} shows the different augmentation techniques, where the main goal is to make real and fake images more challenging to distinguish. This, in a way, forced the model to focus on deep intrinsic characteristics rather than superficial cues.

\begin{table}[t!]
    \centering
    \caption{Techniques for Data augmentation in SimCLR}
    {\fontsize{7pt}{7pt}\selectfont \begin{tabular}{>{\centering\arraybackslash}m{2cm}>{\centering\arraybackslash}m{5.6cm}} \hline
        \textbf{Augmentation Technique} & \textbf{Description} \\ \hline \hline
        Random Cropping & Extracts different patches from the same image
 \\ \hline
        Color Jittering	 & Alters brightness, contrast, and saturation
\\ \hline
        Gaussian Blur	 & Simulates distortions found in compressed deepfakes
 \\ \hline
    Horizontal Flip	 & 	Flips the image to add variation
\\ \hline
    Grayscale Conversion		 & Helps the model learn texture-based features
\\ \hline
    \end{tabular}}
    \label{tab:aug}
\end{table}

Next, SimCLR employed a deep neural network (usually ResNet50 or Vision Transformer (ViT)) to extract features from images. The model processed augmented image pairs and encoded them into high-dimensional feature space. The ResNet50 learned hierarchical feature representations of real and fake images, whereas ViT captured long-range dependencies and textures from deepfake artifacts. These advanced architectures enabled SimCLR to learn distinguishing patterns in real and fake images for previously unseen deepfakes. Thirdly, the extracted features were passed through the projection head, a multi-layer perceptron (MLP) network \cite{Alsmadi_2009}. This head maps features into lower-dimensional embedding \cite{Cunningham} space where the contrastive loss \cite{Zolfaghari_2021} is applied. The non-linear transformation \cite{Lin_2024} helped to separate real and fake clusters, and dimensionality reduction helped reduce complexity and maintain discriminative power. This projection ensured that similar images were mapped closer together while dissimilar ones were pushed apart in the feature space. SimCLR \cite{Migliorelli_2024}, at its last phase, uses the normalized temperature-scaled cross-entropy loss (NT-Xent) for comparing image pairs and optimizing the model's ability to differentiate between deepfakes. The mathematical representation of NT-Xent \cite{Liu_2023} \cite{Migliorelli_2024} loss is shown in Eq. \ref{eq2}.

\begin{equation}
    \ell_1 = -log \frac{exp(\frac{sim(z_i,z_j)}{\tau})}{\sum_{k=1}^{2N}1_{[k \neq 1]}exp(\frac{sim(z_i,z_j)}{\tau})}
    \label{eq2}
\end{equation}

Here, $z_i,z_j$ are feature representations of positive pairs (augmented views of the same image). $sim(z_i,z_j)$ is the cosine similarity between feature vectors. $\tau$ is the temperature parameter, which controls how strictly similar pairs are pulled together. $1_{[k \neq 1]}$ ensured that the denominator includes only negative pairs. This aims to maximize the similarity between positive pairs while minimizing this between negative pairs. SimCLR is a powerful framework for zero-shot deepfake detection \cite{Zhang_2023}, enabling AI models to identify manipulated content without needing labeled data. By leveraging data augmentation \cite{Zhang_2023}, feature extraction, and contrastive loss, SimCLR \cite{Yoon_2024} provides a scalable and generalizable solution against deepfake threats. Future improvements, such as combining SimCLR with Vision Transformers \cite{Yoon_2024} \cite{Song_2024} (ViTs) and adaptive data augmentation, will enhance its robustness further.

\subsubsection{MoCo (Momentum Contrast)}
Momentum Contrast, a self-supervised learning framework, is designed to generate consistent and robust feature representations simultaneously. In contrast with traditional contrastive learning methods that require large batch sizes, MoCo relies on a queue-based memory bank and momentum encoder to maintain stable representations. MoCo is useful for zero-shot deepfake detection because it learns an evolving representation of real and deepfake images. It doesn't require labeled deepfake data but uses a basic contrastive learning principle to differentiate natural and synthetic facial features\cite{He_2020}.

\begin{figure}[t!]
    \centering
    \includegraphics[width=\linewidth]{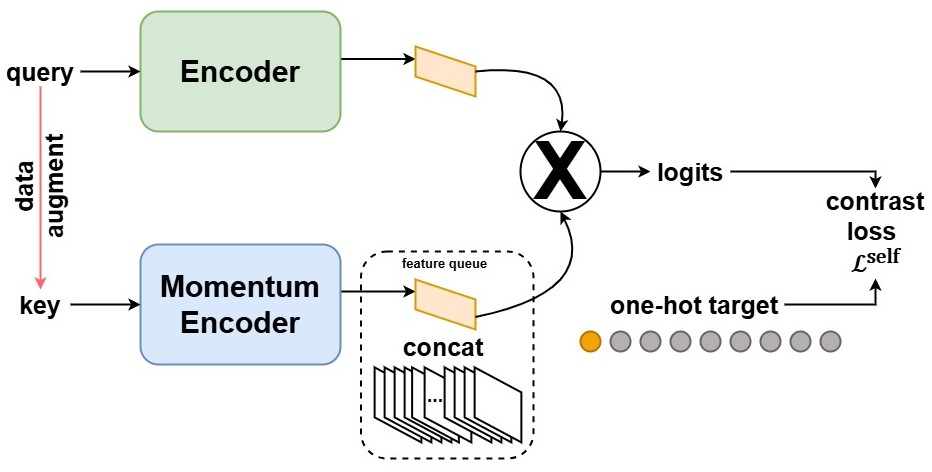}
    \caption{The architectural representation of MoCo (Momentum Contrast))}
    \label{fig:moco}
\end{figure}

MoCo was designed for improvement in contrastive learning by ensuring consistent feature representation over time. The framework consisted of two key networks: Online Network (Query Encoder), which learns feature representations from the input images, and Momentum Network (Key Encoder), which maintains the slowly evolving representation for stability. These two networks interact through a queue-based memory bank to dynamically compare real and deepfake images. The step-by-step process for MoCo is shown in Table \ref{tab:moco}, along with architectural representation in Fig. \ref{fig:moco}.

\begin{table}[t!]
    \centering
    \caption{Step-by-Step Process of MoCo in Zero-shot Deepfake Detection}
    {\fontsize{7pt}{7pt}\selectfont \begin{tabular}{>{\centering\arraybackslash}m{2cm}>{\centering\arraybackslash}m{5.6cm}} \hline
        \textbf{Steps} & \textbf{Description} \\ \hline \hline
        Data Augmentation & Creates multiple variations of the same image (cropping, flipping, color jittering, etc.) to prevent overfitting.
 \\ \hline
        Query Encoder (Online Network)	 & Extracts features from real and deepfake images.
\\ \hline
        Key Encoder (Momentum Network)	 & A slowly updated encoder that provides stable feature representations.
 \\ \hline
    Queue-based Memory Bank		 & 	Stores past embeddings of real and deepfake images, allowing for long-term learning.
\\ \hline
    Contrastive Learning (InfoNCE Loss) & Ensures real images have similar representations, while deepfakes remain distinct.
\\ \hline
    \end{tabular}}
    \label{tab:moco}
\end{table}

The core objective of MoCo is to minimize contrastive loss using similarity metrics. The most common loss function used here is InfoNCE loss, which is defined in Eq. \ref{eq3}.

\begin{equation}
    L = -log \frac{exp(sim(q,k^+)/\tau)}{\sum^K_{i=0}exp(sim(q,k_i)/\tau)}
    \label{eq3}
\end{equation}

Where $q$ is the feature representation of the query image (real or deepfake), $k^+$ is the positive key, $k_i$ is the negative key, $sim(.,.)$ is the similarity function, $\tau$ is the temperature parameter which controlled the contrastive learning strength, $K$ is the number of stored negative keys in the queue. This model maximizes similarity between $q$ and $k^+$ while minimizing similarity with deepfake keys $k_i$.

Momentum Contrast \cite{He_2020} (MoCo) is a powerful self-supervised learning framework for zero-shot deepfake detection. By leveraging a memory bank and momentum encoder, MoCo enables AI models to identify deepfakes without prior labeled data. Its ability to continuously adapt to new deepfake techniques makes it an essential tool for real-world deepfake detection, forensic analysis, and digital content authentication. Future improvements in MoCo-based deepfake detection could reduce computational overhead, enhance feature interpretability, and integrate multi-modal deepfake analysis (e.g., combining visual, audio, and text signals).

\subsubsection{BYOL (Bootstrap Your Own Latent)}
Unlike other contrastive learning, such as SimCLR and MoCo, Bootstrap Your Own Latent (BYOL) was recently proposed as a self-supervised learning framework to learn meaningful feature representations without negative pairs. In zero-shot deepfake detection, BYOL aids in predicting fake content without the need for labeled examples of deepfakes ahead of time. Unlike supervised learning, BYOL has a self-predictive learning mechanism to learn representations from real images and detect deepfake anomalies.

\begin{figure}[t!]
    \centering
    \includegraphics[width=\linewidth]{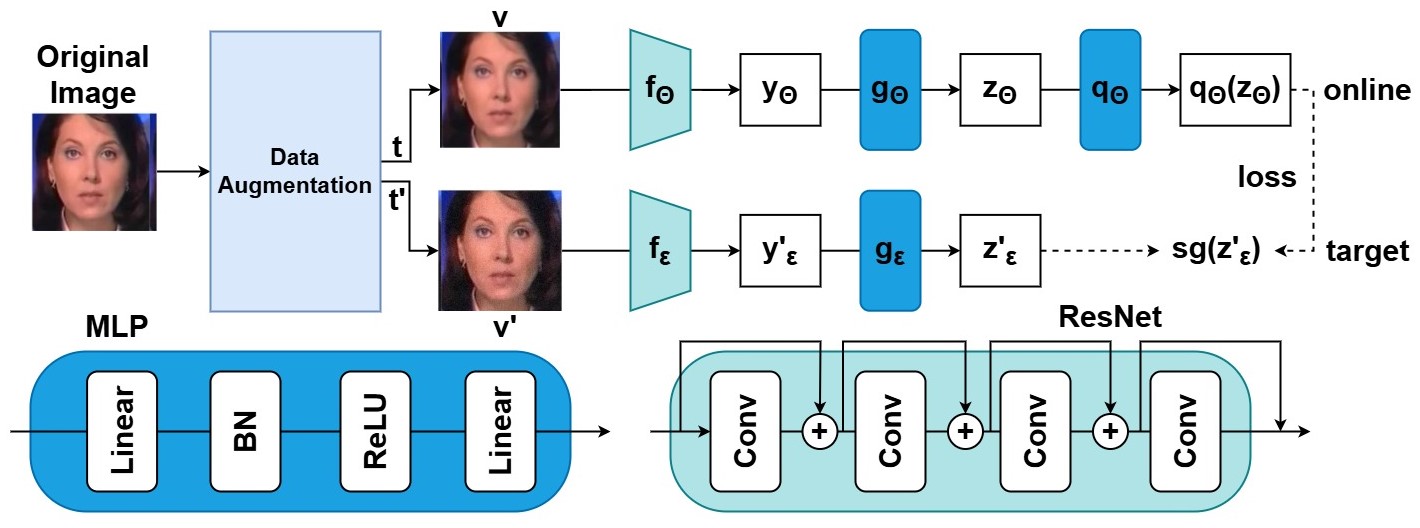}
    \caption{The architectural representation of BYOL (Bootstrap Your Own Latent)}
    \label{fig:byol}
\end{figure}

BYOL \cite {K_t_k__2024} consists of two key networks: the Online network, which learns feature representation dynamically, and the Target network, which provides stable learning signals for training. These networks were trained in a way that the online network predicted the feature representation of the target network, encouraging the model to learn rich representations without explicitly distinguishing between real and fake images. The full architecture of the network is shown in Fig. \ref{fig:byol}. The various components of this architecture and its main function in deepfake detection are also shown in Table \ref{tab:byol}.

\begin{table}[t!]
    \centering
    \caption{Components of BYOL and its function in deepfake detection}
    {\fontsize{7pt}{7pt}\selectfont \begin{tabular}{>{\centering\arraybackslash}m{2cm}>{\centering\arraybackslash}m{5.6cm}} \hline
        \textbf{Components} & \textbf{Function in Deepfake Detection} \\ \hline \hline
        Online Network & Learns dynamic feature representations from real images.
 \\ \hline
        Target Network		 & Serves as a stable reference for training the online network.
\\ \hline
        Projection Head		 & Maps high-dimensional features to a lower-dimensional space.
 \\ \hline
    Prediction Head		 & 	Predicts the feature representation of the target network.
\\ \hline
    \end{tabular}}
    \label{tab:byol}
\end{table}

The first step is applying data augmentation to create different views of real images. These augmentations included random cropping, color jittering, gaussian blur, rotation \& flipping \cite{Bhuse_2021} \cite{Hrga_2022}. This ensured that the model learned robust features that generalized well across different variations of real images. Next, deep neural networks (e.g. ResNet50, EfficientNet, or Vision Transformer) \cite{McNeely_White_2019} \cite{Shah_2020} \cite{Tan_2024} extracted meaningful features from augmented images. The extracted features were then passed to the projection head, which mapped them into lower-dimensional space. The online network processed the first augmented image and tried to predict the representation produced by the target network for the second augmented image. Instead of using explicit negative samples (as in contrastive learning), BYOL minimized the difference between the online network's predicted features and the target network's features \cite{K_t_k__2024}. Over time, the online network learned to generate meaningful representations distinguishing real images from anomalies (potential deepfakes). Next, unlike the online network, the target network was not directly updated using gradient descent. Instead, it followed the momentum update mechanism \cite{Shi_2009} \cite{Xu_2012}, where its parameters were a slow-moving exponential average of the online network's parameters. This ensured stability in learning and prevented overfitting to specific patterns. BYOL has emerged as a powerful zero-shot learning framework for deepfake detection. BYOL learns generalizable deepfake representations without needing explicit fake samples by leveraging self-prediction instead of contrastive loss. This makes it a scalable and future-proof approach as new deepfake techniques evolve \cite{K_t_k__2024}.

\renewcommand{\arraystretch}{1.5}
\begin{table*}[t!]
    \centering
    \caption{Comparison of Contrastive Learning techniques in deepfake detection with Supervised Learning}
    {\fontsize{7pt}{7pt}\selectfont \begin{tabular}{>{\centering\arraybackslash}m{5.3cm}>{\centering\arraybackslash}m{2.5cm}>{\centering\arraybackslash}m{2.5cm}>{\centering\arraybackslash}m{2.5cm}>{\centering\arraybackslash}m{2.5cm}} \hline
        \textbf{Features} & \textbf{Supervised Learning} & \textbf{SimCLR} & \textbf{MoCo} & \textbf{BYOL} \\ \hline \hline
        Required labeled deepfake dataset? & \ding{52} & \XSolidBrush & \XSolidBrush & \XSolidBrush \\ \hline
        Effective in detecting unseen deepfakes? & \XSolidBrush & \ding{52} & \ding{52} & \ding{52} \\ \hline
        Works without predefined deepfake classes? & \XSolidBrush & \ding{52} & \ding{52} & \ding{52} \\ \hline
        Generalizes to new deepfake generation techniques? & \XSolidBrush & \ding{52} & \ding{52} & \ding{52} \\ \hline 
        Learns deep invariant representations? & \XSolidBrush & \ding{52} & \ding{52} & \ding{52} \\ \hline 
        Works with small datasets? & \XSolidBrush & \ding{52} & \ding{52} & \ding{52} \\ \hline 
        Requires expensive data annotation? & \ding{52} & \XSolidBrush & \XSolidBrush & \XSolidBrush \\ \hline
        Scalable to large-scale deepfake datasets? & \XSolidBrush & \ding{52} & \ding{52} & \ding{52} \\ \hline 
        Requires manual feature engineering? & \ding{52} & \XSolidBrush & \XSolidBrush & \XSolidBrush \\ \hline
        Performs well in zero-shot scenarios? & \XSolidBrush & \ding{52} & \ding{52} & \ding{52} \\ \hline 
        Requires large batch sizes for training? & \XSolidBrush & \ding{52} & \XSolidBrush & \XSolidBrush \\ \hline 
        Sensitive to data augmentations? & \XSolidBrush & \ding{52} & \ding{52} & \XSolidBrush \\ \hline 
        Uses a memory queue for learning? & \XSolidBrush & \XSolidBrush & \ding{52} & \XSolidBrush \\ \hline 
        Requires negative pairs for training? & \ding{52} & \ding{52} & \ding{52} & \XSolidBrush \\ \hline
        Uses momentum encoder? & \XSolidBrush & \XSolidBrush & \ding{52} & \XSolidBrush \\ \hline
    \end{tabular}}
    \label{tab:comparative}
\end{table*}

The analysis in Table \ref{tab:comparative} encapsulates the combination of BYOL, other contrastive learning approaches, and supervised learning methods for deepfake detection. Supervised learning models reach very high accuracy when given labeled datasets to train on but are hard to generalize to unseen deepfake techniques. Self-supervised methods obtained through contrastive learning, such as SimCLR, MoCo, and BYOL, provide a lack of labeled deepfake samples alternative, where we achieve robust feature representations. Deepfake detection with contrastive learning has completely removed dependence on labeled datasets \cite{Yeh_2022}. By adopting techniques such as SimCLR, MoCo, and BYOL, AI models can successfully identify deepfakes even without prior exposure and thus serve as highly useful in response to any threats that might be emerging. However, as with deepfake technology, the role of AI-driven defense mechanisms \cite{Zhang_2024} \cite{Zheng_2024}will be contrastive learning in the future.

\subsection{Anomaly Detection via Self-Supervision}
Deepfakes often contain subtle but perceptible distortions \cite{Qureshi_2024} \cite{Le_2023} that depart from natural human features; hence, anomaly detection is crucial to zero-shot deepfake detection. Since they do not require explicit labels, self-supervised anomaly detection \cite{Hojjati_2024} \cite{G_kstorp_2024} is very good at finding new and unseen deepfake methods based on the true difference between real and synthetic content. The second part of this section examines how self-supervised models can discover deepfake anomalies in terms of visual, physiological, and behavioral inconsistency and help in deepfake detection without the dependence on the data set \cite{Lee_2023} \cite{Mandal_2024}. Deepfake videos and images have anomalies caused by the imperfect generative model outputs. In contrast to real human faces, deepfake-generated content is less biometric consistent, has more irregularities in the texture, and cannot realistically keep the facial dynamics. There are three types of these anomalies into which they can be broadly categorized, namely

\begin{enumerate}[wide, labelwidth=!, labelindent=0pt]
    \item Visual artifacts where pixel-level inconsistencies occurred due to imperfect rendering. These are mainly blurred edges, missing reflections, and unnatural skin textures in deepfakes \cite{Kang_2022} \cite{Zhao_2020}.
    \item Physiological irregularities mainly disrupt biometric markers due to AI synthesis. These are improper eye blinking, inconsistent head movements, and abnormal pulse detection \cite{Seha_2019} \cite{Zimmermann_2021} \cite{P_2024}.
    \item Temporal inconsistencies where frame-to-frame mismatches are identified in video deepfakes, such as flickering facial expressions, misaligned speech-lip synchronization, and unnatural posture shifts \cite{Ge_2021} \cite{Datta_2024}.
\end{enumerate}

These anomalies served as key indicators for self-supervised deepfake detection methods. Self-supervised learning (SSL) enabled models for learning patterns from real-world data distributions and identifying deviations in deepfake content. This involved pretraining on real data \cite{Cunha_2024}, where the model learned features from authentic videos/images without deepfake exposure, and anomaly detection through self-supervision, where the model applied predefined learning tasks to uncover hidden irregularities in deepfake samples. The following SSL-based methods have been very effective in zero-shot deepfake anomaly detection \cite{Wang_2024}.

\subsubsection{Autoencoder-Based Anomaly Detection}
\begin{figure}[t!]
    \centering
    \includegraphics[width=\linewidth]{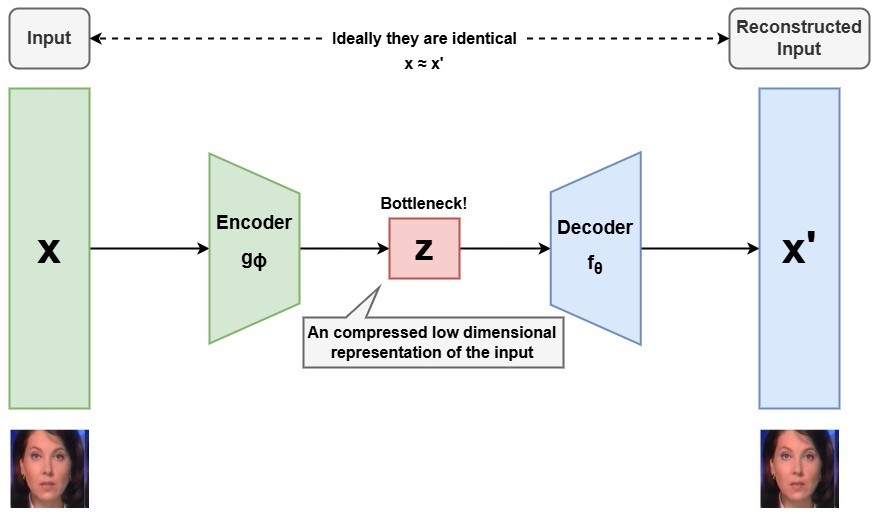}
    \caption{The architectural representation of Autoencoders}
    \label{fig:auto}
\end{figure}

Autoencoder-based anomaly detection is a self-supervised learning framework that detects deepfakes based on reconstruction errors \cite{Wu_2024}. Autoencoders trained on real images are conditioned to encode and decode human faces efficiently. Yet when challenged with deepfake content, its reconstruction quality falls considerably because of hidden inconsistencies in the synthetic faces. The fact that deepfakes produce a different quality of reconstruction compared to real samples indicates that it is a crucial indicator for detecting deepfakes without seeing a fake sample \cite{Hussain_2023}. Autoencoders (AEs) are neural networks that learn compressed latent input data representations. They consisted of two main components: encoder and decoder, as shown in Fig. \ref{fig:auto}. The encoder compressed the input image into a low-dimensional latent representation (feature space). This helps in learning compact representations of real faces. Next, the decoder reconstructs the original image from the encoded representation. This will struggle to reconstruct the deepfake image due to unnatural patterns accurately. Here, the high reconstruction error signals a likely deepfake. If the input image is real, the reconstruction error is low, but for the other case, the error should be high for unexpected distortions.

Autoencoders operate by learning the distribution of real faces and their structural properties. When tested on deepfakes, they failed to reconstruct the synthetic face correctly, which revealed hidden artifacts. The training phase trains the autoencoder exclusively on real human faces to learn the structural and texture distribution \cite{N_2024}. The encoder extracts essential features like skin texture, lighting, and facial symmetry, and the decoder reconstructs the image with minimal distortion \cite{Du_2020}. In the inference phase, a test image is passed through the trained autoencoder, and if the image is real, the model is accurately reconstructed with a low error; otherwise, the reconstruction introduces visible distortions, which results in a high reconstruction error. Different autoencoders are available, enhancing the deepfake detection based on feature learning, compression, and generative capabilities as in Table \ref{tab:auto}. Here, Convolutional Autoencoders are most effective for image-based deepfake detection, while Temporal Autoencoders are best suited for deepfake videos.

\begin{table}[t!]
    \centering
    \caption{Types of autoencoders along with key features and its advantages}
    {\fontsize{7pt}{7pt}\selectfont \begin{tabular}{>{\centering\arraybackslash}m{2cm}>{\centering\arraybackslash}m{2.4cm}>{\centering\arraybackslash}m{2.8cm}} \hline
        \textbf{Types} & \textbf{Key Features} & \textbf{Advantages} \\ \hline \hline
        Vanilla Autoencoder & 	Basic encoder-decoder architecture. & 	Simple, effective for detecting low-quality deepfakes.
 \\ \hline
        Variational Autoencoder (VAE) &	Learns a probabilistic latent space, adding stochasticity. &	Captures subtle variations in deepfake textures.
\\ \hline
        Denoising Autoencoder (DAE)	& Trained to remove noise from images.	& Enhances robustness against adversarial deepfakes.
 \\ \hline
    Convolutional Autoencoder (CAE) & 	Uses CNN layers for better feature extraction. &	Highly effective for detecting image-based deepfakes.
\\ \hline
    Adversarial Autoencoder (AAE) &	Combines autoencoder + GANs for better generalization.	& Handles high-quality deepfakes effectively. \\ \hline
    \end{tabular}}
    \label{tab:auto}
\end{table}

Zero-shot deepfake detection based on an autoencoder is a powerful way of finding previously unseen synthetic content. These models detect deepfake anomalies by combining reconstruction errors \cite{N_2024}, feature extraction, and unsupervised learning without any explicit fake training data. Nevertheless, there is a need to continuously improve the state of the art on hybrid architectures, multimodal analysis \cite{Yu_2024} \cite{Kumar_2024} , and adversarial robustness \cite{Wu_2021} \cite{Chen_2023} to stay ahead of evolving deepfake generation techniques.

\subsubsection{One-Class Support Vector Machines (One-Class SVMs)}
One of the powerful anomaly detection techniques in zero-shot deepfake detection is One-Class Support Vector Machines \cite{Amer_2013} \cite{Yin_2014} (One-Class SVMs). In contrast to more traditional supervised learning techniques, One Class SVMs are trained on real images to approximate a boundary, and a point beyond that boundary is considered a deepfake. This has strong efficacy for zero-shot learning \cite{Al_Machot_2022} \cite{Wang_2019} \cite{Karlos_2021}, as these deepfakes are unnecessary for training. In contrast, it measures deepfakes based on their difference from the learned feature space. One-class SVM is an unsupervised learning approach that learns the distribution of real images and identifies anomalies as outliers. The primary goal is reconstructing a hyperplane or decision boundary that encapsulates real data while excluding anomalous data. The model is trained on real human faces to learn normal feature distributions. A decision boundary is then established around real samples, which minimizes the chances of including deepfake samples. Any new input outside this boundary was flagged as an anomaly (potential deepfake)\cite{Anusha_2025}. 

\begin{figure}[t!]
    \centering
    \includegraphics[width=\linewidth]{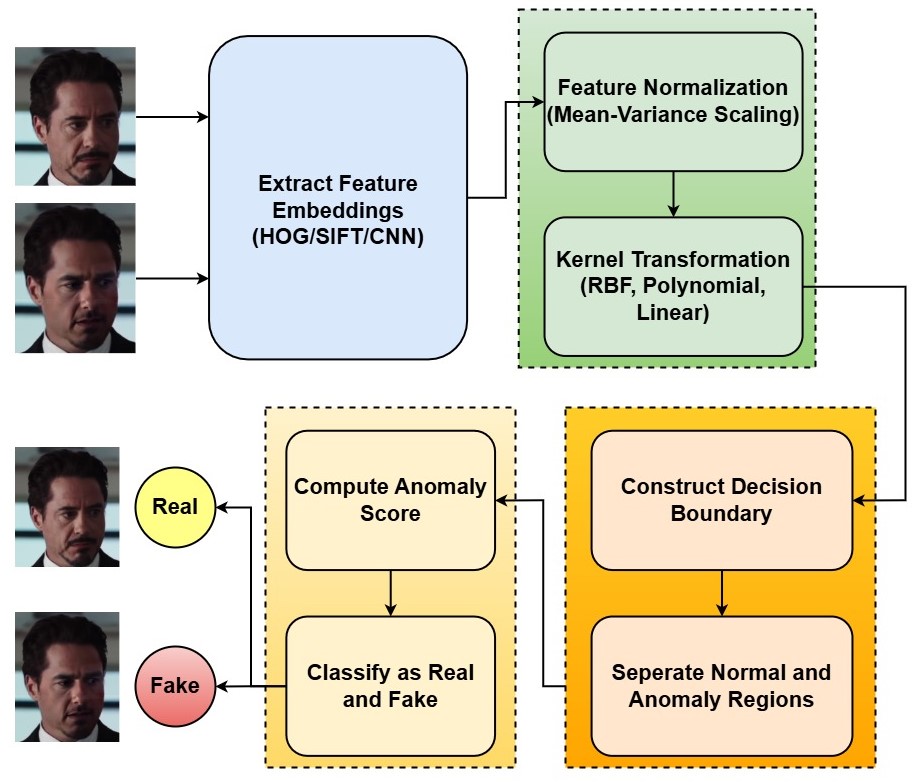}
    \caption{The architectural representation of One-Class Support Vector Machines (One-Class SVMs) in deepfake detection}
    \label{fig:svm}
\end{figure}

\begin{figure*}[t!]
    \centering
    \includegraphics[width=\textwidth]{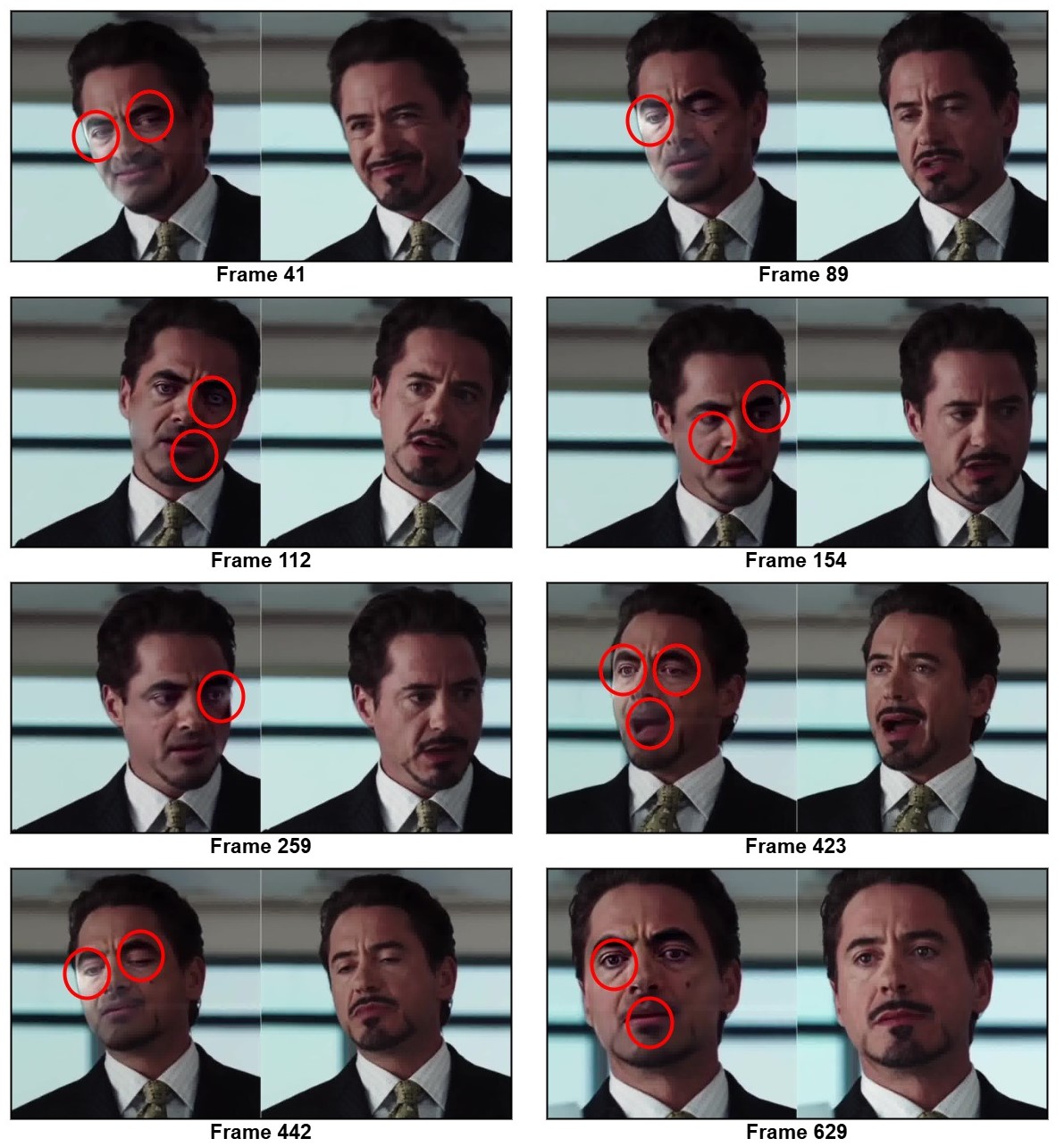}
    \caption{This is a video of Robert Downey Jr., an American actor who was deepfaked by replacing its face with Rowan Atkinson, a well-known figure like Mr. Bean. The right side of the photos shows the original video, and the left side shows the deepfake one. The inconsistent areas are marked with red circles. The video was divided into frames at 26 frames per second with a total of 657 frames, and each frame was manually checked. The best frames with their frame number, where the maximum inconsistency is found, are shown here.}
    \label{fig:incon}
\end{figure*}

Deepfake introduced visual inconsistencies, such as blurred edges, natural textures, or missing reflections, as detailed in Fig. \ref{fig:incon}. These distortions could create statistical anomalies that one-class SVMs could detect. The key features include boundary learning, where the model maps real images into compact feature space. Next, outlier detection detects the deepfakes, being outliers, which fall outside the decision boundary. Zero-shot adaptability eliminates the need for deepfake training samples, making it effective against unseen deepfake types.

The mathematical foundation of one-class SVMs lies in the following optimization problem through Eq. \ref{eq3}.

\begin{equation}
    min_{w, \xi , \rho} \frac{1}{2} ||w||^2 + \frac{1}{vn} \sum ^n_{i=1} \xi_i - \rho
    \label{eq3}
\end{equation}

This is subject to Eq. \ref{eq4}.

\begin{equation}
    (w \cdot \phi(x_i)) \geq \rho - \xi_i, \xi_i \geq 0, \forall_i
    \label{eq4}
\end{equation}

where $w$ is the weight vector defining the decision boundary, $\phi(x)$ is the feature transformation function, $\rho$ is the threshold defining the normal region, $\xi_i$ is the slack variable allowing soft margin violations, and $v$ is the hyperparameter controlling outlier sensitivity. The model maximizes the margin around real images while minimizing violations, ensuring deepfakes are classified as outliers. One-class SVMs use kernel functions \cite{Eshghi_2016} \cite{Elen_2022} \cite{_st_n_2006} to transform input data into higher dimensional space, making complex patterns more distinguishable. Here, the linear kernel\cite{Virmani_2022} separated data with a straight hyperplane, which was effective when deepfakes had simple visual consistencies. The polynomial kernel uses polynomial transformations for complex relationships \cite{Wu_2009}, which capture higher-order patterns in facial distortions. The Radial Basis Function \cite{Tan_2020} (RBF) kernel maps data into an infinite-dimensional space, which is best for detecting subtle deepfake anomalies. The RBF kernel is widely used in deepfake detection because it captures fine-grained differences between real and fake faces \cite{Kafai_2019}.

\subsubsection{Out-of-Distribution (OOD) Detection}

Out-of-Distribution (OOD) \cite{Jo_2024} \cite{Wu_2023} detection was a critical technique in zero-shot deepfake detection that identified anomalies in unseen deepfake content. Since deepfake synthesis methods evolved rapidly, training models on all possible deepfake variants is impractical \cite{Larue_2023}. OOD detection enabled deepfake identification by detecting feature distributions that deviated from real-world data. OOD detection is important for deepfake detection because of its zero-shot capability \cite{Zhao_2023} \cite{Yoon_2024}, which detects deepfakes without prior exposure to specific manipulation techniques. The dataset independence \cite{Jo_2024} avoided reliance on labeled deepfake datasets, which makes the detection scalable. The adaptive nature \cite{Zhu_2022} of this architecture makes it effective in identifying previously unseen deepfake architectures, along with its generalization capability to work from image, video, and audio across different multimedia modalities \cite{_zg_r_2024}. Table \ref{tab:ood} shows the types of distributions in deepfake detection.

\begin{table}[t!]
    \centering
    \caption{Types of distribution in deepfake detection }
    {\fontsize{7pt}{7pt}\selectfont \begin{tabular}{>{\centering\arraybackslash}m{2cm}>{\centering\arraybackslash}m{2.4cm}>{\centering\arraybackslash}m{2.8cm}} \hline
        \textbf{Distribution type} & \textbf{Description} & \textbf{Examples in Deepfake Detection} \\ \hline \hline
        In-distribution (ID)	& Data that follows the expected real-world distribution. & 	Authentic human images and videos with natural facial expressions.
 \\ \hline
        Near-OOD & 	Data that is slightly different from the real-world distribution but still retains some authenticity.	& Partially edited faces (e.g., Snapchat filters, minor retouching).
\\ \hline
        Far-OOD	& Data that is completely unrelated to the learned distribution. &	Cartoon faces, abstract art, synthetic avatars.
 \\ \hline
    \end{tabular}}
    \label{tab:ood}
\end{table}

\begin{figure}[t!]
    \centering
    \includegraphics[width=\linewidth]{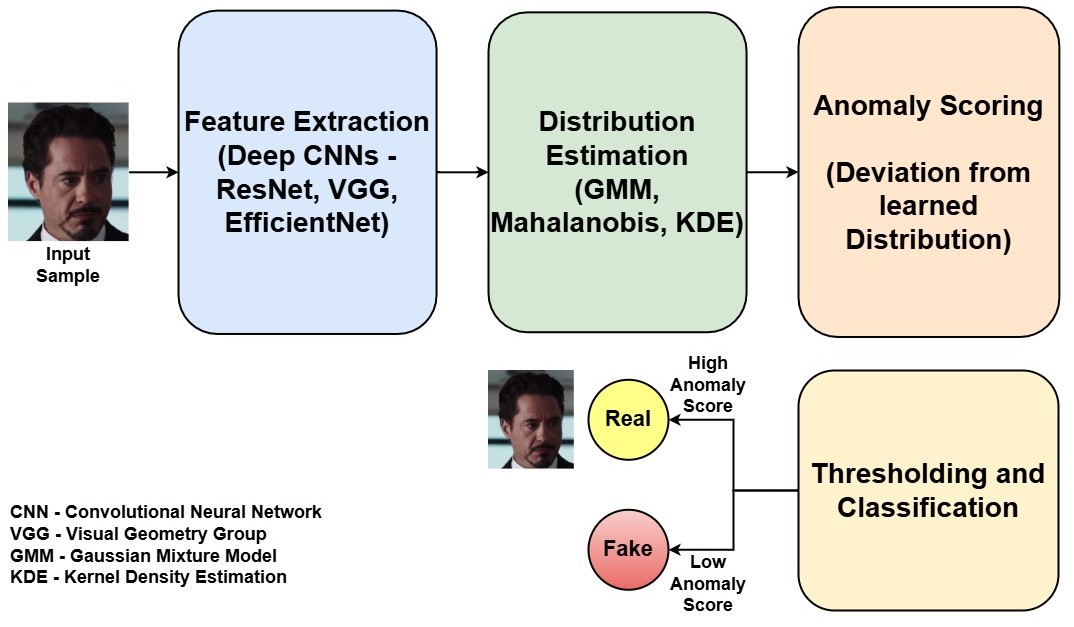}
    \caption{The architectural representation of Out-of-Distribution (OOD) Detection in deepfake detection}
    \label{fig:ood}
\end{figure}

OOD detection was based on the principle that real human images and videos followed a specific distribution in feature space. When deepfakes introduced subtle or drastic changes, they created a distribution shift, allowing detection through statistical and deep learning techniques. The first stage was feature extraction, which extracted embeddings of real and deepfake images using pre-trained deep learning models \cite{Mishra_2023} \cite{Jo_2024}. The distribution estimation helped in learning the distribution of real images using statistical models. The anomaly scoring \cite{Hong_2024} \cite{Deng_2022} assigned an anomaly score for testing samples based on their deviation from the learned distribution. The thresholding and classification differentiated the samples with high anomaly scores as deepfakes. The full architecture is shown in Fig. \ref{fig:ood}.

\begin{table}[t!]
    \centering
    \caption{Methods for OOD Detection in Deepfake Detection}
    {\fontsize{7pt}{7pt}\selectfont \begin{tabular}{>{\centering\arraybackslash}m{1.8cm}>{\centering\arraybackslash}m{2.6cm}>{\centering\arraybackslash}m{2.8cm}} \hline
    \multicolumn{3}{c}{\textbf{Statistical OOD Detection}} \\ \hline
        \textbf{Method} & \textbf{Working Principle} & \textbf{Application in Deepfake Detection} \\ \hline \hline
        Gaussian Mixture Model (GMM)	& Models real data as a mixture of multiple Gaussian distributions. &	Identifies deepfakes as samples falling outside the learned Gaussian clusters.
 \\ \hline
        Mahalanobis Distance &	Measures how far a sample is from the mean of the real data distribution.	 & High distance scores indicate synthetic content.
\\ \hline
        Kernel Density Estimation (KDE) & 	Estimates real data's probability density function (PDF). & 	Deepfake samples show low likelihood in the PDF.
 \\ \hline
    \multicolumn{3}{c}{\textbf{Feature-Based OOD Detection}} \\ \hline   
    Deep Embedding Models & 	Extracts feature vectors from images using deep CNNs. & 	Identifies deepfakes through feature-space distance. \\ \hline
    Contrastive Learning &	Learns to maximize similarity for real images and minimize similarity for deepfakes. &	Creates a discriminative embedding space for real vs. fake. \\ \hline
    t-SNE \& PCA Visualization & 	Reduces high-dimensional embeddings to a lower-dimensional space for anomaly detection. &	Clusters deepfake samples far from real ones. \\ \hline
    \multicolumn{3}{c}{\textbf{Deep Learning-Based OOD Detection}} \\ \hline 
    Energy-based OOD Models & 	Assign an energy score to images; lower energy = real, higher energy = fake. & 	Deepfakes have higher energy values, making them detectable. \\ \hline
    Outlier Exposure (OE) Networks & 	Trained with an auxiliary dataset of fake images to learn an explicit OOD boundary.	& Enhances the ability to reject deepfake samples. \\ \hline
    Self-supervised OOD Detection	& Training on pretext tasks (e.g., rotation, jigsaw, colorization) to learn intrinsic features. &	Learns to distinguish authentic vs. AI-generated images. \\ \hline
    \end{tabular}}
    \label{tab:oods}
\end{table}

OOD detection could be implemented using statistical, feature-based, and deep learning methods \cite{Zheng_2024}. Statistical methods detect deepfakes by measuring distributional differences between the real and fake ones. The main principle behind this is that if a sample does not fit the real data's statistical distribution, it is flagged as a deepfake. Feature-based methods extract high-level representations from deepfake and real images, comparing their embeddings in a learned space. The principle for feature-based OOD is that if the image's features don't align with real image embeddings, it will be classified as a deepfake. Deep learning-based OOD detection employs neural networks to model real data distributions and identify deepfakes based on their energy levels \cite{Cao_2024}, confidence scores\cite{Guan_2023}, and novelty detection mechanisms \cite{Domingues_2018}. Here, the principle is that if a deep learning model fails to classify an image confidently, it will likely be a deepfake. Table \ref{tab:oods} shows the three types of OOD detection techniques.

\subsubsection{Temporal Anomaly Detection for Deepfake Videos}

\begin{figure*}[t!]
    \centering
    \includegraphics[width=\textwidth]{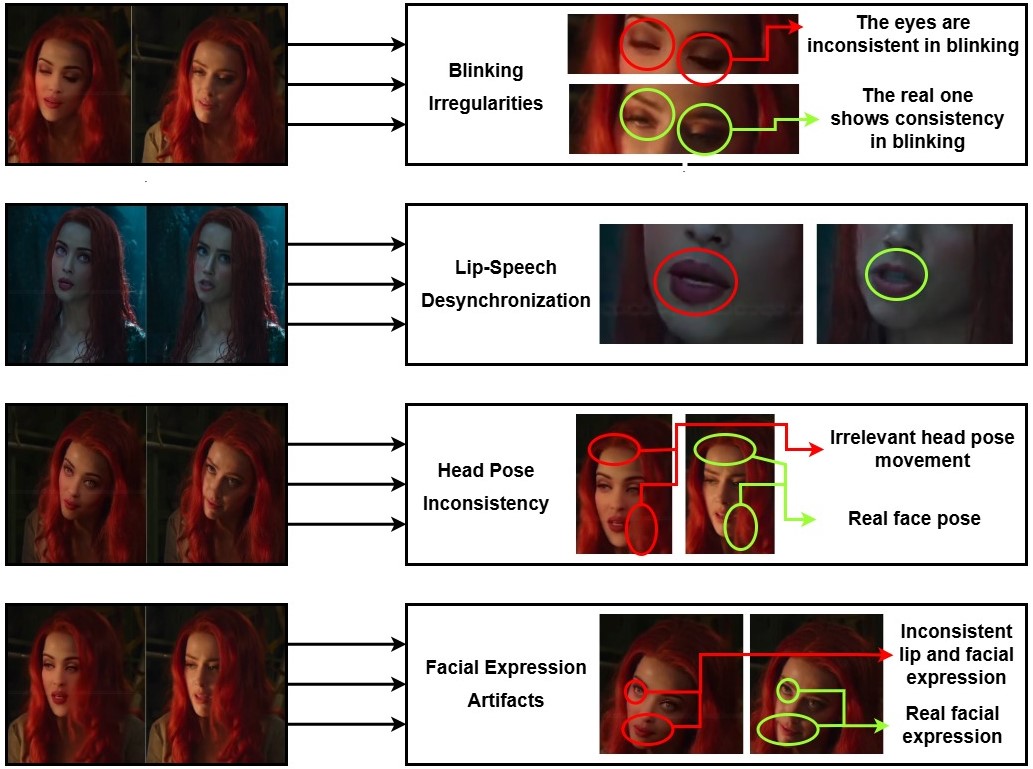}
    \caption{The temporal inconsistencies exhibited by deepfake videos are broadly divided into four major types: blinking irregularities \cite{Yin_2023}, lip-speech syncing \cite{Moufidi_2024}, head pose inconsistency \cite{Lu_2023}, and facial expression \cite{Kopalidis_2024}. The red circle shows the inconsistencies in the deepfake frames, while the green one is its original frame.}
    \label{fig:temp}
\end{figure*}

Deepfake videos often introduce temporal inconsistencies \cite{Zhang_2021} due to imperfections in frame synthesis, facial dynamics, and motion continuity. With still image deepfake detection relying mainly on spatial artifacts, video-based detection focused on temporal coherence, which ensured smooth and natural transitions between frames \cite{Zhu_2024}. Temporal anomaly detection aims to identify irregularities in blinking patterns, speech-lip synchronization, head motion, and facial expressions (see Fig.\ref{fig:temp}) to distinguish real videos from deepfakes. Since many deepfake generation techniques failed to maintain temporal consistency, this approach is crucial for zero-shot deepfake detection.

Eye blinking is an involuntary physiological behavior that occurs at an average rate of 15-20 blinks per minute in real humans \cite{Hoppe_2018}. However, deepfake generation models often fail to replicate natural blink patterns, which results in excessive blinking, lack of blinking, and asymmetrical blinking\cite{Li_2018}. Since deepfake videos struggle to maintain natural blink patterns, the eye blinking anomaly score could be used for the classification of deepfake videos effectively. For lip-speech synchronization analysis, the lip movements synchronize perfectly with spoken words in real videos \cite{Hirishikesh_2023}. However, deepfake synthesis often results in out-of-sync lip motions, where the mouth does not match the phenomena of being spoken \cite{Bohacek_2024}. AI models struggled to generate perfectly synced lip movements \cite{Datta_2024}, leading to misalignment artifacts, which can be effectively detected using audio-visual synchronization models. Head movements \cite{Becattini_2024} in deepfake videos often appeared rigid, unnatural, or desynchronized from body motion. Unlike real humans, deepfake head movements might show jittering motion, unrealistic rotations, and fixed positions \cite{Yang_2023}. Since deepfake videos often have imperfectly aligned head movements \cite{Gr_nquist_2024}, motion-tracking models could detect these anomalies with the highest accuracy possible. Lastly, optical glow refers to the movement of pixels between consecutive video frames. In real videos, facial expressions change smoothly over time, but deepfake videos often exhibit abrupt expression shifts, lack of micro-expressions, and inconsistent motion vectors. Deepfake-generated face lacks natural muscle coordination \cite{Zhao_2020}, which makes optical flow-based analysis highly effective. The detection approach of all the types of temporal analysis with its elaborated steps is shown in Table \ref{tab:temp}.

\begin{table*}[t!]
    \centering
    \caption{Approaches in Detecting Temporal Inconsistency with their Process}
    {\fontsize{7pt}{7pt}\selectfont \begin{tabular}{>{\centering\arraybackslash}m{4cm} >{\centering\arraybackslash}m{4cm}||>{\centering\arraybackslash}m{4cm} >{\centering\arraybackslash}m{4cm}} 
    \hline
    \multicolumn{2}{c}{\textbf{Eye Blink Analysis}} & \multicolumn{2}{c}{\textbf{Lip-Speech Synchronization Analysis}} \\ 
    \hline
    \textbf{Step} & \textbf{Process} & \textbf{Step} & \textbf{Process} \\ \hline
    Face Detection & Detects the face region using a deep learning-based face detector & Audio Extraction & Extracts the audio waveform from the video \\ \hline
    Eye Region Extraction & Extracts the eye region from the face for analysis & Lip Landmark Detection & Tracks lip movement and shape across frames \\ \hline
    Blink Frequency Estimation & Uses a CNN or LSTM model to detect eye closure events over time & Phoneme Alignment & Converts audio into phoneme sequences using speech processing models \\ \hline
    Statistical Analysis & Comparison of Blink Frequency and Duration with Natural Human Patterns & Synchronization Analysis & Compares phoneme timing vs. lip motion to detect mismatches \\ \hline
    Classification & Flags videos as deepfake if blink patterns deviate significantly from real-world data & Classification & Assign a synchronization score—lower scores indicate deepfake content \\ \hline
    \end{tabular}}
    
    \vspace{0.5cm} 

    {\fontsize{7pt}{7pt}\selectfont \begin{tabular}{>{\centering\arraybackslash}m{4cm} >{\centering\arraybackslash}m{4cm}||>{\centering\arraybackslash}m{4cm} >{\centering\arraybackslash}m{4cm}} 
    \hline
    \multicolumn{2}{c}{\textbf{Head Pose Estimation}} & \multicolumn{2}{c}{\textbf{Optical Flow Analysis for Facial Expressions}} \\ 
    \hline
    \textbf{Step} & \textbf{Process} & \textbf{Step} & \textbf{Process} \\ \hline
    Facial Landmark Detection & Extracts key facial points to track motion & Frame Sequence Extraction & Extracts consecutive video frames for analysis \\ \hline
    3D Pose Estimation & Uses machine learning models to estimate head rotation angles & Optical Flow Computation & Uses models like Farneback Optical Flow or Lucas-Kanade to estimate motion vectors \\ \hline
    Motion Analysis & Compares head pose changes over multiple frames & Expression Smoothness Check & Evaluates gradual vs. abrupt expression changes \\ \hline
    Temporal Consistency Check & Flags inconsistencies like abrupt movements or fixed head positions & Motion Consistency Verification & Identifies unrealistic facial region movements \\ \hline
    Classification & Detects deepfakes based on unrealistic motion patterns & Classification & Detects deepfakes based on motion inconsistency scores \\ \hline
    \end{tabular}}
    \label{tab:temp}
\end{table*}

\begin{table*}[h]
    \centering
    \caption{Comparative Analysis of Self-Supervised Anomaly Detection Methods with other deepfake detection methods}
    \renewcommand{\arraystretch}{1.6}
    \setlength{\tabcolsep}{5pt}
    {\fontsize{7pt}{7pt}\selectfont \begin{tabular}{>{\centering\arraybackslash}m{3cm}>{\centering\arraybackslash}m{3cm}>{\centering\arraybackslash}m{3cm}>{\centering\arraybackslash}m{2.5cm}>{\centering\arraybackslash}m{2cm}>{\centering\arraybackslash}m{2cm}}
        \hline
        \textbf{Method} & \textbf{Autoencoder-Based} & \textbf{One-Class SVMs} & \textbf{OOD Detection} & \textbf{Temporal Anomaly Detection} & \textbf{Supervised Learning} \\
        \hline
        \textbf{Requires Labeled Data} & \XSolidBrush & \XSolidBrush & \XSolidBrush & \XSolidBrush & \ding{52} \\
        \hline
        \textbf{Learns Feature Representations Automatically} & \ding{52} & \XSolidBrush & \ding{52} & \ding{52} & \XSolidBrush \\
        \hline
        \textbf{Can Detect Novel Anomalies} & \ding{52} & \ding{52} & \ding{52} & \ding{52} & \XSolidBrush \\
        \hline
        \textbf{Handles High-Dimensional Data Well} & \ding{52} & \XSolidBrush & \ding{52} & \ding{52} & \ding{52} \\
        \hline
        \textbf{Sensitive to Outliers} & \ding{52} & \ding{52} & \ding{52} & \ding{52} & \XSolidBrush \\
        \hline
        \textbf{Requires Negative Samples} & \XSolidBrush & \XSolidBrush & \XSolidBrush & \XSolidBrush & \ding{52} \\
        \hline
        \textbf{Works Well for Temporal Data} & \XSolidBrush & \XSolidBrush & \XSolidBrush & \ding{52} & \ding{52} \\
        \hline
        \textbf{Robust to Adversarial Attacks} & \XSolidBrush & \XSolidBrush & \ding{52} & \ding{52} & \XSolidBrush \\
        \hline
        \textbf{Computationally Efficient} & \ding{52} & \ding{52} & \XSolidBrush & \XSolidBrush & \ding{52} \\
        \hline
        \textbf{Scales Well with Large Datasets} & \ding{52} & \XSolidBrush & \XSolidBrush & \XSolidBrush & \ding{52} \\
        \hline
    \end{tabular}}
    \label{tab:companom}
\end{table*}

Table \ref{tab:companom} reviews other deepfake detection techniques (comparison with self-supervised anomaly detection methods). Other factors the evaluation considers include the need for labeled examples, the capacity to automatically learn a representation of features, the power of detecting new anomalies, and the aptitude for high dimensional. This is notable since autoencoder-based methods \cite{Merrill_2020} \cite{Iqbal_2022} and out-of-distribution (OOD) detection techniques \cite{Graham_2023} are highly effective in learning feature representations and detecting previously unseen anomalies\cite{Naidu_2024}. On the other hand, supervised learning approaches are based on the assumption of well-labeled data and negative samples \cite{Xiang_2010} \cite{Wadekar_2019}. At the same time,
self-supervised anomaly detection methods are more robust against constraints (labeled data and negative samples) of anomalies. In addition, the computational efficiency and scalability of various methods are also addressed, with autoencoder and supervised learning techniques having high computational efficiency and scalability for large data sets. Amongst available methods for temporal anomaly detection, techniques for handling sequential data are the best. This comparative study gives insights into the feasibility of each method in deepfake detection and its strengths and limitations.

\subsection{Generative Model Fingerprinting for Deepfake Attribution}
When dealing with novel (new) or unseen synthetic media, traditional models cannot generalize to previously unseen data and compare relative to the network. Featuring the intrinsic artifacts \cite{Pontorno_2024}, inconsistencies, and fingerprints of deepfake generation models \cite{Yu_2021}, Generative Model Fingerprinting is a promising solution. The fingerprints allow zero-shot detection of both deepfake content and the source model responsible for its creation. Through generative model fingerprints, deepfake detection can transition from a reactive approach (detecting already known deepfakes) to a proactive and forensic viewpoint \cite{Yu_2021} \cite{Lai_2025} (being able to track the origin of unknown deepfakes). This section introduces fundamental concepts, techniques, and applications in deepfake attribution using generative model fingerprinting. Generative models are also just as unique if you look at the outputs of their models because of architectural differences \cite{Liu_2024}, upsampling methods \cite{Shang_2023}, noise injection \cite{Zhang_2024}, and training biases \cite{Korshunov_2022}. The fingerprints for these embeddings are pixel patterns, frequency distortions, or latent space embeddings \cite{muller2022}. Deepfake attribution can be possible even without prior exposure to specific fake content by detecting these artifacts
\cite{cocchi2023unveiling}.

The key sources of generative model fingerprints are shown in Table \ref{tab:gen}. From this table, the key observations in generative model fingerprinting are

\begin{enumerate}[wide, labelwidth=!, labelindent=0pt]
    \item Generative Adversarial Networks (GANs) remained inclined toward pixel-level texture inconsistencies for upsampling layers.
    \item Autoencoders and Variational Autoencoders (VAEs) generated images with blurry edges for compression artifacts.
    \item Deepfake video synthesis models struggled with motion consistency, often leading to temporal flickering between frames.
    \item GAN-generated faces often lack fine skin details, which produces a smooth yet unnatural look when zoomed in.
\end{enumerate}

\begin{table}[t!]
    \centering
    \caption{Key Sources of Generative Model Fingerprints}
    {\fontsize{7pt}{7pt}\selectfont \begin{tabular}{>{\centering\arraybackslash}m{1.7cm}>{\centering\arraybackslash}m{3cm}>{\centering\arraybackslash}m{2.5cm}} \hline
        \textbf{Fingerprint Type} & \textbf{Description} & \textbf{Examples Generative Models Affected} \\ \hline \hline
        Pixel-level artifacts	& Subtle inconsistencies in texture, lighting, and facial alignment were introduced during image synthesis. &	GANs (StyleGAN, ProGAN), VAEs
 \\ \hline
        Frequency Domain Anomalies &	Unnatural spectral patterns, visible in Fourier or wavelet transformations, caused by synthetic upsampling and compression. &	StyleGAN, BigGAN, StarGAN
\\ \hline
        Latent-space representations &	Unique embeddings in the feature space can be traced back to the model’s architecture and training dataset. &	Autoencoders, Transformers (DALL-E, Stable Diffusion)
 \\ \hline
        Upsampling and Noise Patterns	& Interpolation artifacts, and random noise distribution mismatches were introduced during deepfake video synthesis. &	FaceSwap, DeepFaceLab. \\ \hline
    \end{tabular}}
    \label{tab:gen}
\end{table}

Multiple forensic techniques are employed to extract and analyze generative model fingerprints. These techniques capture model-specific patterns in spatial, frequency, or latent space.

\subsubsection{Pixel-Level Fingerprint Analysis}
The basis of deepfake detection is deciding which inconsistencies exist that would differentiate authentic media from synthetic media. Using fundamental pixel-level fingerprint analysis, one investigates the smallest visual units to detect artifacts produced by deep generative models. The cause of these artifacts is upsampling \cite{Pons_2021}, noise injection \cite{Sivabalamurugan_2024}, and imperfect blending and interpolation processes used for artificially generated media \cite{Chen_2023}. This is done because pixel-level patterns apply to zero-shot applicability since these subsist in different deepfake models to detect unseen synthetic content. The nonclassification nature, unlike classification-based detection methods, means that pixel analysis \cite{Lim_2022}is not based on a deepfake model and is, therefore, resistant to upgrades of the deepfake model. Lastly, subtle inconsistency in pixel values, texture, and edge-related errors can affect fine-grain detection even when it is quite subtle \cite{Mi_2024}.

Several image processing and machine learning techniques help extract and analyze pixel-level artifacts to detect deepfake content. The local binary patterns for texture analysis are one of a kind. Texture descriptors capture local pixel variations by comparing each pixel's intensity with its neighbors \cite{Bedi_2018}. This was effective in the detection of over-smoothed textures in deepfake images. The image is first converted to grayscale and divided into small blocks of the same pixels. The LBP \cite{Subash_Kumar_2018} histogram is calculated \cite{Wang_2013} for each block and compared to real images, where deepfake ones exhibit lower texture diversity. This could detect anomalies in AI-generated faces and identify overly smooth or blurry regions. The concept of edge detection \cite{Zhai_2008} identified sharp transitions in pixel intensity, which helped detect unnatural blurring or misalignment at deepfake boundaries. Here, canny edge detection, Sovel and Prewitt filters \cite{Selvakumar_2016}, and Laplacian operation \cite{Sahoo_2016} are vital. This detected boundary mismatches in face-swapped videos and helps identify blending inconsistencies in AI-generated images. Next, deepfake often exhibited unnatural brightness, contrast, or color distributions due to generative model biases. Histogram analysis quantifies these differences by converting the image to HSV \cite{Sucharitha_2023} \cite{Ghimire_2011} (Hue-Saturation-Value) color space and computing histograms for each channel. Then, they compare with real-image histograms \cite{Yu_2009} where deepfakes show irregular spikes in hue and unnatural shifts in saturation/brightness. This helps detect skin-tone inconsistencies in AI-generated faces and helps identify lighting mismatches in face-swapped videos. Generative models also introduced systematic noise patterns that differed from real-world sensor noise. Noise residual analysis \cite{El_Rai_2020} was extracted, and these patterns were compared using different techniques. Some notable of these include Photo Response Non-Uniformity (PRNU) \cite{Amerini_2022} \cite{Lugstein_2021}, where sensor-specific noise patterns absent in deepfake images are captured. Deepfake images show different noise fingerprints compared to real ones so the detection could be done on this basis. Next, Wavelet Transform-based Denoising \cite{Laavanya_2018} \cite{Wahab_2020} separated structured noise from natural textures. The deepfake artifact is detected through AI-generated images with higher unnatural noise components. Lastly, Gaussian Noise Estimation analyzed pixel-level noise distribution where anomalous noise levels were detected in GAN-generated images. Hidden noise artifacts were detected in deepfake images and videos, which were useful for forensic analysis of tampered images \cite{Zhan_2015} \cite{Guo_2019}.

\subsubsection{Frequency Domain Analysis}

\begin{table}[t!]
    \centering
    \caption{Deepfake-Specific Frequency Artifacts}
    {\fontsize{7pt}{7pt}\selectfont \begin{tabular}{>{\centering\arraybackslash}m{2.4cm}>{\centering\arraybackslash}m{3cm}>{\centering\arraybackslash}m{1.8cm}} \hline
        \textbf{Deepfake Type} & \textbf{Characteristic Frequency Artifacts} & \textbf{Detection Method} \\ \hline \hline
        Face-Swaps (DeepFaceLab, FaceSwap)	& Blurry edges, inconsistent texture blending	& High-pass filtering, PSA
 \\ \hline
        GAN-Generated Faces (StyleGAN, ProGAN) &	Checkerboard artifacts, unnatural spectral peaks &	DFT, PSA
\\ \hline
        AI-Generated Videos (First Order Motion Model, Deep Video Portraits) &	Temporal flickering, unnatural motion smoothness &	Wavelet analysis, PSA
 \\ \hline
        Text-to-Image Models (DALL-E, Stable Diffusion)	& Mid-frequency distortions, unnatural fine details & 	DWT, Fourier transform \\ \hline
    \end{tabular}}
    \label{tab:freq}
\end{table}

Deepfake detection traditionally relied on analyzing images and videos in the spatial domain (i.e. pixel-level analysis). However, deepfake generation models often introduce hidden patterns and inconsistencies that are not always visible in the spatial domain but become apparent in the frequency domain \cite{Gao_2024} \cite{Liang_2023}. Frequency domain analysis \cite{Se_k__2017} is a powerful approach that transforms images and videos into their spectral components to uncover artifacts \cite{Vartiainen_2006}, anomalies, and distortions caused by synthetic content generation. This method was particularly useful for zero-shot detection \cite{Yoon_2024}, as it identified fundamental statistical differences between real and fake content without the requirement of prior exposure to specific deepfake models. In the spatial domain, an image was represented as an array of pixels. However, in the frequency domain, an image was decomposed into its constituent sinusoidal frequency components, which revealed patterns that might be invisible to the naked eye. The low-frequency components represented smooth regions (background, uniform textures), crucial for capturing general shape and illumination \cite{Bovik_2009} \cite{Shang_2024}. The high-frequency components represented edges, textures \cite{Li_2024} \cite{Li_2022}, and fine details, which deepfake synthesis, compression, and post-processing altered. The mid-frequency components represented immediate features like gradual shading surface reflections \cite{Hosseinimakarem_2016}, often altered due to GAN upsampling techniques. The model-specific fingerprints allow frequency domain analysis to serve as an effective zero-shot detection method, which even encountered previously unseen deepfake models (see Table \ref{tab:freq}). Since deepfake generation algorithms manipulated specific frequency bands, analyzing their frequency signatures allowed us to detect unseen synthetic content in a zero-shot setting. Several mathematical transformations helped convert images/videos from the spatial domain to the frequency domain for deepfake detection.

\begin{enumerate}[wide, labelwidth=!, labelindent=0pt]
    \item Discrete Fourier Transform (DFT): The main objective behind this is this decomposed image into sinusoidal wave components, which shows how much of each frequency is present. This converted an image $I(x,y)$ into its frequency representation $F(u,v)$, and its frequency components were plotted in a power spectrum, highlighting the deepfake-specific distortions. The key observations include that GAN-generated images exhibited checkerboard artifacts in the frequency spectrum \cite{Jung_2021}. Deepfake videos showed high-frequency inconsistencies, which led to flickering. The compression artifacts in deepfake appeared as unexpected peaks in mid-frequency ranges. 
    \item Discrete Wavelet Transform (DWT): This decomposed an image into multi-resolution frequency components useful for analyzing texture inconsistencies. This broke down an image into four sub-bands, namely LL (Low-Low), which preserved general structure; LH (Low-High), which captured horizontal textures; HL (High-Low), which captured vertical textures; and HH (High-High), which highlighted its fine details \cite{Kolekar_2014} \cite{Carvajal_Gamez_2016}. Here, GAN-generated images often lacked high-frequency details in the HH sub-band, and forgery operations like face-swapping disrupted the HL and LH sub-bands. The post-processing effects, such as blurring and upsampling, also result in distorted wavelet coefficients.
    \item Power Spectrum Analysis (PSA): This detected periodic artifacts and inconsistencies introduced by generative models. This computes the magnitude of frequency components across different orientations and plots the log power spectrum, which reveals patterns invisible to spatial analysis \cite{Lee_2024} \cite{Hsiao_2005}. The deepfake images showed periodic spectral peaks caused by generative upsampling layers, and real images exhibited a natural power-law distribution while deepfake deviated significantly. FFT-based detection algorithms could classify deepfakes with high accuracy using spectral cues.
    \item High-Pass Filtering (HPF): This enhanced high-frequency artifacts \cite{Kaushik_2024} that were often suppressed in deepfake synthesis, and this removed low-frequency components, emphasized edges and fine textures\cite{Stephen_2022}. This detected subtle inconsistencies in facial regions, eye reflections, and skin pores. GANs introduced unnatural textures, visible after high-pass filtering, and deepfake videos had temporal flickering, which was more prominent in filtered frames \cite{Zhu_2024}. The high-pass filtering combined with CNNs enhanced deepfake classification accuracy.
\end{enumerate}

In zero-shot deepfake detection, frequency domain analysis is very useful in detecting hidden artifacts that are often hard to detect in the spatial domain. Forensic deepfake analysis \cite{Kumar_2025} aids in tracing synthetic content connected to the generative model of applications such as forensic fingerprint analysis for law enforcement in cybercrime investigations. Social media platforms use this technique to authenticate content in real time and stop any misinformation being spread by deepfakes. These are also comprised of frequency-based detection to prevent identity fraud and synthetic impersonation attacks and are based on AI. Also, media forensics, financial security, and national defense use frequency domain analysis to validate the integrity of digital content and improve overall cybersecurity.\cite{Hung_2023}

\subsubsection{Latent Space Embedding Analysis}
Latent space embedding analysis is a critical technique for deepfake detection and attribution that involves mapping high-dimensional data (such as images or video frames) into lower-dimensional latent space \cite{Zang_2023} \cite{Pontorno_2024}. This transformation uncovered underlying features not present in the raw data. In the context of generative model fingerprinting \cite{Yu_2021}, these embeddings, after being analyzed, could reveal unique signatures left by deepfake generation models. The latent space is an abstract, compressed representation where high-dimensional inputs (images, audio, etc.) are encoded into features that capture the data's most essential and semantically meaningful characteristics \cite{Ganguly_2024}. With the reduction of dimensionality, the latent space encapsulated the key attributes that defined authentic versus manipulated media. Since different generative models imprinted distinct patterns into the data during synthesis, these unique features could be captured and analyzed in the latent space \cite{Bodria_2022}.

\begin{table}[t!]
    \centering
    \caption{Key Sources of Generative Model Fingerprints}
    {\fontsize{7pt}{7pt}\selectfont \begin{tabular}{>{\centering\arraybackslash}m{1.2cm}>{\centering\arraybackslash}m{2cm}>{\centering\arraybackslash}m{1.8cm}>{\centering\arraybackslash}m{2cm}} \hline
        \textbf{Technique} & \textbf{Description} & \textbf{Advantages} & \textbf{Challenges}\\ \hline \hline
        Pretrained CNN Feature Extractors	& Use models like ResNet or EfficientNet to extract robust high-level features. & 	Leverages proven architectures; high transferability. &	Sensitive to adversarial noise; may require fine-tuning. \\ \hline
        Autoencoders &	Train an autoencoder on authentic data to learn latent representations. &	Highlights reconstruction errors; unsupervised. &	Reconstruction error may be subtle in high-quality fakes. \\ \hline
        t-SNE / UMAP for Visualization &	Non-linear methods to visualize clusters in the latent space. &	Excellent for visual differentiation of clusters. &	Computationally intensive on large datasets; sensitive to parameter settings. \\ \hline
        Clustering Algorithms (K-Means, DBSCAN)	 & Group similar embeddings to identify distinct clusters representing deepfake fingerprints. &	Facilitates clear separation between real and fake data. &	Requires appropriate determination of the number of clusters or density thresholds. \\ \hline
    \end{tabular}}
    \label{tab:latent}
\end{table}

The latent space embedding analysis process starts with feature extraction, where some pretrained models \cite{Jha_2024} are employed to extract high-level features from input images or videos. These features form the basis of the latent representation. Next, autoencoders \cite{Shanckin_2023} are trained on authentic data, where the encoder maps the input into latent space. When decoding back to the original space, the reconstruction error \cite{Fu_2021}can also highlight anomalies present in deepfakes \cite{Peng_2024}. Secondly, dimensionality reduction reduces the high-dimensional features into more manageable latent space, helping visualize and analyze the inherent patterns. Here, techniques include Principal Component Analysis (PCA) \cite{Ballabio_2015} \cite{Bartholomew_2010}, which reduces dimensionality by selecting the components with the highest variance. t-Distributed Stochastic Neighbor Embedding \cite{Rice_2017} \cite{Jung_2024} (t-SNE) is also applied, a non-linear technique that clusters similar data points, enabling visualization of how deepfake samples deviated from authentic ones. Another technique known as Uniform Manifold Approximation and Projection \cite{Myasnikov_2020}(UMAP) is also useful for capturing the global structure \cite{Healy_2024} while preserving local neighborhood relationships. Thirdly, several clustering techniques are used for analysis and clustering \cite{L__2010}. Once the data is embedded into latent space, clustering algorithms (e.g. K-Means or DBSCAN) could group similar representations \cite{Peng_2016}. Differences in the latent space distributions could be highlighted by comparing the clusters of known authentic content with those of suspected deepfakes. The unique signatures or "fingerprints" of generative models might cause deepfake clusters to differentiate from real content separately. Next comes the attribution and anomaly detection phase \cite{Nguyen_2024}, where the latent space representations could be compared against a database of known generative model fingerprints. This matching process helped in attributing a deepfake to a specific source model. If a sample's embedding significantly deviates from the cluster of authentic data, it is flagged as anomalous, suggesting that it might be a deepfake \cite{Ueda_2020}. Table \ref{tab:latent} summarizes the primary techniques used for latent space embedding analysis in the context of deepfake detection.

\subsubsection{Deepfake Attribution via Model Fingerprinting}
Deepfake attribution refers to the process of identification of a generative model that was responsible for creating synthetic content. While traditional deepfake detection focuses on classifying content as real or fake, attribution goes further by tracking the deepfake back to its source model. This was crucial for forensic investigations, regulatory compliance, and proactive countermeasures against misinformation campaigns. Generative models such as StyleGAN, ProGAN, DeepFaceLab, and DALL-E leave unique digital signatures or fingerprints \cite{Song_2024} \cite{Yu_2019}in their outputs due to differences in their architectures, training datasets, and noise distributions. Deepfake attribution enables synthetic content to be tracked back to its generative source by analyzing these fingerprints. This will also help identify the origin of manipulated media \cite{Yoon_2024}. This will, in turn, enhance the zero-shot deepfake detection capabilities. With the growing sophistication of deepfake technology, detecting fakes alone was no longer sufficient. Attribution allowed for 

\begin{enumerate}[wide, labelwidth=!, labelindent=0pt]
    \item Forensic investigations help track down malicious actors spreading synthetic media.
    \item Legal and ethical compliance supported policy enforcement for AI-generated content regulation.
    \item Content authentication enabled social media platforms for verifying content sources.
    \item Model adaptation improved deepfake detectors by learning from emerging generative models.
\end{enumerate}

Traditional deepfake detection models struggled in zero-shot settings when encountering unseen generative models. However, model fingerprinting allowed detection systems to generalize the identification of latent artifacts common across AI-generated content. Deepfake attribution is performed through a systematic pipeline that extracts unique fingerprints from synthetic media and matches them against known generative models \cite{Jeong_2022} \cite{Maho_2023}. The pipeline for deepfake attribution is shown in Table \ref{tab:attr}. Each step ensured high accuracy in detecting and attributing AI-generated content, even in a zero-shot setting.

\begin{table}[t!]
    \centering
    \caption{Pipeline for Deepfake Attribution from start to end}
    {\fontsize{7pt}{7pt}\selectfont \begin{tabular}{>{\centering\arraybackslash}m{1.8cm}>{\centering\arraybackslash}m{3cm}>{\centering\arraybackslash}m{2.4cm}} \hline
        \textbf{Step} & \textbf{Process} & \textbf{Techniques Used} \\ \hline \hline
        Feature Extraction	& Extracts pixel, frequency, and latent space fingerprints from deepfake content.	& CNN Feature Maps, Fourier Transforms, Wavelet Analysis \\ \hline
        Signature Database Creation	& Stores unique generative model patterns for future comparisons. &	Deepfake Signature Databases, Clustering Algorithms \\ \hline
        Signature Matching &	Compares the extracted fingerprint against stored model signatures.	& Cosine Similarity, k-NN Classification, t-SNE \\ \hline
        Source Attribution	& Identifies the specific generative model used to create the deepfake.	& Probabilistic Attribution, Model Comparison Techniques \\ \hline
    \end{tabular}}
    \label{tab:attr}
\end{table}

The first step in deepfake attribution is extracting key forensic features from synthetic media. These features, generative model fingerprints, were derived from three key domains.

\begin{enumerate}[wide, labelwidth=!, labelindent=0pt]
    \item Pixel-level artifacts where texture inconsistencies, unnatural lighting, and alignment errors are found. Examples of models affected are StyleGAN and DeepFaceLab.
    \item Frequency domain anomalies are unnatural spectral distortions from upsampling and compression. Examples of models affected are ProGAN, StarGAN, and BigGAN.
    \item Latent Space Representations consisted of deep feature embeddings unique to different generative models. Examples include DALL-E, Stable Diffusion, and Autoencoders.
\end{enumerate}

These fingerprints are then processed using forensic AI models trained to distinguish between deepfake generators. The key techniques for feature extraction include CNN feature maps, where visual patterns are extracted that are unique to generative models; Fourier transform analysis, which detects anomalies in the frequency domain; and autoencoder residuals, which measure how well an autoencoder trained on real data could reconstruct synthetic media. Once deepfake fingerprints are extracted, they must be stored for future comparisons. The deepfake signature database was created, where unique model fingerprints were indexed based on generative model type (GAN, Transformer, Autoencoder, etc.) \cite{El_Kaddoury_2019} \cite{Laptev_2021} \cite{Han_2025}, Training data characteristics (facial datasets, object datasets, etc.) \cite{Ahmed_2023}, Noise and upsampling artifacts (compression noise, interpolation patterns, etc.). This enabled a centralized repository that detection systems could use for zero-shot identification of emerging deepfake generators \cite{Khan_2024}. When a new deepfake is detected, this is compared against the signature database to determine its likely source. This was done using cosine similarity \cite{Guo_2024}, measuring the distance between extracted fingerprints and stored signatures. K-Nearest Neighbors \cite{2021} (k-NN) could also be a good method that could classify deepfake fingerprints by identifying the closest matching known model. Adding to this, t-SNE \cite{Wu_2022} also clusters generative models based on fingerprint similarity. If a close match is found, the deepfake could be linked back to its generative model with high confidence. If an exact match is not found in the database, probabilistic techniques estimate the most likely generative model responsible for creating the deepfake. Table \ref{tab:attrprob} shows the best probabilistic attribution techniques.

Several deepfake detection architectures have been developed on different architectures, such as Encoder-Decoder architecture \cite{Javaloy_2020}, GAN-based architectures \cite{Sharma_2024}, and Transformer-based architectures \cite{Thuan_2024}. Encoder Decoder models exploit the discrepancies in synthesized images, that is, reconstruction errors, to discern real and fake content. In the case of discriminative features of deepfake generators based on GAN, detection methods are built on robust feature extraction, whereby given artifact patterns can be used to identify the differences between synthesized media and the semantic data they were created from. In contrast, transformer-driven approaches use self-attention mechanisms to discover long-range dependencies \cite{Le_2024} and subtle irregularities of deepfake images and videos. However, these methods are susceptible to adversarial attacks \cite{Li_2022}, prone to overfitting \cite{Kernbach_2021} special datasets, and computationally inefficient \cite{Seifi_2023}. Table \ref{tab:compdeep} provides a detailed summary of these deepfake detection methods, providing a summary of the limitations of each method.

\begin{table}[t!]
    \centering
    \caption{Probabilistic Attribution Techniques}
    {\fontsize{7pt}{7pt}\selectfont \begin{tabular}{>{\centering\arraybackslash}m{1.8cm}>{\centering\arraybackslash}m{3cm}>{\centering\arraybackslash}m{2.4cm}} \hline
        \textbf{Method} & \textbf{Description} & \textbf{Use Case} \\ \hline \hline
        Bayesian Inference	& Compute the probability of a specific model generating a deepfake. &	Used when multiple models share similar artifacts. \\ \hline
        Ensemble Learning &	Combine multiple classification models to improve attribution accuracy. &	Effective in distinguishing between closely related models. \\ \hline
        Deep Metric Learning	& Learns a distance metric between real and synthetic fingerprints.	& Useful for zero-shot detection of new deepfake models. \\ \hline
    \end{tabular}}
    \label{tab:attrprob}
\end{table}

\begin{table*}[t!]
    \centering
    \caption{Summary of Deepfake Detection Methods based on the Encoder-Decoder, GANs and GAN-based-architectures, and Transformer-driven frameworks along with its limitations}
    {\fontsize{7pt}{7pt}\selectfont \begin{tabular}{p{1cm}p{1cm}p{2cm}p{1.4cm}p{1.7cm}p{1.3cm}p{6.6cm}} \hline
        \textbf{Ref} & \textbf{Year} & \textbf{Technique(s) used} & \textbf{Media type(s)} & \textbf{Dataset} & \textbf{Results} & \textbf{Limitations} \\ \hline \hline
    \cite{115} & 2018 & Zero-Shot Encoder-Decoder & Images/Videos & FF and CelebA-HQ & Above 90\% ACC & Deep networks overfit to manipulation-specific artifacts, lacking transferability.Performance declines with new manipulation types, requiring large new datasets. Limited datasets can lead to polarization in training. \\ \hline
    \cite{Huo_2019} & 2019 & Autoencoder and GAN & Videos & FF++, Celeb-DF and DFDC & Above 90\% ACC for 20 projection vectors & MSH-IR has low inter-operability between different acquisition devices. Significant differences exist in iris texture details from various devices. \\ \hline
    \cite{Khalid_2020} & 2020 & One class Autoencoder & Videos & FF++ & 97.50\% ACC & The approach relies on RMSE for reconstruction scoring. Only real and fake images from FaceForensics++ were used. \\ \hline
    \cite{Yang_2021} & 2021 & Encoder-Decoder & Videos & FF++ and DFDC & 99.90\% ACC & Network model training requires large-scale data for optimal results. New authentication networks are needed for unknown face tampering methods. \\ \hline
    \cite{Hu_2021} & 2021 & Encoder-Decoder & Videos & UADFV, FF++, Celeb-DF & 97.60\% AUC & Pixel-level masks for manipulated regions are often unavailable. Existing methods struggle to locate manipulated regions accurately. \\ \hline
    \cite{96} & 2019 & CycleGAN \& StarGAN & Images & CycleGAN and StarGAN & 99\% ACC & The method performs poorly in the 'cityscapes' and 'facades' categories. Original images in certain categories were compressed into JPEG. Non-uniform class distribution affects accuracy in the StarGAN dataset. \\ \hline
    \cite{9010964} & 2019 & Forensics (GAN artifact) \& CNN & Images & CelebA, LSUN & 72.64\% ACC (CelebA) 83.64\% ACC (LSUN) & Attribution by human inspection is no longer feasible due to GAN realism. Existing digital watermarking techniques are impractical for GAN-generated images. \\ \hline
    \cite{8803661} & 2019 & Forensics (GAN artifact) \& SVM & Images & GAN Crop, GAN Full, MFC and CelebA & 0.70 AUC (GAN Crop) 0.61 AUC (GAN Full) & The rapid pace of GAN innovation may outdate current detection methods. Non-GAN regions dilute the method's performance in images. \\ \hline
    \cite{9141516} & 2020 & GAN DeepVision & Videos & Kaggle & 96\% ACC & Continuous and noisy signals lack a universal hard limit for classification. Deep learning detectors are ineffective against realistic generative models. \\ \hline
    \cite{9577744} & 2021 & Recycle-GANs & Videos & Viper Dataset & 99.39\% ACC & Continuous and noisy signals lack a universal hard limit for classification. Deep learning detectors are ineffective against realistic generative models. \\ \hline
    \cite{99} & 2022 & MRI-GAN with SSIM & Videos & DFDC, Celebe, FDF, FFHQ & 74\% Test ACC and 91\%  (Plain frames)  & MRI-GAN accuracy does not surpass plain-frames-based methods tested. Results are preliminary and may improve with better loss functions. \\ \hline
    \cite{131} & 2022 & GAN/3D & Videos & VoxCeleb & ------  & Failures occur with extreme head poses during reenactment. Visual artifacts arise when head poses are outside the estimated distribution. \\ \hline
    \cite{10030936} & 2023 & GAN & Audios/Videos & VoxCeleb & 88\% (ACC) and 97.15\% (AUC)  & The model performs better on seen identities than unseen ones. Existing models require reference sets for identity candidates. \\ \hline
    \cite{Huang_2024} & 2024 & GAN & Videos & CelebA, ProGAN, SNGAN, MMDGAN, CramerGAN & 97.02\% (ACC) & Detectors can be retrained to counter the DeepNotch method. Improved detectors may encourage new attack methods. \\ \hline
    \cite{Gao_2024} & 2024 & GAN & Videos & FF++, Celeb-DF, and OpenForensics & ------ & ------ \\ \hline
    \cite{140} & 2024 & GAN & Videos & FF++, Celeb-DF & 98.22\% (ACC) & The model performs less on DF and FS datasets than the baseline ADD. There is potential overfitting affecting results on the F2F dataset. \\ \hline
    \cite{161} & 2024 & GAN & Videos & ProGAN, FFHQ, LSUN, CelebA, StyleGAN, StyleGAN2, Big-GAN, CycleGAN & 91.50\% (ACC) & Existing detectors overfit to training data artifacts, limiting generalization. Frequency attributes struggle to generalize across diverse sources. \\ \hline
    \cite{236} & 2021 & Transformer-Encoder & Videos & DFDC, Celeb-DF, FaceForensics & ------ & Has no focus on the unseen images manipulated. \\ \hline
    \cite{Lin_2024} & 2024 & Self-supervised transformer & Videos & FSh, DFDCP, DFDC & ------ & The limited generalization power approaches are shown here, which consistently maintained good performances. \\ \hline
    \cite{Kaddar_2024} & 2024 & VIT(Encoder) & Videos & FF++, Celeb-DF(V2), DFDC, DFo & ------ & Generalization was not achieved. \\ \hline
    \end{tabular}}
    \label{tab:compdeep}
\end{table*}

\section{Zero-Shot-Based Prevention Strategies for Deepfake Generation} \label{sec4}
While zero-shot deepfake detection enabled the identification of synthetic content without any prior exposure, there is an equally critical aspect of deepfake defense i.e. prevention of deepfake generation before it occurs \cite{Yang_2024} \cite{Wang_2022}. AI-based prevention strategies aim to proactively mitigate deepfake threats by disrupting its creation pipeline, which embeds authenticity markers and leverages real-time AI monitoring \cite{Yu_2023}. This section explored the various proactive defense mechanisms, focusing on their integration with zero-shot learning strategies for creating a robust multi-layered defense against deepfake manipulation \cite{Priya_2024}.

\subsection{Adversarial Perturbations to Disrupt Deepfake Generators}
Deepfake media generation relies on powerful deep learning models \cite{Cheng_2020} \cite{Mienye_2025}, specifically GANs (Generative Adversarial Networks) and Autoencoders. Well-manipulated images and videos are desired with these models, so the feature needs to be extracted perfectly \cite{Jellali_2024}. The proactive defense is in the form of adversarial perturbations \cite{Nguyen_2018}, subtle, unnoticeable alterations on original media that interfere with the deepfake generation algorithms. Adversarial perturbations strategically distort important aspects of generated deepfakes to render deepfakes ineffective or at least more easily detectable, though not entirely. Nevertheless, most existing adversarial defenses rely on the patterns of pre-trained attacks \cite{Liu_2023} that carry over poorly to the new deepfake models. Against this, zero-shot learning \cite{Yang_2023} \cite{Chemmengath_2021} (ZSL) eliminates this shortcoming by allowing adversarial perturbations to adapt dynamically without prior exposure to specific deepfake architectures. This integration makes it possible to preserve deepfake prevention in the light of developing AI-generated content manipulation techniques \cite{Yang_2023}. 

Adversarial perturbation means imperceptibly modifying an image, video, or audio signal to confuse or disrupt deepfake generators. Similar to the adversarial machine learning techniques used in these attacks, these perturbations leverage adversarial machine learning techniques to attack the fundamental feature extraction and transformation processes used in deepfake models like Generative Adversarial Networks (GANs), Variational Autoencoders (VAEs) and Transformer-based generative models. Adversarial perturbations are, at their core, aimed at manipulating the latent space representation \cite{Cao_2022} \cite{Zeng_2024} of an image in a way that deepfake models are unable to produce a realistic synthetic output. This strategy is particularly effective in a zero-shot setting where no prior knowledge of the deepfake architectures used in generating the images is available \cite{Zeng_2024}.

Let $x$ be the original image, and $f(x)$ represent a deepfake generator model. The main goal is to introduce a perturbation $\delta$ such that the perturbed image $x'$ caused disruption in the ability of the generator to create a deepfake, as in $x' = x + \delta$, where $\delta$ is adversarial perturbation satisfying $||\delta||_p \leq \epsilon$. Here, $|| \cdot ||_p$ denoted the $p$-norm (e.g. $L_2$-norm or $L_{\infty}$-norm), and $\epsilon$ is a small bound which ensured that $\delta$ is imperceptible to human vision. For the ideal deepfake generator $f(x)$, the main goal is maximizing the perturbation loss so that the generated deepfake is visibly corrupted using Eq. \ref{eq6}.

\begin{equation}
    \delta = arg \mathop{max}_{||\delta||_p \leq \epsilon} \mathcal{L}_{adv}(f(x+\delta))
    \label{eq6}
\end{equation}

\begin{figure*}[t!]
    \centering
    \includegraphics[width=\textwidth]{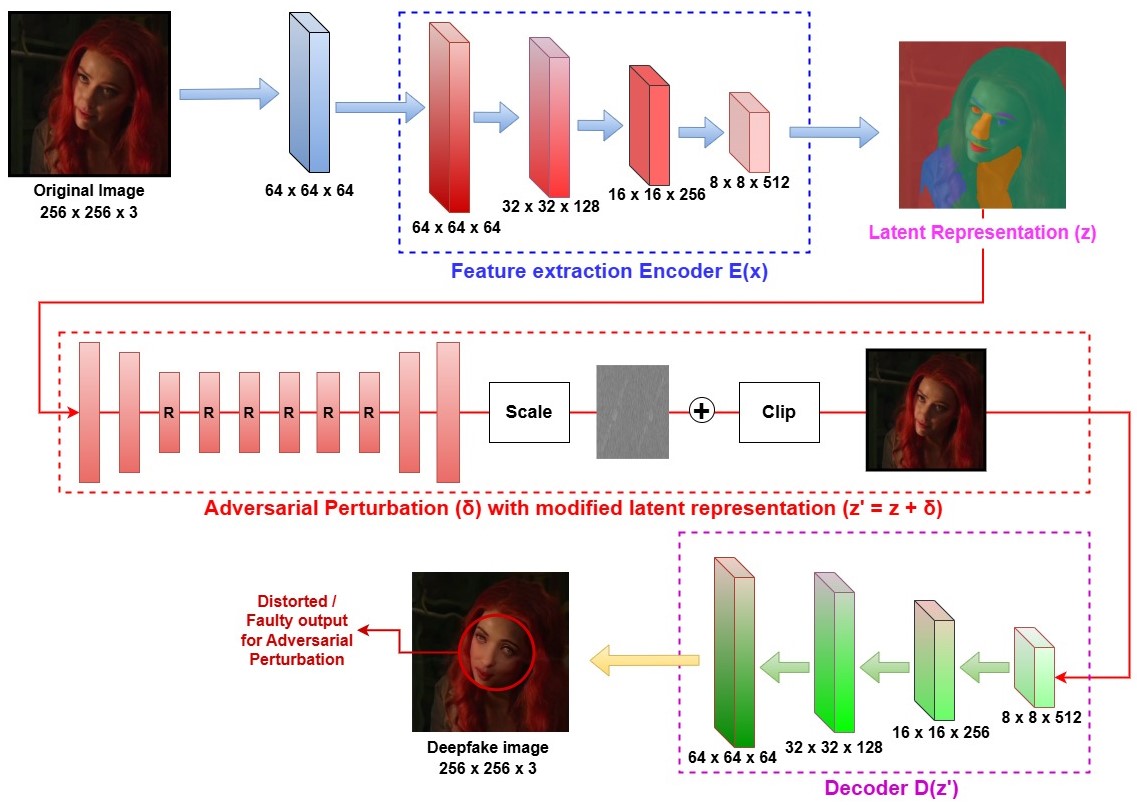}
    \caption{This diagram illustrates generating a deepfake image through adversarial perturbation. An original image is fed into an encoder network, compressing it into a latent representation. This representation is then modified by adding an adversarial perturbation, resulting in a distorted representation. A decoder network reconstructs this distorted representation into a final deepfake image, which is visually altered from the original. The process highlights how adversarial perturbations can manipulate latent representations to create fake images.}
    \label{fig:ap}
\end{figure*}

Here, $\mathcal{L}_{adv}$ is the adversarial loss function designed for breaking the consistency of the generator. A full-proof novel perspective approach is shown in Fig. \ref{fig:ap}, showing how adversarial perturbation could lead to distorted/faulty deepfake outputs. Some of the key techniques that are involved for the successful implementation of this approach are,

\subsubsection{Fast Gradient Sign Method (FGSM)}
The Fast Gradient Sign Method (FGSM) is an adversarial attack technique introduced by Ian J. Goodfellow et al. in his paper, Explaining and Harnessing Adversarial Examples \cite{ap}. This was one of the simplest and most efficient ways to generate adversarial examples that could fool deep neural networks by adding a small, imperceptible perturbation to the input image. FGSM operated in a white-box attach setting, which assumed full access to the target model's parameters, including weights and gradients. The attack was particularly effective because it leveraged gradient information of loss function to an adversarial perturbation that maximally disrupted the model's predictions. Let $x$ be the original input image, $y$ be the true label associated with $x$, $f(x)$ be the deep learning model that predicted the class label from $x$, $\mathcal{L}(f(x),y)$ be the loss function that the model optimized, $\theta$ be the trainable parameters (weights) of the model, $\nabla_x \mathcal{L}(f(x),y)$ be the gradient of loss function with respect to the input image $x$, and $\epsilon$ be the small perturbation factor that controlled the intensity of the attack, the mathematical formulation of FGSM is shown in Eq. \ref{eq7}.

\begin{equation}
    x' = x + \epsilon \cdot sign(\nabla_x \mathcal{L}(f(x),y))
    \label{eq7}
\end{equation}

Here, the adversarial image $x'$ was generated by adding a perturbation to the original image $x$ in the direction of the sign of the gradient of the loss function. With the maximization of the loss function using adversarial perturbation, the model was misled into misclassifying the perturbed image while keeping the perturbation small enough that the change is imperceptible to the human eye.

Deep learning models make decisions based on subtle statistical patterns in the data. FGSM exploited this by making small, targeted modifications to the input image that steered the prediction of the model toward an incorrect class. Consider a neural network with parameters $\theta$ that minimized a loss function $\mathcal{L}$ using gradient descent $\theta \leftarrow \theta - \eta \nabla_{\theta}\mathcal{L}(f(x),y)$. FGSM reversed this process by modifying the input instead of the model parameters. This found a small perturbation that increased the loss function in the direction that maximally confused the model. The implementation of FGSM starts with the computation of loss $\mathcal{L}$ for the given input $x$ and true label $y$ through $\mathcal{L} = - \sum_i y_i log f_i(x)$, where $f_i(x)$ is the predicted probability of class $i$. Next, the gradient is computed using $g = \nabla_x \mathcal{L}(f(x),y)$, and the gradient tells about which direction to modify the pixel values for increasing the model's loss. Thirdly, the adversarial perturbation is computed using $\delta = \epsilon \cdot sign(g)$, where $\epsilon$ is a small scalar that controls how much perturbation to be applied. The final adversarial image is calculated using $x' = x + \delta$, and lastly, to ensure that perturbed image $x'$ remained within valid pixel values (0-255 for images), it is clipped using $x' = clip(x', 0,255)$.

The effectiveness of this algorithm lies in its fast and computationally efficient nature since it only requires a single gradient computation, which is much faster than iterative attacks. FGSM adversarial examples are often transferred to other models, making them useful in black-box attack settings. This can be used to fool state-of-the-art models trained on large datasets. The limitations include its easy defense against adversarial training, which could improve the robustness of the model against FGSM, and defensive distillation and gradient masking, which could make FGSM less effective. This also produced high-confidence misclassification, which revealed a major vulnerability in deep networks. FGSM does not adapt itself based on the model's defense mechanisms, unlike other stronger iterative methods such as Projected Gradient Descent (PGD).

\subsubsection{Projected Gradient Descent (PGD)}
Projected Gradient Descent (PGD) is an iterative adversarial attack that extends the Fast Gradient Sign Method (FGSM) by refining the adversarial perturbation over multiple steps. This stronger and more flexible attack could generate highly effective examples, making it widely used in adversarial machine learning, particularly in robust deepfake prevention and defensive AI research. This is more powerful than FGSM and guarantees constraint satisfaction. For given input image $x$, a target model $f(x)$, and ground truth label $y$, the goal is to create an adversarial example $x'$ such that:

\begin{enumerate}[wide, labelwidth=!, labelindent=0pt]
    \item The perturbation remains small using $x' = x + \delta$ where $||\delta||_p \leq \epsilon$. Here, $\epsilon$ is the small bound controlling the amount of adversarial perturbation that could change $x$.
    \item The modified input fools the model through $f(x') \neq y$ meaning the model incorrectly classified $x'$.
\end{enumerate}

PGD iteratively updates the input image in the direction of the loss function's gradient, ensuring that each update remains within the $L_p$-norm constraint. This first initialized perturbed image using $x^{(0)} = x + random noise$. This small random perturbation was added at the start to prevent gradient masking effects. Secondly, the iterative gradient update took place at each iteration $t$, the adversarial perturbation is updated using the gradient sign method using Eq. \ref{eq8}.

\begin{equation}
    x^{(t+1)} = \prod_{\mathcal{B}(x, \epsilon)} (x^{(t)} + \alpha \cdot sign(\nabla_x \mathcal{L}(f(x^{(t)}),y)))
    \label{eq8}
\end{equation}

Here, $\mathcal{L}(f(x),y)$ is the loss function, $\nabla_x \mathcal{L}$ is the gradient of the loss function with respect to input $x$, $sign(\cdot)$ extracted the direction of the gradient, $\alpha$ is the step size for each iteration, $\prod_{\mathcal{B}(x, \epsilon)} (\cdot)$ projected the perturbed image back into the allowed $L_p$-norm perturbation ball. The process was repeated for a predefined number of steps $T$ until the attack converges. Since adversarial perturbation must remain within a specific norm constraint, $||\delta||_p \leq \epsilon, \quad PGD$ ensures this using the projection step in EQ. \ref{eq9}.

\begin{equation}
    \prod_{\mathcal{B}(x, \epsilon)}(x') = min(max(x', x - \epsilon),x + \epsilon)
    \label{eq9}
\end{equation}

This clips the perturbed image to remain within the valid pixel range and the predefined perturbation bound. If the perturbation $\delta$ exceeds the allowed $L_p$-norm ball, it will be projected back into the valid space. This prevented excessive distortion, which ensured that the adversarial perturbation remained imperceptible to human vision.

Deepfake detection and prevention mechanisms typically relied on supervised learning, which required extensive labeled datasets of fake and real media for training models \cite{Guo_2017} \cite{Yan_2024}. However, zero-shot learning (ZSL) enabled the model to generalize beyond its training data, which made it particularly effective for adversarial perturbation generation \cite{Zhao_2022} \cite{Zhou_2023} against unseen deepfake models. Zero-shot adversarial perturbations could prevent deepfake generation by dynamically disrupting deepfake models without requiring prior exposure to specific architectures or training datasets. This perspective approach discussed contrastive learning, self-supervised adversarial training, and latent space disruption for introducing robust defense applicable to known and unknown deepfake models.

Let $X$ be the input image, and $G$ represent a deepfake generator. The goal of zero-shot adversarial perturbations is to introduce a perturbation $\delta$ such that the deepfake model failed to generate realistic output using $G(X + \delta) \not\approx \widetilde{X}$, where $\widetilde{X}$ is the expected manipulated image. The perturbation $\delta$ is crafted for maximizing the generative loss using $\delta = \arg \max_{\delta} \mathcal{L}_{gen}(G(X + \delta))$, where $\mathcal{L}_{gen}$ is the generative loss function of the deepfake model. This determines how well the generated image matches the real image distribution. By maximizing the loss, adversarial perturbations confused the deepfake generator, which led to poor-quality outputs.

\subsubsection{Contrastive Learning-Based Zero-Shot Perturbations}
Contrastive learning \cite{Hu_2024} was widely used to learn generalizable feature representations without explicit labels. In zero-shot adversarial perturbations, contrastive learning \cite{Vito_2022} ensured that perturbations remained effective across different deepfake models by identifying common vulnerabilities in deepfake feature spaces. Given a set of input images $\{X_i\}$, a set of positive pairs (real images) is constructed along with negative pairs (deepfakes). The model learns a representation $f\{X\}$ using the contrastive loss function InfoNCE (Noise Contrastive Estimation) using Eq. \ref{eq10}.

\begin{equation}
    \mathcal{L}_{contrastive} = - \sum_i \log \frac{\exp (sim (f(X_i), f(X_i^+))/ \tau)}{\sum_j \exp (sim(f(X_i), f(X_j^-))/\tau)}
    \label{eq10}
\end{equation}

where, $X_i^+$ is the positive pair (real image), $X_i^-$ is the negative pair (deepfake), $sim(.,.)$ is the similarity function and $\tau$ is the temparature scaling parameter. The learned feature space ensures that real and fake images are distinguishable, which allows perturbations to target deepfake-prone areas across unseen deepfake architectures.

\subsubsection{Self-Supervised Adversarial Training}
Self-supervised adversarial training enabled a model for generating adversarial perturbations without requiring labeled fake images. This method involved training perturbation generator $P$ to maximize deepfake distortion without explicit supervision. The adversarial perturbation model was trained using the meta-learning framework, where a perturbation function $P$ was optimized to increase deepfake error rates using Eq. \ref{eq11}.

\begin{equation}
    min_P\mathbb{E}_{X \sim \mathcal{D}} [max_{\delta \in \Delta} \mathcal{L_{adv}}(G(X + P(X)))]
    \label{eq11}
\end{equation}

Here, $\mathcal{L}_{adv}$ is the adversarial loss function that penalized deepfake realism, $\Delta$ is the set of allowable perturbations, and $\mathbb{E}_{X \sim \mathcal{D}}$ represented the expectation over the data distribution. This training strategy allowed perturbations to be self-adjusting, which ensured that they remained effective even when new deepfake models emerged.

\subsubsection{Latent Space Disruption via Zero-Shot Adaptation}
Deepfake models are operated by mapping input images into latent feature spaces before reconstructing manipulated output. Introducing perturbations directly into the latent space could degrade the quality of deepfake transformations without prior knowledge of the internal structure of the generator \cite{Guo_2024}. Let $Z$ be the latent representation of input image $X$ under deepfake generator $G$, then $Z = G_{enc}(X)$, where $G_{enc}$ is the encoder of the deepfake model. The main goal is to introduce a perturbation $\delta_z$ such that the decoded output is distorted as in $G_{dec} (Z + \delta_z) \not\approx \widetilde{X}$, where $G_{dec}$ is the decoder function of the deepfake generator. The perturbation is learned using latent-space adversarial loss through Eq. \ref{eq11}.

\begin{equation}
    \delta_z = \arg \max_{\delta_z} \mathcal{L}_{latent}(G_{dec}(Z + \delta_z))
    \label{eq11}
\end{equation}

Here, $\mathcal{L}_{latent}$ measured the divergence between deepfake features and real images, which forced the generator to low-quality results.

Consider a real-time deepfake prevention system integrated into social media platform. When a user uploads an image, 

\begin{enumerate}[wide, labelwidth=!, labelindent=0pt]
    \item The system first analyzes the image using a zero-shot feature extraction model (e.g. contrastive learning-based deepfake detector).
    \item If the image is highly susceptible to deepfake attacks, the system applies an adversarial perturbation $\delta$ before storing the image.
    \item If the deepfake model later attempts to modify the image, the perturbation disrupts the latent feature encoding, which results in poor-quality deepfake outputs.
\end{enumerate}

This method ensured that even newly developed deepfake models struggled to generate realistic fakes, which protected users from identifying threats and misinformation. Although adversarial perturbations provide a promising zero-shot strategy for removing the power of deepfakes generation, some challenges lie ahead for ensuring that the adversarial perturbations \cite{Chen_2021} \cite{Wyawahare_2025} will be an effective means of disruption in the future. Another problem is balancing effectiveness and perceptibility since perturbations have to be strong enough to fool deepfake models. Still, perceptible by human observers otherwise, or else they fail to be effective while noticeable, making them impractical for deployment to tools in the real world. Meanwhile, adaptive deepfakes might also answer adversarial attacks through adversarially robust training \cite{Y_ksel_2020}, where generative networks \cite{Liu_2019} learn to evade or pull over adversarial perturbations, for which updated methods imposed by adversarial strategies must be in continuous process. The computational overhead of creating real-time, personalized perturbation is another key challenge \cite{Li_2019}, which requires a lot of processing power, and it becomes hard to implement for resources on large-scale social media platforms and digital content providers. Adversarial techniques generally work on images and videos, and deepfakes threaten audio, text, and multimodal content, demanding cross-modality adversarial \cite{Wang_2023} \cite{Wang_2024} noise to disrupt deep fake synthesis in different modality domains. These challenges notwithstanding, AI models can enhance the resilience of adversarial perturbations with personalized perturbation generation, which uses AI to educate adversarial perturbations that prevent unauthorized synthetic media manipulation for a given unique individual \cite{Barach_2025}. The strength of defenses in federated adversarial training, in addition to privacy, can be further boosted as each organization may defend against the deepfake technique on a massive scale without exposing their users' privacy \cite{Lee_2024} \cite{Lu_2024}. Quantum-inspired adversarial defenses \cite{Hou_2023} can also introduce highly unpredictable perturbations, where the perturbations are quantum-inspired and virtually impossible for deepfake models to adapt\cite{Akter_2024}. The other direction that is quite promising and something that might happen is the real-time deepfake resilience scoring \cite{Alrowais_2024}, where an AI-powered system will be able to assess an image or a video and see if it's susceptible to being a deepfake and then warn and alert the user and allow them to take protective measures in case that is required. Integrating adversarial AI \cite{Li_2024}, federated learning \cite{Awan_2021}, and quantum computing \cite{Das_2023} \cite{Kania_2017} will be crucial in establishing robust, future-proof defenses against synthetic media manipulation as deepfake technology evolves \cite{de_Wolf_2017}

\subsection{Digital Watermarking and AI-Embedded Content Authentication}
With the development of deepfake technology, robust content authentication mechanisms have become necessary. Digital watermarking \cite{Ambadekar_2018} \cite{Singh_2013} is one of the most effective solutions with digital media, a type of digital watermarking in which invisible or semi-invisible signatures are contained in the digital media to verify the use authenticity and detect tampering \cite{Radharapu_2024}. Unlike conventional detection, where after content creation it is detected, digital watermarking helps prevent content dishonesty from the time it is created. Integrating the watermarking film with AI-powered techniques and zero-seen learning allows authentication to be more adaptive, scalable, and not prone to deepfake attacks \cite{AL_ardhi_2019}.

Digital watermarking is a security measure that embeds a unique, often imperceptible digital signature into multimedia content, such as images, videos, and audio files. The primary objectives of digital watermarking include:

\begin{enumerate}[wide, labelwidth=!, labelindent=0pt]
    \item Authenticity verification where it is ensured that the content has not been altered from its original form.
    \item Copyright protection where media is marked with an identifier that proves ownership and prevents unauthorized usage.
    \item Tamper detection where an image, video, or audio file is identified as modified and manipulated.
    \item Deepfake prevention where AI-generated content is prevented from being misrepresented as real.
\end{enumerate}

\begin{figure*}[t!]
    \centering
    \includegraphics[width=\textwidth]{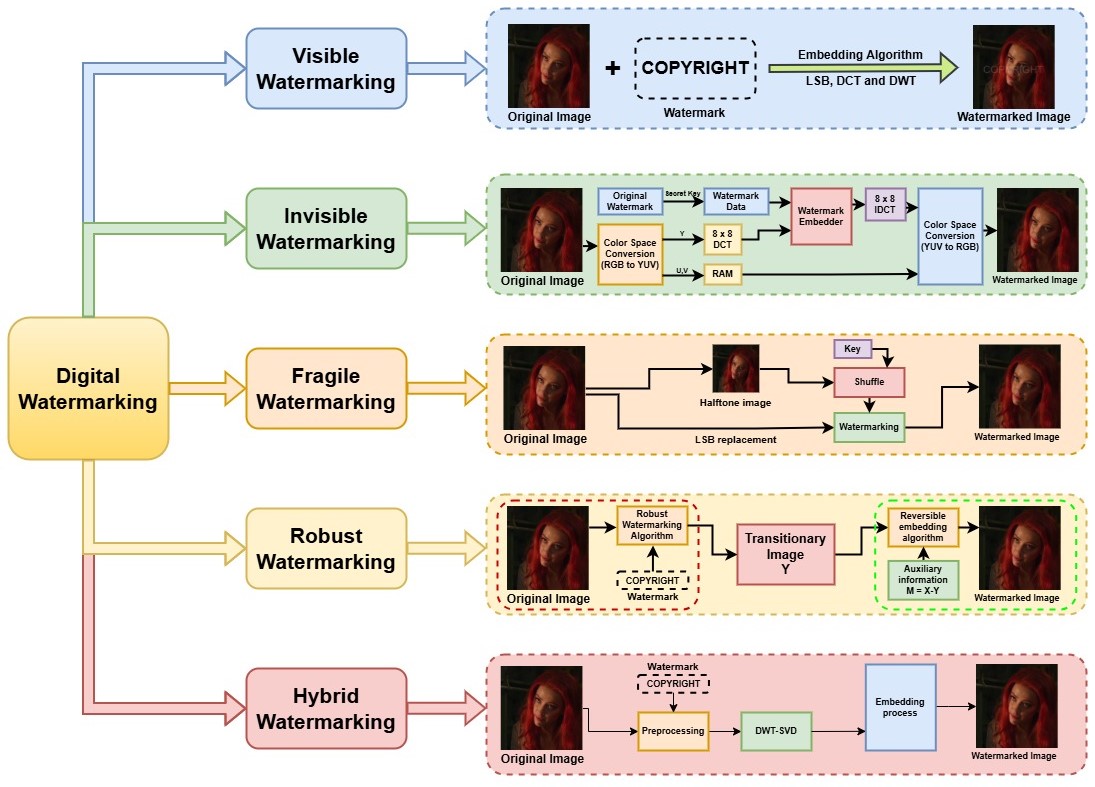}
    \caption{This diagram illustrates the various types of digital watermarking techniques, categorized into visible, invisible, fragile, robust, and hybrid approaches, each showcasing a distinct methodology for embedding information within an image. Each method results in a watermarked image, demonstrating the diverse applications of digital watermarking in copyright protection and data authentication.}
    \label{fig:watermarking}
\end{figure*}

With the integration of AI-enhanced watermarking mechanisms, digital watermarking \cite{Xie_2024} could evolve into an advanced zero-shot detection strategy capable of identifying previously unseen deepfake techniques without extensive training on synthetic datasets. Digital watermarking methods can be categorized into several types based on visibility, robustness \cite{Liao_2022}, and intended application. The different types of watermarking with its implementation are shown in Fig. \ref{fig:watermarking}.

\begin{enumerate}[wide, labelwidth=!, labelindent=0pt]
    \item \textbf{Visible Watermarks}: These watermarks are overt and easily recognizable. Logos, text overlays, or semi-transparent patterns are visible watermarks indicating media ownership \cite{Thakkar_2024}. News agencies and stock image platforms overlay visible watermarks on images to prevent unauthorized use. Visible watermarks could be cropped or blurred, which makes them ineffective for deepfake prevention \cite{Iacobici_2020}.
    \item \textbf{Invisible Watermarks}: These watermarks are hidden within the media and are not perceptible to the human eye. They could only be detected using AI-driven analysis or specialized software \cite{Lai_2025} \cite{Wang_2024}. AI-generated signals embedded in the pixel values of an image remain unchanged despite compression or minor editing. Deepfake algorithms struggled to remove these watermarks without degrading the quality of media.
    \item \textbf{Fragile Watermarks}: Fragile watermarks are sensitive to modifications and disappear when any part of the media is altered. A watermark that vanished if an image was edited or compressed, which signaled possible tampering. The main use of this technique lies in the prevention of any kind of post-processing on government-issued documents or critical medical images\cite{Schirripa_Spagnolo_2006}.
    \item \textbf{Robust Watermarks}: These are designed for withstanding modifications, including compression, resizing, or filtering \cite{Kakkirala_2015} \cite{Guo_2024}. This watermark persisted even after a deepfake model attempted to reconstruct an image using Gen AI. This ensured long-term content authentication, which made it useful for forensic deepfake detection \cite{Xia_2014}.
    \item \textbf{Dual Watermarking (Hybrid Approach)}: A combination of visible and invisible watermarking techniques that provided double-layer security against deepfake alterations. A real use case is a news broadcast embedding an invisible forensic watermark while displaying a visible logo overlay \cite{Nadimpalli_2024} \cite{Zhao_2023}.
\end{enumerate}

Watermarking was a passive security measure and actively combatted deepfake generation when integrated with AI. This happened through

\begin{enumerate}[wide, labelwidth=!, labelindent=0pt]
    \item \textbf{Embedding Watermarks During Content Creation}: AI-powered watermarking tools embed unique identification markers during media creation. These markers helped verify whether the content was synthetically modified or not. A real example is AI-generated news reports, which are watermarked to prevent false attribution.
    \item \textbf{AI-Based Deepfake Identification Through Watermark Analysis}: Machine learning models analyzed media for watermark inconsistencies. If an image or video lacks the expected watermark, this is flagged as potentially synthetic. An example is a deepfake video of a political leader uploaded without the expected watermark, which triggers an AI authentication check\cite{K_2024}.
    \item \textbf{Tamper-Detection Using Watermarking Techniques}: Invisible watermarks get altered slightly when content is modified (e.g. using GAN-based deepfake models manipulated video frames) \cite{Ramesh_2009}. AI-powered algorithms analyzed minute distortions in watermark integrity to detect deepfake manipulations. Forensic tools that analyze celebrity deepfakes use watermarking to verify authenticity \cite{Hydara_2024}.
    \item \textbf{Zero-Shot Learning in Watermark-Based Authentication}: Deepfake models evolved rapidly, making pre-training AI detectors impossible on all possible manipulations. Zero-shot learning \cite{Li_2024} enabled AI models to detect anomalies in watermark structures without prior exposure to specific deepfake techniques \cite{Sun_2023} \cite{Sun_2023}. If an AI-based watermarking system was trained on conventional deepfakes but encounters a new AI-generated synthetic voice, this could still infer fake content through watermark inconsistencies.
\end{enumerate}

Deepfake detection traditionally relied on supervised learning, where models were trained on large datasets that contained real and fake samples. However, this approach had a fundamental weakness: its dependence on prior knowledge \cite{Moon_2024} of deepfake types. With deepfake models evolving rapidly, new types of manipulations appeared that the model had never encountered before, making it effective. To overcome this limitation, zero-shot learning was introduced into watermark-based authentication systems. ZSL \cite{Liu_2023} \cite{Guo_2024} enabled AI models to detect manipulations without specific training on newly emerging deepfake models. In the context of digital watermarking, ZSL enhanced the ability to validate content authenticity, identifying tampering and preventing deepfake-generated modifications \cite{Sharma_2020}. Traditional watermarking methods used static patterns to verify authenticity. However, adversarial deepfake techniques could forge or remove these static watermarks \cite{Bashardoost_2015}. To counter this, an AI-driven watermarking system used ZSL-powered dynamic watermarks, which change based on the context of the content. These watermarks are adaptive, which could alter their structure based on video, image, or audio characteristics. These also used semantic attributes \cite{Zhang_2024} to make deepfake manipulation detectable even when no prior data on that specific deepfake type existed. These encoded high-dimensional feature representations that deepfake generators struggled for replication of accuracy \cite{Kapre_2020}. A real use case would be a news agency embedded a unique AI-generated watermark in every news video. If someone tries to alter the face of the speaker using a deepfake generator, the AI model would detect subtle inconsistencies in the watermark pattern, and no deepfake algorithm could perfectly reproduce this.

\subsection{Blockchain for Secure Content Verification}
Blockchain technology is a decentralized and immutable digital ledger system that records tamper-proof and verifiable transactions. Originally designed for cryptocurrencies, blockchain has found applications in various fields, including content authentication and deepfake prevention. The fundamental principle behind blockchain-based secure content verification is that once a piece of media - such as image, video, or audio - was created, its unique digital fingerprint (hash) was recorded on a blockchain ledger, which ensured that any subsequent modifications could be detected. A blockchain-based content verification system typically involves the following key steps:

\begin{enumerate}[wide, labelwidth=!, labelindent=0pt]
    \item \textbf{Content hashing and digital signatures}: When an original media file (image, video, or audio) is created, a cryptographic hash function generates a unique digital fingerprint of the file. The hash and metadata, such as timestamp, creator, and source device identity, are stored on a blockchain network \cite{DEKA_2019} \cite{Huang_2007}.
    \item \textbf{Decentralized storage and verification}: Unlike centralized systems (such as Google Drive or YouTube's content verification tools), blockchain stores media fingerprints across multiple distributed nodes \cite{Thilakavathy_2023}. Since the implemented ledger is immutable, tampering with the original file will straightly result in a different hash value, which signals manipulation \cite{Jain_2021}.
    \item \textbf{Real-time verification of media integrity}: When content is uploaded or shared online, AI-based forensic tools extract its hash and compare it with the registered hash stored on the blockchain \cite{Ali_2020}. If there is a discrepancy, the system flags the content as potentially manipulated and alerts relevant stakeholders (e.g. social media platform admins, forensic investigators, or law enforcement) \cite{Picha_Edwardsson_2024}.
    \item \textbf{Smart Contracts for Automated Enforcement}: Smart contracts could automatically trigger verification mechanisms when content is uploaded to digital platforms. If deepfake detection algorithms determine that a file has been tampered with, smart contracts could prevent its distribution or label it as a synthetic content \cite{Wang_2020} \cite{Fisher_2024}.
\end{enumerate}

\begin{figure*}[t!]
    \centering
    \includegraphics[width=\textwidth]{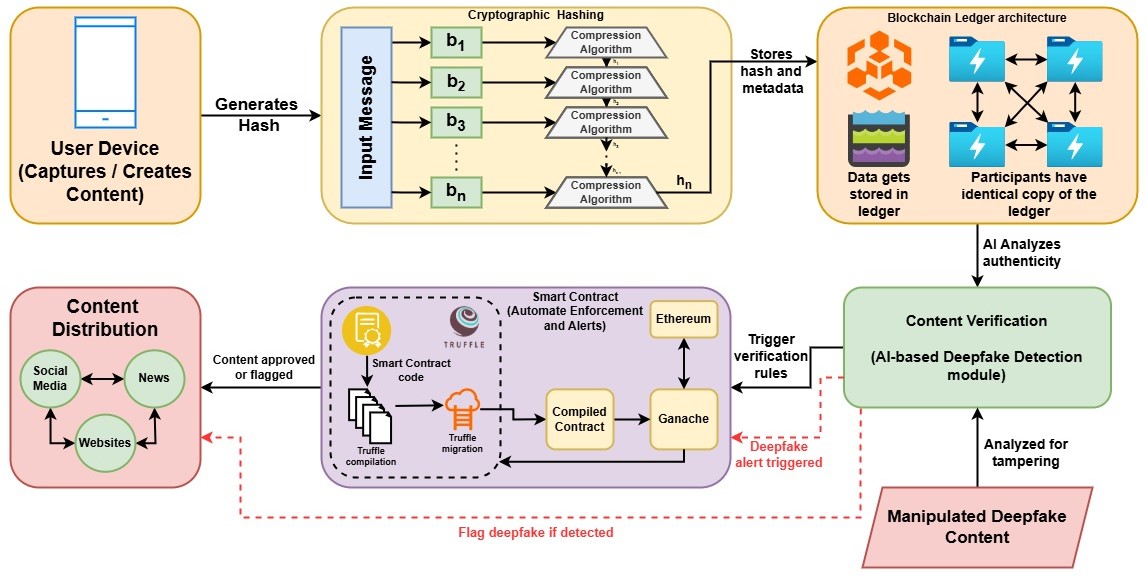}
    \caption{This diagram illustrates a blockchain-based system for content authentication and distribution, employing cryptographic hashing and smart contracts to ensure data integrity and provenance. User-generated content is hashed and stored on a distributed ledger, accessible to all participants, with metadata. Smart contracts on the Ethereum blockchain automate content verification, including AI-driven deepfake detection, triggering alerts for manipulated content, and facilitating secure distribution across social media, news platforms, and websites. This process guarantees content authenticity and transparency, mitigating the risks of tampering and misinformation.}
    \label{fig:blockchain}
\end{figure*}

Thus, blockchain acts as an immutable ledger that tracks media provenance, ensuring that content modifications can be traced, verified, and flagged before misleading information spreads. A proposed perspective approach of how blockchain could be used for deepfake prevention through zero-shot learning is shown in Fig. \ref{fig:blockchain}.

\subsubsection{Zero-Shot Integration}
While blockchain ensured content integrity, this traditionally required a predefined dataset of hatches for verification. Deepfake variants \cite{Sun_2025} created using new, unseen AI models could bypass conventional blockchain-based systems. Zero-shot learning (ZSL) \cite{Fu_2014} addressed this challenge by enabling real-time anomaly detection and dynamic adaptation for novel deepfake threats \cite{Kirk_2023}. Zero-shot learning enhanced blockchain for deepfake prevention in the following ways:

\begin{enumerate}[wide, labelwidth=!, labelindent=0pt]
    \item \textbf{Detecting Anomalies in Blockchain Logs}: Zero-shot learning models could analyze blockchain-stored metadata and identify inconsistencies or previously unseen deepfake pattern . Instead of requiring labeled deepfake datasets, ZSL-based classifiers could detect out-of-distribution changes in media characteristics by comparing content to historical authenticity markers.
    \item \textbf{Classifying Media Authenticity Without Prior Exposure}: Traditional blockchain verification only checks for an exact match between a file's hash and its stored record. However, deepfake generation techniques often introduce subtle modifications (e.g. slight distortions in facial expressions, voice tone shifts). Zero-shot models leveraged semantic similarity analysis (e.g., ViTs, Contrastive Language-Image Pretraining) for detecting deepfake patterns without needing pre-existing labeled data \cite{Lokhande_2024}.
    \item \textbf{Adaptive deepfake detection in real-time}: Since deepfake technologies evolved rapidly, blockchain ledger alone might not be sufficient. Zero-shot models used feature-based learning to analyze media content before its addition to blockchain, which ensured that even newly emerging deepfake architectures could be dynamically flagged \cite{Priya_2024}. AI-driven self-learning forensic models could enhance the ability of blockchain to identify fake content before it is registered as authentic media.
\end{enumerate}

For example, here, if a new deepfake generator (e.g., AI-based face-swapping model) produced content that had never been detected before, a zero-shot learning model could extract key features from the deepfake content, compare it with genuine media stored on the blockchain \cite{Satone_2024} \cite{Lokhande_2024}, and predict whether the content is real or synthetic when without prior labeled data. This fusion of blockchain \cite{Costales_2023} and zero-shot learning ensured that even previously unseen deepfake manipulations could be detected, which made content verification systems more robust and future-proof.

Although blockchain-based safe content verification promises to revolutionize the future, numerous obstacles prevent widespread use. However, the computational overhead \cite{Kim_2025} of blockchain transactions in large-scale multimedia verification is too high and can cause latency issues \cite{Lao_2024}. It also poses challenges in handling legitimate content updates \cite{Zhou_2023} and removals due to the immutability of blockchain\cite{Chunli_Wang_2023}. Complicating deployment further, there is a great need for high storage capacity to store cryptographic hashes and metadata \cite{Zhu_2020}. Additionally, ensuring that this integration benefits seamlessly with AI-driven deepfake detection models is difficult, and it is done in a decentralized manner \cite{Jain_2024}. While there are limitations, future improvements in lightweight blockchain architectures, quantum-resistant cryptographic hashing \cite{Fathalla_2024}, and AI-optimal smart contracts \cite{Tyagi_2023} offer hope of addressing these limitations. Fine integration of zero-knowledge proofs for privacy-preserving verification \cite{Handayani_2023} and federated blockchain networks can improve efficiency yet maintain trustworthiness. With the evolution of AI and blockchain technology, these two technologies may lead to an autonomous self-verifying content ecosystem far less prone to spreading deepfakes and, therefore, supports more authentic digital media \cite{Jain_2024}.

\subsection{AI-Powered Real-Time Monitoring and Content Moderation}

\begin{figure*}[t!]
    \centering
    \includegraphics[width=\textwidth]{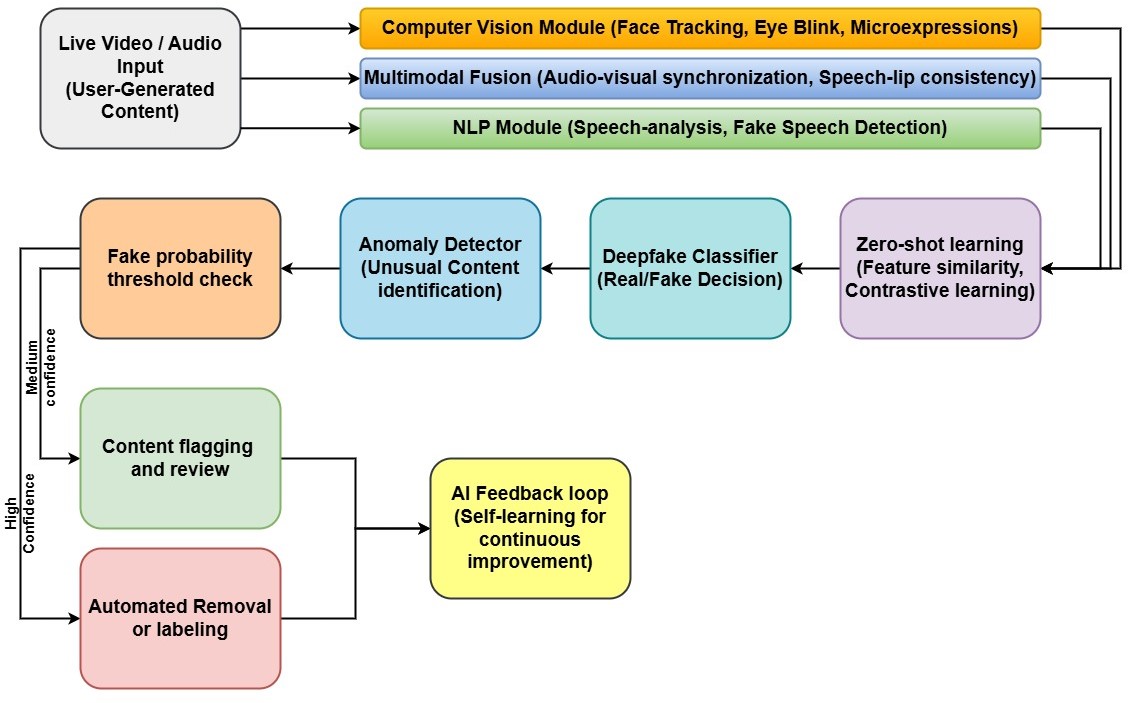}
    \caption{This diagram illustrates a comprehensive deepfake detection pipeline, processing live video/audio input through a series of modules: computer vision for facial analysis, multimodal fusion for audio-visual consistency, and NLP for speech analysis and fake speech detection. An anomaly detector identifies unusual content, while a deepfake classifier, aided by zero-shot learning, makes a real/fake decision. The system employs a fake probability threshold to determine confidence levels, triggering content flagging and review for medium confidence cases and automated removal or labeling for high confidence cases. An AI feedback loop ensures continuous improvement through self-learning.}
    \label{fig:cm}
\end{figure*}

As deepfake technology advances, the ability to detect and mitigate manipulated content in real time \cite{Javed_2024} becomes a crucial aspect of digital security. AI-powered real-time monitoring systems are an active defense mechanism to analyze media content created, modified, or shared online. Unlike traditional deepfake detection approaches that completely relied on offline forensic analysis, real-time systems functioned at the point of content generation or distribution, which prevented harmful synthetic content from reaching a wider audience. Integrating zero-shot learning (ZSL) \cite{Sariyildiz_2019} into real-time monitoring enhanced its ability to identify novel deepfake variations without requiring pre-existing labeled training data. This ensured that even newly developed, unseen deepfake techniques could be detected as soon as they emerged. Real-time monitoring \cite{Suresh_2024} and moderation rely on deep learning models that continuously analyze content streams to identify manipulated or synthetic media. These systems leveraged multiple AI techniques, including computer vision and deep neural networks, which analyzed video frames and facial movements for inconsistencies. Techniques also included natural language processing (NLP), which detected synthetic speech patterns and scripted deepfake-generated conversation, along with multimodal analysis, which combined audio, video, and textural cues to detect deepfake content comprehensively, as shown in Fig. \ref{fig:cm}. This works through:

\begin{enumerate}[wide, labelwidth=!, labelindent=0pt]
    \item Content ingestion, where a user uploads or streams a video, where AI models begin analyzing individual frames, voice modulation, and facial expressions in real-time.
    \item Feature extraction where the system extracted key features such as blink rate, facial texture inconsistencies, unnatural lip-syncing, and abrupt changes in voice tone.
    \item Deepfake probability estimate is a probabilistic model that assigns a score indicating whether the content is likely synthetic.
    \item Decision-making and flagging occur when a video surpasses a predefined deepfake risk threshold; it is either flagged for further review, automatically removed, or labeled with a warning.
    \item Feedback loop for continuous improvements is needed where real-time monitoring systems learn from false positives and negatives, which refined detection strategies over time.
\end{enumerate}

AI-based monitoring could be deployed in various digital platforms, including social media networks, to detect manipulated political or misleading content before it spreads, and live streaming services to prevent real-time identity impersonation using AI-generated avatars \cite{Kim_2023}. This also includes online verification systems for combating deepfake fraud in banking and legal proceedings. While real-time monitoring already plays a role in content moderation policies, integrating zero-shot learning could improve its adaptability and accuracy \cite{Singh_2025} \cite{Kanamori_2022}.

\subsubsection{How Zero-Shot Learning Enhances Real-Time Monitoring}
Traditional deepfake detection models required large labeled datasets for training, which limited their ability to detect new types of deepfakes. Zero-shot learning (ZSL) eliminated this limitation by enabling AI to identify fake content it had never encountered before. The key zero-shot strategies in real-time monitoring include:

\begin{enumerate}[wide, labelwidth=!, labelindent=0pt]
    \item \textbf{Feature similarity matching}: Instead of memorizing deepfake characteristics, ZSL models compare live content against high-level features of real human behavior and natural speech patterns \cite{Kaur_2018} \cite{Winkler_2021}.
    \item \textbf{Anomaly detection without prior training}: AI models use out-of-distribution detection techniques for identifying content that deviates from normal human-like behavior, even if the specific manipulation method is unknown \cite{Jo_2024} \cite{}.
    \item \textbf{Contrastive learning (CLIP-based approaches)}: AI models, such as OpenAI's CLIP (contrastive Language-Image Pretraining), could associate deepfake content with textual descriptions (e.g., "unnatural facial expressions," "lip desynchronization") which enabled zero-shot detection \cite{Dong_2022} \cite{Khan_2024}.
    \item \textbf{Vision Transformers (ViTs)}: Instead of manually learning specific deepfake features, ViTs analyzed spatial and temporal inconsistencies in video frames for detecting previously unseen manipulation methods.
\end{enumerate}

Using the potential of zero-shot techniques with real-time monitoring systems could detect deepfake patterns that had never been explicitly trained on, adapt instantly to new deepfake-generation architectures, and scale across different languages and regions without requiring localized datasets. A real use case could be a political figure live-streaming \cite{Yoon_2024} an event and an adversary attempting to replace their face with a digitally altered deepfake version. The zero-shot enabled real-time monitoring AI \cite{Mittal_2024} could analyze facial motion and speech alignment. At the same time, the stream is ongoing, and it can be compared against any historical footage of a real person to identify subtle inconsistencies. This could also help detect anomalies in lip-syncing \cite{Liu_2024} and head movements \cite{Datta_2024} that indicate synthetic alterations and automatically flag or pause the live stream for human verification before misinformation spreads. This proactive approach \cite{Wang_2013} prevented real-time misinformation attacks, fraud, and AI-generated impersonations from being widely disseminated \cite{Mittal_2024}.

Despite the many technical, ethical, and implementation challenges, real-time deepfake detection \cite{Chetty_2010} and moderation are potential applications. However, high computational costs make the high-scale deployment expensive, and false positives and negatives \cite{Singh_2024} make removing content unnecessary or undetected a deepfake threat. The volume of live content \cite{Wang_2024} that needs to be monitored is too large for scalable solutions, and yet, lightweight AI models \cite{Javed_2024} for monitoring this live content need to be efficient. Ethical dilemmas \cite{Adiwijaya_2023} include that legitimate content may get flagged as such, and there may be biases in detection models \cite{Yang_2022} that could adversely impact others more than others. Using AI to monitor users' content means it needs to be exposed to them, posing the security and privacy risk of disclosing and being surveilled on content. Future work will be continued to advance the techniques, such as federated learning \cite{Mohammadi_2024}, that would preserve privacy and increase detection accuracy \cite{Jagarlamudi_2023}. Content moderation can benefit from explainable AI (XAI); edge AI can enable low latency detection on user devices, offloading as much as possible from centralized cloud processing. The reliability can be further improved by adopting hybrid human-AI moderation that allows for automated detection and human verification by fact-checkers. These challenges will be central to developing robust, fair, and scalable real-time deepfake prevention systems.

\subsection{Federated Learning for Privacy-Preserving Deepfake Prevention}

\begin{figure*}[t!]
    \centering
    \includegraphics[width=\textwidth]{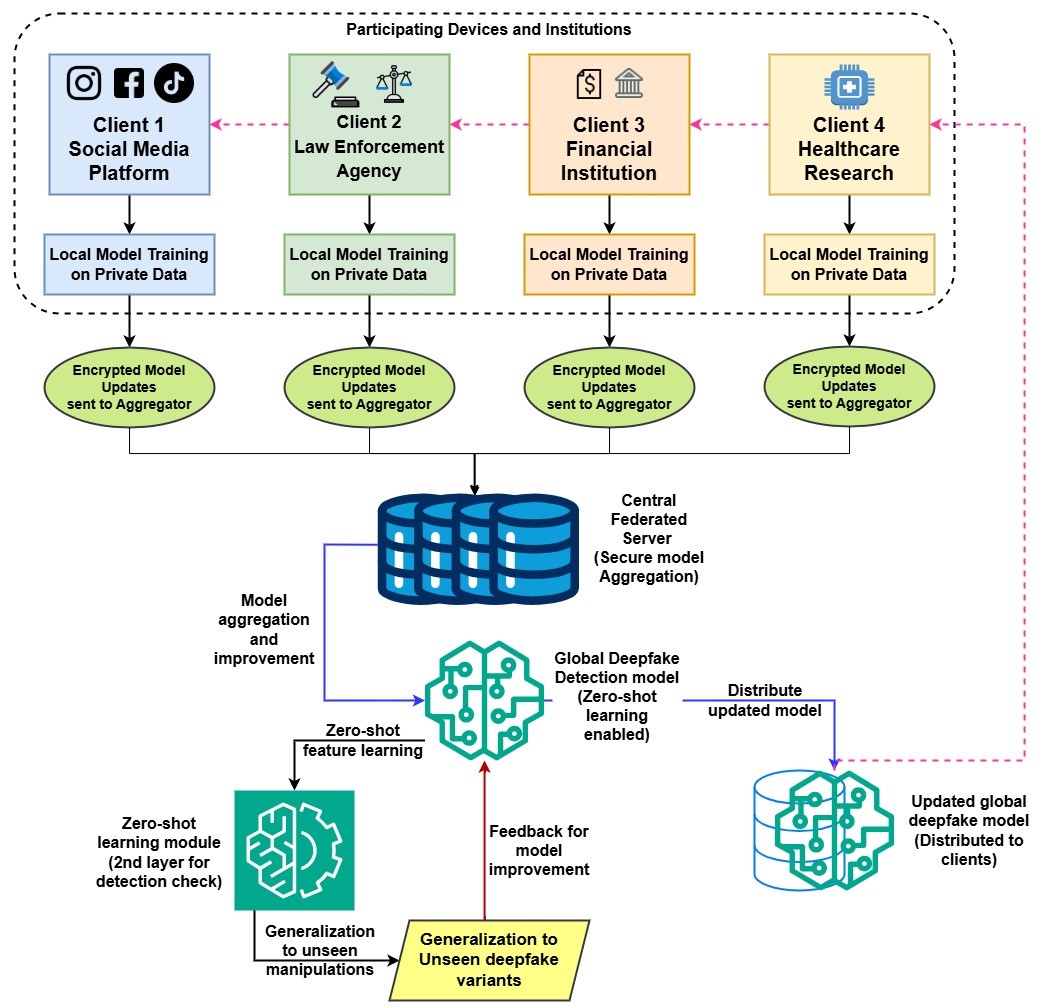}
    \caption{This diagram illustrates a federated learning framework for deepfake detection, where multiple clients (social media, law enforcement, finance, and healthcare) collaboratively train a global model without sharing raw data. Each client trains a local model on private data and sends encrypted model updates to a central server. The server securely aggregates these updates to improve the global deepfake detection model, which is then distributed back to the clients. A zero-shot learning module is integrated to enhance generalization to unseen manipulation techniques and deepfake variants, with feedback loops further refining the model's performance.}
    \label{fig:fl}
\end{figure*}

As deepfake techniques continued to evolve, traditional centralized AI-based defenses faced data privacy, scalability, and adaptability challenges. Federated learning (FL) \cite{Asif_2024} offers a decentralized, privacy-preserving approach to deepfake prevention by allowing multiple entities to collaboratively train AI models without sharing sensitive data \cite{Yin_2024}. Using the potential of zero-shot learning strategies, FL could enhance deepfake prevention by ensuring that the detection models are generalized for previously unseen deepfake variants across different datasets and domains. Federated learning is a decentralized machine learning approach that allows AI models to be trained across multiple devices or institutions without transferring raw data to a central server. Instead, only model updates (such as gradients and parameters) are exchanged between participants, which ensures data privacy while enabling collaborative learning. The workflow of FL in deepfake prevention could be done through these steps:

\begin{enumerate}[wide, labelwidth=!, labelindent=0pt]
    \item Local model training where each participating device in the framework trains a deepfake detection model using its private dataset \cite{Karathanasis_2025}.
    \item Secure model update sharing could be done where each device sends encrypted model updates to the central coordinator instead of sharing raw data \cite{Yan_2024}.
    \item Global model aggregation, where the central server aggregates updates from multiple devices for improving the global model \cite{Li_2022}.
        \item The updated global model is sent back to all devices, allowing them to refine their local models through iterative refinement \cite{Bhatti_2023}.
    \item The trained model is deployed across multiple locations for identifying and preventing deepfake content at scale.
\end{enumerate}

This collaborative and privacy-focused training enabled institutions to train high-quality deepfake detection models while complying with strict data protection regulations (e.g., GDPR, HIPAA) (see Fig. \ref{fig:fl}).

\subsubsection{Integration of Zero-Shot Learning (ZSL) in Federated Deepfake Prevention}
Traditional federated learning models relied on labeled datasets from different sources for training deepfake detection algorithms. However, new deepfake techniques emerged frequently, which made it impractical to pre-train models on every possible manipulation method. This is where zero-shot learning \cite{Zhang_2020} (ZSL) played a crucial role. ZSL allowed federated learning models for detecting unseen deepfake variants by learning generalized feature representations, classification of new deepfake content without requiring labeled examples of every manipulation type, and adapting dynamically to emerging deepfake techniques across different datasets and institutions \cite{Hao_2021}. Some perspectives on how zero-shot learning could enhance federated deepfake detection,

\begin{enumerate}[wide, labelwidth=!, labelindent=0pt]
    \item \textbf{Generalization across domains}: Here, zero-shot models used semantic embedding techniques (e.g., CLIP, contrastive learning) \cite{Liu_2024} for detecting deepfakes based on fundamental manipulation characteristics rather than specific examples. This allowed federated deepfake detection systems to identify fake content across different media formats, languages, and styles without retaining each variation \cite{Ba_2023}.
    \item \textbf{Efficient Resource Utilization}: Since ZSL enabled models to infer deepfake features without labeled data, federated learning could train efficient deepfake prevention models without extensive dataset collection\cite{Le_Cacheux_2012}. This reduced computational overhead and accelerated the model update process in federated learning networks.
    \item \textbf{Robust against novel deepfake attacks}: Zero-shot models leveraged pre-trained embeddings and feature representations from diverse datasets \cite{Li_2023}, making them more resilient to new and evolving deepfake techniques \cite{Nagarhalli_2024} \cite{Awan_2021}. This ensured that federated deepfake detection systems remained effective even when attackers introduced novel deepfake architectures.
\end{enumerate}

Together with zero-shot learning, federated learning has many applications across different industries, allowing deepfake detection without compromising data privacy \cite{Yurdem_2024}. Federated deepfake detection in social media platforms enables companies like Meta, Twitter, and TikTok \cite{Joshi_2022} to train AI models against evolving deepfake threats with real-time protection by collaborating without sharing raw data. Because zero-shot learning greatly facilitated the detection of unseen deepfake variants across diverse content sources, federated learning models can be used by law enforcement agencies and forensic investigators to deploy across jurisdictions \cite{Sulaiman_2024} to identify manipulated media in cybercrime cases \cite{Gupta_2023}. Federated learning is used by financial institutions to identify synthetic identity frauds, thereby allowing global banks to detect deepfake-generated fraud identities without exposing the privacy of their customers \cite{Li_2024}. Federated deepfake detection protects biomedical data in the healthcare sector by protecting patient confidentiality; hospitals and research institutions can detect AI-generated forgeries in medical imaging \cite{B__2023}. These applications show how federated learning enables deepfake prevention in critical domains at once, maximizing model adaptability by allowing zero-shot learning is also shown in Fig. \ref{fig:fl}.

However, federated learning for deepfake prevention still has several problems to overcome before scaling. Federated systems need frequent updates of the models over devices, giving rise to communication overhead and scalability problems, which hierarchical aggregation strategies can alleviate \cite{Ye_2024}. Adversarial participants \cite{Godavarthi_2024} \cite{Qammar_2021} can corrupt the global model or reconstruct private data using model inversion techniques, which raises security and privacy risks, and SMPC and DP \cite{Zheng_2024} \cite{Fotohi_2024} measures are required. The second challenge is how to deal with non-IID data (i.e., data that is not independent and identically distributed) since deepfake detection models trained on specific demographics or on given content types are not always easily generalizable to different datasets, and such a challenge can be overcome by using adaptive aggregation methods. Further, zero-shot learning allows for the detection of unseen deepfake manipulations. However, efficient zero-shot models in the federated setting are computationally demanding \cite{Sarma_2023} and require lightweight transformer-based architectures that run on device optimization. To support privacy-preserving, artificial intelligence (AI) \cite{Jagarlamudi_2023} based media authentication of media in the face of growing sophistication in synthetic media threats, overcoming these challenges will be necessary for scalable, secure, and efficient federated deepfake prevention.

\section{Challenges in Zero-Shot Deepfake Detection and Prevention} \label{sec5}
However, despite the great progress in zero-shot deepfake detection and prevention, many severe issues and limitations still obstruct the progress of this field. In addition, all these challenges arise from the complexity of deepfake generation, evolving attack techniques, adversarial robustness, scalability constraints, and ethical concerns. Therefore, it is important to put efforts into addressing these challenges to realize the practical application of Zero-Shot AI models in the real world.

\subsection{Generalization and Adaptability Issues}
An extremely important yet hard problem in zero-shot deepfake detection is generalization and adaptability across a fast-evolving space of methods for deepfake generation. In contrast to traditional supervised learning models that rely on training data of labeled deepfake types, zero-shot models must make inferences of the previously unseen deepfake manipulations through learned feature representations and high-level semantic relationships. However, this approach has a fundamental drawback in the form of the sustaining productivity gains made in generative adversarial networks (GANs) and diffusion models that constantly push the quality and realism of synthetic content. The modern deepfake generators, such as StyleGAN3, DALL·E 3, and Stable Diffusion \cite{Daneshfar_2024} \cite{Das_2023} \cite{Keita_2025}, allow one to synthesize human faces, expressions, and even human bodies to a high level of convincing detail and with less of the usual artifacts that earlier detection models depended on. This means a zero-shot detector trained to samples of the old datasets will be at a loss in identifying real content from AI-generated deepfakes, especially when confronted with sophisticated blend techniques, fine-grained texture synthesis, and photorealistic lighting adjustments.

The high dimensionality and the multimodal aspect of deepfakes are other major concerns. Early work on deepfake detection largely explored visual forgery in images or videos, with the latest advances adding a modal dimension of multimodal deepfakes, where visual and auditory contents are both being modified. One example is lip-syncing with speech generated by AI-powered models, which makes it virtually impossible to distinguish between human lip movements and the corresponding speech. These cross-modal inconsistencies \cite{Zhao_2023} are not well captured by zero-shot models that usually depend on semantic embedding or contrastive learning, resulting in misclassification errors. Furthermore, neural voice cloning and text-based video synthesis add depth to the detection problem and make it even harder, as their output is both subtle and highly deceptive changes that will evade the feature extraction aperture of all existing zero-shot models. A primary open research problem of multimodal learning architectures is integrating architectures capable of analyzing visual, audio, and textual deepfake patterns.

Another big barrier concerning generalization is domain adaptation dataset bias. In particular, most of the zero-shot deepfake detectors are trained on public datasets such as FaceForensics++, Celeb-DF, DFDC (Deepfake Detection Challenge), and DeeperForensics-1.0 \cite{Mao_2023} \cite{Huang_2022} \cite{Mancini_2020}, the majority of which are celebrity faces in controlled settings. Real-world deepfake attacks are targeted toward various people with varying lighting conditions, backgrounds, and camera angles, giving rise to the domain shift phenomenon \cite{Liu_2024}. Therefore, this implies a model trained on high-quality datasets with ample lighting may yield poor results in identifying deepfakes in videos appearing in social media misinformation campaigns and other cybercriminal endeavors. Furthermore, culturally different datasets restrict the adaptability of zero-shot models because variations in facial structure, skin tone, and linguistic cues can affect detection accuracy. Moreover, real-time forensic applications call for zero-shot detectors to wield a low false positive and false negative rate so that the system is reliable. Therefore, without extensive domain adaptation techniques, these detectors are prone to high false positive and false negative rates, rendering the system unreliable in real-time forensic applications.

Additionally, self-improving deepfake models create a particularly worrisome danger for zero-shot generalization. Unlike the static deepfake datasets used in training, real-world deepfake generators now use the combination of reinforcement learning optimized based on adversarial training to circumvent detection mechanisms actively. They are powerful enough to analyze the detection output and iterate, refining the synthetic output until they pass as real content \cite{Liu_2022}. As such, zero-shot detection frameworks must continuously adapt to these changes to their feature representations, making it a hard problem to learn adaptive behavior using no supervised retraining. An open problem in research is to develop self-evolving zero-shot detection algorithms that can dynamically self-train from deepfakes that have been encountered for the first time \cite{Cao_2022}.

\subsection{Adversarial Vulnerabilities}
A recent and serious threat to deepfake detection is that of adversarial attacks against an AI model. The sophistication of deepfake creators is getting increasingly sophisticated to avoid being detectable, using highly advanced methods for manipulating synthetic content's visual and latent representations. However, zero-shot detection models are not exposed to future attack strategies and can be more vulnerable to adversarial deception. The main adversarial challenges for zero-shot deepfake detection are adversarial attacks over the detection model and data poisoning through model manipulation.

\subsubsection{Attacks on Zero-Shot Learning Models}
The main ways of zero-shot deepfake detection are based on semantic similarity, anomaly detection, and feature-based classification. However, as these methods inherently suffer from weakness, attackers can use the inherent weakness that introduces adversarial perturbations in deepfake images or videos, making them indistinguishable from legitimate content. In particular, adversarial examples are carefully crafted modifications of an image or video that can be paired with the pixel values or with the internal feature representations of the image or video while retaining perceptual realism. Deepfakes generated by generative adversarial networks (GANs) \cite{Mishra_2024} contain imperceptible noise patterns that interfere with the zero-shot detection model's feature extraction process \cite{Ziyadinov_2023}. Adversaries, for example, seek to add high-frequency perturbations (Fourier-based adversarial attacks) to change the underlying texture and edge details of a fake image \cite{Huang_2022}. Since zero-shot models learn from representations instead of pixel comparisons, fakes can be perturbed to appear real to the model. A more sophisticated attack consists of style transfer-based deepfakes that utilize real videos to blend the style and texture onto synthetic ones to hide digital artifacts that zero-shot models can detect. Attackers may also use generative noise injections to introduce subtle artifacts to a deepfake signal to create an adversarially misleading signal that will fail the AI model.

Finally, it is possible for an adversary to use iterative adaptive adversarial attacks where the adversary and deepfake generations monitor the model’s predictions are refined continuously to evade detection. For example, the deepfake generator may exploit reinforcement learning-based adversarial adaptation, where the deepfake generator is trained that is indistinguishable from real data against different state-of-the-art deepfake detectors in a competitive scenario until it finds the configuration that is indistinguishable from real data \cite{Anbalagan_2022}. By running this GAN-detector co-evolution, we obtain deepfakes that learn to improve and improve at eluding detection. As a result, the zero-shot approach becomes much more challenging in the long run \cite{Larue_2023}.

\subsubsection{Model Poisoning and Data Manipulation}
Data poisoning is another major adversarial attack challenge where attackers manipulate the dataset used for training zero-shot models to misclassify deepfakes as authentic. Zero-shot models leverage a lot of work based on self-supervised learning and large-scale pre-trained networks, so the adversaries can inject maliciously altered training samples into the zero-shot models and make them unable to differentiate between real and fake. However, one way of poisoning a model is by putting “trojans" or “backdoors" into deepfake datasets \cite{Gao_2019}. These backdoors make hidden patterns in the form of a trigger, forcing the model to always classify any deepfakes input with this trigger pattern as genuine. By data augmentation attacks \cite{Wang_2023}, adversaries may manipulate a subset of the training images or videos to induce an imperceptible but undesirable adversarial signal. Meta-learning-based adversarial training involves attackers pretrain generative models using their learned representations to create deepfakes with distributions similar to those learned representations. Therefore, the zero-shot model starts recognizing fake facial structures, motion patterns, or speech signals as real, contaminating its detection performance. It is particularly dangerous in cases where the zero-shot deepfake detectors have to be somehow based on transformer-based models like Vision Transformers (ViTs) or CLIP-like architectures, which heavily depend on the pre-learned semantic space \cite{Yoon_2024} \cite{2025}. An attacker can manipulate this semantic space during model training. In such cases, while learning associations, the zero-shot detector can misclass, resulting in highly realistic deepfakes being considered real.

In addition, overcoming data poisoning can be lifted to the setting of federated learning, where deepfake detection models are trained in multiple, decentralized sources. Model inversion attacks, where attackers can extract internal feature representations from a federated model for subsequent use with deepfake generation pipelines, can be done by attackers. This setting allows the crafting of gradient-based adversarial deepfake from generative models, which learn to manipulate specific latent features to which the zero-shot detector is sensitive and becomes ineffective. Adversarial vulnerabilities in zero-shot deepfake detection are severe and call for robust adversarial defense mechanisms \cite{Li_2023}. Adversarial training \cite{Gong_2020}, which pre-exposes models to adversarial deepfakes, is a potential countermeasure, and differential privacy-based model training \cite{Miguelez_Tercero_2022} prevents attackers from reverse engineering the learned feature space. Furthermore, secure model validation frameworks like blockchain-based provenance tracking could also be introduced to prevent the training datasets from being tampered with and adversarial contaminated.

\subsection{Scalability and Computational Constraints}
Existing zero-shot deepfake detection models are either based on transformers, contrastive learning, or self-supervision methods, they are highly computation intensive and hard to deploy in real-time, in resource-constrained environments, for instance, in mobile devices and IoT systems. Processing time and energy consumption can be increased with the complexity of analyzing high-resolution videos \cite{Chamot_2022}, multimodal inputs (audio-visual data), and large-scale datasets. In addition, real-time detection is another task because deepfake generation techniques are becoming more capable of generating live manipulation of video calls and streaming media, making detectors impractical and needing minimal latency but still very high accuracy. To scale, the researchers need to develop lightweight, efficient architectures, optimize hardware acceleration (e.g., GPU/TPU-based inference), and investigate edge AI solutions to improve the speed and accuracy, neither of which takes a heavy load on the cloud resources.

\subsection{Lack of Standardized Evaluation Metrics and Datasets}
Evaluation of zero-shot deepfake detection based on existing standardized metrics is one of the major challenges in this task due to the lack of a variety of datasets that cover a wide range of real-world attack scenarios. While the traditional deepfake detectors trained on particular datasets (FaceForensics++, DFDC, and Celeb-DF) exhibit specificity to a specific deepfake technique, zero-shot models require continually large datasets with divergent distribution so that they can adapt when the deepfakes are emerging. Unfortunately, most datasets available to the public are biased and subjected to domain shift, being largely controlled environments, that is, celebrity faces, making them ill-suited for the real world, where deepfakes attack regular people on the street under various circumstances. Furthermore, while existing benchmarks do not include cross-modal manipulations, adversarial perturbations, or real-time deepfake attacks, they cannot test the robustness of zero-shot learning models \cite{Coccomini_2022}. A lack of a fundamental, universal benchmark informed by the fast evolution of deepfake technology creates a significant open problem in the foundation of comparison between different options for detecting deepfakes in zero-shot settings, which motivates the need for open and adversarially robust deepfake datasets with large variation capable of distinguishing different detection models \cite{Yoon_2024}.

\section{Future Research Directions} \label{sec6}
\begin{figure*}[t!]
    \centering
    \includegraphics[width=\textwidth]{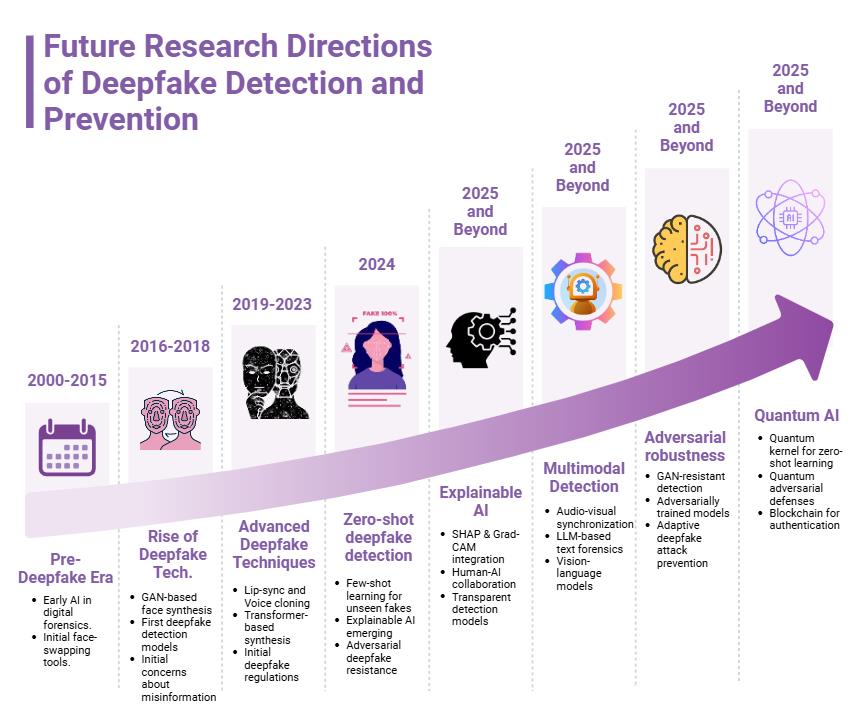}
    \caption{This diagram presents a timeline of future research directions in deepfake detection and prevention, highlighting key advancements from the pre-deepfake era to projected developments beyond 2025. It illustrates the evolution from early AI in digital forensics to the rise of deepfake technology, followed by advancements in detection techniques, the emergence of zero-shot deepfake detection, and the focus on explainable AI. The diagram anticipates significant strides in multimodal detection, adversarial robustness, and the integration of quantum AI for enhanced security and detection capabilities.}
    \label{fig:fd}
\end{figure*}

As deepfake technology evolves, zero-shot deepfake detection and prevention strategies must advance to counter increasingly sophisticated synthetic content. While current zero-shot approaches offer promising adaptability and robustness, several key challenges remain regarding generalization, adversarial resilience, real-time efficiency, and ethical considerations. Fig. \ref{fig:fd} outlined four major future research directions that could enhance the efficacy and scalability of zero-shot deepfake detection and prevention systems.

\subsection{Explainable AI (XAI) for Deepfake Detection}
One of the primary concerns with deepfake detection models, which are particularly based on deep neural networks (DNNs) and transformers, is their black-box nature. Current models lack interpretability, making understanding why a specific video or image is classified as a deepfake. The key research challenges are

\begin{enumerate}[wide, labelwidth=!, labelindent=0pt]
    \item Existing models provided binary outputs (real or fake) without explaining the decision.
    \item The lack of transparency hinders item Trust and adoption of AI-based detection systems.
    \item Ethical concerns arose when decisions impacted media authenticity in legal or journalistic contexts.
\end{enumerate}

Some promising future directions could be utilizing methods like SHAP (Shapley Additive Explanations), LIME (Local Interpretable Model-Agnostic Explanations), and Grad-CAM (Gradient-weighted Class Activation Mapping) for visualizing the specific regions or patterns in an image or video that led to its classification as a deepfake. The human-AI collaboration developed hybrid systems where explainable AI assisted fact-checkers, forensic analysts, and policymakers in properly understanding deepfake manipulations, which could be another direction. The interpretable zero-shot learning models incorporated prototype-based learning, attention mechanisms, or symbolic reasoning to make detection decisions more interpretable.

\subsection{Multimodal Deepfake Detection: Integrating Image, Audio, and Text}
Most current deepfake detection models focus only on visual features, which neglect audio and textual cues that could improve detection accuracy. Advanced deepfake attacks now manipulated facial expressions, lip synchronization, and speech synthesis, making unimodal detection approaches ineffective. The key research challenges here consisted of:

\begin{enumerate}[wide, labelwidth=!, labelindent=0pt]
    \item Visual-only detection models failed against audio-visual deepfakes (e.g., manipulated voice synchronization in deepfake videos).
    \item Text-based misinformation combined with deepfakes increased the risk of synthetic propaganda.
\end{enumerate}

Here, future research could focus on audio-visual synchronization analysis, where inconsistencies in lip movements and speech patterns are found using wav2vec, Whisper, or Mel-spectrogram-based CNNs. The text-driven deepfake detection, which employed Large Language Models (LLMs) for analyzing metadata, captions, and transcripts for inconsistencies in deepfake videos, could also be a promising future direction. Along with these, fusion models for multimodal analysis could be developed with transformer-based architectures like Vision-Language Models (VLMs) for jointly processing image, text, and audio cues for holistic deepfake detection. 

\subsection{Adversarial Robustness: Defending Against Deepfake Evasion Tactics}
Deepfake generators continuously evolve, incorporating adversarial learning techniques for bypassing detection models. Attackers could introduce subtle noise, adversarial perturbations, or GAN-based camouflage techniques to fool AI-based detection systems. The key research challenges consisted of:

\begin{enumerate}[wide, labelwidth=!, labelindent=0pt]
    \item Current detection models struggled against adversarially enhanced deepfakes.
    \item Deepfake generators could use adversarial training to produce "undetectable" fakes.
    \item The lack of standardized benchmark adversarial deepfake datasets limited research progress.
\end{enumerate}

Here, the key future research focus could be on adversarially trained zero-shot detectors to enhance these with adversarial robustness to detect stealthy deepfake variations. Developing and using defensive GANs to actively counter deepfake generation techniques could also be an option. The robust benchmarking frameworks could create adversarial deepfake datasets with perturbed attacks for rigorous testing of deepfake detection models.

\subsection{Quantum AI for Deepfake Detection and Prevention}
Integrating quantum computing with deep learning offers a revolutionary approach to pattern recognition and anomaly detection in deepfake forensics. Quantum algorithms could efficiently process high-dimensional data and detect subtle artifacts in synthetic content that classical AI models missed. The key challenges noticed here are Quantum AI, which is still in its early stages, and practical quantum computing hardware, not yet widely available. Also, quantum-based models required novel training paradigms and quantum-ready datasets. From this perspective, future research could focus on quantum-assisted zero-shot learning, where quantum kernel methods and hybrid quantum-classical neural networks could be used to enhance zero-shot generalization for unseen deepfake types. Quantum adversarial learning explored quantum-resistant defenses to protect deepfake detection models from adversarial attacks. Also, quantum blockchain could be another way to prevent deepfakes, where quantum-secure cryptographic techniques could enhance blockchain-based authentication.

\section{Conclusion} \label{sec7}
The growing sophistication of deepfake tech is quite an intriguing double-edged situation in the digital realm. On the other hand, deepfakes are a powerful means in fields such as entertainment, accessibility, and creative media for real-life simulations and digital enhancements. On the other hand, however, there are serious questions regarding the integrity of digital information due to its misuse for malicious purposes, such as misinformation campaigns and political propaganda, identity fraud, and financial scams. Traditional defection mechanisms lag behind as deepfake generation models evolve, often falling victim to the new, more sophisticated synthetic media. As far as the model, zero-shot deepfake detection has become a revolutionary proposal for the identification of manipulated content completely without the need to have previously seen the techniques of manipulation, and thus adaptable to the constant evolution of the deepfake.

Finally, this research has deeply covered a technological approach to zero-shot deepfake detection, detection strategies, protection strategies, issues, and prospects. This concludes that deepfake detection models have greatly benefited from developing self-supervised learning, contrastive learning, and transformer-based architectures. In contrast to conventional detection approaches that need high-scale annotated datasets for training, zero-shot methods detect deepfakes with latent representations, anomaly detection techniques, and domain invariant features in a more flexible, scalable fashion. Because of this adaptability, it is particularly useful in real-world cases where the manipulation techniques do not exist in the present dataset and continuously evolve, leading to the obsoleteness of dataset-based approaches.

Regarding prevention strategies for deepfake-generated misinformation, we have proposed many of the most cutting-edge solutions to curb its spread. However, blockchain-based authentication systems have also ensured media integrity by providing verifiable cryptographic proof of originality. Like watermarking techniques, AI-generated watermarking techniques can introduce undetectable signatures in multimedia content for forensic tracking of synthetic media. It also introduces federated learning architectures into privacy by making collaborative deepfake detection possible without centralized data storage, addressing data security concerns, and making models more robust. While advancements in such prevention mechanisms can cut down on the deepfake threat, prevention could never be so comprehensive—if it had to do that, there would be no need for detection algorithms in the first place.

So far, the continuous fight between forensic AI models and adversarial deepfake generators is one of the most pressing issues in zero-shot deepfake detection. These deepfake generation models now introduce perturbations to trick forensic networks and avoid detection. This challenge highlights the need to design zero-shot detection frameworks that operate in the presence of adversarial inputs and unseen data distributions. Furthermore, interpretability remains a serious issue, as most deep learning-based detection models remain black boxes which impedes forensic analysts and policymakers from verifying the credibility of deepfakes information. Since deepfake detection systems can be susceptible to misuse in the legal and judicial fields where trust is key, improving the model transparency by applying explainable AI (XAI) techniques is highly relevant.

Several transformations in the research areas are expected to mold the future direction of zero-shot deepfake detection. The ability to integrate multimodal learning in which audio, video, textual, and behavioral cues are combined to improve detection accuracy has the potential to improve detection accuracy by exploiting cross-modality inconsistencies. As with deepfake forensics, using quantum computing could provide previously unattainable computational power through which AI models could more efficiently process high-dimensional deepfake data. At the same time, federated learning supports decentralized deepfake detection across various devices without compromising user privacy. Federated models can become more resilient to adversarial attacks that break the principle of ethical AI while localizing the learning process.

However, the arms race between the deepfake generators and the detection models continues to exist, thus requiring continuous research and innovation. It is necessary to form global AI governance rules and ethical regulatory standards to minimize the chances of deepfake detection failing in the long term. So, governments, research institutions, and technology companies should think clearly and define concrete guidelines for the ethical creation and dissemination of such AI-generated content. Public awareness campaigns and educational initiatives should be conducted to improve the ability of individuals to spot and critically analyze deepfake media.

Finally, we conclude that zero-shot deepfake detection is a paradigm change in the war against AI-created misinformation. This approach mitigates the deepfake crisis in two ways: without the need for ‘tons of labeled datasets’ and with adaptability against unknown deepfake manipulations. While AI research has made great advances, technical, ethical, and legal issues still need to be solved through interdisciplinary work – opening a gap between AI research, cybersecurity, and the work of policymakers. With the proliferation of synthetic content drawing ever closer to the realm where it can not be told apart from real-world content, the importance of online media that provides the necessary credibility, fidelity, and security has never been higher. As AI develops more and more AI-driven forensic techniques, strong ethical frameworks and public debates will all be crucial to maintaining truth, transparency, and trust in the era of generative AI.


\bibliographystyle{IEEEtran}
\bibliography{ieee}

\begin{thebibliography}{100}
\providecommand{\url}[1]{#1}
\csname url@samestyle\endcsname
\providecommand{\newblock}{\relax}
\providecommand{\bibinfo}[2]{#2}
\providecommand{\BIBentrySTDinterwordspacing}{\spaceskip=0pt\relax}
\providecommand{\BIBentryALTinterwordstretchfactor}{4}
\providecommand{\BIBentryALTinterwordspacing}{\spaceskip=\fontdimen2\font plus
\BIBentryALTinterwordstretchfactor\fontdimen3\font minus \fontdimen4\font\relax}
\providecommand{\BIBforeignlanguage}[2]{{%
\expandafter\ifx\csname l@#1\endcsname\relax
\typeout{** WARNING: IEEEtran.bst: No hyphenation pattern has been}%
\typeout{** loaded for the language `#1'. Using the pattern for}%
\typeout{** the default language instead.}%
\else
\language=\csname l@#1\endcsname
\fi
#2}}
\providecommand{\BIBdecl}{\relax}
\BIBdecl

\bibitem{Mishra_2024}
\BIBentryALTinterwordspacing
A.~Mishra, A.~Bharwaj, A.~K. Yadav, K.~Batra, and N.~Mishra, ``Deepfakes - generating synthetic images, and detecting artificially generated fake visuals using deep learning,'' in \emph{2024 14th International Conference on Cloud Computing, Data Science \& Engineering (Confluence)}.\hskip 1em plus 0.5em minus 0.4em\relax IEEE, Jan. 2024, p. 587–592. [Online]. Available: \url{http://dx.doi.org/10.1109/confluence60223.2024.10463337}
\BIBentrySTDinterwordspacing

\bibitem{Sharma_2024}
\BIBentryALTinterwordspacing
P.~Sharma, M.~Kumar, and H.~K. Sharma, ``Robust gan-based cnn model as generative ai application for deepfake detection,'' \emph{EAI Endorsed Transactions on Internet of Things}, vol.~10, Apr. 2024. [Online]. Available: \url{http://dx.doi.org/10.4108/eetiot.5637}
\BIBentrySTDinterwordspacing

\bibitem{Dagar_2022}
\BIBentryALTinterwordspacing
D.~Dagar and D.~K. Vishwakarma, ``A literature review and perspectives in deepfakes: generation, detection, and applications,'' \emph{International Journal of Multimedia Information Retrieval}, vol.~11, no.~3, p. 219–289, Jul. 2022. [Online]. Available: \url{http://dx.doi.org/10.1007/s13735-022-00241-w}
\BIBentrySTDinterwordspacing

\bibitem{Chawki_2024}
\BIBentryALTinterwordspacing
M.~Chawki, ``Navigating legal challenges of deepfakes in the american context: a call to action,'' \emph{Cogent Engineering}, vol.~11, no.~1, Feb. 2024. [Online]. Available: \url{http://dx.doi.org/10.1080/23311916.2024.2320971}
\BIBentrySTDinterwordspacing

\bibitem{Rathi_2024}
\BIBentryALTinterwordspacing
P.~Rathi, S.~K. Budhani, G.~Murari~Upadhyay, P.~Vats, R.~Kaur, and A.~K. Saini, ``Unmasking deepfakes: Understanding the technology, risks, and countermeasures,'' in \emph{2024 4th International Conference on Innovative Practices in Technology and Management (ICIPTM)}.\hskip 1em plus 0.5em minus 0.4em\relax IEEE, Feb. 2024, p. 1–6. [Online]. Available: \url{http://dx.doi.org/10.1109/iciptm59628.2024.10563353}
\BIBentrySTDinterwordspacing

\bibitem{Tchaptchet_2025}
\BIBentryALTinterwordspacing
E.~Tchaptchet, E.~Fute~Tagne, J.~Acosta, D.~B. Rawat, and C.~Kamhoua, ``Deepfakes detection by iris analysis,'' \emph{IEEE Access}, vol.~13, p. 8977–8987, 2025. [Online]. Available: \url{http://dx.doi.org/10.1109/access.2025.3527868}
\BIBentrySTDinterwordspacing

\bibitem{AL_KHAZRAJI_2023}
\BIBentryALTinterwordspacing
S.~H. AL-KHAZRAJI, H.~H. SALEH, A.~I. KHALID, and I.~A. MISHKHAL, ``Impact of deepfake technology on social media: Detection, misinformation and societal implications,'' \emph{The Eurasia Proceedings of Science Technology Engineering and Mathematics}, vol.~23, p. 429–441, Oct. 2023. [Online]. Available: \url{http://dx.doi.org/10.55549/epstem.1371792}
\BIBentrySTDinterwordspacing

\bibitem{Maniyal_2024}
\BIBentryALTinterwordspacing
V.~Maniyal and V.~Kumar, ``Unveiling the deepfake dilemma: Framework, classification, and future trajectories,'' \emph{IT Professional}, vol.~26, no.~2, p. 32–38, Mar. 2024. [Online]. Available: \url{http://dx.doi.org/10.1109/mitp.2024.3369948}
\BIBentrySTDinterwordspacing

\bibitem{Samoilenko_2023}
\BIBentryALTinterwordspacing
S.~A. Samoilenko and I.~Suvorova, \emph{Artificial Intelligence and Deepfakes in Strategic Deception Campaigns: The U.S. and Russian Experiences}.\hskip 1em plus 0.5em minus 0.4em\relax Springer International Publishing, 2023, p. 507–529. [Online]. Available: \url{http://dx.doi.org/10.1007/978-3-031-22552-9_19}
\BIBentrySTDinterwordspacing

\bibitem{Gamb_n_2024}
\BIBentryALTinterwordspacing
Ã.~F. Gambín, A.~Yazidi, A.~Vasilakos, H.~Haugerud, and Y.~Djenouri, ``Deepfakes: current and future trends,'' \emph{Artificial Intelligence Review}, vol.~57, no.~3, Feb. 2024. [Online]. Available: \url{http://dx.doi.org/10.1007/s10462-023-10679-x}
\BIBentrySTDinterwordspacing

\bibitem{Barrientos_B_ez_2024}
\BIBentryALTinterwordspacing
A.~Barrientos-Báez, M.~T. Piñeiro~Otero, and D.~Porto~Renó, ``Imágenes falsas, efectos reales. deepfakes como manifestaciones de la violencia política de género,'' \emph{Revista Latina de Comunicación Social}, no.~82, p. 1–30, May 2024. [Online]. Available: \url{http://dx.doi.org/10.4185/rlcs-2024-2278}
\BIBentrySTDinterwordspacing

\bibitem{Nagarhalli_2024}
\BIBentryALTinterwordspacing
T.~P. Nagarhalli, A.~Save, S.~Patil, and U.~Aswalekar, ``A comprehensive review of deepfake and its detection techniques,'' \emph{International Journal of Electrical and Electronics Engineering}, vol.~11, no.~8, p. 121–133, Aug. 2024. [Online]. Available: \url{http://dx.doi.org/10.14445/23488379/ijeee-v11i8p111}
\BIBentrySTDinterwordspacing

\bibitem{Babaei_2025}
\BIBentryALTinterwordspacing
R.~Babaei, S.~Cheng, R.~Duan, and S.~Zhao, ``Generative artificial intelligence and the evolving challenge of deepfake detection: A systematic analysis,'' \emph{Journal of Sensor and Actuator Networks}, vol.~14, no.~1, p.~17, Feb. 2025. [Online]. Available: \url{http://dx.doi.org/10.3390/jsan14010017}
\BIBentrySTDinterwordspacing

\bibitem{Ramadhani_2020}
\BIBentryALTinterwordspacing
K.~N. Ramadhani and R.~Munir, ``A comparative study of deepfake video detection method,'' in \emph{2020 3rd International Conference on Information and Communications Technology (ICOIACT)}.\hskip 1em plus 0.5em minus 0.4em\relax IEEE, Nov. 2020, p. 394–399. [Online]. Available: \url{http://dx.doi.org/10.1109/icoiact50329.2020.9331963}
\BIBentrySTDinterwordspacing

\bibitem{Rana_2024}
\BIBentryALTinterwordspacing
P.~Rana and S.~Bansal, ``Exploring deepfake detection: Techniques, datasets and challenges,'' \emph{International Journal of Computing and Digital Systems}, vol.~15, no.~1, p. 769–781, Aug. 2024. [Online]. Available: \url{http://dx.doi.org/10.12785/ijcds/160156}
\BIBentrySTDinterwordspacing

\bibitem{key}
``{D}eep{F}ake {A}{I} {M}arket --- market.us,'' \url{https://market.us/report/deepfake-ai-market/}, [Accessed 11-03-2025].

\bibitem{Le_2023}
\BIBentryALTinterwordspacing
B.~Le, S.~Tariq, A.~Abuadbba, K.~Moore, and S.~Woo, ``Why do facial deepfake detectors fail?'' in \emph{The 2nd Workshop on the security implications of Deepfakes and Cheapfakes}, ser. ASIA CCS '23.\hskip 1em plus 0.5em minus 0.4em\relax ACM, Jul. 2023, p. 24–28. [Online]. Available: \url{http://dx.doi.org/10.1145/3595353.3595882}
\BIBentrySTDinterwordspacing

\bibitem{Zhalgasbayev_2024}
\BIBentryALTinterwordspacing
A.~Zhalgasbayev, T.~Aiteni, and N.~Khaimuldin, ``Using the cnn architecture based on the efficientnetb4 model to efficiently detect deepfake images,'' in \emph{2024 IEEE AITU: Digital Generation}.\hskip 1em plus 0.5em minus 0.4em\relax IEEE, Apr. 2024, p. 14–19. [Online]. Available: \url{http://dx.doi.org/10.1109/ieeeconf61558.2024.10585385}
\BIBentrySTDinterwordspacing

\bibitem{Kaur_2024}
\BIBentryALTinterwordspacing
A.~Kaur, A.~Noori~Hoshyar, V.~Saikrishna, S.~Firmin, and F.~Xia, ``Deepfake video detection: challenges and opportunities,'' \emph{Artificial Intelligence Review}, vol.~57, no.~6, May 2024. [Online]. Available: \url{http://dx.doi.org/10.1007/s10462-024-10810-6}
\BIBentrySTDinterwordspacing

\bibitem{Chamot_2022}
\BIBentryALTinterwordspacing
F.~Chamot, Z.~Geradts, and E.~Haasdijk, ``Deepfake forensics: Cross-manipulation robustness of feedforward- and recurrent convolutional forgery detection methods,'' \emph{Forensic Science International: Digital Investigation}, vol.~40, p. 301374, Mar. 2022. [Online]. Available: \url{http://dx.doi.org/10.1016/j.fsidi.2022.301374}
\BIBentrySTDinterwordspacing

\bibitem{Pang_2023}
\BIBentryALTinterwordspacing
C.~Pang, Y.~Pan, and H.~Bai, ``Metamorphic testing for the deepfake detection model,'' in \emph{2023 IEEE 23rd International Conference on Software Quality, Reliability, and Security Companion (QRS-C)}.\hskip 1em plus 0.5em minus 0.4em\relax IEEE, Oct. 2023, p. 1–8. [Online]. Available: \url{http://dx.doi.org/10.1109/qrs-c60940.2023.00012}
\BIBentrySTDinterwordspacing

\bibitem{10924158}
A.~Sar, S.~Sati, T.~Choudhury, P.~Joshi, R.~Sille, K.~Srihari, and K.~Bansal, ``A unified neural framework for real-time deepfake detection across multimedia modalities to combat misleading content,'' \emph{IEEE Access}, vol.~13, pp. 48\,683--48\,702, 2025.

\bibitem{Khormali_2024}
\BIBentryALTinterwordspacing
A.~Khormali and J.-S. Yuan, ``Self-supervised graph transformer for deepfake detection,'' \emph{IEEE Access}, vol.~12, p. 58114–58127, 2024. [Online]. Available: \url{http://dx.doi.org/10.1109/access.2024.3392512}
\BIBentrySTDinterwordspacing

\bibitem{Atamna_2023}
\BIBentryALTinterwordspacing
M.~Atamna, I.~Tkachenko, and S.~Miguet, ``Improving generalization in facial manipulation detection using image noise residuals and temporal features,'' in \emph{2023 IEEE International Conference on Image Processing (ICIP)}.\hskip 1em plus 0.5em minus 0.4em\relax IEEE, Oct. 2023, p. 3424–3428. [Online]. Available: \url{http://dx.doi.org/10.1109/icip49359.2023.10222043}
\BIBentrySTDinterwordspacing

\bibitem{Agarwal_2024}
\BIBentryALTinterwordspacing
A.~Agarwal and N.~Ratha, ``Deepfake: Classifiers, fairness, and demographically robust algorithm,'' in \emph{2024 IEEE 18th International Conference on Automatic Face and Gesture Recognition (FG)}.\hskip 1em plus 0.5em minus 0.4em\relax IEEE, May 2024, p. 1–9. [Online]. Available: \url{http://dx.doi.org/10.1109/fg59268.2024.10581915}
\BIBentrySTDinterwordspacing

\bibitem{Stanciu_2024}
\BIBentryALTinterwordspacing
D.-C. Stanciu and B.~Ionescu, ``Improving generalization in deepfake detection via augmentation with recurrent adversarial attacks,'' in \emph{3rd ACM International Workshop on Multimedia AI against Disinformation}, ser. ICMR '24.\hskip 1em plus 0.5em minus 0.4em\relax ACM, Jun. 2024, p. 46–54. [Online]. Available: \url{http://dx.doi.org/10.1145/3643491.3660291}
\BIBentrySTDinterwordspacing

\bibitem{Li_2023}
\BIBentryALTinterwordspacing
M.~Li, X.~Li, K.~Yu, C.~Deng, H.~Huang, F.~Mao, H.~Xue, and M.~Li, ``Spatio-temporal catcher: A self-supervised transformer for deepfake video detection,'' in \emph{Proceedings of the 31st ACM International Conference on Multimedia}, ser. MM '23.\hskip 1em plus 0.5em minus 0.4em\relax ACM, Oct. 2023, p. 8707–8718. [Online]. Available: \url{http://dx.doi.org/10.1145/3581783.3613842}
\BIBentrySTDinterwordspacing

\bibitem{Karathanasis_2025}
\BIBentryALTinterwordspacing
A.~Karathanasis, J.~Violos, and I.~Kompatsiaris, ``A comparative analysis of compression and transfer learning techniques in deepfake detection models,'' \emph{Mathematics}, vol.~13, no.~5, p. 887, Mar. 2025. [Online]. Available: \url{http://dx.doi.org/10.3390/math13050887}
\BIBentrySTDinterwordspacing

\bibitem{Mohzary_2023}
\BIBentryALTinterwordspacing
M.~Mohzary, K.~J. Almalki, B.-Y. Choi, and S.~Song, ``Mobideep: Mobile deepfake detection through machine learning-based corneal-specular backscattering,'' in \emph{2023 IEEE 20th Consumer Communications \& Networking Conference (CCNC)}.\hskip 1em plus 0.5em minus 0.4em\relax IEEE, Jan. 2023, p. 1104–1109. [Online]. Available: \url{http://dx.doi.org/10.1109/ccnc51644.2023.10059841}
\BIBentrySTDinterwordspacing

\bibitem{Sridevi_2022}
\BIBentryALTinterwordspacing
K.~Sridevi, K.~S. Kumar, D.~Sameera, Y.~Garapati, D.~Krishnamadhuri, and S.~Bethu, ``Iot based application designing of deep fake test for face animation,'' in \emph{Proceedings of the 2022 6th International Conference on Cloud and Big Data Computing}, ser. ICCBDC 2022.\hskip 1em plus 0.5em minus 0.4em\relax ACM, Aug. 2022, p. 24–30. [Online]. Available: \url{http://dx.doi.org/10.1145/3555962.3555967}
\BIBentrySTDinterwordspacing

\bibitem{Gandhi_2020}
\BIBentryALTinterwordspacing
A.~Gandhi and S.~Jain, ``Adversarial perturbations fool deepfake detectors,'' in \emph{2020 International Joint Conference on Neural Networks (IJCNN)}.\hskip 1em plus 0.5em minus 0.4em\relax IEEE, Jul. 2020. [Online]. Available: \url{http://dx.doi.org/10.1109/ijcnn48605.2020.9207034}
\BIBentrySTDinterwordspacing

\bibitem{Ma_2019}
\BIBentryALTinterwordspacing
C.~Ma, C.~Zhao, H.~Shi, L.~Chen, J.~Yong, and D.~Zeng, ``Metaadvdet: Towards robust detection of evolving adversarial attacks,'' in \emph{Proceedings of the 27th ACM International Conference on Multimedia}, ser. MM '19.\hskip 1em plus 0.5em minus 0.4em\relax ACM, Oct. 2019, p. 692–701. [Online]. Available: \url{http://dx.doi.org/10.1145/3343031.3350887}
\BIBentrySTDinterwordspacing

\bibitem{Bagaria_2024}
\BIBentryALTinterwordspacing
U.~Bagaria, V.~Kumar, T.~Rajesh, V.~Deepak, and S.~S~S, ``Disrupting deepfakes: A survey on adversarial perturbation techniques and prevention strategies,'' in \emph{Proceedings of the 2024 10th International Conference on Computing and Artificial Intelligence}, ser. ICCAI 2024.\hskip 1em plus 0.5em minus 0.4em\relax ACM, Apr. 2024, p. 301–306. [Online]. Available: \url{http://dx.doi.org/10.1145/3669754.3669799}
\BIBentrySTDinterwordspacing

\bibitem{Guo_2023}
\BIBentryALTinterwordspacing
J.~Guo, S.~Guo, Q.~Zhou, Z.~Liu, X.~Lu, and F.~Huo, ``Graph knows unknowns: Reformulate zero-shot learning as sample-level graph recognition,'' \emph{Proceedings of the AAAI Conference on Artificial Intelligence}, vol.~37, no.~6, p. 7775–7783, Jun. 2023. [Online]. Available: \url{http://dx.doi.org/10.1609/aaai.v37i6.25942}
\BIBentrySTDinterwordspacing

\bibitem{Chen_2021}
\BIBentryALTinterwordspacing
J.~Chen, Y.~Geng, Z.~Chen, I.~Horrocks, J.~Z.~Pan, and H.~Chen, ``Knowledge-aware zero-shot learning: Survey and perspective,'' in \emph{Proceedings of the Thirtieth International Joint Conference on Artificial Intelligence}, ser. IJCAI-2021.\hskip 1em plus 0.5em minus 0.4em\relax International Joint Conferences on Artificial Intelligence Organization, Aug. 2021, p. 4366–4373. [Online]. Available: \url{http://dx.doi.org/10.24963/ijcai.2021/597}
\BIBentrySTDinterwordspacing

\bibitem{Verma_2024}
\BIBentryALTinterwordspacing
V.~Verma, N.~Mehta, K.~J. Liang, A.~Mishra, and L.~Carin, ``Meta-learned attribute self-interaction network for continual and generalized zero-shot learning,'' in \emph{2024 IEEE/CVF Winter Conference on Applications of Computer Vision (WACV)}.\hskip 1em plus 0.5em minus 0.4em\relax IEEE, Jan. 2024, p. 2709–2719. [Online]. Available: \url{http://dx.doi.org/10.1109/wacv57701.2024.00270}
\BIBentrySTDinterwordspacing

\bibitem{guo2018zero}
Y.~Guo, G.~Ding, J.~Han, and S.~Tang, ``Zero-shot learning with attribute selection,'' in \emph{Proceedings of the AAAI Conference on Artificial Intelligence}, vol.~32, no.~1, 2018.

\bibitem{Cao_2023}
\BIBentryALTinterwordspacing
W.~Cao, Y.~Wu, Y.~Sun, H.~Zhang, J.~Ren, D.~Gu, and X.~Wang, ``A review on multimodal zero‐shot learning,'' \emph{WIREs Data Mining and Knowledge Discovery}, vol.~13, no.~2, Jan. 2023. [Online]. Available: \url{http://dx.doi.org/10.1002/widm.1488}
\BIBentrySTDinterwordspacing

\bibitem{Parida_2020}
\BIBentryALTinterwordspacing
K.~K. Parida, N.~Matiyali, T.~Guha, and G.~Sharma, ``Coordinated joint multimodal embeddings for generalized audio-visual zero-shot classification and retrieval of videos,'' in \emph{2020 IEEE Winter Conference on Applications of Computer Vision (WACV)}.\hskip 1em plus 0.5em minus 0.4em\relax IEEE, Mar. 2020, p. 3240–3249. [Online]. Available: \url{http://dx.doi.org/10.1109/wacv45572.2020.9093438}
\BIBentrySTDinterwordspacing

\bibitem{Hong_2023}
\BIBentryALTinterwordspacing
J.~Hong, Z.~Hayder, J.~Han, P.~Fang, M.~Harandi, and L.~Petersson, ``Hyperbolic audio-visual zero-shot learning,'' in \emph{2023 IEEE/CVF International Conference on Computer Vision (ICCV)}.\hskip 1em plus 0.5em minus 0.4em\relax IEEE, Oct. 2023, p. 7839–7849. [Online]. Available: \url{http://dx.doi.org/10.1109/iccv51070.2023.00724}
\BIBentrySTDinterwordspacing

\bibitem{Khare_2020}
\BIBentryALTinterwordspacing
V.~Khare, D.~Mahajan, H.~Bharadhwaj, V.~K. Verma, and P.~Rai, ``A generative framework for zero-shot learning with adversarial domain adaptation,'' in \emph{2020 IEEE Winter Conference on Applications of Computer Vision (WACV)}.\hskip 1em plus 0.5em minus 0.4em\relax IEEE, Mar. 2020, p. 3090–3099. [Online]. Available: \url{http://dx.doi.org/10.1109/wacv45572.2020.9093348}
\BIBentrySTDinterwordspacing

\bibitem{NIPS2014_f033ed80}
\BIBentryALTinterwordspacing
I.~J. Goodfellow, J.~Pouget-Abadie, M.~Mirza, B.~Xu, D.~Warde-Farley, S.~Ozair, A.~Courville, and Y.~Bengio, ``Generative adversarial nets,'' in \emph{Advances in Neural Information Processing Systems}, Z.~Ghahramani, M.~Welling, C.~Cortes, N.~Lawrence, and K.~Weinberger, Eds., vol.~27.\hskip 1em plus 0.5em minus 0.4em\relax Curran Associates, Inc., 2014. [Online]. Available: \url{https://proceedings.neurips.cc/paper_files/paper/2014/file/f033ed80deb0234979a61f95710dbe25-Paper.pdf}
\BIBentrySTDinterwordspacing

\bibitem{Trevisan_de_Souza_2023}
\BIBentryALTinterwordspacing
V.~L. Trevisan~de Souza, B.~A.~D. Marques, H.~C. Batagelo, and J.~P. Gois, ``A review on generative adversarial networks for image generation,'' \emph{Computers \& Graphics}, vol. 114, p. 13–25, Aug. 2023. [Online]. Available: \url{http://dx.doi.org/10.1016/j.cag.2023.05.010}
\BIBentrySTDinterwordspacing

\bibitem{bhat2025review}
R.~Bhat and R.~Nanjundegowda, ``A review on comparative analysis of generative adversarial networks' architectures and applications,'' \emph{Journal of Robotics and Control (JRC)}, vol.~6, no.~1, pp. 53--64, 2025.

\bibitem{J_2024}
\BIBentryALTinterwordspacing
B.~J, J.~R, T.~Richard, R.~Chowdhury, P.~Kumar, and K.~Radhika, ``Evaluating generative adversarial networks performance in image synthesis with graphical analysis of loss functions and quality metrics,'' in \emph{2024 International Conference on Integrated Intelligence and Communication Systems (ICIICS)}.\hskip 1em plus 0.5em minus 0.4em\relax IEEE, Nov. 2024, p. 1–8. [Online]. Available: \url{http://dx.doi.org/10.1109/iciics63763.2024.10859959}
\BIBentrySTDinterwordspacing

\bibitem{Cao_2019}
\BIBentryALTinterwordspacing
Y.-J. Cao, L.-L. Jia, Y.-X. Chen, N.~Lin, C.~Yang, B.~Zhang, Z.~Liu, X.-X. Li, and H.-H. Dai, ``Recent advances of generative adversarial networks in computer vision,'' \emph{IEEE Access}, vol.~7, p. 14985–15006, 2019. [Online]. Available: \url{http://dx.doi.org/10.1109/access.2018.2886814}
\BIBentrySTDinterwordspacing

\bibitem{Aarti_2021}
\BIBentryALTinterwordspacing
E.~Aarti, \emph{Generative adversarial networks and their variants}.\hskip 1em plus 0.5em minus 0.4em\relax Elsevier, 2021, p. 59–80. [Online]. Available: \url{http://dx.doi.org/10.1016/b978-0-12-823519-5.00003-8}
\BIBentrySTDinterwordspacing

\bibitem{Purwins_2019}
\BIBentryALTinterwordspacing
H.~Purwins, B.~Sturm, B.~Li, J.~Nam, and A.~Alwan, ``Introduction to the issue on data science: Machine learning for audio signal processing,'' \emph{IEEE Journal of Selected Topics in Signal Processing}, vol.~13, no.~2, p. 203–205, May 2019. [Online]. Available: \url{http://dx.doi.org/10.1109/jstsp.2019.2914321}
\BIBentrySTDinterwordspacing

\bibitem{Kang_2020}
\BIBentryALTinterwordspacing
S.~Kang, D.~Han, J.~Lee, D.~Im, S.~Kim, S.~Kim, and H.-J. Yoo, ``7.4 ganpu: A 135tflops/w multi-dnn training processor for gans with speculative dual-sparsity exploitation,'' in \emph{2020 IEEE International Solid- State Circuits Conference - (ISSCC)}.\hskip 1em plus 0.5em minus 0.4em\relax IEEE, Feb. 2020. [Online]. Available: \url{http://dx.doi.org/10.1109/isscc19947.2020.9062989}
\BIBentrySTDinterwordspacing

\bibitem{Prasanthi_2024}
\BIBentryALTinterwordspacing
S.~Prasanthi, M.~Rayavarapu, S.~Rao, and R.~Goswami, ``Investigation of different generative adversarial networks techniques for image restoration,'' \emph{International Journal of Computing and Digital Systems}, vol.~15, no.~1, p. 673–683, Aug. 2024. [Online]. Available: \url{http://dx.doi.org/10.12785/ijcds/160150}
\BIBentrySTDinterwordspacing

\bibitem{radford2016unsupervisedrepresentationlearningdeep}
\BIBentryALTinterwordspacing
A.~Radford, L.~Metz, and S.~Chintala, ``Unsupervised representation learning with deep convolutional generative adversarial networks,'' 2016. [Online]. Available: \url{https://arxiv.org/abs/1511.06434}
\BIBentrySTDinterwordspacing

\bibitem{karras2018progressivegrowinggansimproved}
\BIBentryALTinterwordspacing
T.~Karras, T.~Aila, S.~Laine, and J.~Lehtinen, ``Progressive growing of gans for improved quality, stability, and variation,'' 2018. [Online]. Available: \url{https://arxiv.org/abs/1710.10196}
\BIBentrySTDinterwordspacing

\bibitem{karras2019stylebasedgeneratorarchitecturegenerative}
\BIBentryALTinterwordspacing
T.~Karras, S.~Laine, and T.~Aila, ``A style-based generator architecture for generative adversarial networks,'' 2019. [Online]. Available: \url{https://arxiv.org/abs/1812.04948}
\BIBentrySTDinterwordspacing

\bibitem{karras2020analyzingimprovingimagequality}
\BIBentryALTinterwordspacing
T.~Karras, S.~Laine, M.~Aittala, J.~Hellsten, J.~Lehtinen, and T.~Aila, ``Analyzing and improving the image quality of stylegan,'' 2020. [Online]. Available: \url{https://arxiv.org/abs/1912.04958}
\BIBentrySTDinterwordspacing

\bibitem{Karras2021}
T.~Karras, M.~Aittala, S.~Laine, E.~H\"ark\"onen, J.~Hellsten, J.~Lehtinen, and T.~Aila, ``Alias-free generative adversarial networks,'' in \emph{Proc. NeurIPS}, 2021.

\bibitem{zhu2020unpairedimagetoimagetranslationusing}
\BIBentryALTinterwordspacing
J.-Y. Zhu, T.~Park, P.~Isola, and A.~A. Efros, ``Unpaired image-to-image translation using cycle-consistent adversarial networks,'' 2020. [Online]. Available: \url{https://arxiv.org/abs/1703.10593}
\BIBentrySTDinterwordspacing

\bibitem{choi2018starganunifiedgenerativeadversarial}
\BIBentryALTinterwordspacing
Y.~Choi, M.~Choi, M.~Kim, J.-W. Ha, S.~Kim, and J.~Choo, ``Stargan: Unified generative adversarial networks for multi-domain image-to-image translation,'' 2018. [Online]. Available: \url{https://arxiv.org/abs/1711.09020}
\BIBentrySTDinterwordspacing

\bibitem{Zhang_2025}
\BIBentryALTinterwordspacing
Q.~Zhang, ``Exploring the frontier: recent advances and evaluations in vision transformers,'' \emph{IET Conference Proceedings}, vol. 2024, no.~24, p. 184–188, Jan. 2025. [Online]. Available: \url{http://dx.doi.org/10.1049/icp.2024.4478}
\BIBentrySTDinterwordspacing

\bibitem{Arshed_2024}
\BIBentryALTinterwordspacing
M.~A. Arshed, S.~Mumtaz, M.~Ibrahim, C.~Dewi, M.~Tanveer, and S.~Ahmed, ``Multiclass ai-generated deepfake face detection using patch-wise deep learning model,'' \emph{Computers}, vol.~13, no.~1, p.~31, Jan. 2024. [Online]. Available: \url{http://dx.doi.org/10.3390/computers13010031}
\BIBentrySTDinterwordspacing

\bibitem{Fnu_2024}
\BIBentryALTinterwordspacing
N.~Fnu and A.~Bansal, ``Understanding the architecture of vision transformer and its variants: A review,'' in \emph{2024 1st International Conference on Innovative Engineering Sciences and Technological Research (ICIESTR)}.\hskip 1em plus 0.5em minus 0.4em\relax IEEE, May 2024, p. 1–6. [Online]. Available: \url{http://dx.doi.org/10.1109/iciestr60916.2024.10798341}
\BIBentrySTDinterwordspacing

\bibitem{vaswani2023attentionneed}
\BIBentryALTinterwordspacing
A.~Vaswani, N.~Shazeer, N.~Parmar, J.~Uszkoreit, L.~Jones, A.~N. Gomez, L.~Kaiser, and I.~Polosukhin, ``Attention is all you need,'' 2023. [Online]. Available: \url{https://arxiv.org/abs/1706.03762}
\BIBentrySTDinterwordspacing

\bibitem{Adriana_Mercioni_2024}
\BIBentryALTinterwordspacing
M.~Adriana~Mercioni and C.~Daniel~Căleanu, ``Computer aided diagnosis for contrast-enhanced ultrasound using a small hybrid transformer neural network,'' in \emph{2024 International Symposium on Electronics and Telecommunications (ISETC)}.\hskip 1em plus 0.5em minus 0.4em\relax IEEE, Nov. 2024, p. 1–4. [Online]. Available: \url{http://dx.doi.org/10.1109/isetc63109.2024.10797437}
\BIBentrySTDinterwordspacing

\bibitem{Khan_2022}
\BIBentryALTinterwordspacing
S.~Khan, M.~Naseer, M.~Hayat, S.~W. Zamir, F.~S. Khan, and M.~Shah, ``Transformers in vision: A survey,'' \emph{ACM Computing Surveys}, vol.~54, no. 10s, p. 1–41, Jan. 2022. [Online]. Available: \url{http://dx.doi.org/10.1145/3505244}
\BIBentrySTDinterwordspacing

\bibitem{Naik_2024}
\BIBentryALTinterwordspacing
D.~Naik, I.~Naik, and N.~Naik, \emph{Decoder-Only Transformers: The Brains Behind Generative AI, Large Language Models and Large Multimodal Models}.\hskip 1em plus 0.5em minus 0.4em\relax Springer Nature Switzerland, 2024, p. 315–331. [Online]. Available: \url{http://dx.doi.org/10.1007/978-3-031-74443-3_19}
\BIBentrySTDinterwordspacing

\bibitem{Sudha_2024}
\BIBentryALTinterwordspacing
L.~Sudha, K.~B. Aruna, V.~Sureka, M.~Niveditha, and S.~Prema, ``Semantic image synthesis from text: Current trends and future horizons in text-to-image generation,'' \emph{EAI Endorsed Transactions on Internet of Things}, vol.~11, Nov. 2024. [Online]. Available: \url{http://dx.doi.org/10.4108/eetiot.5336}
\BIBentrySTDinterwordspacing

\bibitem{Xu_2024}
\BIBentryALTinterwordspacing
S.~Xu, L.~Sun, G.~Liu, and Z.~Wei, \emph{Controlling Attention Map Better for Text-Guided Image Editing Diffusion Models}.\hskip 1em plus 0.5em minus 0.4em\relax Springer Nature Singapore, 2024, p. 54–65. [Online]. Available: \url{http://dx.doi.org/10.1007/978-981-97-5615-5_5}
\BIBentrySTDinterwordspacing

\bibitem{Saha_2024}
\BIBentryALTinterwordspacing
M.~Saha, P.~K. Pagadala, and M.~T. Basu, ``Cyber deceptions: Unveiling the threat of deepfakes in digital realms,'' in \emph{2024 IEEE 6th International Conference on Cybernetics, Cognition and Machine Learning Applications (ICCCMLA)}.\hskip 1em plus 0.5em minus 0.4em\relax IEEE, Oct. 2024, p. 113–116. [Online]. Available: \url{http://dx.doi.org/10.1109/icccmla63077.2024.10871909}
\BIBentrySTDinterwordspacing

\bibitem{Gonzales_2023}
\BIBentryALTinterwordspacing
N.~H. Gonzales, A.~Louis~Lobian, F.~M. Hengky, M.~Rendra~Andhika, S.~Achmad, and R.~Sutoyo, ``Deepfake technology: Negative impacts, mitigation methods, and preventive algorithms,'' in \emph{2023 IEEE 8th International Conference on Recent Advances and Innovations in Engineering (ICRAIE)}.\hskip 1em plus 0.5em minus 0.4em\relax IEEE, Dec. 2023, p. 1–5. [Online]. Available: \url{http://dx.doi.org/10.1109/icraie59459.2023.10468370}
\BIBentrySTDinterwordspacing

\bibitem{Qi_2024}
\BIBentryALTinterwordspacing
H.~Qi, Y.~Qiu, X.~Luo, and Z.~Jin, ``An efficient latent style guided transformer-cnn framework for face super-resolution,'' \emph{IEEE Transactions on Multimedia}, vol.~26, p. 1589–1599, 2024. [Online]. Available: \url{http://dx.doi.org/10.1109/tmm.2023.3283856}
\BIBentrySTDinterwordspacing

\bibitem{Liu_2022}
\BIBentryALTinterwordspacing
J.~Liu, P.~Chen, X.~Wang, X.~Fu, J.~Dai, and J.~Han, ``Critical review of human face reenactment methods,'' \emph{Journal of Image and Graphics}, vol.~27, no.~9, p. 2629–2651, 2022. [Online]. Available: \url{http://dx.doi.org/10.11834/jig.211243}
\BIBentrySTDinterwordspacing

\bibitem{Wu_2018}
\BIBentryALTinterwordspacing
W.~Wu, Y.~Zhang, C.~Li, C.~Qian, and C.~C. Loy, \emph{ReenactGAN: Learning to Reenact Faces via Boundary Transfer}.\hskip 1em plus 0.5em minus 0.4em\relax Springer International Publishing, 2018, p. 622–638. [Online]. Available: \url{http://dx.doi.org/10.1007/978-3-030-01246-5_37}
\BIBentrySTDinterwordspacing

\bibitem{Chapagain_2024}
\BIBentryALTinterwordspacing
D.~Chapagain, N.~Kshetri, and B.~Aryal, ``Deepfake disasters: A comprehensive review of technology, ethical concerns, countermeasures, and societal implications,'' in \emph{2024 International Conference on Emerging Trends in Networks and Computer Communications (ETNCC)}.\hskip 1em plus 0.5em minus 0.4em\relax IEEE, Jul. 2024, p. 1–9. [Online]. Available: \url{http://dx.doi.org/10.1109/etncc63262.2024.10767452}
\BIBentrySTDinterwordspacing

\bibitem{Xu_2023}
\BIBentryALTinterwordspacing
Y.~Xu, Z.~Yang, T.~Chen, K.~Li, and C.~Qing, ``Progressive transformer machine for natural character reenactment,'' \emph{ACM Transactions on Multimedia Computing, Communications, and Applications}, vol.~19, no.~2s, p. 1–22, Feb. 2023. [Online]. Available: \url{http://dx.doi.org/10.1145/3559107}
\BIBentrySTDinterwordspacing

\bibitem{Wang_2023}
\BIBentryALTinterwordspacing
L.~Wang, X.~Zhao, J.~Sun, Y.~Zhang, H.~Zhang, T.~Yu, and Y.~Liu, ``Styleavatar: Real-time photo-realistic portrait avatar from a single video,'' in \emph{Special Interest Group on Computer Graphics and Interactive Techniques Conference Conference Proceedings}, ser. SIGGRAPH '23.\hskip 1em plus 0.5em minus 0.4em\relax ACM, Jul. 2023, p. 1–10. [Online]. Available: \url{http://dx.doi.org/10.1145/3588432.3591517}
\BIBentrySTDinterwordspacing

\bibitem{Brey_2004}
\BIBentryALTinterwordspacing
P.~Brey, ``Ethical aspects of facial recognition systems in public places,'' \emph{Journal of Information, Communication and Ethics in Society}, vol.~2, no.~2, p. 97–109, May 2004. [Online]. Available: \url{http://dx.doi.org/10.1108/14779960480000246}
\BIBentrySTDinterwordspacing

\bibitem{Hu_2021}
\BIBentryALTinterwordspacing
J.~Hu, S.~Wang, and X.~Li, ``Improving the generalization ability of deepfake detection via disentangled representation learning,'' in \emph{2021 IEEE International Conference on Image Processing (ICIP)}.\hskip 1em plus 0.5em minus 0.4em\relax IEEE, Sep. 2021, p. 3577–3581. [Online]. Available: \url{http://dx.doi.org/10.1109/icip42928.2021.9506730}
\BIBentrySTDinterwordspacing

\bibitem{Rafiei_Oskooei_2024}
\BIBentryALTinterwordspacing
A.~Rafiei~Oskooei, E.~Yahsi, M.~Sungur, and M.~S.~Aktas, \emph{Can One Model Fit All? An Exploration of Wav2Lip's Lip-Syncing Generalizability Across Culturally Distinct Languages}.\hskip 1em plus 0.5em minus 0.4em\relax Springer Nature Switzerland, 2024, p. 149–164. [Online]. Available: \url{http://dx.doi.org/10.1007/978-3-031-65282-0_10}
\BIBentrySTDinterwordspacing

\bibitem{Chen_2024}
\BIBentryALTinterwordspacing
Z.~Chen, Z.~Ai, Y.~Ma, X.~Li, and S.~Xu, ``Optimizing feature fusion for improved zero-shot adaptation in text-to-speech synthesis,'' \emph{EURASIP Journal on Audio, Speech, and Music Processing}, vol. 2024, no.~1, May 2024. [Online]. Available: \url{http://dx.doi.org/10.1186/s13636-024-00351-9}
\BIBentrySTDinterwordspacing

\bibitem{Ujoodha_2024}
\BIBentryALTinterwordspacing
K.~Ujoodha, G.~R. Baby, and N.~G. Sahib, ``Training tacotron 2 for text-to-speech synthesis in kreole language: Methodology and audio processing,'' in \emph{2024 International Conference on Next Generation Computing Applications (NextComp)}.\hskip 1em plus 0.5em minus 0.4em\relax IEEE, Oct. 2024, p. 1–6. [Online]. Available: \url{http://dx.doi.org/10.1109/nextcomp63004.2024.10779830}
\BIBentrySTDinterwordspacing

\bibitem{Sang_2021}
\BIBentryALTinterwordspacing
D.~V. Sang and L.~X. Thu, ``Fasttacotron: A fast, robust and controllable method for speech synthesis,'' in \emph{2021 International Conference on Multimedia Analysis and Pattern Recognition (MAPR)}.\hskip 1em plus 0.5em minus 0.4em\relax IEEE, Oct. 2021, p. 1–5. [Online]. Available: \url{http://dx.doi.org/10.1109/mapr53640.2021.9585267}
\BIBentrySTDinterwordspacing

\bibitem{Galyashina_2022}
\BIBentryALTinterwordspacing
E.~I. Galyashina and V.~D. Nikishin, ``The protection of megascience projects from deepfake technologies threats: information law aspects,'' \emph{Journal of Physics: Conference Series}, vol. 2210, no.~1, p. 012007, Mar. 2022. [Online]. Available: \url{http://dx.doi.org/10.1088/1742-6596/2210/1/012007}
\BIBentrySTDinterwordspacing

\bibitem{Borrelli_2021}
\BIBentryALTinterwordspacing
C.~Borrelli, P.~Bestagini, F.~Antonacci, A.~Sarti, and S.~Tubaro, ``Synthetic speech detection through short-term and long-term prediction traces,'' \emph{EURASIP Journal on Information Security}, vol. 2021, no.~1, Apr. 2021. [Online]. Available: \url{http://dx.doi.org/10.1186/s13635-021-00116-3}
\BIBentrySTDinterwordspacing

\bibitem{Wang_2024}
\BIBentryALTinterwordspacing
X.~Wang, Z.~He, and X.~Peng, ``Artificial-intelligence-generated content with diffusion models: A literature review,'' \emph{Mathematics}, vol.~12, no.~7, p. 977, Mar. 2024. [Online]. Available: \url{http://dx.doi.org/10.3390/math12070977}
\BIBentrySTDinterwordspacing

\bibitem{Kumar_2024}
\BIBentryALTinterwordspacing
A.~Kumar, S.~Soni, S.~Chauhan, S.~Kaur, R.~Sharma, P.~Kalsi, R.~Chauhan, and A.~Birla, \emph{Navigating the Realm of Generative Models: GANs, Diffusion, Limitations, and Future Prospects—A Review}.\hskip 1em plus 0.5em minus 0.4em\relax Springer Nature Singapore, 2024, p. 301–319. [Online]. Available: \url{http://dx.doi.org/10.1007/978-981-97-2550-2_23}
\BIBentrySTDinterwordspacing

\bibitem{Solano_Carrillo_2023}
\BIBentryALTinterwordspacing
E.~Solano-Carrillo, A.~B. Rodriguez, B.~Carrillo-Perez, Y.~Steiniger, and J.~Stoppe, ``Look atme: The discriminator mean entropy needs attention,'' in \emph{2023 IEEE/CVF Conference on Computer Vision and Pattern Recognition Workshops (CVPRW)}.\hskip 1em plus 0.5em minus 0.4em\relax IEEE, Jun. 2023, p. 787–796. [Online]. Available: \url{http://dx.doi.org/10.1109/cvprw59228.2023.00086}
\BIBentrySTDinterwordspacing

\bibitem{Preeti_2023}
\BIBentryALTinterwordspacing
Preeti and S.~Bansal, ``Artifact based deepfake detection methods,'' in \emph{2023 Second International Conference on Informatics (ICI)}.\hskip 1em plus 0.5em minus 0.4em\relax IEEE, Nov. 2023, p. 1–6. [Online]. Available: \url{http://dx.doi.org/10.1109/ici60088.2023.10421629}
\BIBentrySTDinterwordspacing

\bibitem{Hu_2023}
\BIBentryALTinterwordspacing
X.~Hu, Y.~Jin, J.~Liang, J.~Liu, R.~Luo, M.~Li, and T.~Peng, ``Diffusion model for image generation - a survey,'' in \emph{2023 2nd International Conference on Artificial Intelligence, Human-Computer Interaction and Robotics (AIHCIR)}.\hskip 1em plus 0.5em minus 0.4em\relax IEEE, Dec. 2023, p. 416–424. [Online]. Available: \url{http://dx.doi.org/10.1109/aihcir61661.2023.00073}
\BIBentrySTDinterwordspacing

\bibitem{Mirza_2023}
\BIBentryALTinterwordspacing
M.~U. Mirza and T.~Çukur, ``Super-resolution diffusion model for accelerated mri reconstruction,'' in \emph{2023 31st Signal Processing and Communications Applications Conference (SIU)}.\hskip 1em plus 0.5em minus 0.4em\relax IEEE, Jul. 2023, p. 1–4. [Online]. Available: \url{http://dx.doi.org/10.1109/siu59756.2023.10223786}
\BIBentrySTDinterwordspacing

\bibitem{Duan_2023}
\BIBentryALTinterwordspacing
Z.~Duan, C.~Wang, C.~Chen, J.~Huang, and W.~Qian, ``Optimal linear subspace search: Learning to construct fast and high-quality schedulers for diffusion models,'' in \emph{Proceedings of the 32nd ACM International Conference on Information and Knowledge Management}, ser. CIKM '23.\hskip 1em plus 0.5em minus 0.4em\relax ACM, Oct. 2023, p. 463–472. [Online]. Available: \url{http://dx.doi.org/10.1145/3583780.3614999}
\BIBentrySTDinterwordspacing

\bibitem{Shin_2022}
\BIBentryALTinterwordspacing
S.~Y. Shin and J.~Lee, ``The effect of deepfake video on news credibility and corrective influence of cost-based knowledge about deepfakes,'' \emph{Digital Journalism}, vol.~10, no.~3, p. 412–432, Mar. 2022. [Online]. Available: \url{http://dx.doi.org/10.1080/21670811.2022.2026797}
\BIBentrySTDinterwordspacing

\bibitem{Plazas_Olmedo_2022}
\BIBentryALTinterwordspacing
M.~Plazas-Olmedo and P.~López-Rabadán, ``Nuevos formatos del vídeo electoral en redes. la estrategia multiplataforma de los partidos españoles en las elecciones de 2019,'' \emph{index.comunicación}, vol.~12, no.~2, p. 305–331, Jul. 2022. [Online]. Available: \url{http://dx.doi.org/10.33732/ixc/12/02nuevos}
\BIBentrySTDinterwordspacing

\bibitem{Lalancette_2017}
\BIBentryALTinterwordspacing
M.~Lalancette and V.~Raynauld, ``The power of political image: Justin trudeau, instagram, and celebrity politics,'' \emph{American Behavioral Scientist}, vol.~63, no.~7, p. 888–924, Nov. 2017. [Online]. Available: \url{http://dx.doi.org/10.1177/0002764217744838}
\BIBentrySTDinterwordspacing

\bibitem{Shen_2024}
\BIBentryALTinterwordspacing
Z.~Shen, M.~Mao, and P.~Fan, ``A primary comparison of diffusion models and generative adversarial networks for image synthesis,'' in \emph{Proceedings of the 2024 7th International Conference on Machine Learning and Machine Intelligence (MLMI)}, ser. MLMI 2024.\hskip 1em plus 0.5em minus 0.4em\relax ACM, Aug. 2024, p. 225–234. [Online]. Available: \url{http://dx.doi.org/10.1145/3696271.3696307}
\BIBentrySTDinterwordspacing

\bibitem{Ali_Raza_2024}
\BIBentryALTinterwordspacing
S.~Ali~Raza, U.~Habib, M.~Usman, A.~Ashraf~Cheema, and M.~Sajid~Khan, ``Mmganguard: A robust approach for detecting fake images generated by gans using multi-model techniques,'' \emph{IEEE Access}, vol.~12, p. 104153–104164, 2024. [Online]. Available: \url{http://dx.doi.org/10.1109/access.2024.3393842}
\BIBentrySTDinterwordspacing

\bibitem{Tan_2024}
\BIBentryALTinterwordspacing
C.~Tan, H.~Liu, Y.~Zhao, S.~Wei, G.~Gu, P.~Liu, and Y.~Wei, ``Rethinking the up-sampling operations in cnn-based generative network for generalizable deepfake detection,'' in \emph{2024 IEEE/CVF Conference on Computer Vision and Pattern Recognition (CVPR)}.\hskip 1em plus 0.5em minus 0.4em\relax IEEE, Jun. 2024, p. 28130–28139. [Online]. Available: \url{http://dx.doi.org/10.1109/cvpr52733.2024.02657}
\BIBentrySTDinterwordspacing

\bibitem{Liu_2024}
\BIBentryALTinterwordspacing
R.~Liu, S.~Zhang, Y.~Xu, W.~Xu, and X.~He, ``High-resolution network-based multi-feature fusion for generalized forgery detection,'' \emph{Multimedia Systems}, vol.~31, no.~1, Dec. 2024. [Online]. Available: \url{http://dx.doi.org/10.1007/s00530-024-01626-z}
\BIBentrySTDinterwordspacing

\bibitem{Phung_2023}
\BIBentryALTinterwordspacing
H.~Phung, Q.~Dao, and A.~Tran, ``Wavelet diffusion models are fast and scalable image generators,'' in \emph{2023 IEEE/CVF Conference on Computer Vision and Pattern Recognition (CVPR)}.\hskip 1em plus 0.5em minus 0.4em\relax IEEE, Jun. 2023, p. 10199–10208. [Online]. Available: \url{http://dx.doi.org/10.1109/cvpr52729.2023.00983}
\BIBentrySTDinterwordspacing

\bibitem{Arshad_2024}
\BIBentryALTinterwordspacing
T.~Arshad, M.~H. Khan, and M.~S. Farid, ``An efficient framework to recognize deepfake faces using a light-weight cnn,'' in \emph{Proceedings of the 2024 9th International Conference on Multimedia Systems and Signal Processing (ICMSSP)}, ser. ICMSSP 2024.\hskip 1em plus 0.5em minus 0.4em\relax ACM, May 2024, p. 24–29. [Online]. Available: \url{http://dx.doi.org/10.1145/3690063.3690064}
\BIBentrySTDinterwordspacing

\bibitem{Singh_2023}
\BIBentryALTinterwordspacing
A.~Singh, ``A survey of ai text-to-image and ai text-to-video generators,'' in \emph{2023 4th International Conference on Artificial Intelligence, Robotics and Control (AIRC)}.\hskip 1em plus 0.5em minus 0.4em\relax IEEE, May 2023, p. 32–36. [Online]. Available: \url{http://dx.doi.org/10.1109/airc57904.2023.10303174}
\BIBentrySTDinterwordspacing

\bibitem{Hasan_Fuad_2023}
\BIBentryALTinterwordspacing
M.~T. Hasan~Fuad, F.~Bin~Amin, and S.~M. Masudul~Ahsan, ``Deepfake detection from face-swapped videos using transfer learning approach,'' in \emph{2023 26th International Conference on Computer and Information Technology (ICCIT)}.\hskip 1em plus 0.5em minus 0.4em\relax IEEE, Dec. 2023, p. 1–6. [Online]. Available: \url{http://dx.doi.org/10.1109/iccit60459.2023.10441067}
\BIBentrySTDinterwordspacing

\bibitem{Kang_2022}
\BIBentryALTinterwordspacing
J.~Kang, S.-K. Ji, S.~Lee, D.~Jang, and J.-U. Hou, ``Detection enhancement for various deepfake types based on residual noise and manipulation traces,'' \emph{IEEE Access}, vol.~10, p. 69031–69040, 2022. [Online]. Available: \url{http://dx.doi.org/10.1109/access.2022.3185121}
\BIBentrySTDinterwordspacing

\bibitem{Huang_2020}
\BIBentryALTinterwordspacing
Y.~Huang, F.~Juefei-Xu, R.~Wang, Q.~Guo, L.~Ma, X.~Xie, J.~Li, W.~Miao, Y.~Liu, and G.~Pu, ``Fakepolisher: Making deepfakes more detection-evasive by shallow reconstruction,'' in \emph{Proceedings of the 28th ACM International Conference on Multimedia}, ser. MM '20.\hskip 1em plus 0.5em minus 0.4em\relax ACM, Oct. 2020, p. 1217–1226. [Online]. Available: \url{http://dx.doi.org/10.1145/3394171.3413732}
\BIBentrySTDinterwordspacing

\bibitem{Liu_2023}
\BIBentryALTinterwordspacing
H.~Liu, P.~Bestagini, L.~Huang, W.~Zhou, S.~Tubaro, W.~Zhang, and N.~Yu, ``It wasn't me: Irregular identity in deepfake videos,'' in \emph{2023 IEEE International Conference on Image Processing (ICIP)}.\hskip 1em plus 0.5em minus 0.4em\relax IEEE, Oct. 2023, p. 2770–2774. [Online]. Available: \url{http://dx.doi.org/10.1109/icip49359.2023.10222654}
\BIBentrySTDinterwordspacing

\bibitem{Khan_2024}
\BIBentryALTinterwordspacing
S.~A. Khan and D.-T. Dang-Nguyen, ``Clipping the deception: Adapting vision-language models for universal deepfake detection,'' in \emph{Proceedings of the 2024 International Conference on Multimedia Retrieval}, ser. ICMR '24.\hskip 1em plus 0.5em minus 0.4em\relax ACM, May 2024, p. 1006–1015. [Online]. Available: \url{http://dx.doi.org/10.1145/3652583.3658035}
\BIBentrySTDinterwordspacing

\bibitem{Zhang_2024}
\BIBentryALTinterwordspacing
Y.~Zhang, S.~Xu, and H.~Zhang, \emph{Enhancing the Transferability and Stealth of Deepfake Detection Attacks Through Latent Diffusion Models}.\hskip 1em plus 0.5em minus 0.4em\relax Springer Nature Singapore, Nov. 2024, p. 269–282. [Online]. Available: \url{http://dx.doi.org/10.1007/978-981-97-8505-6_19}
\BIBentrySTDinterwordspacing

\bibitem{Lu_2024}
\BIBentryALTinterwordspacing
Y.~Lu and T.~Ebrahimi, ``Towards the detection of ai-synthesized human face images,'' in \emph{2024 IEEE International Conference on Image Processing (ICIP)}.\hskip 1em plus 0.5em minus 0.4em\relax IEEE, Oct. 2024, p. 3778–3784. [Online]. Available: \url{http://dx.doi.org/10.1109/icip51287.2024.10647451}
\BIBentrySTDinterwordspacing

\bibitem{Yoon_2024}
\BIBentryALTinterwordspacing
J.~Yoon, A.~Panizo-LLedot, D.~Camacho, and C.~Choi, ``Triple-modality interaction for deepfake detection on zero-shot identity,'' \emph{Information Fusion}, vol. 109, p. 102424, Sep. 2024. [Online]. Available: \url{http://dx.doi.org/10.1016/j.inffus.2024.102424}
\BIBentrySTDinterwordspacing

\bibitem{Morris_2024}
\BIBentryALTinterwordspacing
K.~W. Morris, \emph{Deepfake Sockpuppets: The Toxic "Realities" of a Weaponised Internet}.\hskip 1em plus 0.5em minus 0.4em\relax Springer International Publishing, 2024, p. 61–79. [Online]. Available: \url{http://dx.doi.org/10.1007/978-3-031-43852-3_5}
\BIBentrySTDinterwordspacing

\bibitem{Singh_2021}
\BIBentryALTinterwordspacing
R.~K. Singh, P.~V. Sarda, S.~Aggarwal, and D.~K. Vishwakarma, ``Demystifying deepfakes using deep learning,'' in \emph{2021 5th International Conference on Computing Methodologies and Communication (ICCMC)}.\hskip 1em plus 0.5em minus 0.4em\relax IEEE, Apr. 2021, p. 1290–1298. [Online]. Available: \url{http://dx.doi.org/10.1109/iccmc51019.2021.9418477}
\BIBentrySTDinterwordspacing

\bibitem{Choquette_2021}
\BIBentryALTinterwordspacing
J.~Choquette, E.~Lee, R.~Krashinsky, V.~Balan, and B.~Khailany, ``3.2 the a100 datacenter gpu and ampere architecture,'' in \emph{2021 IEEE International Solid- State Circuits Conference (ISSCC)}.\hskip 1em plus 0.5em minus 0.4em\relax IEEE, Feb. 2021. [Online]. Available: \url{http://dx.doi.org/10.1109/isscc42613.2021.9365803}
\BIBentrySTDinterwordspacing

\bibitem{Gschwind_2017}
\BIBentryALTinterwordspacing
M.~Gschwind, T.~Kaldewey, and D.~K. Tam, ``Optimizing the efficiency of deep learning through accelerator virtualization,'' \emph{IBM Journal of Research and Development}, vol.~61, no. 4/5, pp. 12:1--12:11, Jul. 2017. [Online]. Available: \url{http://dx.doi.org/10.1147/jrd.2017.2716598}
\BIBentrySTDinterwordspacing

\bibitem{Kwak_2020}
\BIBentryALTinterwordspacing
J.~g. Kwak and H.~Ko, ``Unsupervised generation and synthesis of facial images via an auto-encoder-based deep generative adversarial network,'' \emph{Applied Sciences}, vol.~10, no.~6, p. 1995, Mar. 2020. [Online]. Available: \url{http://dx.doi.org/10.3390/app10061995}
\BIBentrySTDinterwordspacing

\bibitem{Sun_2022}
\BIBentryALTinterwordspacing
Y.~Sun, H.~Zhou, K.~Wang, Q.~Wu, Z.~Hong, J.~Liu, E.~Ding, J.~Wang, Z.~Liu, and K.~Hideki, ``Masked lip-sync prediction by audio-visual contextual exploitation in transformers,'' in \emph{SIGGRAPH Asia 2022 Conference Papers}, ser. SA '22.\hskip 1em plus 0.5em minus 0.4em\relax ACM, Nov. 2022, p. 1–9. [Online]. Available: \url{http://dx.doi.org/10.1145/3550469.3555393}
\BIBentrySTDinterwordspacing

\bibitem{Yang_2023}
\BIBentryALTinterwordspacing
S.~Yang, Y.~Bao, and Y.~Qi, ``Local context and dimensional relation aware transformer network for continuous affect estimation,'' in \emph{2023 IEEE International Conference on Image Processing (ICIP)}.\hskip 1em plus 0.5em minus 0.4em\relax IEEE, Oct. 2023, p. 1315–1319. [Online]. Available: \url{http://dx.doi.org/10.1109/icip49359.2023.10222702}
\BIBentrySTDinterwordspacing

\bibitem{Han_2024}
\BIBentryALTinterwordspacing
Y.~Han, J.~Zhu, K.~He, X.~Chen, Y.~Ge, W.~Li, X.~Li, J.~Zhang, C.~Wang, and Y.~Liu, \emph{Face-Adapter for Pre-trained Diffusion Models with Fine-Grained ID and Attribute Control}.\hskip 1em plus 0.5em minus 0.4em\relax Springer Nature Switzerland, Nov. 2024, p. 20–36. [Online]. Available: \url{http://dx.doi.org/10.1007/978-3-031-72973-7_2}
\BIBentrySTDinterwordspacing

\bibitem{Hao_2021}
\BIBentryALTinterwordspacing
C.~Hao, J.~Dotzel, J.~Xiong, L.~Benini, Z.~Zhang, and D.~Chen, ``Enabling design methodologies and future trends for edge ai: Specialization and codesign,'' \emph{IEEE Design \& Test}, vol.~38, no.~4, p. 7–26, Aug. 2021. [Online]. Available: \url{http://dx.doi.org/10.1109/mdat.2021.3069952}
\BIBentrySTDinterwordspacing

\bibitem{Gill_2024}
\BIBentryALTinterwordspacing
S.~S. Gill, M.~Golec, J.~Hu, M.~Xu, J.~Du, H.~Wu, G.~K. Walia, S.~S. Murugesan, B.~Ali, M.~Kumar, K.~Ye, P.~Verma, S.~Kumar, F.~Cuadrado, and S.~Uhlig, ``Edge ai: A taxonomy, systematic review and future directions,'' \emph{Cluster Computing}, vol.~28, no.~1, Oct. 2024. [Online]. Available: \url{http://dx.doi.org/10.1007/s10586-024-04686-y}
\BIBentrySTDinterwordspacing

\bibitem{Wu_2021}
\BIBentryALTinterwordspacing
C.~Wu, V.~Fresse, B.~Suffran, and H.~Konik, ``Accelerating dnns from local to virtualized fpga in the cloud: A survey of trends,'' \emph{Journal of Systems Architecture}, vol. 119, p. 102257, Oct. 2021. [Online]. Available: \url{http://dx.doi.org/10.1016/j.sysarc.2021.102257}
\BIBentrySTDinterwordspacing

\bibitem{Shrestha_2023}
\BIBentryALTinterwordspacing
R.~Shrestha, R.~Bajracharya, A.~Mishra, and S.~Kim, \emph{AI Accelerators for Cloud and Server Applications}.\hskip 1em plus 0.5em minus 0.4em\relax Springer International Publishing, 2023, p. 95–125. [Online]. Available: \url{http://dx.doi.org/10.1007/978-3-031-22170-5_3}
\BIBentrySTDinterwordspacing

\bibitem{den_Uyl_2015}
\BIBentryALTinterwordspacing
T.~M. den Uyl, H.~E. Tasli, P.~Ivan, and M.~Snijdewind, ``Who do you want to be? real-time face swap,'' in \emph{2015 11th IEEE International Conference and Workshops on Automatic Face and Gesture Recognition (FG)}.\hskip 1em plus 0.5em minus 0.4em\relax IEEE, May 2015, p. 1–1. [Online]. Available: \url{http://dx.doi.org/10.1109/fg.2015.7163172}
\BIBentrySTDinterwordspacing

\bibitem{Priya_2023}
\BIBentryALTinterwordspacing
K.~Priya and M.~Maanesh, ``Enabling global communication through automated real-time video dubbing,'' in \emph{2023 IEEE Technology \& Engineering Management Conference - Asia Pacific (TEMSCON-ASPAC)}.\hskip 1em plus 0.5em minus 0.4em\relax IEEE, Dec. 2023, p. 1–5. [Online]. Available: \url{http://dx.doi.org/10.1109/temscon-aspac59527.2023.10531326}
\BIBentrySTDinterwordspacing

\bibitem{Datta_2024}
\BIBentryALTinterwordspacing
S.~K. Datta, S.~Jia, and S.~Lyu, ``Exposing lip-syncing deepfakes from mouth inconsistencies,'' in \emph{2024 IEEE International Conference on Multimedia and Expo (ICME)}.\hskip 1em plus 0.5em minus 0.4em\relax IEEE, Jul. 2024, p. 1–6. [Online]. Available: \url{http://dx.doi.org/10.1109/icme57554.2024.10687902}
\BIBentrySTDinterwordspacing

\bibitem{Mittal_2024}
\BIBentryALTinterwordspacing
S.~Mittal, M.~Joshi, P.~Vats, G.~M. Upadhayay, S.~K. Vats, and S.~Kumar, ``Virtual illusions: Unleashing deepfake expertise for enhanced visual effects in film production,'' in \emph{2024 11th International Conference on Reliability, Infocom Technologies and Optimization (Trends and Future Directions) (ICRITO)}.\hskip 1em plus 0.5em minus 0.4em\relax IEEE, Mar. 2024, p. 1–6. [Online]. Available: \url{http://dx.doi.org/10.1109/icrito61523.2024.10522334}
\BIBentrySTDinterwordspacing

\bibitem{Sardana_2024}
\BIBentryALTinterwordspacing
F.~Sardana, K.~K. Mishra, A.~Singh, and N.~Saini, \emph{Transforming Social Media Marketing Through Deepfake Technology}.\hskip 1em plus 0.5em minus 0.4em\relax IGI Global, Jul. 2024, p. 431–453. [Online]. Available: \url{http://dx.doi.org/10.4018/979-8-3693-5298-4.ch022}
\BIBentrySTDinterwordspacing

\bibitem{Nasar_2020}
\BIBentryALTinterwordspacing
B.~F. Nasar, S.~T, and E.~R. Lason, ``Deepfake detection in media files - audios, images and videos,'' in \emph{2020 IEEE Recent Advances in Intelligent Computational Systems (RAICS)}.\hskip 1em plus 0.5em minus 0.4em\relax IEEE, Dec. 2020, p. 74–79. [Online]. Available: \url{http://dx.doi.org/10.1109/raics51191.2020.9332516}
\BIBentrySTDinterwordspacing

\bibitem{Teresa_2023}
\BIBentryALTinterwordspacing
L.~A. Teresa, N.~M. Sunil, S.~R. Andrews, T.~T. Thengumpallil, S.~Thomas, and B.~V~A, ``Enhancing children's learning experience: Interactive and personalized video learning with ai technology,'' in \emph{2023 IEEE International Conference on Recent Advances in Systems Science and Engineering (RASSE)}.\hskip 1em plus 0.5em minus 0.4em\relax IEEE, Nov. 2023, p. 1–5. [Online]. Available: \url{http://dx.doi.org/10.1109/rasse60029.2023.10363506}
\BIBentrySTDinterwordspacing

\bibitem{Regondi_2025}
\BIBentryALTinterwordspacing
S.~Regondi, G.~Donvito, E.~Frontoni, M.~Kostovic, F.~Minazzi, S.~Bratières, M.~Filosto, and R.~Pugliese, ``Artificial intelligence empowered voice generation for amyotrophic lateral sclerosis patients,'' \emph{Scientific Reports}, vol.~15, no.~1, Jan. 2025. [Online]. Available: \url{http://dx.doi.org/10.1038/s41598-024-84728-y}
\BIBentrySTDinterwordspacing

\bibitem{Fink_2024}
\BIBentryALTinterwordspacing
M.~C. Fink, S.~A. Robinson, and B.~Ertl, ``Ai-based avatars are changing the way we learn and teach: benefits and challenges,'' \emph{Frontiers in Education}, vol.~9, Jul. 2024. [Online]. Available: \url{http://dx.doi.org/10.3389/feduc.2024.1416307}
\BIBentrySTDinterwordspacing

\bibitem{de_Ruiter_2021}
\BIBentryALTinterwordspacing
A.~de~Ruiter, ``The distinct wrong of deepfakes,'' \emph{Philosophy \& Technology}, vol.~34, no.~4, p. 1311–1332, Jun. 2021. [Online]. Available: \url{http://dx.doi.org/10.1007/s13347-021-00459-2}
\BIBentrySTDinterwordspacing

\bibitem{Matli_2024}
\BIBentryALTinterwordspacing
W.~Matli, ``Extending the theory of information poverty to deepfake technology,'' \emph{International Journal of Information Management Data Insights}, vol.~4, no.~2, p. 100286, Nov. 2024. [Online]. Available: \url{http://dx.doi.org/10.1016/j.jjimei.2024.100286}
\BIBentrySTDinterwordspacing

\bibitem{Lundberg_2024}
\BIBentryALTinterwordspacing
E.~Lundberg and P.~Mozelius, ``The potential effects of deepfakes on news media and entertainment,'' \emph{AI \& SOCIETY}, Oct. 2024. [Online]. Available: \url{http://dx.doi.org/10.1007/s00146-024-02072-1}
\BIBentrySTDinterwordspacing

\bibitem{Malik_2024}
\BIBentryALTinterwordspacing
S.~Malik, A.~Surbhi, and D.~Roy, ``Blurring boundaries between truth and illusion: Analysis of human rights and regulatory concerns arising from abuse of deepfake technology,'' in \emph{ETLTC2024 INTERNATIONAL CONFERENCE SERIES ON ICT, ENTERTAINMENT TECHNOLOGIES, AND INTELLIGENT INFORMATION MANAGEMENT IN EDUCATION AND INDUSTRY}, vol. 3220.\hskip 1em plus 0.5em minus 0.4em\relax AIP Publishing, 2024, p. 050016. [Online]. Available: \url{http://dx.doi.org/10.1063/5.0234995}
\BIBentrySTDinterwordspacing

\bibitem{Kopecky_2024}
\BIBentryALTinterwordspacing
S.~Kopecky, \emph{Challenges of Deepfakes}.\hskip 1em plus 0.5em minus 0.4em\relax Springer Nature Switzerland, 2024, p. 158–166. [Online]. Available: \url{http://dx.doi.org/10.1007/978-3-031-62281-6_11}
\BIBentrySTDinterwordspacing

\bibitem{Shakil_2024}
\BIBentryALTinterwordspacing
M.~Shakil and F.~Mekuria, ``Balancing the risks and rewards of deepfake and synthetic media technology: A regulatory framework for emerging economies,'' in \emph{2024 International Conference on Information and Communication Technology for Development for Africa (ICT4DA)}.\hskip 1em plus 0.5em minus 0.4em\relax IEEE, Nov. 2024, p. 114–119. [Online]. Available: \url{http://dx.doi.org/10.1109/ict4da62874.2024.10777194}
\BIBentrySTDinterwordspacing

\bibitem{Volkova_2023}
\BIBentryALTinterwordspacing
S.~S. Volkova, ``A method for deepfake detection using convolutional neural networks,'' \emph{Scientific and Technical Information Processing}, vol.~50, no.~5, p. 475–485, Dec. 2023. [Online]. Available: \url{http://dx.doi.org/10.3103/s0147688223050143}
\BIBentrySTDinterwordspacing

\bibitem{Do_2021}
\BIBentryALTinterwordspacing
T.-L. Do, M.-K. Tran, H.~H. Nguyen, and M.-T. Tran, \emph{Potential Threat of Face Swapping to eKYC with Face Registration and Augmented Solution with Deepfake Detection}.\hskip 1em plus 0.5em minus 0.4em\relax Springer International Publishing, 2021, p. 293–307. [Online]. Available: \url{http://dx.doi.org/10.1007/978-3-030-91387-8_19}
\BIBentrySTDinterwordspacing

\bibitem{Gilbert_2024}
\BIBentryALTinterwordspacing
A.~Gilbert and Z.~Gong, \emph{Digital Identity Theft Using Deepfakes}.\hskip 1em plus 0.5em minus 0.4em\relax CRC Press, Apr. 2024, p. 307–314. [Online]. Available: \url{http://dx.doi.org/10.1201/9781003264415-47}
\BIBentrySTDinterwordspacing

\bibitem{Agarwal_2023}
\BIBentryALTinterwordspacing
A.~Agarwal and N.~Ratha, \emph{Manipulating faces for identity theft via morphing and deepfake: Digital privacy}.\hskip 1em plus 0.5em minus 0.4em\relax Elsevier, 2023, p. 223–241. [Online]. Available: \url{http://dx.doi.org/10.1016/bs.host.2022.12.003}
\BIBentrySTDinterwordspacing

\bibitem{de_Rancourt_Raymond_2022}
\BIBentryALTinterwordspacing
A.~de~Rancourt-Raymond and N.~Smaili, ``The unethical use of deepfakes,'' \emph{Journal of Financial Crime}, vol.~30, no.~4, p. 1066–1077, May 2022. [Online]. Available: \url{http://dx.doi.org/10.1108/jfc-04-2022-0090}
\BIBentrySTDinterwordspacing

\bibitem{Stupp_2019}
\BIBentryALTinterwordspacing
C.~Stupp, ``\BIBforeignlanguage{en_US}{Fraudsters used ai to mimic ceo's voice in unusual cybercrime case},'' \emph{\BIBforeignlanguage{en_US}{WSJ}}, Aug 2019. [Online]. Available: \url{https://shorturl.at/6JvLe}
\BIBentrySTDinterwordspacing

\bibitem{TheHackerNews}
\BIBentryALTinterwordspacing
T.~H. News, ``Chinese hackers using deepfakes in advanced mobile banking malware attacks.'' [Online]. Available: \url{https://thehackernews.com/2024/02/chinese-hackers-using-deepfakes-in.html}
\BIBentrySTDinterwordspacing

\bibitem{Allyn_2022}
\BIBentryALTinterwordspacing
B.~Allyn, ``Deepfake video of zelenskyy could be "tip of the iceberg" in info war, experts warn,'' \emph{NPR}, Mar. 2022. [Online]. Available: \url{https://www.npr.org/2022/03/16/1087062648/deepfake-video-zelenskyy-experts-war-manipulation-ukraine-russia}
\BIBentrySTDinterwordspacing

\bibitem{Alavi_Achom_2020}
\BIBentryALTinterwordspacing
M.~Alavi and D.~Achom, ``\BIBforeignlanguage{en}{Bjp shared deepfake video on whatsapp during delhi campaign},'' Feb. 2020. [Online]. Available: \url{https://shorturl.at/y3DLm}
\BIBentrySTDinterwordspacing

\bibitem{Harwell_2018}
\BIBentryALTinterwordspacing
D.~Harwell, ``Scarlett johansson on fake ai-generated sex videos: 'nothing can stop someone from cutting and pasting my image','' \emph{The Washington Post}, Dec. 2018. [Online]. Available: \url{https://tinyurl.com/2hkums8n}
\BIBentrySTDinterwordspacing

\bibitem{K_t_k__2024}
\BIBentryALTinterwordspacing
Y.~E. Kütükçü and H.~Polat, ``Byol yaklasimini kullanarak mtcnn ile yakalanan yüzlerde deepfake tespiti deepfake detection on faces captured with mtcnn, by using the byol approach,'' in \emph{2024 Innovations in Intelligent Systems and Applications Conference (ASYU)}.\hskip 1em plus 0.5em minus 0.4em\relax IEEE, Oct. 2024, p. 1–6. [Online]. Available: \url{http://dx.doi.org/10.1109/asyu62119.2024.10756986}
\BIBentrySTDinterwordspacing

\bibitem{Lin_2022}
\BIBentryALTinterwordspacing
Y.~Lin, H.~Chen, B.~Li, and J.~Wu, ``Towards generalizable deepfake face forgery detection with semi-supervised learning and knowledge distillation,'' in \emph{2022 IEEE International Conference on Image Processing (ICIP)}.\hskip 1em plus 0.5em minus 0.4em\relax IEEE, Oct. 2022. [Online]. Available: \url{http://dx.doi.org/10.1109/icip46576.2022.9897792}
\BIBentrySTDinterwordspacing

\bibitem{Patil_2024}
\BIBentryALTinterwordspacing
M.~B. Patil, V.~A. Sangolgi, V.~V. Bag, A.~B. Patwegar, R.~Koli, A.~Naikwadi, and A.~G. Shaikh, ``Genconvit+: Advanced hybrid framework for deepfake detection for safeguarding digital media integrity,'' \emph{Journal of Integrated Science and Technology}, vol.~12, no.~5, May 2024. [Online]. Available: \url{http://dx.doi.org/10.62110/sciencein.jist.2024.v12.820}
\BIBentrySTDinterwordspacing

\bibitem{Pontorno_2024}
\BIBentryALTinterwordspacing
O.~Pontorno, L.~Guarnera, and S.~Battiato, ``On the exploitation of dct-traces in the generative-ai domain,'' in \emph{2024 IEEE International Conference on Image Processing (ICIP)}.\hskip 1em plus 0.5em minus 0.4em\relax IEEE, Oct. 2024, p. 3806–3812. [Online]. Available: \url{http://dx.doi.org/10.1109/icip51287.2024.10648013}
\BIBentrySTDinterwordspacing

\bibitem{Fung_2021}
\BIBentryALTinterwordspacing
S.~Fung, X.~Lu, C.~Zhang, and C.-T. Li, ``Deepfakeucl: Deepfake detection via unsupervised contrastive learning,'' in \emph{2021 International Joint Conference on Neural Networks (IJCNN)}.\hskip 1em plus 0.5em minus 0.4em\relax IEEE, Jul. 2021, p. 1–8. [Online]. Available: \url{http://dx.doi.org/10.1109/ijcnn52387.2021.9534089}
\BIBentrySTDinterwordspacing

\bibitem{Dai_2023}
\BIBentryALTinterwordspacing
X.~Dai, C.~Wang, H.~Li, S.~Lin, L.~Dong, J.~Wu, and J.~Wang, ``Synthetic feature assessment for zero-shot object detection,'' in \emph{2023 IEEE International Conference on Multimedia and Expo (ICME)}.\hskip 1em plus 0.5em minus 0.4em\relax IEEE, Jul. 2023, p. 444–449. [Online]. Available: \url{http://dx.doi.org/10.1109/icme55011.2023.00083}
\BIBentrySTDinterwordspacing

\bibitem{Moskalev_2022}
\BIBentryALTinterwordspacing
A.~Moskalev, I.~Sosnovik, V.~Fischer, and A.~Smeulders, \emph{Contrasting Quadratic Assignments for Set-Based Representation Learning}.\hskip 1em plus 0.5em minus 0.4em\relax Springer Nature Switzerland, 2022, p. 88–104. [Online]. Available: \url{http://dx.doi.org/10.1007/978-3-031-19812-0_6}
\BIBentrySTDinterwordspacing

\bibitem{John_2022}
\BIBentryALTinterwordspacing
J.~John and B.~V. Sherif, ``Comparative analysis on different deepfake detection methods and semi supervised gan architecture for deepfake detection,'' in \emph{2022 Sixth International Conference on I-SMAC (IoT in Social, Mobile, Analytics and Cloud) (I-SMAC)}.\hskip 1em plus 0.5em minus 0.4em\relax IEEE, Nov. 2022, p. 516–521. [Online]. Available: \url{http://dx.doi.org/10.1109/i-smac55078.2022.9987265}
\BIBentrySTDinterwordspacing

\bibitem{Pan_2023}
\BIBentryALTinterwordspacing
Y.~Pan, H.~Cheng, Y.~Fang, Y.~Liu, and T.~Liu, ``Hard-negatives focused self-supervised learning,'' in \emph{2023 International Conference on Cyber-Enabled Distributed Computing and Knowledge Discovery (CyberC)}.\hskip 1em plus 0.5em minus 0.4em\relax IEEE, Nov. 2023, p. 68–77. [Online]. Available: \url{http://dx.doi.org/10.1109/cyberc58899.2023.00022}
\BIBentrySTDinterwordspacing

\bibitem{Yu_Diong_2024}
\BIBentryALTinterwordspacing
Z.~Yu~Diong, W.~Y. Lim, and C.~P. Goh, ``Self-supervised learning in medical diagnostics: An examination of simclr and byol in image classification,'' in \emph{2024 3rd International Conference on Digital Transformation and Applications (ICDXA)}.\hskip 1em plus 0.5em minus 0.4em\relax IEEE, Jan. 2024, p. 210–214. [Online]. Available: \url{http://dx.doi.org/10.1109/icdxa61007.2024.10470922}
\BIBentrySTDinterwordspacing

\bibitem{Alsmadi_2009}
\BIBentryALTinterwordspacing
M.~k. Alsmadi, K.~B. Omar, S.~A. Noah, and I.~Almarashdah, ``Performance comparison of multi-layer perceptron (back propagation, delta rule and perceptron) algorithms in neural networks,'' in \emph{2009 IEEE International Advance Computing Conference}.\hskip 1em plus 0.5em minus 0.4em\relax IEEE, Mar. 2009, p. 296–299. [Online]. Available: \url{http://dx.doi.org/10.1109/iadcc.2009.4809024}
\BIBentrySTDinterwordspacing

\bibitem{Cunningham}
\BIBentryALTinterwordspacing
P.~Cunningham, \emph{Dimension Reduction}.\hskip 1em plus 0.5em minus 0.4em\relax Springer Berlin Heidelberg, p. 91–112. [Online]. Available: \url{http://dx.doi.org/10.1007/978-3-540-75171-7_4}
\BIBentrySTDinterwordspacing

\bibitem{Zolfaghari_2021}
\BIBentryALTinterwordspacing
M.~Zolfaghari, Y.~Zhu, P.~Gehler, and T.~Brox, ``Crossclr: Cross-modal contrastive learning for multi-modal video representations,'' in \emph{2021 IEEE/CVF International Conference on Computer Vision (ICCV)}.\hskip 1em plus 0.5em minus 0.4em\relax IEEE, Oct. 2021, p. 1430–1439. [Online]. Available: \url{http://dx.doi.org/10.1109/iccv48922.2021.00148}
\BIBentrySTDinterwordspacing

\bibitem{Lin_2024}
\BIBentryALTinterwordspacing
K.~Lin, W.~Han, S.~Li, Z.~Gu, H.~Zhao, and Y.~Mei, ``Detecting deepfake videos using spatiotemporal trident network,'' \emph{ACM Transactions on Multimedia Computing, Communications, and Applications}, vol.~20, no.~11, p. 1–20, Sep. 2024. [Online]. Available: \url{http://dx.doi.org/10.1145/3623639}
\BIBentrySTDinterwordspacing

\bibitem{Migliorelli_2024}
\BIBentryALTinterwordspacing
G.~Migliorelli, M.~C. Fiorentino, M.~Di~Cosmo, F.~P. Villani, A.~Mancini, and S.~Moccia, ``On the use of contrastive learning for standard-plane classification in fetal ultrasound imaging,'' \emph{Computers in Biology and Medicine}, vol. 174, p. 108430, May 2024. [Online]. Available: \url{http://dx.doi.org/10.1016/j.compbiomed.2024.108430}
\BIBentrySTDinterwordspacing

\bibitem{Zhang_2023}
\BIBentryALTinterwordspacing
J.~Zhang, T.~Wang, J.~Wang, C.~Li, Y.~Fu, and H.~Soussi, ``A study of contrastive self-supervised learning generalization based on augmented data,'' in \emph{2023 38th Youth Academic Annual Conference of Chinese Association of Automation (YAC)}.\hskip 1em plus 0.5em minus 0.4em\relax IEEE, Aug. 2023, p. 659–664. [Online]. Available: \url{http://dx.doi.org/10.1109/yac59482.2023.10401602}
\BIBentrySTDinterwordspacing

\bibitem{Song_2024}
\BIBentryALTinterwordspacing
Q.~Song, Y.~Dai, H.~Lu, and G.~Jin, ``High-throughput systolic array-based accelerator for hybrid transformer-cnn networks,'' \emph{Journal of King Saud University - Computer and Information Sciences}, vol.~36, no.~8, p. 102194, Oct. 2024. [Online]. Available: \url{http://dx.doi.org/10.1016/j.jksuci.2024.102194}
\BIBentrySTDinterwordspacing

\bibitem{He_2020}
\BIBentryALTinterwordspacing
K.~He, H.~Fan, Y.~Wu, S.~Xie, and R.~Girshick, ``Momentum contrast for unsupervised visual representation learning,'' in \emph{2020 IEEE/CVF Conference on Computer Vision and Pattern Recognition (CVPR)}.\hskip 1em plus 0.5em minus 0.4em\relax IEEE, Jun. 2020. [Online]. Available: \url{http://dx.doi.org/10.1109/cvpr42600.2020.00975}
\BIBentrySTDinterwordspacing

\bibitem{Bhuse_2021}
\BIBentryALTinterwordspacing
P.~Bhuse, B.~Singh, and P.~Raut, \emph{Effect of Data Augmentation on the Accuracy of Convolutional Neural Networks}.\hskip 1em plus 0.5em minus 0.4em\relax Springer Singapore, Jul. 2021, p. 337–348. [Online]. Available: \url{http://dx.doi.org/10.1007/978-981-16-0739-4_33}
\BIBentrySTDinterwordspacing

\bibitem{Hrga_2022}
\BIBentryALTinterwordspacing
I.~Hrga and M.~Ivasic-Kos, ``Effect of data augmentation methods on face image classification results,'' in \emph{Proceedings of the 11th International Conference on Pattern Recognition Applications and Methods}.\hskip 1em plus 0.5em minus 0.4em\relax SCITEPRESS - Science and Technology Publications, 2022, p. 660–667. [Online]. Available: \url{http://dx.doi.org/10.5220/0010883800003122}
\BIBentrySTDinterwordspacing

\bibitem{McNeely_White_2019}
\BIBentryALTinterwordspacing
D.~G. McNeely-White, J.~R. Beveridge, and B.~A. Draper, \emph{Inception and ResNet: Same Training, Same Features}.\hskip 1em plus 0.5em minus 0.4em\relax Springer International Publishing, Jul. 2019, p. 352–357. [Online]. Available: \url{http://dx.doi.org/10.1007/978-3-030-25719-4_45}
\BIBentrySTDinterwordspacing

\bibitem{Shah_2020}
\BIBentryALTinterwordspacing
P.~Shah and M.~El-Sharkawy, ``A-mnasnet: Augmented mnasnet for computer vision,'' in \emph{2020 IEEE 63rd International Midwest Symposium on Circuits and Systems (MWSCAS)}.\hskip 1em plus 0.5em minus 0.4em\relax IEEE, Aug. 2020, p. 1044–1047. [Online]. Available: \url{http://dx.doi.org/10.1109/mwscas48704.2020.9184619}
\BIBentrySTDinterwordspacing

\bibitem{Shi_2009}
\BIBentryALTinterwordspacing
H.~Shi, ``Evolving artificial neural networks using ga and momentum,'' in \emph{2009 Second International Symposium on Electronic Commerce and Security}.\hskip 1em plus 0.5em minus 0.4em\relax IEEE, 2009, p. 475–478. [Online]. Available: \url{http://dx.doi.org/10.1109/isecs.2009.132}
\BIBentrySTDinterwordspacing

\bibitem{Xu_2012}
\BIBentryALTinterwordspacing
D.~Xu, H.~Shao, and H.~Zhang, ``A new adaptive momentum algorithm for split-complex recurrent neural networks,'' \emph{Neurocomputing}, vol.~93, p. 133–136, Sep. 2012. [Online]. Available: \url{http://dx.doi.org/10.1016/j.neucom.2012.03.013}
\BIBentrySTDinterwordspacing

\bibitem{Yeh_2022}
\BIBentryALTinterwordspacing
C.-H. Yeh, C.-Y. Hong, Y.-C. Hsu, T.-L. Liu, Y.~Chen, and Y.~LeCun, \emph{Decoupled Contrastive Learning}.\hskip 1em plus 0.5em minus 0.4em\relax Springer Nature Switzerland, 2022, p. 668–684. [Online]. Available: \url{http://dx.doi.org/10.1007/978-3-031-19809-0_38}
\BIBentrySTDinterwordspacing

\bibitem{Zheng_2024}
\BIBentryALTinterwordspacing
J.~Zheng, X.~Hu, C.~Chen, Y.~Zhou, D.~Gao, and Z.~Tang, ``A new deepfake detection model for responding to perception attacks in embodied artificial intelligence,'' \emph{Image and Vision Computing}, vol. 151, p. 105279, Nov. 2024. [Online]. Available: \url{http://dx.doi.org/10.1016/j.imavis.2024.105279}
\BIBentrySTDinterwordspacing

\bibitem{Qureshi_2024}
\BIBentryALTinterwordspacing
S.~M. Qureshi, A.~Saeed, S.~H. Almotiri, F.~Ahmad, and M.~A. Al~Ghamdi, ``Deepfake forensics: a survey of digital forensic methods for multimodal deepfake identification on social media,'' \emph{PeerJ Computer Science}, vol.~10, p. e2037, May 2024. [Online]. Available: \url{http://dx.doi.org/10.7717/peerj-cs.2037}
\BIBentrySTDinterwordspacing

\bibitem{Hojjati_2024}
\BIBentryALTinterwordspacing
H.~Hojjati, T.~K.~K. Ho, and N.~Armanfard, ``Self-supervised anomaly detection in computer vision and beyond: A survey and outlook,'' \emph{Neural Networks}, vol. 172, p. 106106, Apr. 2024. [Online]. Available: \url{http://dx.doi.org/10.1016/j.neunet.2024.106106}
\BIBentrySTDinterwordspacing

\bibitem{G_kstorp_2024}
\BIBentryALTinterwordspacing
S.~Gökstorp, J.~Nyberg, Y.~Kim, P.~Johnson, and G.~Dán, ``Anomaly detection in security logs using sequence modeling,'' in \emph{NOMS 2024-2024 IEEE Network Operations and Management Symposium}.\hskip 1em plus 0.5em minus 0.4em\relax IEEE, May 2024, p. 1–9. [Online]. Available: \url{http://dx.doi.org/10.1109/noms59830.2024.10575561}
\BIBentrySTDinterwordspacing

\bibitem{Lee_2023}
\BIBentryALTinterwordspacing
E.-G. Lee, I.~Lee, and S.-B. Yoo, ``Cluecatcher: Catching domain-wise independent clues for deepfake detection,'' \emph{Mathematics}, vol.~11, no.~18, p. 3952, Sep. 2023. [Online]. Available: \url{http://dx.doi.org/10.3390/math11183952}
\BIBentrySTDinterwordspacing

\bibitem{Mandal_2024}
\BIBentryALTinterwordspacing
S.~Mandal, B.~Ghosh, S.~Chakraborty, and R.~Naskar, ``Can deepfakes mimic human emotions? a perspective on synthesia videos,'' in \emph{TENCON 2024 - 2024 IEEE Region 10 Conference (TENCON)}.\hskip 1em plus 0.5em minus 0.4em\relax IEEE, Dec. 2024, p. 306–309. [Online]. Available: \url{http://dx.doi.org/10.1109/tencon61640.2024.10902983}
\BIBentrySTDinterwordspacing

\bibitem{Zhao_2020}
\BIBentryALTinterwordspacing
Z.~Zhao, P.~Wang, and W.~Lu, ``Detecting deepfake video by learning two-level features with two-stream convolutional neural network,'' in \emph{Proceedings of the 2020 6th International Conference on Computing and Artificial Intelligence}, ser. ICCAI '20.\hskip 1em plus 0.5em minus 0.4em\relax ACM, Apr. 2020, p. 291–297. [Online]. Available: \url{http://dx.doi.org/10.1145/3404555.3404564}
\BIBentrySTDinterwordspacing

\bibitem{Seha_2019}
\BIBentryALTinterwordspacing
S.~Seha, G.~Papangelakis, D.~Hatzinakos, A.~S. Zandi, and F.~J. Comeau, ``Improving eye movement biometrics using remote registration of eye blinking patterns,'' in \emph{ICASSP 2019 - 2019 IEEE International Conference on Acoustics, Speech and Signal Processing (ICASSP)}.\hskip 1em plus 0.5em minus 0.4em\relax IEEE, May 2019. [Online]. Available: \url{http://dx.doi.org/10.1109/icassp.2019.8683757}
\BIBentrySTDinterwordspacing

\bibitem{Zimmermann_2021}
\BIBentryALTinterwordspacing
E.~Zimmermann, ``Sensorimotor serial dependencies in head movements,'' \emph{Journal of Neurophysiology}, vol. 126, no.~3, p. 913–923, Sep. 2021. [Online]. Available: \url{http://dx.doi.org/10.1152/jn.00231.2021}
\BIBentrySTDinterwordspacing

\bibitem{P_2024}
\BIBentryALTinterwordspacing
V.~P, M.~M, K.~A. Ivaturi, and P.~Takur, ``Improving biometric security using pulseprint: Real-time defense against fingerprint spoofing,'' in \emph{2024 2nd International Conference on Networking and Communications (ICNWC)}.\hskip 1em plus 0.5em minus 0.4em\relax IEEE, Apr. 2024, p. 1–7. [Online]. Available: \url{http://dx.doi.org/10.1109/icnwc60771.2024.10537386}
\BIBentrySTDinterwordspacing

\bibitem{Ge_2021}
\BIBentryALTinterwordspacing
S.~Ge, F.~Lin, C.~Li, D.~Zhang, J.~Tan, W.~Wang, and D.~Zeng, ``Latent pattern sensing: Deepfake video detection via predictive representation learning,'' in \emph{ACM Multimedia Asia}, ser. MMAsia '21.\hskip 1em plus 0.5em minus 0.4em\relax ACM, Dec. 2021, p. 1–7. [Online]. Available: \url{http://dx.doi.org/10.1145/3469877.3490586}
\BIBentrySTDinterwordspacing

\bibitem{Cunha_2024}
\BIBentryALTinterwordspacing
Y.~M. B.~G. Cunha, B.~R. Gomes, J.~M.~C. Boaro, D.~d.~S. Moraes, A.~J.~G. Busson, J.~C. Duarte, and S.~Colcher, ``Learning self-distilled features for facial deepfake detection using visual foundation models: General results and demographic analysis,'' \emph{Journal on Interactive Systems}, vol.~15, no.~1, p. 682–694, Jul. 2024. [Online]. Available: \url{http://dx.doi.org/10.5753/jis.2024.4120}
\BIBentrySTDinterwordspacing

\bibitem{Wu_2024}
\BIBentryALTinterwordspacing
H.~Wu, L.~Leng, and P.~Yu, ``Learning local reconstruction errors for face forgery detection,'' \emph{International Journal of Advanced Computer Science and Applications}, vol.~15, no.~11, 2024. [Online]. Available: \url{http://dx.doi.org/10.14569/ijacsa.2024.01511120}
\BIBentrySTDinterwordspacing

\bibitem{Hussain_2023}
\BIBentryALTinterwordspacing
M.~Hussain, J.-W. Suh, B.-S. Seo, and J.-E. Hong, ``How reliable are the deep learning-based anomaly detectors? a comprehensive reliability analysis of autoencoder-based anomaly detectors,'' in \emph{2023 Fourteenth International Conference on Ubiquitous and Future Networks (ICUFN)}.\hskip 1em plus 0.5em minus 0.4em\relax IEEE, Jul. 2023, p. 317–322. [Online]. Available: \url{http://dx.doi.org/10.1109/icufn57995.2023.10199315}
\BIBentrySTDinterwordspacing

\bibitem{N_2024}
\BIBentryALTinterwordspacing
A.~N, K.~K, Madhumitha, M.~Shet, and P.~Alekhya, ``Deepfake detection using deep learning: A two-pronged approach with cnns and autoencoders,'' in \emph{2024 2nd International Conference on Recent Advances in Information Technology for Sustainable Development (ICRAIS)}.\hskip 1em plus 0.5em minus 0.4em\relax IEEE, Nov. 2024, p. 24–29. [Online]. Available: \url{http://dx.doi.org/10.1109/icrais62903.2024.10811712}
\BIBentrySTDinterwordspacing

\bibitem{Du_2020}
\BIBentryALTinterwordspacing
M.~Du, S.~Pentyala, Y.~Li, and X.~Hu, ``Towards generalizable deepfake detection with locality-aware autoencoder,'' in \emph{Proceedings of the 29th ACM International Conference on Information \& Knowledge Management}, ser. CIKM '20.\hskip 1em plus 0.5em minus 0.4em\relax ACM, Oct. 2020. [Online]. Available: \url{http://dx.doi.org/10.1145/3340531.3411892}
\BIBentrySTDinterwordspacing

\bibitem{Yu_2024}
\BIBentryALTinterwordspacing
Y.~Yu, X.~Liu, R.~Ni, S.~Yang, Y.~Zhao, and A.~C. Kot, ``Pvass-mdd: Predictive visual-audio alignment self-supervision for multimodal deepfake detection,'' \emph{IEEE Transactions on Circuits and Systems for Video Technology}, vol.~34, no.~8, p. 6926–6936, Aug. 2024. [Online]. Available: \url{http://dx.doi.org/10.1109/tcsvt.2023.3309899}
\BIBentrySTDinterwordspacing

\bibitem{Chen_2023}
\BIBentryALTinterwordspacing
P.~Chen, M.~Xu, and J.~Qi, ``Deepfake detection against adversarial examples based on d‐vaegan,'' \emph{IET Image Processing}, vol.~18, no.~3, p. 615–626, Nov. 2023. [Online]. Available: \url{http://dx.doi.org/10.1049/ipr2.12973}
\BIBentrySTDinterwordspacing

\bibitem{Amer_2013}
\BIBentryALTinterwordspacing
M.~Amer, M.~Goldstein, and S.~Abdennadher, ``Enhancing one-class support vector machines for unsupervised anomaly detection,'' in \emph{Proceedings of the ACM SIGKDD Workshop on Outlier Detection and Description}, ser. KDD' 13.\hskip 1em plus 0.5em minus 0.4em\relax ACM, Aug. 2013, p. 8–15. [Online]. Available: \url{http://dx.doi.org/10.1145/2500853.2500857}
\BIBentrySTDinterwordspacing

\bibitem{Yin_2014}
\BIBentryALTinterwordspacing
S.~Yin, X.~Zhu, and C.~Jing, ``Fault detection based on a robust one class support vector machine,'' \emph{Neurocomputing}, vol. 145, p. 263–268, Dec. 2014. [Online]. Available: \url{http://dx.doi.org/10.1016/j.neucom.2014.05.035}
\BIBentrySTDinterwordspacing

\bibitem{Al_Machot_2022}
\BIBentryALTinterwordspacing
F.~Al~Machot, M.~Ullah, and H.~Ullah, ``Hfm: A hybrid feature model based on conditional auto encoders for zero-shot learning,'' \emph{Journal of Imaging}, vol.~8, no.~6, p. 171, Jun. 2022. [Online]. Available: \url{http://dx.doi.org/10.3390/jimaging8060171}
\BIBentrySTDinterwordspacing

\bibitem{Wang_2019}
\BIBentryALTinterwordspacing
W.~Wang, V.~W. Zheng, H.~Yu, and C.~Miao, ``A survey of zero-shot learning: Settings, methods, and applications,'' \emph{ACM Transactions on Intelligent Systems and Technology}, vol.~10, no.~2, p. 1–37, Jan. 2019. [Online]. Available: \url{http://dx.doi.org/10.1145/3293318}
\BIBentrySTDinterwordspacing

\bibitem{Karlos_2021}
\BIBentryALTinterwordspacing
S.~Karlos, N.~Mylonas, and G.~Tsoumakas, ``Instance-based zero-shot learning for semi-automatic mesh indexing,'' \emph{Pattern Recognition Letters}, vol. 151, p. 62–68, Nov. 2021. [Online]. Available: \url{http://dx.doi.org/10.1016/j.patrec.2021.08.009}
\BIBentrySTDinterwordspacing

\bibitem{Anusha_2025}
\BIBentryALTinterwordspacing
T.~Anusha and A.~Srinagesh, ``Deepfake video detection: A comprehensive survey of advanced machine learning and deep learning techniques to combat synthetic video manipulation,'' in \emph{2025 International Conference on Multi-Agent Systems for Collaborative Intelligence (ICMSCI)}.\hskip 1em plus 0.5em minus 0.4em\relax IEEE, Jan. 2025, p. 1033–1041. [Online]. Available: \url{http://dx.doi.org/10.1109/icmsci62561.2025.10894187}
\BIBentrySTDinterwordspacing

\bibitem{Eshghi_2016}
\BIBentryALTinterwordspacing
K.~Eshghi and M.~Kafai, ``The cro kernel: Using concomitant rank order hashes for sparse high dimensional randomized feature maps,'' in \emph{2016 IEEE 32nd International Conference on Data Engineering (ICDE)}.\hskip 1em plus 0.5em minus 0.4em\relax IEEE, May 2016, p. 721–730. [Online]. Available: \url{http://dx.doi.org/10.1109/icde.2016.7498284}
\BIBentrySTDinterwordspacing

\bibitem{Elen_2022}
\BIBentryALTinterwordspacing
A.~Elen, S.~Baş, and C.~Közkurt, ``An adaptive gaussian kernel for support vector machine,'' \emph{Arabian Journal for Science and Engineering}, vol.~47, no.~8, p. 10579–10588, Mar. 2022. [Online]. Available: \url{http://dx.doi.org/10.1007/s13369-022-06654-3}
\BIBentrySTDinterwordspacing

\bibitem{_st_n_2006}
\BIBentryALTinterwordspacing
B.~Üstün, W.~Melssen, and L.~Buydens, ``Facilitating the application of support vector regression by using a universal pearson vii function based kernel,'' \emph{Chemometrics and Intelligent Laboratory Systems}, vol.~81, no.~1, p. 29–40, Mar. 2006. [Online]. Available: \url{http://dx.doi.org/10.1016/j.chemolab.2005.09.003}
\BIBentrySTDinterwordspacing

\bibitem{Virmani_2022}
\BIBentryALTinterwordspacing
D.~Virmani and H.~Pandey, \emph{Comparative Analysis on Effect of Different SVM Kernel Functions for Classification}.\hskip 1em plus 0.5em minus 0.4em\relax Springer Nature Singapore, Nov. 2022, p. 657–670. [Online]. Available: \url{http://dx.doi.org/10.1007/978-981-19-3679-1_56}
\BIBentrySTDinterwordspacing

\bibitem{Wu_2009}
\BIBentryALTinterwordspacing
D.~Wu and F.~Cao, ``Learning rates for svm classifiers with polynomial kernels,'' in \emph{2009 International Conference on Machine Learning and Cybernetics}.\hskip 1em plus 0.5em minus 0.4em\relax IEEE, Jul. 2009, p. 1111–1116. [Online]. Available: \url{http://dx.doi.org/10.1109/icmlc.2009.5212388}
\BIBentrySTDinterwordspacing

\bibitem{Tan_2020}
\BIBentryALTinterwordspacing
R.~Tan, J.~R. Ottewill, and N.~F. Thornhill, ``Monitoring statistics and tuning of kernel principal component analysis with radial basis function kernels,'' \emph{IEEE Access}, vol.~8, p. 198328–198342, 2020. [Online]. Available: \url{http://dx.doi.org/10.1109/access.2020.3034550}
\BIBentrySTDinterwordspacing

\bibitem{Kafai_2019}
\BIBentryALTinterwordspacing
M.~Kafai and K.~Eshghi, ``Croification: Accurate kernel classification with the efficiency of sparse linear svm,'' \emph{IEEE Transactions on Pattern Analysis and Machine Intelligence}, vol.~41, no.~1, p. 34–48, Jan. 2019. [Online]. Available: \url{http://dx.doi.org/10.1109/tpami.2017.2785313}
\BIBentrySTDinterwordspacing

\bibitem{Jo_2024}
\BIBentryALTinterwordspacing
H.~Jo, S.~Park, and S.~I. Cho, ``A survey of unsupervised learning-based out-of-distribution detection,'' in \emph{2024 IEEE International Conference on Consumer Electronics-Asia (ICCE-Asia)}.\hskip 1em plus 0.5em minus 0.4em\relax IEEE, Nov. 2024, p. 1–4. [Online]. Available: \url{http://dx.doi.org/10.1109/icce-asia63397.2024.10773891}
\BIBentrySTDinterwordspacing

\bibitem{Wu_2023}
\BIBentryALTinterwordspacing
X.~Wu, J.~Lu, Z.~Fang, and G.~Zhang, ``Meta ood learning for continuously adaptive ood detection,'' in \emph{2023 IEEE/CVF International Conference on Computer Vision (ICCV)}.\hskip 1em plus 0.5em minus 0.4em\relax IEEE, Oct. 2023, p. 19296–19307. [Online]. Available: \url{http://dx.doi.org/10.1109/iccv51070.2023.01773}
\BIBentrySTDinterwordspacing

\bibitem{Larue_2023}
\BIBentryALTinterwordspacing
N.~Larue, N.-S. Vu, V.~Struc, P.~Peer, and V.~Christophides, ``Seeable: Soft discrepancies and bounded contrastive learning for exposing deepfakes,'' in \emph{2023 IEEE/CVF International Conference on Computer Vision (ICCV)}.\hskip 1em plus 0.5em minus 0.4em\relax IEEE, Oct. 2023, p. 20954–20964. [Online]. Available: \url{http://dx.doi.org/10.1109/iccv51070.2023.01921}
\BIBentrySTDinterwordspacing

\bibitem{Zhao_2023}
\BIBentryALTinterwordspacing
X.~Zhao and Q.~Dou, ``Multi-level feature networks for out-of-distribution image detection,'' in \emph{International Conference on Computer, Artificial Intelligence, and Control Engineering (CAICE 2023)}, A.~Bhattacharjya and X.~Feng, Eds.\hskip 1em plus 0.5em minus 0.4em\relax SPIE, May 2023, p. 197. [Online]. Available: \url{http://dx.doi.org/10.1117/12.2681328}
\BIBentrySTDinterwordspacing

\bibitem{Zhu_2022}
\BIBentryALTinterwordspacing
Q.~Zhu, G.~Zheng, J.~Shen, and R.~Wang, ``Out-of-distribution detection based on feature fusion in neural network classifier pre-trained by pedcc-loss,'' \emph{IEEE Access}, vol.~10, p. 66190–66197, 2022. [Online]. Available: \url{http://dx.doi.org/10.1109/access.2022.3184694}
\BIBentrySTDinterwordspacing

\bibitem{_zg_r_2024}
\BIBentryALTinterwordspacing
O.~Özgür and Ä.~M. Baytas¸, ``Out-of-distribution detection with prototype similarity,'' in \emph{2024 IEEE International Conference on Knowledge Graph (ICKG)}.\hskip 1em plus 0.5em minus 0.4em\relax IEEE, Dec. 2024, p. 235–243. [Online]. Available: \url{http://dx.doi.org/10.1109/ickg63256.2024.00037}
\BIBentrySTDinterwordspacing

\bibitem{Mishra_2023}
\BIBentryALTinterwordspacing
D.~Mishra, H.~Zhao, P.~Saha, A.~T. Papageorghiou, and J.~A. Noble, \emph{Dual Conditioned Diffusion Models for Out-of-Distribution Detection: Application to Fetal Ultrasound Videos}.\hskip 1em plus 0.5em minus 0.4em\relax Springer Nature Switzerland, 2023, p. 216–226. [Online]. Available: \url{http://dx.doi.org/10.1007/978-3-031-43907-0_21}
\BIBentrySTDinterwordspacing

\bibitem{Hong_2024}
\BIBentryALTinterwordspacing
J.~Hong and S.~Kang, ``Score distillation for anomaly detection,'' \emph{Knowledge-Based Systems}, vol. 295, p. 111842, Jul. 2024. [Online]. Available: \url{http://dx.doi.org/10.1016/j.knosys.2024.111842}
\BIBentrySTDinterwordspacing

\bibitem{Deng_2022}
\BIBentryALTinterwordspacing
A.~Deng, A.~Goodge, L.~Y. Ang, and B.~Hooi, ``Cadet: Calibrated anomaly detection for mitigating hardness bias,'' in \emph{Proceedings of the Thirty-First International Joint Conference on Artificial Intelligence}, ser. IJCAI-2022.\hskip 1em plus 0.5em minus 0.4em\relax International Joint Conferences on Artificial Intelligence Organization, Jul. 2022, p. 2002–2008. [Online]. Available: \url{http://dx.doi.org/10.24963/ijcai.2022/278}
\BIBentrySTDinterwordspacing

\bibitem{Cao_2024}
\BIBentryALTinterwordspacing
Z.~Cao, Y.~Li, and B.-S. Shin, \emph{Attention-Guided Energy-Based Model for Out-of-Distribution Data Detection}.\hskip 1em plus 0.5em minus 0.4em\relax Springer Nature Switzerland, Dec. 2024, p. 1–15. [Online]. Available: \url{http://dx.doi.org/10.1007/978-3-031-78395-1_1}
\BIBentrySTDinterwordspacing

\bibitem{Guan_2023}
\BIBentryALTinterwordspacing
X.~Guan, Z.~Liu, W.-S. Zheng, Y.~Zhou, and R.~Wang, ``Revisit pca-based technique for out-of-distribution detection,'' in \emph{2023 IEEE/CVF International Conference on Computer Vision (ICCV)}.\hskip 1em plus 0.5em minus 0.4em\relax IEEE, Oct. 2023, p. 19374–19382. [Online]. Available: \url{http://dx.doi.org/10.1109/iccv51070.2023.01780}
\BIBentrySTDinterwordspacing

\bibitem{Domingues_2018}
\BIBentryALTinterwordspacing
R.~Domingues, P.~Michiardi, J.~Zouaoui, and M.~Filippone, ``Deep gaussian process autoencoders for novelty detection,'' \emph{Machine Learning}, vol. 107, no. 8–10, p. 1363–1383, Jun. 2018. [Online]. Available: \url{http://dx.doi.org/10.1007/s10994-018-5723-3}
\BIBentrySTDinterwordspacing

\bibitem{Yin_2023}
\BIBentryALTinterwordspacing
Q.~Yin, W.~Lu, B.~Li, and J.~Huang, ``Dynamic difference learning with spatio–temporal correlation for deepfake video detection,'' \emph{IEEE Transactions on Information Forensics and Security}, vol.~18, p. 4046–4058, 2023. [Online]. Available: \url{http://dx.doi.org/10.1109/tifs.2023.3290752}
\BIBentrySTDinterwordspacing

\bibitem{Moufidi_2024}
\BIBentryALTinterwordspacing
A.~Moufidi, D.~Rousseau, and P.~Rasti, ``Multimodal deepfake detection for short videos,'' in \emph{Proceedings of the 4th International Conference on Image Processing and Vision Engineering}.\hskip 1em plus 0.5em minus 0.4em\relax SCITEPRESS - Science and Technology Publications, 2024, p. 67–73. [Online]. Available: \url{http://dx.doi.org/10.5220/0012557300003720}
\BIBentrySTDinterwordspacing

\bibitem{Lu_2023}
\BIBentryALTinterwordspacing
Y.~Lu, C.~Liu, F.~Chang, H.~Liu, and H.~Huan, ``Jhpfa-net: Joint head pose and facial action network for driver yawning detection across arbitrary poses in videos,'' \emph{IEEE Transactions on Intelligent Transportation Systems}, vol.~24, no.~11, p. 11850–11863, Nov. 2023. [Online]. Available: \url{http://dx.doi.org/10.1109/tits.2023.3285923}
\BIBentrySTDinterwordspacing

\bibitem{Kopalidis_2024}
\BIBentryALTinterwordspacing
T.~Kopalidis, V.~Solachidis, N.~Vretos, and P.~Daras, ``Advances in facial expression recognition: A survey of methods, benchmarks, models, and datasets,'' \emph{Information}, vol.~15, no.~3, p. 135, Feb. 2024. [Online]. Available: \url{http://dx.doi.org/10.3390/info15030135}
\BIBentrySTDinterwordspacing

\bibitem{Zhang_2021}
\BIBentryALTinterwordspacing
D.~Zhang, C.~Li, F.~Lin, D.~Zeng, and S.~Ge, ``Detecting deepfake videos with temporal dropout 3dcnn,'' in \emph{Proceedings of the Thirtieth International Joint Conference on Artificial Intelligence}, ser. IJCAI-2021.\hskip 1em plus 0.5em minus 0.4em\relax International Joint Conferences on Artificial Intelligence Organization, Aug. 2021, p. 1288–1294. [Online]. Available: \url{http://dx.doi.org/10.24963/ijcai.2021/178}
\BIBentrySTDinterwordspacing

\bibitem{Zhu_2024}
\BIBentryALTinterwordspacing
Y.~Zhu, C.~Zhang, J.~Gao, X.~Sun, Z.~Rui, and X.~Zhou, ``High-compressed deepfake video detection with contrastive spatiotemporal distillation,'' \emph{Neurocomputing}, vol. 565, p. 126872, Jan. 2024. [Online]. Available: \url{http://dx.doi.org/10.1016/j.neucom.2023.126872}
\BIBentrySTDinterwordspacing

\bibitem{Hoppe_2018}
\BIBentryALTinterwordspacing
D.~Hoppe, S.~Helfmann, and C.~A. Rothkopf, ``Humans quickly learn to blink strategically in response to environmental task demands,'' \emph{Proceedings of the National Academy of Sciences}, vol. 115, no.~9, p. 2246–2251, Feb. 2018. [Online]. Available: \url{http://dx.doi.org/10.1073/pnas.1714220115}
\BIBentrySTDinterwordspacing

\bibitem{Li_2018}
\BIBentryALTinterwordspacing
Y.~Li, M.-C. Chang, and S.~Lyu, ``In ictu oculi: Exposing ai created fake videos by detecting eye blinking,'' in \emph{2018 IEEE International Workshop on Information Forensics and Security (WIFS)}.\hskip 1em plus 0.5em minus 0.4em\relax IEEE, Dec. 2018, p. 1–7. [Online]. Available: \url{http://dx.doi.org/10.1109/wifs.2018.8630787}
\BIBentrySTDinterwordspacing

\bibitem{Hirishikesh_2023}
\BIBentryALTinterwordspacing
P.~Hirishikesh, M.~Yaswanth, and H.~V. A, ``Speech to lip sync generation using deep learning algorithm,'' in \emph{2023 OITS International Conference on Information Technology (OCIT)}.\hskip 1em plus 0.5em minus 0.4em\relax IEEE, Dec. 2023, p. 426–431. [Online]. Available: \url{http://dx.doi.org/10.1109/ocit59427.2023.10430635}
\BIBentrySTDinterwordspacing

\bibitem{Bohacek_2024}
\BIBentryALTinterwordspacing
M.~Bohacek and H.~Farid, ``Lost in translation: Lip-sync deepfake detection from audio-video mismatch,'' in \emph{2024 IEEE/CVF Conference on Computer Vision and Pattern Recognition Workshops (CVPRW)}.\hskip 1em plus 0.5em minus 0.4em\relax IEEE, Jun. 2024, p. 4315–4323. [Online]. Available: \url{http://dx.doi.org/10.1109/cvprw63382.2024.00435}
\BIBentrySTDinterwordspacing

\bibitem{Becattini_2024}
\BIBentryALTinterwordspacing
F.~Becattini, C.~Bisogni, V.~Loia, C.~Pero, and F.~Hao, ``Head pose estimation patterns as deepfake detectors,'' \emph{ACM Transactions on Multimedia Computing, Communications, and Applications}, vol.~20, no.~11, p. 1–24, Sep. 2024. [Online]. Available: \url{http://dx.doi.org/10.1145/3612928}
\BIBentrySTDinterwordspacing

\bibitem{Gr_nquist_2024}
\BIBentryALTinterwordspacing
P.~Grönquist, Y.~Ren, Q.~He, A.~Verardo, and S.~Süsstrunk, ``Efficient temporally-aware deepfake detection using h.264 motion vectors,'' \emph{Electronic Imaging}, vol.~36, no.~4, pp. 335--1--335–9, Jan. 2024. [Online]. Available: \url{http://dx.doi.org/10.2352/ei.2024.36.4.mwsf-335}
\BIBentrySTDinterwordspacing

\bibitem{Merrill_2020}
\BIBentryALTinterwordspacing
N.~Merrill and A.~Eskandarian, ``Modified autoencoder training and scoring for robust unsupervised anomaly detection in deep learning,'' \emph{IEEE Access}, vol.~8, p. 101824–101833, 2020. [Online]. Available: \url{http://dx.doi.org/10.1109/access.2020.2997327}
\BIBentrySTDinterwordspacing

\bibitem{Iqbal_2022}
\BIBentryALTinterwordspacing
T.~Iqbal and S.~Qureshi, ``Reconstruction probability-based anomaly detection using variational auto-encoders,'' \emph{International Journal of Computers and Applications}, vol.~45, no.~3, p. 231–237, Nov. 2022. [Online]. Available: \url{http://dx.doi.org/10.1080/1206212x.2022.2143026}
\BIBentrySTDinterwordspacing

\bibitem{Graham_2023}
\BIBentryALTinterwordspacing
M.~S. Graham, W.~H.~L. Pinaya, P.-D. Tudosiu, P.~Nachev, S.~Ourselin, and M.~J. Cardoso, ``Denoising diffusion models for out-of-distribution detection,'' in \emph{2023 IEEE/CVF Conference on Computer Vision and Pattern Recognition Workshops (CVPRW)}.\hskip 1em plus 0.5em minus 0.4em\relax IEEE, Jun. 2023, p. 2948–2957. [Online]. Available: \url{http://dx.doi.org/10.1109/cvprw59228.2023.00296}
\BIBentrySTDinterwordspacing

\bibitem{Naidu_2024}
\BIBentryALTinterwordspacing
S.~M. Naidu and N.~Xiong, ``S2devfmap: Self-supervised learning framework with dual ensemble voting fusion for maximizing anomaly prediction in timeseries,'' in \emph{2024 9th International Conference on Machine Learning Technologies (ICMLT)}, ser. ICMLT 2024.\hskip 1em plus 0.5em minus 0.4em\relax ACM, May 2024, p. 49–57. [Online]. Available: \url{http://dx.doi.org/10.1145/3674029.3674038}
\BIBentrySTDinterwordspacing

\bibitem{Xiang_2010}
\BIBentryALTinterwordspacing
G.~Xiang and W.~Min, ``Applying semi-supervised cluster algorithm for anomaly detection,'' in \emph{2010 Third International Symposium on Information Processing}.\hskip 1em plus 0.5em minus 0.4em\relax IEEE, Oct. 2010, p. 43–45. [Online]. Available: \url{http://dx.doi.org/10.1109/isip.2010.68}
\BIBentrySTDinterwordspacing

\bibitem{Wadekar_2019}
\BIBentryALTinterwordspacing
A.~Wadekar, T.~Gupta, R.~Vijan, and F.~Kazi, ``Hybrid cae-vae for unsupervised anomaly detection in log file systems,'' in \emph{2019 10th International Conference on Computing, Communication and Networking Technologies (ICCCNT)}.\hskip 1em plus 0.5em minus 0.4em\relax IEEE, Jul. 2019, p. 1–7. [Online]. Available: \url{http://dx.doi.org/10.1109/icccnt45670.2019.8944863}
\BIBentrySTDinterwordspacing

\bibitem{Yu_2021}
\BIBentryALTinterwordspacing
N.~Yu, V.~Skripniuk, S.~Abdelnabi, and M.~Fritz, ``Artificial fingerprinting for generative models: Rooting deepfake attribution in training data,'' in \emph{2021 IEEE/CVF International Conference on Computer Vision (ICCV)}.\hskip 1em plus 0.5em minus 0.4em\relax IEEE, Oct. 2021. [Online]. Available: \url{http://dx.doi.org/10.1109/iccv48922.2021.01418}
\BIBentrySTDinterwordspacing

\bibitem{Lai_2025}
\BIBentryALTinterwordspacing
Z.~Lai, S.~Arif, C.~Feng, G.~Liao, and C.~Wang, ``Enhancing deepfake detection: Proactive forensics techniques using digital watermarking,'' \emph{Computers, Materials \& Continua}, vol.~82, no.~1, p. 73–102, 2025. [Online]. Available: \url{http://dx.doi.org/10.32604/cmc.2024.059370}
\BIBentrySTDinterwordspacing

\bibitem{Shang_2023}
\BIBentryALTinterwordspacing
Z.~Shang, H.~Zhang, P.~Zhang, L.~Wang, and T.~Li, ``Analysis and solution to aliasing artifacts in neural waveform generation models,'' \emph{Applied Acoustics}, vol. 203, p. 109183, Feb. 2023. [Online]. Available: \url{http://dx.doi.org/10.1016/j.apacoust.2022.109183}
\BIBentrySTDinterwordspacing

\bibitem{Korshunov_2022}
\BIBentryALTinterwordspacing
P.~Korshunov, A.~Jain, and S.~Marcel, ``Custom attribution loss for improving generalization and interpretability of deepfake detection,'' in \emph{ICASSP 2022 - 2022 IEEE International Conference on Acoustics, Speech and Signal Processing (ICASSP)}.\hskip 1em plus 0.5em minus 0.4em\relax IEEE, May 2022, p. 8972–8976. [Online]. Available: \url{http://dx.doi.org/10.1109/icassp43922.2022.9747628}
\BIBentrySTDinterwordspacing

\bibitem{muller2022}
\BIBentryALTinterwordspacing
N.~M. Müller, F.~Dieckmann, and J.~Williams, ``Attacker attribution of audio deepfakes,'' 2022. [Online]. Available: \url{https://arxiv.org/abs/2203.15563}
\BIBentrySTDinterwordspacing

\bibitem{cocchi2023unveiling}
F.~Cocchi, L.~Baraldi, S.~Poppi, M.~Cornia, L.~Baraldi, and R.~Cucchiara, ``Unveiling the impact of image transformations on deepfake detection: An experimental analysis,'' in \emph{International Conference on Image Analysis and Processing}.\hskip 1em plus 0.5em minus 0.4em\relax Springer, 2023, pp. 345--356.

\bibitem{Pons_2021}
\BIBentryALTinterwordspacing
J.~Pons, S.~Pascual, G.~Cengarle, and J.~Serra, ``Upsampling artifacts in neural audio synthesis,'' in \emph{ICASSP 2021 - 2021 IEEE International Conference on Acoustics, Speech and Signal Processing (ICASSP)}.\hskip 1em plus 0.5em minus 0.4em\relax IEEE, Jun. 2021, p. 3005–3009. [Online]. Available: \url{http://dx.doi.org/10.1109/icassp39728.2021.9414913}
\BIBentrySTDinterwordspacing

\bibitem{Sivabalamurugan_2024}
\BIBentryALTinterwordspacing
M.~Sivabalamurugan and T.~R. Swapna, ``Deepfake detection and classification using local surface geometrical features,'' in \emph{2024 IEEE International Conference on Computer Vision and Machine Intelligence (CVMI)}.\hskip 1em plus 0.5em minus 0.4em\relax IEEE, Oct. 2024, p. 1–6. [Online]. Available: \url{http://dx.doi.org/10.1109/cvmi61877.2024.10782175}
\BIBentrySTDinterwordspacing

\bibitem{Lim_2022}
\BIBentryALTinterwordspacing
N.~T. Lim, M.~Yi~Kuan, M.~Pu, M.~K. Lim, and C.~Yong~Chong, ``Metamorphic testing-based adversarial attack to fool deepfake detectors,'' in \emph{2022 26th International Conference on Pattern Recognition (ICPR)}.\hskip 1em plus 0.5em minus 0.4em\relax IEEE, Aug. 2022, p. 2503–2509. [Online]. Available: \url{http://dx.doi.org/10.1109/icpr56361.2022.9956543}
\BIBentrySTDinterwordspacing

\bibitem{Mi_2024}
\BIBentryALTinterwordspacing
Z.~Mi, X.~Jiang, T.~Sun, K.~Xu, Q.~Xu, and L.~Meng, \emph{Low-Quality Deepfake Video Detection Model Targeting Compression-Degraded Spatiotemporal Inconsistencies}.\hskip 1em plus 0.5em minus 0.4em\relax Springer Nature Singapore, 2024, p. 267–280. [Online]. Available: \url{http://dx.doi.org/10.1007/978-981-97-5606-3_23}
\BIBentrySTDinterwordspacing

\bibitem{Bedi_2018}
\BIBentryALTinterwordspacing
A.~K. Bedi, R.~K. Sunkaria, and S.~K. Randhawa, ``Local binary pattem variants: A review,'' in \emph{2018 First International Conference on Secure Cyber Computing and Communication (ICSCCC)}.\hskip 1em plus 0.5em minus 0.4em\relax IEEE, Dec. 2018, p. 234–237. [Online]. Available: \url{http://dx.doi.org/10.1109/icsccc.2018.8703326}
\BIBentrySTDinterwordspacing

\bibitem{Subash_Kumar_2018}
\BIBentryALTinterwordspacing
T.~G. Subash~Kumar and V.~Nagarajan, ``Local curve pattern for content-based image retrieval,'' \emph{Pattern Analysis and Applications}, vol.~22, no.~3, p. 1233–1242, Jul. 2018. [Online]. Available: \url{http://dx.doi.org/10.1007/s10044-018-0724-1}
\BIBentrySTDinterwordspacing

\bibitem{Wang_2013}
\BIBentryALTinterwordspacing
S.~Wang, X.~He, Q.~Wu, and J.~Yang, ``Generalized local n-ary patterns for texture classification,'' in \emph{2013 10th IEEE International Conference on Advanced Video and Signal Based Surveillance}.\hskip 1em plus 0.5em minus 0.4em\relax IEEE, Aug. 2013, p. 324–329. [Online]. Available: \url{http://dx.doi.org/10.1109/avss.2013.6636660}
\BIBentrySTDinterwordspacing

\bibitem{Zhai_2008}
\BIBentryALTinterwordspacing
L.~Zhai, S.~Dong, and H.~Ma, ``Recent methods and applications on image edge detection,'' in \emph{2008 International Workshop on Education Technology and Training \& 2008 International Workshop on Geoscience and Remote Sensing}.\hskip 1em plus 0.5em minus 0.4em\relax IEEE, Dec. 2008, p. 332–335. [Online]. Available: \url{http://dx.doi.org/10.1109/ettandgrs.2008.39}
\BIBentrySTDinterwordspacing

\bibitem{Selvakumar_2016}
\BIBentryALTinterwordspacing
P.~Selvakumar and S.~Hariganesh, ``The performance analysis of edge detection algorithms for image processing,'' in \emph{2016 International Conference on Computing Technologies and Intelligent Data Engineering (ICCTIDE'16)}.\hskip 1em plus 0.5em minus 0.4em\relax IEEE, Jan. 2016, p. 1–5. [Online]. Available: \url{http://dx.doi.org/10.1109/icctide.2016.7725371}
\BIBentrySTDinterwordspacing

\bibitem{Sahoo_2016}
\BIBentryALTinterwordspacing
T.~Sahoo and S.~Pine, ``Design and simulation of various edge detection techniques using matlab simulink,'' in \emph{2016 International Conference on Signal Processing, Communication, Power and Embedded System (SCOPES)}.\hskip 1em plus 0.5em minus 0.4em\relax IEEE, Oct. 2016, p. 1224–1228. [Online]. Available: \url{http://dx.doi.org/10.1109/scopes.2016.7955636}
\BIBentrySTDinterwordspacing

\bibitem{Sucharitha_2023}
\BIBentryALTinterwordspacing
G.~Sucharitha, B.~J.~D. Kalyani, G.~Chandra~Sekhar, and C.~Srividya, \emph{Efficient Image Retrieval Technique with Local Edge Binary Pattern Using Combined Color and Texture Features}.\hskip 1em plus 0.5em minus 0.4em\relax Springer Nature Singapore, 2023, p. 261–276. [Online]. Available: \url{http://dx.doi.org/10.1007/978-981-19-8493-8_21}
\BIBentrySTDinterwordspacing

\bibitem{Ghimire_2011}
\BIBentryALTinterwordspacing
D.~Ghimire and J.~Lee, \emph{Nonlinear Transfer Function-Based Image Detail Preserving Dynamic Range Compression for Color Image Enhancement}.\hskip 1em plus 0.5em minus 0.4em\relax Springer Berlin Heidelberg, 2011, p. 1–12. [Online]. Available: \url{http://dx.doi.org/10.1007/978-3-642-25367-6_1}
\BIBentrySTDinterwordspacing

\bibitem{Yu_2009}
\BIBentryALTinterwordspacing
H.~Yu, J.~Cao, Y.~Liu, and W.~Luo, ``Non-equal spacing division of hsv components for wood image retrieval,'' in \emph{2009 2nd International Congress on Image and Signal Processing}.\hskip 1em plus 0.5em minus 0.4em\relax IEEE, Oct. 2009, p. 1–3. [Online]. Available: \url{http://dx.doi.org/10.1109/cisp.2009.5303915}
\BIBentrySTDinterwordspacing

\bibitem{El_Rai_2020}
\BIBentryALTinterwordspacing
M.~C. El~Rai, H.~Al~Ahmad, O.~Gouda, D.~Jamal, M.~A. Talib, and Q.~Nasir, ``Fighting deepfake by residual noise using convolutional neural networks,'' in \emph{2020 3rd International Conference on Signal Processing and Information Security (ICSPIS)}.\hskip 1em plus 0.5em minus 0.4em\relax IEEE, Nov. 2020, p. 1–4. [Online]. Available: \url{http://dx.doi.org/10.1109/icspis51252.2020.9340138}
\BIBentrySTDinterwordspacing

\bibitem{Amerini_2022}
\BIBentryALTinterwordspacing
I.~Amerini, M.~Conti, P.~Giacomazzi, and L.~Pajola, ``Prana: Prnu-based technique to tell real and deepfake videos apart,'' in \emph{2022 International Joint Conference on Neural Networks (IJCNN)}.\hskip 1em plus 0.5em minus 0.4em\relax IEEE, Jul. 2022, p. 1–7. [Online]. Available: \url{http://dx.doi.org/10.1109/ijcnn55064.2022.9892413}
\BIBentrySTDinterwordspacing

\bibitem{Lugstein_2021}
\BIBentryALTinterwordspacing
F.~Lugstein, S.~Baier, G.~Bachinger, and A.~Uhl, ``Prnu-based deepfake detection,'' in \emph{Proceedings of the 2021 ACM Workshop on Information Hiding and Multimedia Security}, ser. IH\&MMSec '21.\hskip 1em plus 0.5em minus 0.4em\relax ACM, Jun. 2021. [Online]. Available: \url{http://dx.doi.org/10.1145/3437880.3460400}
\BIBentrySTDinterwordspacing

\bibitem{Laavanya_2018}
\BIBentryALTinterwordspacing
M.~Laavanya and M.~Karthikeyan, ``Dual tree complex wavelet transform incorporating svd and bilateral filter for image denoising,'' \emph{International Journal of Biomedical Engineering and Technology}, vol.~26, no. 3/4, p. 266, 2018. [Online]. Available: \url{http://dx.doi.org/10.1504/ijbet.2018.089956}
\BIBentrySTDinterwordspacing

\bibitem{Wahab_2020}
\BIBentryALTinterwordspacing
M.~F. Wahab and T.~C. O'Haver, ``Wavelet transforms in separation science for denoising and peak overlap detection,'' \emph{Journal of Separation Science}, vol.~43, no. 9–10, p. 1998–2010, Mar. 2020. [Online]. Available: \url{http://dx.doi.org/10.1002/jssc.202000013}
\BIBentrySTDinterwordspacing

\bibitem{Zhan_2015}
\BIBentryALTinterwordspacing
L.~Zhan and Y.~Zhu, ``Passive forensics for image splicing based on pca noise estimation,'' in \emph{2015 10th International Conference for Internet Technology and Secured Transactions (ICITST)}.\hskip 1em plus 0.5em minus 0.4em\relax IEEE, Dec. 2015, p. 78–83. [Online]. Available: \url{http://dx.doi.org/10.1109/icitst.2015.7412062}
\BIBentrySTDinterwordspacing

\bibitem{Guo_2019}
\BIBentryALTinterwordspacing
T.~Guo, C.~Xu, B.~Shi, C.~Xu, and D.~Tao, ``Smooth deep image generator from noises,'' \emph{Proceedings of the AAAI Conference on Artificial Intelligence}, vol.~33, no.~01, p. 3731–3738, Jul. 2019. [Online]. Available: \url{http://dx.doi.org/10.1609/aaai.v33i01.33013731}
\BIBentrySTDinterwordspacing

\bibitem{Gao_2024}
\BIBentryALTinterwordspacing
J.~Gao, Z.~Xia, G.~L. Marcialis, C.~Dang, J.~Dai, and X.~Feng, ``Deepfake detection based on high-frequency enhancement network for highly compressed content,'' \emph{Expert Systems with Applications}, vol. 249, p. 123732, Sep. 2024. [Online]. Available: \url{http://dx.doi.org/10.1016/j.eswa.2024.123732}
\BIBentrySTDinterwordspacing

\bibitem{Liang_2023}
\BIBentryALTinterwordspacing
W.~Liang, Y.~Wu, J.~Wu, and J.~Xu, ``Faclue: Exploring frequency clues by adaptive frequency-attention for deepfake detection,'' in \emph{2023 42nd Chinese Control Conference (CCC)}.\hskip 1em plus 0.5em minus 0.4em\relax IEEE, Jul. 2023, p. 7621–7626. [Online]. Available: \url{http://dx.doi.org/10.23919/ccc58697.2023.10240940}
\BIBentrySTDinterwordspacing

\bibitem{Se_k__2017}
\BIBentryALTinterwordspacing
P.~Sečkář and D.~Svoboda, ``Ftutor: An interactive guide to the fundamentals of frequency analysis,'' \emph{Computer Applications in Engineering Education}, vol.~25, no.~3, p. 508–520, Apr. 2017. [Online]. Available: \url{http://dx.doi.org/10.1002/cae.21817}
\BIBentrySTDinterwordspacing

\bibitem{Vartiainen_2006}
\BIBentryALTinterwordspacing
J.~Vartiainen, A.~Sadovnikov, L.~Lensu, J.-K. Kamarainen, and H.~Kalviainen, ``Detecting irregularities in regular patterns,'' in \emph{18th International Conference on Pattern Recognition (ICPR'06)}.\hskip 1em plus 0.5em minus 0.4em\relax IEEE, 2006, p. 926–929. [Online]. Available: \url{http://dx.doi.org/10.1109/icpr.2006.432}
\BIBentrySTDinterwordspacing

\bibitem{Bovik_2009}
\BIBentryALTinterwordspacing
A.~C. Bovik, \emph{Basic Tools for Image Fourier Analysis}.\hskip 1em plus 0.5em minus 0.4em\relax Elsevier, 2009, p. 97–121. [Online]. Available: \url{http://dx.doi.org/10.1016/b978-0-12-374457-9.00005-6}
\BIBentrySTDinterwordspacing

\bibitem{Shang_2024}
\BIBentryALTinterwordspacing
K.~Shang, M.~Shao, Y.~Qiao, and H.~Liu, ``Frequency-aware network for low-light image enhancement,'' \emph{Computers \& Graphics}, vol. 118, p. 210–219, Feb. 2024. [Online]. Available: \url{http://dx.doi.org/10.1016/j.cag.2023.12.014}
\BIBentrySTDinterwordspacing

\bibitem{Li_2024}
\BIBentryALTinterwordspacing
X.~Li, H.~Yang, C.~Sun, and H.~Li, ``A high-low frequency guided generative adversarial network for shadow removal,'' in \emph{Fifth International Conference on Computer Vision and Data Mining (ICCVDM 2024)}, X.~Zhang and M.~Yin, Eds.\hskip 1em plus 0.5em minus 0.4em\relax SPIE, Oct. 2024, p.~50. [Online]. Available: \url{http://dx.doi.org/10.1117/12.3048153}
\BIBentrySTDinterwordspacing

\bibitem{Li_2022}
\BIBentryALTinterwordspacing
Z.~Li, P.~Xia, X.~Rui, and B.~Li, ``Exploring the effect of high-frequency components in gans training,'' \emph{ACM Transactions on Multimedia Computing, Communications, and Applications}, vol.~19, no.~5, p. 1–22, Sep. 2022. [Online]. Available: \url{http://dx.doi.org/10.1145/3578585}
\BIBentrySTDinterwordspacing

\bibitem{Hosseinimakarem_2016}
\BIBentryALTinterwordspacing
Z.~Hosseinimakarem, A.~D. Davies, and C.~J. Evans, ``Zernike polynomials for mid-spatial frequency representation on optical surfaces,'' in \emph{Reflection, Scattering, and Diffraction from Surfaces V}, L.~M. Hanssen, Ed., vol. 9961.\hskip 1em plus 0.5em minus 0.4em\relax SPIE, Sep. 2016, p. 99610P. [Online]. Available: \url{http://dx.doi.org/10.1117/12.2236520}
\BIBentrySTDinterwordspacing

\bibitem{Jung_2021}
\BIBentryALTinterwordspacing
S.~Jung and M.~Keuper, ``Spectral distribution aware image generation,'' \emph{Proceedings of the AAAI Conference on Artificial Intelligence}, vol.~35, no.~2, p. 1734–1742, May 2021. [Online]. Available: \url{http://dx.doi.org/10.1609/aaai.v35i2.16267}
\BIBentrySTDinterwordspacing

\bibitem{Kolekar_2014}
\BIBentryALTinterwordspacing
M.~K.~H. Kolekar, G.~L. Raja, and S.~Sengupta, \emph{An Introduction to Wavelet-Based Image Processing and its Applications}.\hskip 1em plus 0.5em minus 0.4em\relax IGI Global, 2014, p. 38–53. [Online]. Available: \url{http://dx.doi.org/10.4018/978-1-4666-4558-5.ch003}
\BIBentrySTDinterwordspacing

\bibitem{Carvajal_Gamez_2016}
\BIBentryALTinterwordspacing
B.~E. Carvajal-Gamez, A.~E. Moreno-Cervantes, F.~J. Gallegos-Funes, and V.~Ponomaryov, ``Image reconstruction in complex media by the fourth moment wavelet,'' in \emph{2016 9th International Kharkiv Symposium on Physics and Engineering of Microwaves, Millimeter and Submillimeter Waves (MSMW)}.\hskip 1em plus 0.5em minus 0.4em\relax IEEE, Jun. 2016, p. 1–3. [Online]. Available: \url{http://dx.doi.org/10.1109/msmw.2016.7538151}
\BIBentrySTDinterwordspacing

\bibitem{Lee_2024}
\BIBentryALTinterwordspacing
S.~Lee, S.-W. Jung, and H.~Seo, ``Spectrum translation for refinement of image generation (stig) based on contrastive learning and spectral filter profile,'' \emph{Proceedings of the AAAI Conference on Artificial Intelligence}, vol.~38, no.~4, p. 2929–2937, Mar. 2024. [Online]. Available: \url{http://dx.doi.org/10.1609/aaai.v38i4.28074}
\BIBentrySTDinterwordspacing

\bibitem{Hsiao_2005}
\BIBentryALTinterwordspacing
W.~H. Hsiao and R.~P. Millane, ``Effects of occlusion, edges, and scaling on the power spectra of natural images,'' \emph{Journal of the Optical Society of America A}, vol.~22, no.~9, p. 1789, Sep. 2005. [Online]. Available: \url{http://dx.doi.org/10.1364/josaa.22.001789}
\BIBentrySTDinterwordspacing

\bibitem{Kaushik_2024}
\BIBentryALTinterwordspacing
B.~Kaushik, K.~Vinay, and M.~Singh, \emph{Unveiling Deepfakes: Convolutional Neural Networks for Detection}.\hskip 1em plus 0.5em minus 0.4em\relax Springer Nature Switzerland, 2024, p. 68–77. [Online]. Available: \url{http://dx.doi.org/10.1007/978-3-031-64836-6_7}
\BIBentrySTDinterwordspacing

\bibitem{Stephen_2022}
\BIBentryALTinterwordspacing
D.~Stephen and T.~Mantoro, ``Usage of convolutional neural network for deepfake video detection with face-swapping technique,'' in \emph{2022 5th International Conference of Computer and Informatics Engineering (IC2IE)}.\hskip 1em plus 0.5em minus 0.4em\relax IEEE, Sep. 2022, p. 22–28. [Online]. Available: \url{http://dx.doi.org/10.1109/ic2ie56416.2022.9970025}
\BIBentrySTDinterwordspacing

\bibitem{Kumar_2025}
\BIBentryALTinterwordspacing
A.~B. Kumar and K.~Sanjaya, \emph{Ethics, Algorithms, and the Rules of Evidence: New Era of AI-Driven Forensics}.\hskip 1em plus 0.5em minus 0.4em\relax IGI Global, Apr. 2025, p. 103–124. [Online]. Available: \url{http://dx.doi.org/10.4018/979-8-3693-9405-2.ch006}
\BIBentrySTDinterwordspacing

\bibitem{Hung_2023}
\BIBentryALTinterwordspacing
R.-J. Hung, C.-C. Hsu, and J.-H. Ho, ``Malicious traffic blocking mechanism and protection based on dns,'' in \emph{2023 IEEE 5th Eurasia Conference on IOT, Communication and Engineering (ECICE)}.\hskip 1em plus 0.5em minus 0.4em\relax IEEE, Oct. 2023, p. 88–91. [Online]. Available: \url{http://dx.doi.org/10.1109/ecice59523.2023.10383095}
\BIBentrySTDinterwordspacing

\bibitem{Zang_2023}
\BIBentryALTinterwordspacing
Z.~Zang, Y.~Xu, L.~Lu, Y.~Geng, S.~Yang, and S.~Z. Li, ``Udrn: Unified dimensional reduction neural network for feature selection and feature projection,'' \emph{Neural Networks}, vol. 161, p. 626–637, Apr. 2023. [Online]. Available: \url{http://dx.doi.org/10.1016/j.neunet.2023.02.018}
\BIBentrySTDinterwordspacing

\bibitem{Ganguly_2024}
\BIBentryALTinterwordspacing
R.~Ganguly, M.~D. Bah, and M.~Dahmane, \emph{Diffusion Models as a Representation Learner for Deepfake Image Detection}.\hskip 1em plus 0.5em minus 0.4em\relax Springer Nature Switzerland, Dec. 2024, p. 228–241. [Online]. Available: \url{http://dx.doi.org/10.1007/978-3-031-78305-0_15}
\BIBentrySTDinterwordspacing

\bibitem{Bodria_2022}
\BIBentryALTinterwordspacing
F.~Bodria, R.~Guidotti, F.~Giannotti, and D.~Pedreschi, \emph{Interpretable Latent Space to Enable Counterfactual Explanations}.\hskip 1em plus 0.5em minus 0.4em\relax Springer Nature Switzerland, 2022, p. 525–540. [Online]. Available: \url{http://dx.doi.org/10.1007/978-3-031-18840-4_37}
\BIBentrySTDinterwordspacing

\bibitem{Jha_2024}
\BIBentryALTinterwordspacing
R.~Jha, M.~K. Khirwal, and V.~Bhattacharjee, ``Feature vector learning for images with pretrained models,'' in \emph{2024 First International Conference on Innovations in Communications, Electrical and Computer Engineering (ICICEC)}.\hskip 1em plus 0.5em minus 0.4em\relax IEEE, Oct. 2024, p. 1–5. [Online]. Available: \url{http://dx.doi.org/10.1109/icicec62498.2024.10808740}
\BIBentrySTDinterwordspacing

\bibitem{Shanckin_2023}
\BIBentryALTinterwordspacing
S.~Shanckin, Mayank, A.~Singh, and R.~Patwadi, ``Unveiling latent spaces with variational autoencoders,'' in \emph{2023 IEEE International Conference on Service Operations and Logistics, and Informatics (SOLI)}.\hskip 1em plus 0.5em minus 0.4em\relax IEEE, Dec. 2023, p. 1–5. [Online]. Available: \url{http://dx.doi.org/10.1109/soli60636.2023.10425383}
\BIBentrySTDinterwordspacing

\bibitem{Fu_2021}
\BIBentryALTinterwordspacing
S.~Fu, S.~Zhong, L.~Lin, and M.~Zhao, ``A re-optimized deep auto-encoder for gas turbine unsupervised anomaly detection,'' \emph{Engineering Applications of Artificial Intelligence}, vol. 101, p. 104199, May 2021. [Online]. Available: \url{http://dx.doi.org/10.1016/j.engappai.2021.104199}
\BIBentrySTDinterwordspacing

\bibitem{Peng_2024}
\BIBentryALTinterwordspacing
C.~Peng, T.~Chen, D.~Liu, Y.~Zheng, and N.~Wang, \emph{Spatial-Frequency Dual-Stream Reconstruction for Deepfake Detection}.\hskip 1em plus 0.5em minus 0.4em\relax Springer Nature Singapore, Nov. 2024, p. 473–487. [Online]. Available: \url{http://dx.doi.org/10.1007/978-981-97-8795-1_32}
\BIBentrySTDinterwordspacing

\bibitem{Ballabio_2015}
\BIBentryALTinterwordspacing
D.~Ballabio, ``A matlab toolbox for principal component analysis and unsupervised exploration of data structure,'' \emph{Chemometrics and Intelligent Laboratory Systems}, vol. 149, p. 1–9, Dec. 2015. [Online]. Available: \url{http://dx.doi.org/10.1016/j.chemolab.2015.10.003}
\BIBentrySTDinterwordspacing

\bibitem{Bartholomew_2010}
\BIBentryALTinterwordspacing
D.~Bartholomew, \emph{Principal Components Analysis}.\hskip 1em plus 0.5em minus 0.4em\relax Elsevier, 2010, p. 374–377. [Online]. Available: \url{http://dx.doi.org/10.1016/b978-0-08-044894-7.01358-0}
\BIBentrySTDinterwordspacing

\bibitem{Rice_2017}
\BIBentryALTinterwordspacing
I.~Rice, ``Γ-stochastic neighbour embedding for feed-forward data visualization,'' \emph{Information Visualization}, vol.~17, no.~4, p. 306–315, Jul. 2017. [Online]. Available: \url{http://dx.doi.org/10.1177/1473871617715212}
\BIBentrySTDinterwordspacing

\bibitem{Jung_2024}
\BIBentryALTinterwordspacing
S.~Jung, T.~Dagobert, J.-M. Morel, and G.~Facciolo, ``A review of t-sne,'' \emph{Image Processing On Line}, vol.~14, p. 250–270, Oct. 2024. [Online]. Available: \url{http://dx.doi.org/10.5201/ipol.2024.528}
\BIBentrySTDinterwordspacing

\bibitem{Myasnikov_2020}
\BIBentryALTinterwordspacing
E.~Myasnikov, ``Using umap for dimensionality reduction of hyperspectral data,'' in \emph{2020 International Multi-Conference on Industrial Engineering and Modern Technologies (FarEastCon)}.\hskip 1em plus 0.5em minus 0.4em\relax IEEE, Oct. 2020, p. 1–5. [Online]. Available: \url{http://dx.doi.org/10.1109/fareastcon50210.2020.9271656}
\BIBentrySTDinterwordspacing

\bibitem{Healy_2024}
\BIBentryALTinterwordspacing
J.~Healy and L.~McInnes, ``Uniform manifold approximation and projection,'' \emph{Nature Reviews Methods Primers}, vol.~4, no.~1, Nov. 2024. [Online]. Available: \url{http://dx.doi.org/10.1038/s43586-024-00363-x}
\BIBentrySTDinterwordspacing

\bibitem{L__2010}
\BIBentryALTinterwordspacing
S.~L., J.~B. Simha, and R.~Veluru, ``Generating optimum number of clusters using median search and projection algorithms,'' in \emph{2010 IEEE 24th International Conference on Advanced Information Networking and Applications Workshops}.\hskip 1em plus 0.5em minus 0.4em\relax IEEE, 2010, p. 97–102. [Online]. Available: \url{http://dx.doi.org/10.1109/waina.2010.196}
\BIBentrySTDinterwordspacing

\bibitem{Peng_2016}
\BIBentryALTinterwordspacing
C.~Peng, S.~Guoyou, L.~Shuang, and Y.~Jian, ``An incremental density based spatial clustering of application with noise algorithm based on partition index,'' \emph{Journal of Computational and Theoretical Nanoscience}, vol.~13, no.~12, p. 10273–10280, Dec. 2016. [Online]. Available: \url{http://dx.doi.org/10.1166/jctn.2016.6104}
\BIBentrySTDinterwordspacing

\bibitem{Nguyen_2024}
\BIBentryALTinterwordspacing
V.~Q. Nguyen, L.~T. Ngo, L.~M. Nguyen, V.~H. Nguyen, and N.~Shone, ``Deep clustering hierarchical latent representation for anomaly-based cyber-attack detection,'' \emph{Knowledge-Based Systems}, vol. 301, p. 112366, Oct. 2024. [Online]. Available: \url{http://dx.doi.org/10.1016/j.knosys.2024.112366}
\BIBentrySTDinterwordspacing

\bibitem{Ueda_2020}
\BIBentryALTinterwordspacing
T.~Ueda, Y.~Tohsato, and I.~Nishikawa, \emph{Temporal Anomaly Detection by Deep Generative Models with Applications to Biological Data}.\hskip 1em plus 0.5em minus 0.4em\relax Springer International Publishing, 2020, p. 553–565. [Online]. Available: \url{http://dx.doi.org/10.1007/978-3-030-61609-0_44}
\BIBentrySTDinterwordspacing

\bibitem{Yu_2019}
\BIBentryALTinterwordspacing
N.~Yu, L.~Davis, and M.~Fritz, ``Attributing fake images to gans: Learning and analyzing gan fingerprints,'' in \emph{2019 IEEE/CVF International Conference on Computer Vision (ICCV)}.\hskip 1em plus 0.5em minus 0.4em\relax IEEE, Oct. 2019, p. 7555–7565. [Online]. Available: \url{http://dx.doi.org/10.1109/iccv.2019.00765}
\BIBentrySTDinterwordspacing

\bibitem{Jeong_2022}
\BIBentryALTinterwordspacing
Y.~Jeong, D.~Kim, Y.~Ro, P.~Kim, and J.~Choi, \emph{FingerprintNet: Synthesized Fingerprints for Generated Image Detection}.\hskip 1em plus 0.5em minus 0.4em\relax Springer Nature Switzerland, 2022, p. 76–94. [Online]. Available: \url{http://dx.doi.org/10.1007/978-3-031-19781-9_5}
\BIBentrySTDinterwordspacing

\bibitem{Maho_2023}
\BIBentryALTinterwordspacing
T.~Maho, T.~Furon, and E.~L. Merrer, ``Fingerprinting classifiers with benign inputs,'' \emph{IEEE Transactions on Information Forensics and Security}, vol.~18, p. 5459–5472, 2023. [Online]. Available: \url{http://dx.doi.org/10.1109/tifs.2023.3301268}
\BIBentrySTDinterwordspacing

\bibitem{El_Kaddoury_2019}
\BIBentryALTinterwordspacing
M.~El-Kaddoury, A.~Mahmoudi, and M.~M. Himmi, \emph{Deep Generative Models for Image Generation: A Practical Comparison Between Variational Autoencoders and Generative Adversarial Networks}.\hskip 1em plus 0.5em minus 0.4em\relax Springer International Publishing, 2019, p. 1–8. [Online]. Available: \url{http://dx.doi.org/10.1007/978-3-030-22885-9_1}
\BIBentrySTDinterwordspacing

\bibitem{Laptev_2021}
\BIBentryALTinterwordspacing
V.~V. Laptev, O.~M. Gerget, and N.~A. Markova, \emph{Generative Models Based on VAE and GAN for New Medical Data Synthesis}.\hskip 1em plus 0.5em minus 0.4em\relax Springer International Publishing, 2021, p. 217–226. [Online]. Available: \url{http://dx.doi.org/10.1007/978-3-030-63563-3_17}
\BIBentrySTDinterwordspacing

\bibitem{Han_2025}
\BIBentryALTinterwordspacing
D.~Han and A.~P. Chandrakasan, ``Mega.mini: A universal generative ai processor with a new big/little core architecture for npu,'' in \emph{2025 IEEE International Solid-State Circuits Conference (ISSCC)}.\hskip 1em plus 0.5em minus 0.4em\relax IEEE, Feb. 2025, p. 1–3. [Online]. Available: \url{http://dx.doi.org/10.1109/isscc49661.2025.10904514}
\BIBentrySTDinterwordspacing

\bibitem{Ahmed_2023}
\BIBentryALTinterwordspacing
O.~Ahmed, S.~Fekry, A.~Metwally, A.~Abdelwahid, D.~Khattab, A.~Abdulelmagid, and R.~Hossieny, ``Deepfake detection system using deep learning,'' in \emph{2023 Eleventh International Conference on Intelligent Computing and Information Systems (ICICIS)}.\hskip 1em plus 0.5em minus 0.4em\relax IEEE, Nov. 2023, p. 502–508. [Online]. Available: \url{http://dx.doi.org/10.1109/icicis58388.2023.10391147}
\BIBentrySTDinterwordspacing

\bibitem{Guo_2024}
\BIBentryALTinterwordspacing
H.~Guo, C.~Sun, J.~Zhang, W.~Zhang, and N.~Zhang, ``Mmyfnet: Multi-modality yolo fusion network for object detection in remote sensing images,'' \emph{Remote Sensing}, vol.~16, no.~23, p. 4451, Nov. 2024. [Online]. Available: \url{http://dx.doi.org/10.3390/rs16234451}
\BIBentrySTDinterwordspacing

\bibitem{2021}
\BIBentryALTinterwordspacing
N.~G. Bach, L.~H. Hoang, and T.~H. Hai, ``Improvement of k-nearest neighbors (knn) algorithm for network intrusion detection using shannon-entropy,'' \emph{Journal of Communications}, p. 347–354, 2021. [Online]. Available: \url{http://dx.doi.org/10.12720/jcm.16.8.347-354}
\BIBentrySTDinterwordspacing

\bibitem{Wu_2022}
\BIBentryALTinterwordspacing
C.~Wu, B.~Wang, Z.~Yang, W.~Nai, Y.~Xing, Z.~Wang, and Y.~Lin, ``t-sne based on sobol sequence initialized exchange market algorithm,'' in \emph{2022 IEEE 10th Joint International Information Technology and Artificial Intelligence Conference (ITAIC)}.\hskip 1em plus 0.5em minus 0.4em\relax IEEE, Jun. 2022, p. 2498–2502. [Online]. Available: \url{http://dx.doi.org/10.1109/itaic54216.2022.9836613}
\BIBentrySTDinterwordspacing

\bibitem{Javaloy_2020}
\BIBentryALTinterwordspacing
A.~Javaloy and G.~García-Mateos, ``Preliminary results on different text processing tasks using encoder-decoder networks and the causal feature extractor,'' \emph{Applied Sciences}, vol.~10, no.~17, p. 5772, Aug. 2020. [Online]. Available: \url{http://dx.doi.org/10.3390/app10175772}
\BIBentrySTDinterwordspacing

\bibitem{Thuan_2024}
\BIBentryALTinterwordspacing
P.~M. Thuan, B.~T. Lam, and P.~D. Trung, ``Spatial vision transformer: A novel approach to deepfake video detection,'' in \emph{2024 1st International Conference On Cryptography And Information Security (VCRIS)}.\hskip 1em plus 0.5em minus 0.4em\relax IEEE, Dec. 2024, p. 1–6. [Online]. Available: \url{http://dx.doi.org/10.1109/vcris63677.2024.10813391}
\BIBentrySTDinterwordspacing

\bibitem{Le_2024}
\BIBentryALTinterwordspacing
D.~P.~C. Le, D.~Wang, and V.-T. Le, ``A comprehensive survey of recent transformers in image, video and diffusion models,'' \emph{Computers, Materials \& Continua}, vol.~80, no.~1, p. 37–60, 2024. [Online]. Available: \url{http://dx.doi.org/10.32604/cmc.2024.050790}
\BIBentrySTDinterwordspacing

\bibitem{Kernbach_2021}
\BIBentryALTinterwordspacing
J.~M. Kernbach and V.~E. Staartjes, \emph{Foundations of Machine Learning-Based Clinical Prediction Modeling: Part II—Generalization and Overfitting}.\hskip 1em plus 0.5em minus 0.4em\relax Springer International Publishing, Dec. 2021, p. 15–21. [Online]. Available: \url{http://dx.doi.org/10.1007/978-3-030-85292-4_3}
\BIBentrySTDinterwordspacing

\bibitem{Seifi_2023}
\BIBentryALTinterwordspacing
F.~Seifi and S.~T.~A. Niaki, ``Extending the hypergradient descent technique to reduce the time of optimal solution achieved in hyperparameter optimization algorithms,'' \emph{International Journal of Industrial Engineering Computations}, vol.~14, no.~3, p. 501–510, 2023. [Online]. Available: \url{http://dx.doi.org/10.5267/j.ijiec.2023.4.004}
\BIBentrySTDinterwordspacing

\bibitem{115}
\BIBentryALTinterwordspacing
D.~Cozzolino, J.~Thies, A.~Rössler, C.~Riess, M.~Nießner, and L.~Verdoliva, ``Forensictransfer: Weakly-supervised domain adaptation for forgery detection,'' 2019. [Online]. Available: \url{https://arxiv.org/abs/1812.02510}
\BIBentrySTDinterwordspacing

\bibitem{Huo_2019}
\BIBentryALTinterwordspacing
G.~Huo, Q.~Zhang, H.~Guo, W.~Li, and Y.~Zhang, \emph{Multi-source Heterogeneous Iris Recognition Using Locality Preserving Projection}.\hskip 1em plus 0.5em minus 0.4em\relax Springer International Publishing, 2019, p. 304–311. [Online]. Available: \url{http://dx.doi.org/10.1007/978-3-030-31456-9_34}
\BIBentrySTDinterwordspacing

\bibitem{Khalid_2020}
\BIBentryALTinterwordspacing
H.~Khalid and S.~S. Woo, ``Oc-fakedect: Classifying deepfakes using one-class variational autoencoder,'' in \emph{2020 IEEE/CVF Conference on Computer Vision and Pattern Recognition Workshops (CVPRW)}.\hskip 1em plus 0.5em minus 0.4em\relax IEEE, Jun. 2020, p. 2794–2803. [Online]. Available: \url{http://dx.doi.org/10.1109/cvprw50498.2020.00336}
\BIBentrySTDinterwordspacing

\bibitem{Yang_2021}
\BIBentryALTinterwordspacing
J.~Yang, S.~Xiao, A.~Li, G.~Lan, and H.~Wang, ``Detecting fake images by identifying potential texture difference,'' \emph{Future Generation Computer Systems}, vol. 125, p. 127–135, Dec. 2021. [Online]. Available: \url{http://dx.doi.org/10.1016/j.future.2021.06.043}
\BIBentrySTDinterwordspacing

\bibitem{96}
\BIBentryALTinterwordspacing
L.~Nataraj, T.~M. Mohammed, S.~Chandrasekaran, A.~Flenner, J.~H. Bappy, A.~K. Roy-Chowdhury, and B.~S. Manjunath, ``Detecting gan generated fake images using co-occurrence matrices,'' 2019. [Online]. Available: \url{https://arxiv.org/abs/1903.06836}
\BIBentrySTDinterwordspacing

\bibitem{9010964}
N.~Yu, L.~Davis, and M.~Fritz, ``Attributing fake images to gans: Learning and analyzing gan fingerprints,'' in \emph{2019 IEEE/CVF International Conference on Computer Vision (ICCV)}, 2019, pp. 7555--7565.

\bibitem{8803661}
S.~McCloskey and M.~Albright, ``Detecting gan-generated imagery using saturation cues,'' in \emph{2019 IEEE International Conference on Image Processing (ICIP)}, 2019, pp. 4584--4588.

\bibitem{9141516}
U.~A. Ciftci, I.~Demir, and L.~Yin, ``Fakecatcher: Detection of synthetic portrait videos using biological signals,'' \emph{IEEE Transactions on Pattern Analysis and Machine Intelligence}, pp. 1--1, 2020.

\bibitem{9577744}
H.~Liu, X.~Li, W.~Zhou, Y.~Chen, Y.~He, H.~Xue, W.~Zhang, and N.~Yu, ``Spatial-phase shallow learning: Rethinking face forgery detection in frequency domain,'' in \emph{2021 IEEE/CVF Conference on Computer Vision and Pattern Recognition (CVPR)}, 2021, pp. 772--781.

\bibitem{99}
\BIBentryALTinterwordspacing
P.~Prajapati and C.~Pollett, ``Mri-gan: A generalized approach to detect deepfakes using perceptual image assessment,'' 2022. [Online]. Available: \url{https://arxiv.org/abs/2203.00108}
\BIBentrySTDinterwordspacing

\bibitem{131}
\BIBentryALTinterwordspacing
S.~Bounareli, V.~Argyriou, and G.~Tzimiropoulos, ``Finding directions in gan's latent space for neural face reenactment,'' 2022. [Online]. Available: \url{https://arxiv.org/abs/2202.00046}
\BIBentrySTDinterwordspacing

\bibitem{10030936}
B.~Liu, B.~Liu, M.~Ding, T.~Zhu, and X.~Yu, ``Ti2net: Temporal identity inconsistency network for deepfake detection,'' in \emph{2023 IEEE/CVF Winter Conference on Applications of Computer Vision (WACV)}, 2023, pp. 4680--4689.

\bibitem{Huang_2024}
\BIBentryALTinterwordspacing
Y.~Huang, F.~Juefei-Xu, Q.~Guo, Y.~Liu, and G.~Pu, ``Dodging deepfake detection via implicit spatial-domain notch filtering,'' \emph{IEEE Transactions on Circuits and Systems for Video Technology}, vol.~34, no.~8, p. 6949–6962, Aug. 2024. [Online]. Available: \url{http://dx.doi.org/10.1109/tcsvt.2023.3325427}
\BIBentrySTDinterwordspacing

\bibitem{140}
\BIBentryALTinterwordspacing
B.~Wang, X.~Wu, F.~Wang, Y.~Zhang, F.~Wei, and Z.~Song, ``Spatial-frequency feature fusion based deepfake detection through knowledge distillation,'' \emph{Engineering Applications of Artificial Intelligence}, vol. 133, p. 108341, Jul. 2024. [Online]. Available: \url{http://dx.doi.org/10.1016/j.engappai.2024.108341}
\BIBentrySTDinterwordspacing

\bibitem{161}
\BIBentryALTinterwordspacing
C.~Tan, Y.~Zhao, S.~Wei, G.~Gu, P.~Liu, and Y.~Wei, ``Frequency-aware deepfake detection: Improving generalizability through frequency space learning,'' 2024. [Online]. Available: \url{https://arxiv.org/abs/2403.07240}
\BIBentrySTDinterwordspacing

\bibitem{236}
\BIBentryALTinterwordspacing
M.~Bonomi, C.~Pasquini, and G.~Boato, ``Dynamic texture analysis for detecting fake faces in video sequences,'' 2020. [Online]. Available: \url{https://arxiv.org/abs/2007.15271}
\BIBentrySTDinterwordspacing

\bibitem{Kaddar_2024}
\BIBentryALTinterwordspacing
B.~Kaddar, S.~A. Fezza, Z.~Akhtar, W.~Hamidouche, A.~Hadid, and J.~Serra-Sagristá, ``Deepfake detection using spatiotemporal transformer,'' \emph{ACM Transactions on Multimedia Computing, Communications, and Applications}, vol.~20, no.~11, p. 1–21, Sep. 2024. [Online]. Available: \url{http://dx.doi.org/10.1145/3643030}
\BIBentrySTDinterwordspacing

\bibitem{Yang_2024}
\BIBentryALTinterwordspacing
Y.~Yang, N.~B. Idris, C.~Liu, H.~Wu, and D.~Yu, ``A destructive active defense algorithm for deepfake face images,'' \emph{PeerJ Computer Science}, vol.~10, p. e2356, Oct. 2024. [Online]. Available: \url{http://dx.doi.org/10.7717/peerj-cs.2356}
\BIBentrySTDinterwordspacing

\bibitem{Wang_2022}
\BIBentryALTinterwordspacing
R.~Wang, Z.~Huang, Z.~Chen, L.~Liu, J.~Chen, and L.~Wang, ``Anti-forgery: Towards a stealthy and robust deepfake disruption attack via adversarial perceptual-aware perturbations,'' in \emph{Proceedings of the Thirty-First International Joint Conference on Artificial Intelligence}, ser. IJCAI-2022.\hskip 1em plus 0.5em minus 0.4em\relax International Joint Conferences on Artificial Intelligence Organization, Jul. 2022, p. 761–767. [Online]. Available: \url{http://dx.doi.org/10.24963/ijcai.2022/107}
\BIBentrySTDinterwordspacing

\bibitem{Yu_2023}
\BIBentryALTinterwordspacing
Z.~Yu, S.~Zhai, and N.~Zhang, ``Antifake: Using adversarial audio to prevent unauthorized speech synthesis,'' in \emph{Proceedings of the 2023 ACM SIGSAC Conference on Computer and Communications Security}, ser. CCS '23.\hskip 1em plus 0.5em minus 0.4em\relax ACM, Nov. 2023, p. 460–474. [Online]. Available: \url{http://dx.doi.org/10.1145/3576915.3623209}
\BIBentrySTDinterwordspacing

\bibitem{Priya_2024}
\BIBentryALTinterwordspacing
M.~Priya, J.~Murugesan, P.~Bhuvaneswari, M.~Rubigha, S.~Lalithambikai, and B.~Mohanraj, ``Preserving visual authenticity: Block chain-augmented ai frameworks for advanced digital deception recognition and mitigation,'' in \emph{2024 5th International Conference on Smart Electronics and Communication (ICOSEC)}.\hskip 1em plus 0.5em minus 0.4em\relax IEEE, Sep. 2024, p. 707–713. [Online]. Available: \url{http://dx.doi.org/10.1109/icosec61587.2024.10722740}
\BIBentrySTDinterwordspacing

\bibitem{Cheng_2020}
\BIBentryALTinterwordspacing
K.~Cheng, R.~Tahir, L.~K. Eric, and M.~Li, ``An analysis of generative adversarial networks and variants for image synthesis on mnist dataset,'' \emph{Multimedia Tools and Applications}, vol.~79, no. 19–20, p. 13725–13752, Feb. 2020. [Online]. Available: \url{http://dx.doi.org/10.1007/s11042-019-08600-2}
\BIBentrySTDinterwordspacing

\bibitem{Mienye_2025}
\BIBentryALTinterwordspacing
I.~D. Mienye and T.~G. Swart, ``Deep autoencoder neural networks: A comprehensive review and new perspectives,'' \emph{Archives of Computational Methods in Engineering}, Mar. 2025. [Online]. Available: \url{http://dx.doi.org/10.1007/s11831-025-10260-5}
\BIBentrySTDinterwordspacing

\bibitem{Jellali_2024}
\BIBentryALTinterwordspacing
A.~Jellali, I.~Ben~Fredj, and K.~Ouni, ``Pushing the boundaries of deepfake audio detection with a hybrid mfcc and spectral contrast approach,'' \emph{Multimedia Tools and Applications}, Jul. 2024. [Online]. Available: \url{http://dx.doi.org/10.1007/s11042-024-19819-z}
\BIBentrySTDinterwordspacing

\bibitem{Nguyen_2018}
\BIBentryALTinterwordspacing
L.~Nguyen, S.~Wang, and A.~Sinha, \emph{A Learning and Masking Approach to Secure Learning}.\hskip 1em plus 0.5em minus 0.4em\relax Springer International Publishing, 2018, p. 453–464. [Online]. Available: \url{http://dx.doi.org/10.1007/978-3-030-01554-1_26}
\BIBentrySTDinterwordspacing

\bibitem{Chemmengath_2021}
\BIBentryALTinterwordspacing
S.~A. Chemmengath, S.~Paul, S.~Bharadwaj, S.~Samanta, and K.~Sankaranarayanan, ``Addressing target shift in zero-shot learning using grouped adversarial learning,'' in \emph{2021 IEEE/CVF International Conference on Computer Vision Workshops (ICCVW)}.\hskip 1em plus 0.5em minus 0.4em\relax IEEE, Oct. 2021, p. 2368–2377. [Online]. Available: \url{http://dx.doi.org/10.1109/iccvw54120.2021.00268}
\BIBentrySTDinterwordspacing

\bibitem{Cao_2022}
\BIBentryALTinterwordspacing
Y.~Cao, C.~Zhu, H.~Wang, and Y.~Zhuang, ``An adversarial attack algorithm based on edge-sketched feature from latent space,'' in \emph{2022 2nd International Conference on Consumer Electronics and Computer Engineering (ICCECE)}.\hskip 1em plus 0.5em minus 0.4em\relax IEEE, Jan. 2022, p. 723–728. [Online]. Available: \url{http://dx.doi.org/10.1109/iccece54139.2022.9712755}
\BIBentrySTDinterwordspacing

\bibitem{Zeng_2024}
\BIBentryALTinterwordspacing
S.~Zeng, W.~Wang, F.~Huang, and Y.~Fang, ``Loft: Latent space optimization and generator fine-tuning for defending against deepfakes,'' in \emph{ICASSP 2024 - 2024 IEEE International Conference on Acoustics, Speech and Signal Processing (ICASSP)}.\hskip 1em plus 0.5em minus 0.4em\relax IEEE, Apr. 2024, p. 4750–4754. [Online]. Available: \url{http://dx.doi.org/10.1109/icassp48485.2024.10447890}
\BIBentrySTDinterwordspacing

\bibitem{ap}
\BIBentryALTinterwordspacing
I.~J. Goodfellow, J.~Shlens, and C.~Szegedy, ``Explaining and harnessing adversarial examples,'' 2015. [Online]. Available: \url{https://arxiv.org/abs/1412.6572}
\BIBentrySTDinterwordspacing

\bibitem{Guo_2017}
\BIBentryALTinterwordspacing
Y.~Guo, G.~Ding, J.~Han, and Y.~Gao, ``Synthesizing samples for zero-shot learning,'' in \emph{Proceedings of the Twenty-Sixth International Joint Conference on Artificial Intelligence}, ser. IJCAI-2017.\hskip 1em plus 0.5em minus 0.4em\relax International Joint Conferences on Artificial Intelligence Organization, Aug. 2017, p. 1774–1780. [Online]. Available: \url{http://dx.doi.org/10.24963/ijcai.2017/246}
\BIBentrySTDinterwordspacing

\bibitem{Yan_2024}
\BIBentryALTinterwordspacing
S.~Yan, Q.~Zeng, Y.~Qi, L.~Lu, W.~Dong, and L.~Yu, \emph{Research on Zero-Shot Learning Based on Generative Model}.\hskip 1em plus 0.5em minus 0.4em\relax IOS Press, Feb. 2024. [Online]. Available: \url{http://dx.doi.org/10.3233/faia231364}
\BIBentrySTDinterwordspacing

\bibitem{Zhao_2022}
\BIBentryALTinterwordspacing
P.~Zhao, P.~Ram, S.~Lu, Y.~Yao, D.~Bouneffouf, X.~Lin, and S.~Liu, ``Learning to generate image source-agnostic universal adversarial perturbations,'' in \emph{Proceedings of the Thirty-First International Joint Conference on Artificial Intelligence}, ser. IJCAI-2022.\hskip 1em plus 0.5em minus 0.4em\relax International Joint Conferences on Artificial Intelligence Organization, Jul. 2022, p. 1714–1720. [Online]. Available: \url{http://dx.doi.org/10.24963/ijcai.2022/239}
\BIBentrySTDinterwordspacing

\bibitem{Zhou_2023}
\BIBentryALTinterwordspacing
X.~Zhou, N.~Yang, and O.~Wu, ``Combining adversaries with anti-adversaries in training,'' \emph{Proceedings of the AAAI Conference on Artificial Intelligence}, vol.~37, no.~9, p. 11435–11442, Jun. 2023. [Online]. Available: \url{http://dx.doi.org/10.1609/aaai.v37i9.26352}
\BIBentrySTDinterwordspacing

\bibitem{Hu_2024}
\BIBentryALTinterwordspacing
H.~Hu, X.~Wang, Y.~Zhang, Q.~Chen, and Q.~Guan, ``A comprehensive survey on contrastive learning,'' \emph{Neurocomputing}, vol. 610, p. 128645, Dec. 2024. [Online]. Available: \url{http://dx.doi.org/10.1016/j.neucom.2024.128645}
\BIBentrySTDinterwordspacing

\bibitem{Vito_2022}
\BIBentryALTinterwordspacing
V.~Vito and L.~Y. Stefanus, ``An asymmetric contrastive loss for handling imbalanced datasets,'' \emph{Entropy}, vol.~24, no.~9, p. 1303, Sep. 2022. [Online]. Available: \url{http://dx.doi.org/10.3390/e24091303}
\BIBentrySTDinterwordspacing

\bibitem{Wyawahare_2025}
\BIBentryALTinterwordspacing
M.~Wyawahare, S.~Bhorge, M.~Rane, V.~Parkhi, M.~Jha, and N.~Muhal, ``Comparative analysis of deepfake detection models on diverse gan-generated images,'' \emph{International journal of electrical and computer engineering systems}, vol.~16, no.~1, p. 9–18, Jan. 2025. [Online]. Available: \url{http://dx.doi.org/10.32985/ijeces.16.1.2}
\BIBentrySTDinterwordspacing

\bibitem{Y_ksel_2020}
\BIBentryALTinterwordspacing
O.~K. Yüksel and Ä.~M. Baytaş, ``Adversarial training with orthogonal regularization,'' in \emph{2020 28th Signal Processing and Communications Applications Conference (SIU)}.\hskip 1em plus 0.5em minus 0.4em\relax IEEE, Oct. 2020, p. 1–4. [Online]. Available: \url{http://dx.doi.org/10.1109/siu49456.2020.9302247}
\BIBentrySTDinterwordspacing

\bibitem{Liu_2019}
\BIBentryALTinterwordspacing
X.~Liu and C.-J. Hsieh, ``Rob-gan: Generator, discriminator, and adversarial attacker,'' in \emph{2019 IEEE/CVF Conference on Computer Vision and Pattern Recognition (CVPR)}.\hskip 1em plus 0.5em minus 0.4em\relax IEEE, Jun. 2019. [Online]. Available: \url{http://dx.doi.org/10.1109/cvpr.2019.01149}
\BIBentrySTDinterwordspacing

\bibitem{Li_2019}
\BIBentryALTinterwordspacing
S.~Li, A.~Neupane, S.~Paul, C.~Song, S.~V. Krishnamurthy, A.~K.~R. Chowdhury, and A.~Swami, ``Stealthy adversarial perturbations against real-time video classification systems,'' in \emph{Proceedings 2019 Network and Distributed System Security Symposium}, ser. NDSS 2019.\hskip 1em plus 0.5em minus 0.4em\relax Internet Society, 2019. [Online]. Available: \url{http://dx.doi.org/10.14722/ndss.2019.23202}
\BIBentrySTDinterwordspacing

\bibitem{Barach_2025}
\BIBentryALTinterwordspacing
J.~Barach, \emph{Cross-Domain Adversarial Attacks and Robust Defense Mechanisms for Multimodal Neural Networks}.\hskip 1em plus 0.5em minus 0.4em\relax Springer Nature Switzerland, 2025, p. 345–362. [Online]. Available: \url{http://dx.doi.org/10.1007/978-3-031-83796-8_23}
\BIBentrySTDinterwordspacing

\bibitem{Hou_2023}
\BIBentryALTinterwordspacing
X.~Hou, ``Adversarial attack generation in quantum data classification,'' in \emph{Proceedings of the 2023 4th International Conference on Computer Science and Management Technology}, ser. ICCSMT 2023.\hskip 1em plus 0.5em minus 0.4em\relax ACM, Oct. 2023, p. 548–551. [Online]. Available: \url{http://dx.doi.org/10.1145/3644523.3644623}
\BIBentrySTDinterwordspacing

\bibitem{Akter_2024}
\BIBentryALTinterwordspacing
M.~S. Akter, H.~Shahriar, A.~Cuzzocrea, and F.~Wu, ``Quantum adversarial attacks: Developing quantum fgsm algorithm,'' in \emph{2024 IEEE 48th Annual Computers, Software, and Applications Conference (COMPSAC)}.\hskip 1em plus 0.5em minus 0.4em\relax IEEE, Jul. 2024, p. 1073–1079. [Online]. Available: \url{http://dx.doi.org/10.1109/compsac61105.2024.00145}
\BIBentrySTDinterwordspacing

\bibitem{Alrowais_2024}
\BIBentryALTinterwordspacing
F.~Alrowais, A.~Abbas~Hassan, W.~Sulaiman~Almukadi, M.~H. Alanazi, R.~Marzouk, and A.~Mahmud, ``Boosting deep feature fusion-based detection model for fake faces generated by generative adversarial networks for consumer space environment,'' \emph{IEEE Access}, vol.~12, p. 147680–147693, 2024. [Online]. Available: \url{http://dx.doi.org/10.1109/access.2024.3470128}
\BIBentrySTDinterwordspacing

\bibitem{Awan_2021}
\BIBentryALTinterwordspacing
S.~Awan, B.~Luo, and F.~Li, \emph{CONTRA: Defending Against Poisoning Attacks in Federated Learning}.\hskip 1em plus 0.5em minus 0.4em\relax Springer International Publishing, 2021, p. 455–475. [Online]. Available: \url{http://dx.doi.org/10.1007/978-3-030-88418-5_22}
\BIBentrySTDinterwordspacing

\bibitem{Das_2023}
\BIBentryALTinterwordspacing
S.~Das, A.~Chatterjee, and S.~Ghosh, ``Sok: A first order survey of quantum supply dynamics and threat landscapes,'' in \emph{Proceedings of the 12th International Workshop on Hardware and Architectural Support for Security and Privacy}, ser. HASP '23.\hskip 1em plus 0.5em minus 0.4em\relax ACM, Oct. 2023, p. 82–90. [Online]. Available: \url{http://dx.doi.org/10.1145/3623652.3623664}
\BIBentrySTDinterwordspacing

\bibitem{Kania_2017}
\BIBentryALTinterwordspacing
E.~B. Kania and J.~K. Costello, ``Quantum technologies, u.s.-china strategic competition, and future dynamics of cyber stability,'' in \emph{2017 International Conference on Cyber Conflict (CyCon U.S.)}.\hskip 1em plus 0.5em minus 0.4em\relax IEEE, Nov. 2017, p. 89–96. [Online]. Available: \url{http://dx.doi.org/10.1109/cyconus.2017.8167502}
\BIBentrySTDinterwordspacing

\bibitem{de_Wolf_2017}
\BIBentryALTinterwordspacing
R.~de~Wolf, ``The potential impact of quantum computers on society,'' \emph{Ethics and Information Technology}, vol.~19, no.~4, p. 271–276, Sep. 2017. [Online]. Available: \url{http://dx.doi.org/10.1007/s10676-017-9439-z}
\BIBentrySTDinterwordspacing

\bibitem{Ambadekar_2018}
\BIBentryALTinterwordspacing
S.~P. Ambadekar, J.~Jain, and J.~Khanapuri, \emph{Digital Image Watermarking Through Encryption and DWT for Copyright Protection}.\hskip 1em plus 0.5em minus 0.4em\relax Springer Singapore, May 2018, p. 187–195. [Online]. Available: \url{http://dx.doi.org/10.1007/978-981-10-8863-6_19}
\BIBentrySTDinterwordspacing

\bibitem{Singh_2013}
\BIBentryALTinterwordspacing
A.~K. Singh, M.~Dave, and A.~Mohan, ``Performance comparison of wavelet filters against signal processing attacks,'' in \emph{2013 IEEE Second International Conference on Image Information Processing (ICIIP-2013)}.\hskip 1em plus 0.5em minus 0.4em\relax IEEE, Dec. 2013, p. 695–698. [Online]. Available: \url{http://dx.doi.org/10.1109/iciip.2013.6707685}
\BIBentrySTDinterwordspacing

\bibitem{Radharapu_2024}
\BIBentryALTinterwordspacing
B.~Radharapu and H.~Krishna, ``Realseal: Revolutionizing media authentication with real-time realism scoring,'' in \emph{International Conference on Multimodel Interaction}, ser. ICMI '24.\hskip 1em plus 0.5em minus 0.4em\relax ACM, Nov. 2024, p. 585–590. [Online]. Available: \url{http://dx.doi.org/10.1145/3678957.3678960}
\BIBentrySTDinterwordspacing

\bibitem{AL_ardhi_2019}
\BIBentryALTinterwordspacing
S.~AL-ardhi, V.~Thayananthan, and A.~Basuhail, ``A watermarking system architecture using the cellular automata transform for 2d vector map,'' \emph{International Journal of Advanced Computer Science and Applications}, vol.~10, no.~6, 2019. [Online]. Available: \url{http://dx.doi.org/10.14569/ijacsa.2019.0100652}
\BIBentrySTDinterwordspacing

\bibitem{Xie_2024}
\BIBentryALTinterwordspacing
S.~Xie, C.~Zhao, N.~Sun, W.~Li, and H.~Ling, ``Picking watermarks from noise (pwfn): an improved robust watermarking model against intensive distortions,'' in \emph{2024 IEEE International Conference on Multimedia and Expo (ICME)}.\hskip 1em plus 0.5em minus 0.4em\relax IEEE, Jul. 2024, p. 1–6. [Online]. Available: \url{http://dx.doi.org/10.1109/icme57554.2024.10687631}
\BIBentrySTDinterwordspacing

\bibitem{Liao_2022}
\BIBentryALTinterwordspacing
C.-Y. Liao, C.-H. Huang, J.-C. Chen, and J.-L. Wu, ``Enhancing the robustness of deep learning based fingerprinting to improve deepfake attribution,'' in \emph{Proceedings of the 4th ACM International Conference on Multimedia in Asia}, ser. MMAsia '22.\hskip 1em plus 0.5em minus 0.4em\relax ACM, Dec. 2022, p. 1–7. [Online]. Available: \url{http://dx.doi.org/10.1145/3551626.3564981}
\BIBentrySTDinterwordspacing

\bibitem{Thakkar_2024}
\BIBentryALTinterwordspacing
J.~J. Thakkar and A.~Kaur, ``From deepfakes to digital truths: The role of watermarking in ai-generated image verification,'' in \emph{2024 47th International Conference on Telecommunications and Signal Processing (TSP)}.\hskip 1em plus 0.5em minus 0.4em\relax IEEE, Jul. 2024, p. 216–222. [Online]. Available: \url{http://dx.doi.org/10.1109/tsp63128.2024.10605975}
\BIBentrySTDinterwordspacing

\bibitem{Iacobici_2020}
\BIBentryALTinterwordspacing
N.~L. Iacobici, M.~Frigura-Iliasa, H.~E. Filipescu, M.~Nen, F.~M. Frigura-Iliasa, and M.~Iorga, ``Digital imaging processing and reconstruction for general applications,'' in \emph{2020 IEEE 18th World Symposium on Applied Machine Intelligence and Informatics (SAMI)}.\hskip 1em plus 0.5em minus 0.4em\relax IEEE, Jan. 2020, p. 231–234. [Online]. Available: \url{http://dx.doi.org/10.1109/sami48414.2020.9108753}
\BIBentrySTDinterwordspacing

\bibitem{Schirripa_Spagnolo_2006}
\BIBentryALTinterwordspacing
G.~Schirripa~Spagnolo and M.~De~Santis, ``Potentiality of holographic technique in fragile watermarking,'' in \emph{Optical Security and Counterfeit Deterrence Techniques VI}, R.~L. van Renesse, Ed., vol. 6075.\hskip 1em plus 0.5em minus 0.4em\relax SPIE, Feb. 2006, p. 607509. [Online]. Available: \url{http://dx.doi.org/10.1117/12.652315}
\BIBentrySTDinterwordspacing

\bibitem{Kakkirala_2015}
\BIBentryALTinterwordspacing
K.~R. Kakkirala, S.~R. Chalamala, and B.~M. Garlapati, ``An audio/speech watermarking method for copyright protection,'' in \emph{2015 3rd International Conference on Artificial Intelligence, Modelling and Simulation (AIMS)}.\hskip 1em plus 0.5em minus 0.4em\relax IEEE, Dec. 2015, p. 338–342. [Online]. Available: \url{http://dx.doi.org/10.1109/aims.2015.61}
\BIBentrySTDinterwordspacing

\bibitem{Xia_2014}
\BIBentryALTinterwordspacing
S.~Xia, S.~Ge, and C.~Tang, ``Overview of digital watermarking,'' in \emph{Future Information Engineering}, ser. ICIE2013, vol.~1.\hskip 1em plus 0.5em minus 0.4em\relax WIT Press, Mar. 2014, p. 457–464. [Online]. Available: \url{http://dx.doi.org/10.2495/icie130531}
\BIBentrySTDinterwordspacing

\bibitem{Nadimpalli_2024}
\BIBentryALTinterwordspacing
A.~V. Nadimpalli and A.~Rattani, ``Social media authentication and combating deepfakes using semi-fragile invisible image watermarking,'' \emph{Digital Threats: Research and Practice}, vol.~5, no.~4, p. 1–30, Dec. 2024. [Online]. Available: \url{http://dx.doi.org/10.1145/3700146}
\BIBentrySTDinterwordspacing

\bibitem{K_2024}
\BIBentryALTinterwordspacing
K.~K, S.~R, D.~S, and D.~S, ``Guardian ai: Synthetic media forensics through multimodal fusion and advanced machine learning,'' in \emph{2024 International Conference on Cognitive Robotics and Intelligent Systems (ICC - ROBINS)}.\hskip 1em plus 0.5em minus 0.4em\relax IEEE, Apr. 2024, p. 226–232. [Online]. Available: \url{http://dx.doi.org/10.1109/icc-robins60238.2024.10533980}
\BIBentrySTDinterwordspacing

\bibitem{Ramesh_2009}
\BIBentryALTinterwordspacing
S.~Ramesh and M.~M.~I. Majeed, ``Implementation of a visible watermarking in a secure still digital camera using vlsi design,'' in \emph{AFRICON 2009}.\hskip 1em plus 0.5em minus 0.4em\relax IEEE, Sep. 2009, p. 1–4. [Online]. Available: \url{http://dx.doi.org/10.1109/afrcon.2009.5308417}
\BIBentrySTDinterwordspacing

\bibitem{Hydara_2024}
\BIBentryALTinterwordspacing
E.~Hydara, M.~Kikuchi, and T.~Ozono, ``Empirical assessment of deepfake detection: Advancing judicial evidence verification through artificial intelligence,'' \emph{IEEE Access}, vol.~12, p. 151188–151203, 2024. [Online]. Available: \url{http://dx.doi.org/10.1109/access.2024.3480320}
\BIBentrySTDinterwordspacing

\bibitem{Sun_2023}
\BIBentryALTinterwordspacing
C.~Sun, S.~Jia, S.~Hou, and S.~Lyu, ``Ai-synthesized voice detection using neural vocoder artifacts,'' in \emph{2023 IEEE/CVF Conference on Computer Vision and Pattern Recognition Workshops (CVPRW)}.\hskip 1em plus 0.5em minus 0.4em\relax IEEE, Jun. 2023, p. 904–912. [Online]. Available: \url{http://dx.doi.org/10.1109/cvprw59228.2023.00097}
\BIBentrySTDinterwordspacing

\bibitem{Moon_2024}
\BIBentryALTinterwordspacing
K.-H. Moon, S.-Y. Ok, and S.-H. Lee, ``Supcon-mpl-dp: Supervised contrastive learning with meta pseudo labels for deepfake image detection,'' \emph{Applied Sciences}, vol.~14, no.~8, p. 3249, Apr. 2024. [Online]. Available: \url{http://dx.doi.org/10.3390/app14083249}
\BIBentrySTDinterwordspacing

\bibitem{Sharma_2020}
\BIBentryALTinterwordspacing
S.~S. Sharma and V.~Chandrasekaran, ``A robust hybrid digital watermarking technique against a powerful cnn-based adversarial attack,'' \emph{Multimedia Tools and Applications}, vol.~79, no. 43–44, p. 32769–32790, Aug. 2020. [Online]. Available: \url{http://dx.doi.org/10.1007/s11042-020-09555-5}
\BIBentrySTDinterwordspacing

\bibitem{Bashardoost_2015}
\BIBentryALTinterwordspacing
M.~Bashardoost, M.~S. Mohd~Rahim, and N.~Hadipour, ``A novel zero-watermarking scheme for text document authentication,'' \emph{Jurnal Teknologi}, vol.~75, no.~4, Jul. 2015. [Online]. Available: \url{http://dx.doi.org/10.11113/jt.v75.5066}
\BIBentrySTDinterwordspacing

\bibitem{Kapre_2020}
\BIBentryALTinterwordspacing
B.~S. Kapre and A.~M. Rajurkar, \emph{Robust and Secure Lucas Sequence-Based Video Watermarking}.\hskip 1em plus 0.5em minus 0.4em\relax Springer Singapore, 2020, p. 300–307. [Online]. Available: \url{http://dx.doi.org/10.1007/978-981-15-4029-5_30}
\BIBentrySTDinterwordspacing

\bibitem{DEKA_2019}
\BIBentryALTinterwordspacing
R.~DEKA, C.~GALDI, and J.-L. DUGELAY, ``Hybrid g-prnu: Optimal parameter selection for scale-invariant asymmetric source smartphone identification,'' \emph{Electronic Imaging}, vol.~31, no.~5, pp. 546--1--546–7, Jan. 2019. [Online]. Available: \url{http://dx.doi.org/10.2352/issn.2470-1173.2019.5.mwsf-546}
\BIBentrySTDinterwordspacing

\bibitem{Huang_2007}
\BIBentryALTinterwordspacing
W.~Huang and I.~M. Panahi, ``Anti-collusion fingerprinting scheme based on error correction ability of nonlinear combinatorial code,'' in \emph{Wireless Sensing and Processing II}, R.~M. Rao, S.~A. Dianat, and M.~D. Zoltowski, Eds., vol. 6577.\hskip 1em plus 0.5em minus 0.4em\relax SPIE, Apr. 2007, p. 65770I. [Online]. Available: \url{http://dx.doi.org/10.1117/12.724260}
\BIBentrySTDinterwordspacing

\bibitem{Thilakavathy_2023}
\BIBentryALTinterwordspacing
P.~Thilakavathy, S.~Jayachitra, A.~Aeron, N.~Kumar, S.~S. Ali, and M.~Malathy, ``Investigating blockchain security mechanisms for tamper-proof data storage,'' in \emph{2023 International Conference on Communication, Security and Artificial Intelligence (ICCSAI)}.\hskip 1em plus 0.5em minus 0.4em\relax IEEE, Nov. 2023, p. 926–930. [Online]. Available: \url{http://dx.doi.org/10.1109/iccsai59793.2023.10421006}
\BIBentrySTDinterwordspacing

\bibitem{Jain_2021}
\BIBentryALTinterwordspacing
A.~Jain, S.~Das, A.~Singh~Kushwah, T.~Rajora, and S.~Saboo, ``Blockchain-based criminal record database management,'' in \emph{2021 Asian Conference on Innovation in Technology (ASIANCON)}.\hskip 1em plus 0.5em minus 0.4em\relax IEEE, Aug. 2021, p. 1–5. [Online]. Available: \url{http://dx.doi.org/10.1109/asiancon51346.2021.9544655}
\BIBentrySTDinterwordspacing

\bibitem{Ali_2020}
\BIBentryALTinterwordspacing
N.~H.~M. Ali and M.~E. Mahdi, ``Detecting similarity in color images based on perceptual image hash algorithm,'' \emph{IOP Conference Series: Materials Science and Engineering}, vol. 737, no.~1, p. 012244, Feb. 2020. [Online]. Available: \url{http://dx.doi.org/10.1088/1757-899x/737/1/012244}
\BIBentrySTDinterwordspacing

\bibitem{Picha_Edwardsson_2024}
\BIBentryALTinterwordspacing
M.~Picha~Edwardsson and W.~Al-Saqaf, ``Blockchain solutions for generative ai challenges in journalism,'' \emph{Frontiers in Blockchain}, vol.~7, Nov. 2024. [Online]. Available: \url{http://dx.doi.org/10.3389/fbloc.2024.1440355}
\BIBentrySTDinterwordspacing

\bibitem{Wang_2020}
\BIBentryALTinterwordspacing
S.~Wang, X.~Zhang, W.~Yu, K.~Hu, and J.~Zhu, ``Smart contract microservitization,'' in \emph{2020 IEEE 44th Annual Computers, Software, and Applications Conference (COMPSAC)}.\hskip 1em plus 0.5em minus 0.4em\relax IEEE, Jul. 2020, p. 1569–1574. [Online]. Available: \url{http://dx.doi.org/10.1109/compsac48688.2020.00-31}
\BIBentrySTDinterwordspacing

\bibitem{Fisher_2024}
\BIBentryALTinterwordspacing
S.~A. Fisher, ``Something <scp>ai</scp> should tell you – the case for labelling synthetic content,'' \emph{Journal of Applied Philosophy}, vol.~42, no.~1, p. 272–286, Aug. 2024. [Online]. Available: \url{http://dx.doi.org/10.1111/japp.12758}
\BIBentrySTDinterwordspacing

\bibitem{Sun_2025}
\BIBentryALTinterwordspacing
Z.~Sun, N.~Ruan, and J.~Li, ``Ddl: Effective and comprehensible interpretation framework for diverse deepfake detectors,'' \emph{IEEE Transactions on Information Forensics and Security}, p. 1–1, 2025. [Online]. Available: \url{http://dx.doi.org/10.1109/tifs.2025.3553803}
\BIBentrySTDinterwordspacing

\bibitem{Fu_2014}
\BIBentryALTinterwordspacing
Y.~Fu, T.~M. Hospedales, T.~Xiang, Z.~Fu, and S.~Gong, \emph{Transductive Multi-view Embedding for Zero-Shot Recognition and Annotation}.\hskip 1em plus 0.5em minus 0.4em\relax Springer International Publishing, 2014, p. 584–599. [Online]. Available: \url{http://dx.doi.org/10.1007/978-3-319-10605-2_38}
\BIBentrySTDinterwordspacing

\bibitem{Kirk_2023}
\BIBentryALTinterwordspacing
R.~Kirk, A.~Zhang, E.~Grefenstette, and T.~Rocktäschel, ``A survey of zero-shot generalisation in deep reinforcement learning,'' \emph{Journal of Artificial Intelligence Research}, vol.~76, p. 201–264, Jan. 2023. [Online]. Available: \url{http://dx.doi.org/10.1613/jair.1.14174}
\BIBentrySTDinterwordspacing

\bibitem{Lokhande_2024}
\BIBentryALTinterwordspacing
M.~Lokhande, P.~Raut, K.~Gawali, M.~Ahirrao, and A.~Bhande, ``Artificial intelligence for detecting cyber attacks in deepfake \& identity theft,'' in \emph{2024 8th International Conference on Computing, Communication, Control and Automation (ICCUBEA)}.\hskip 1em plus 0.5em minus 0.4em\relax IEEE, Aug. 2024, p. 1–6. [Online]. Available: \url{http://dx.doi.org/10.1109/iccubea61740.2024.10774771}
\BIBentrySTDinterwordspacing

\bibitem{Satone_2024}
\BIBentryALTinterwordspacing
K.~Satone and S.~Y. Amdani, ``Preserving video authenticity in the age of synthetic media using blockchain,'' in \emph{INTERNATIONAL CONFERENCE ON INTELLIGENT TECHNOLOGIES FOR SUSTAINABLE ENERGY MANAGEMENT AND CONTROL 2023: ITSEMC2023}, vol. 3188.\hskip 1em plus 0.5em minus 0.4em\relax AIP Publishing, 2024, p. 080008. [Online]. Available: \url{http://dx.doi.org/10.1063/5.0240462}
\BIBentrySTDinterwordspacing

\bibitem{Costales_2023}
\BIBentryALTinterwordspacing
J.~A. Costales, S.~Shiromani, and M.~Devaraj, ``The impact of blockchain technology to protect image and video integrity from identity theft using deepfake analyzer,'' in \emph{2023 International Conference on Innovative Data Communication Technologies and Application (ICIDCA)}.\hskip 1em plus 0.5em minus 0.4em\relax IEEE, Mar. 2023, p. 730–733. [Online]. Available: \url{http://dx.doi.org/10.1109/icidca56705.2023.10099668}
\BIBentrySTDinterwordspacing

\bibitem{Kim_2025}
\BIBentryALTinterwordspacing
D.~Kim, S.~Yun, S.~Lee, J.~Lee, and D.~Niyato, ``Intelligent transaction generation control for permissioned blockchain-based services,'' \emph{IEEE Transactions on Services Computing}, p. 1–12, 2025. [Online]. Available: \url{http://dx.doi.org/10.1109/tsc.2025.3528318}
\BIBentrySTDinterwordspacing

\bibitem{Lao_2024}
\BIBentryALTinterwordspacing
C.~K. Lao, S.~Zhou, L.~Zhang, F.~Zhang, and K.~Y. Wang, ``Unpacking long-latency transactions in ethereum,'' in \emph{Proceedings of the Workshop on Decentralized Finance and Security}, ser. CCS '24.\hskip 1em plus 0.5em minus 0.4em\relax ACM, Nov. 2024, p. 11–20. [Online]. Available: \url{http://dx.doi.org/10.1145/3689931.3694909}
\BIBentrySTDinterwordspacing

\bibitem{Chunli_Wang_2023}
\BIBentryALTinterwordspacing
C.~W. Chunli~Wang, Y.~C. Chunli~Wang, and W.~J. Yuling~Chen, ``Constraints-based and one-time modification redactable blockchain,'' \emph{網際網路技術學刊}, vol.~24, no.~7, p. 1437–1446, Dec. 2023. [Online]. Available: \url{http://dx.doi.org/10.53106/160792642023122407005}
\BIBentrySTDinterwordspacing

\bibitem{Zhu_2020}
\BIBentryALTinterwordspacing
L.~Zhu, C.~Xiao, and X.~Gong, ``Keyword search in decentralized storage systems,'' \emph{Electronics}, vol.~9, no.~12, p. 2041, Dec. 2020. [Online]. Available: \url{http://dx.doi.org/10.3390/electronics9122041}
\BIBentrySTDinterwordspacing

\bibitem{Jain_2024}
\BIBentryALTinterwordspacing
A.~Jain, A.~Gaur, G.~Gupta, S.~Mishra, R.~Johari, and D.~P. Vidyarthi, \emph{Securing Digital Integrity: Proposed Comprehensive Framework for Deepfake Detection and Blockchain Validation}.\hskip 1em plus 0.5em minus 0.4em\relax Springer Nature Singapore, Dec. 2024, p. 579–589. [Online]. Available: \url{http://dx.doi.org/10.1007/978-981-97-7371-8_45}
\BIBentrySTDinterwordspacing

\bibitem{Fathalla_2024}
\BIBentryALTinterwordspacing
E.~Fathalla and M.~Azab, ``Beyond classical cryptography: A systematic review of post-quantum hash-based signature schemes, security, and optimizations,'' \emph{IEEE Access}, vol.~12, p. 175969–175987, 2024. [Online]. Available: \url{http://dx.doi.org/10.1109/access.2024.3485602}
\BIBentrySTDinterwordspacing

\bibitem{Tyagi_2023}
\BIBentryALTinterwordspacing
A.~K. Tyagi and S.~Tiwari, \emph{The Future of Artificial Intelligence in Blockchain Applications}.\hskip 1em plus 0.5em minus 0.4em\relax IGI Global, Dec. 2023, p. 346–373. [Online]. Available: \url{http://dx.doi.org/10.4018/978-1-6684-8531-6.ch018}
\BIBentrySTDinterwordspacing

\bibitem{Handayani_2023}
\BIBentryALTinterwordspacing
A.~D. Handayani, S.~S. Carita, and N.~Yulianti, ``A modified zero-knowledge scheme from bilinear pairing for securing multiple secret values,'' in \emph{2023 IEEE International Conference on Cryptography, Informatics, and Cybersecurity (ICoCICs)}.\hskip 1em plus 0.5em minus 0.4em\relax IEEE, Aug. 2023, p. 195–198. [Online]. Available: \url{http://dx.doi.org/10.1109/icocics58778.2023.10276959}
\BIBentrySTDinterwordspacing

\bibitem{Javed_2024}
\BIBentryALTinterwordspacing
M.~Javed, Z.~Zhang, F.~H. Dahri, and A.~A. Laghari, ``Real-time deepfake video detection using eye movement analysis with a hybrid deep learning approach,'' \emph{Electronics}, vol.~13, no.~15, p. 2947, Jul. 2024. [Online]. Available: \url{http://dx.doi.org/10.3390/electronics13152947}
\BIBentrySTDinterwordspacing

\bibitem{Sariyildiz_2019}
\BIBentryALTinterwordspacing
M.~B. Sariyildiz and R.~G. Cinbis, ``Gradient matching generative networks for zero-shot learning,'' in \emph{2019 IEEE/CVF Conference on Computer Vision and Pattern Recognition (CVPR)}.\hskip 1em plus 0.5em minus 0.4em\relax IEEE, Jun. 2019, p. 2163–2173. [Online]. Available: \url{http://dx.doi.org/10.1109/cvpr.2019.00227}
\BIBentrySTDinterwordspacing

\bibitem{Suresh_2024}
\BIBentryALTinterwordspacing
M.~Suresh, J.~Raghavan, A.~Roy, and S.~S. Das, ``Hybrid xception-resnext bilstm model with attention and visual synchronization for robust deepfake detection,'' in \emph{2024 International Conference on System, Computation, Automation and Networking (ICSCAN)}.\hskip 1em plus 0.5em minus 0.4em\relax IEEE, Dec. 2024, p. 1–6. [Online]. Available: \url{http://dx.doi.org/10.1109/icscan62807.2024.10894583}
\BIBentrySTDinterwordspacing

\bibitem{Kim_2023}
\BIBentryALTinterwordspacing
S.-O. Kim, D.-W. Jeong, and S.-Y. Lee, \emph{Method of Facial De-identification Using Machine Learning in Real-Time Video}.\hskip 1em plus 0.5em minus 0.4em\relax Springer Nature Switzerland, 2023, p. 201–208. [Online]. Available: \url{http://dx.doi.org/10.1007/978-3-031-35836-4_22}
\BIBentrySTDinterwordspacing

\bibitem{Singh_2025}
\BIBentryALTinterwordspacing
B.~Singh, A.~Raghav, S.~Ahmed, M.~K. Arora, and S.~Lal, \emph{Smearing Machine Learning and Deep Learning in E-Commerce Transactions for Monetary Justice: Crushing Financial Frauds and Fostering Strong Financial Institutions}.\hskip 1em plus 0.5em minus 0.4em\relax IGI Global, Mar. 2025, p. 303–320. [Online]. Available: \url{http://dx.doi.org/10.4018/979-8-3693-9395-6.ch014}
\BIBentrySTDinterwordspacing

\bibitem{Kanamori_2022}
\BIBentryALTinterwordspacing
S.~Kanamori, T.~Abe, T.~Ito, K.~Emura, L.~Wang, S.~Yamamoto, L.~T. Phong, K.~Abe, S.~Kim, R.~Nojima, S.~Ozawa, and S.~Moriai, ``Privacy-preserving federated learning for detecting fraudulent financial transactions in japanese banks,'' \emph{Journal of Information Processing}, vol.~30, no.~0, p. 789–795, 2022. [Online]. Available: \url{http://dx.doi.org/10.2197/ipsjjip.30.789}
\BIBentrySTDinterwordspacing

\bibitem{Kaur_2018}
\BIBentryALTinterwordspacing
J.~Kaur, K.~C. Juglan, and V.~Sharma, ``Voice stress analysis for punjabi and hindi database: Detection of deception,'' in \emph{AIP Conference Proceedings}, vol. 2006.\hskip 1em plus 0.5em minus 0.4em\relax Author(s), 2018, p. 030022. [Online]. Available: \url{http://dx.doi.org/10.1063/1.5051278}
\BIBentrySTDinterwordspacing

\bibitem{Winkler_2021}
\BIBentryALTinterwordspacing
S.~Winkler, W.~Chen, A.~Dhall, and P.~Korshunov, ``Adgd'21: 1st workshop on synthetic multimedia - audiovisual deepfake generation and detection,'' in \emph{Proceedings of the 29th ACM International Conference on Multimedia}, ser. MM '21.\hskip 1em plus 0.5em minus 0.4em\relax ACM, Oct. 2021, p. 5698–5699. [Online]. Available: \url{http://dx.doi.org/10.1145/3474085.3478578}
\BIBentrySTDinterwordspacing

\bibitem{Dong_2022}
\BIBentryALTinterwordspacing
X.~Dong, J.~Guo, A.~Li, W.-T. Ting, C.~Liu, and H.~Kung, ``Neural mean discrepancy for efficient out-of-distribution detection,'' in \emph{2022 IEEE/CVF Conference on Computer Vision and Pattern Recognition (CVPR)}.\hskip 1em plus 0.5em minus 0.4em\relax IEEE, Jun. 2022, p. 19195–19205. [Online]. Available: \url{http://dx.doi.org/10.1109/cvpr52688.2022.01862}
\BIBentrySTDinterwordspacing

\bibitem{Chetty_2010}
\BIBentryALTinterwordspacing
G.~Chetty, \emph{Robust Audio Visual Biometric Person Authentication with Liveness Verification}.\hskip 1em plus 0.5em minus 0.4em\relax Springer Berlin Heidelberg, 2010, p. 59–78. [Online]. Available: \url{http://dx.doi.org/10.1007/978-3-642-11756-5_3}
\BIBentrySTDinterwordspacing

\bibitem{Singh_2024}
\BIBentryALTinterwordspacing
R.~P. Singh, N.~H. Sree, K.~L. S.~P. Reddy, and K.~Jashwanth, ``Convergence of deep learning and forensic methodologies using self-attention integrated efficientnet model for deep fake detection,'' \emph{SN Computer Science}, vol.~5, no.~8, Dec. 2024. [Online]. Available: \url{http://dx.doi.org/10.1007/s42979-024-03455-3}
\BIBentrySTDinterwordspacing

\bibitem{Adiwijaya_2023}
\BIBentryALTinterwordspacing
J.~Adiwijaya, V.~R. Tanaya, Anderies, and A.~Chowanda, ``Federated learning and differential privacy in ai-based surveillance systems model,'' in \emph{2023 14th International Conference on Information \& Communication Technology and System (ICTS)}.\hskip 1em plus 0.5em minus 0.4em\relax IEEE, Oct. 2023, p. 283–288. [Online]. Available: \url{http://dx.doi.org/10.1109/icts58770.2023.10330863}
\BIBentrySTDinterwordspacing

\bibitem{Yang_2022}
\BIBentryALTinterwordspacing
Y.~Yang, A.~Gupta, J.~Feng, P.~Singhal, V.~Yadav, Y.~Wu, P.~Natarajan, V.~Hedau, and J.~Joo, ``Enhancing fairness in face detection in computer vision systems by demographic bias mitigation,'' in \emph{Proceedings of the 2022 AAAI/ACM Conference on AI, Ethics, and Society}, ser. AIES '22.\hskip 1em plus 0.5em minus 0.4em\relax ACM, Jul. 2022. [Online]. Available: \url{http://dx.doi.org/10.1145/3514094.3534153}
\BIBentrySTDinterwordspacing

\bibitem{Mohammadi_2024}
\BIBentryALTinterwordspacing
S.~Mohammadi, A.~Balador, S.~Sinaei, and F.~Flammini, ``Balancing privacy and performance in federated learning: A systematic literature review on methods and metrics,'' \emph{Journal of Parallel and Distributed Computing}, vol. 192, p. 104918, Oct. 2024. [Online]. Available: \url{http://dx.doi.org/10.1016/j.jpdc.2024.104918}
\BIBentrySTDinterwordspacing

\bibitem{Jagarlamudi_2023}
\BIBentryALTinterwordspacing
G.~K. Jagarlamudi, A.~Yazdinejad, R.~M. Parizi, and S.~Pouriyeh, ``Exploring privacy measurement in federated learning,'' \emph{The Journal of Supercomputing}, vol.~80, no.~8, p. 10511–10551, Dec. 2023. [Online]. Available: \url{http://dx.doi.org/10.1007/s11227-023-05846-4}
\BIBentrySTDinterwordspacing

\bibitem{Asif_2024}
\BIBentryALTinterwordspacing
M.~Asif, S.~Naz, F.~Ali, A.~Alabrah, A.~Salam, F.~Amin, and F.~Ullah, ``Advanced zero-shot learning (azsl) framework for secure model generalization in federated learning,'' \emph{IEEE Access}, vol.~12, p. 184393–184407, 2024. [Online]. Available: \url{http://dx.doi.org/10.1109/access.2024.3510756}
\BIBentrySTDinterwordspacing

\bibitem{Yin_2024}
\BIBentryALTinterwordspacing
C.~Yin and Q.~Zeng, ``Defending against data poisoning attack in federated learning with non-iid data,'' \emph{IEEE Transactions on Computational Social Systems}, vol.~11, no.~2, p. 2313–2325, Apr. 2024. [Online]. Available: \url{http://dx.doi.org/10.1109/tcss.2023.3296885}
\BIBentrySTDinterwordspacing

\bibitem{Bhatti_2023}
\BIBentryALTinterwordspacing
D.~M.~S. Bhatti and H.~Nam, ``A robust aggregation approach for heterogeneous federated learning,'' in \emph{2023 Fourteenth International Conference on Ubiquitous and Future Networks (ICUFN)}.\hskip 1em plus 0.5em minus 0.4em\relax IEEE, Jul. 2023, p. 300–304. [Online]. Available: \url{http://dx.doi.org/10.1109/icufn57995.2023.10201227}
\BIBentrySTDinterwordspacing

\bibitem{Zhang_2020}
\BIBentryALTinterwordspacing
Z.~Zhang, J.~Zhoa, and X.~Liang, ``Zero-shot learning based on semantic embedding for ship detection,'' in \emph{2020 3rd International Conference on Unmanned Systems (ICUS)}.\hskip 1em plus 0.5em minus 0.4em\relax IEEE, Nov. 2020, p. 1152–1156. [Online]. Available: \url{http://dx.doi.org/10.1109/icus50048.2020.9274981}
\BIBentrySTDinterwordspacing

\bibitem{Ba_2023}
\BIBentryALTinterwordspacing
Z.~Ba, Q.~Wen, P.~Cheng, Y.~Wang, F.~Lin, L.~Lu, and Z.~Liu, ``Transferring audio deepfake detection capability across languages,'' in \emph{Proceedings of the ACM Web Conference 2023}, ser. WWW '23.\hskip 1em plus 0.5em minus 0.4em\relax ACM, Apr. 2023, p. 2033–2044. [Online]. Available: \url{http://dx.doi.org/10.1145/3543507.3583222}
\BIBentrySTDinterwordspacing

\bibitem{Le_Cacheux_2012}
\BIBentryALTinterwordspacing
Y.~Le~Cacheux, H.~Le~Borgne, and M.~Crucianu, \emph{Zero-Shot Learning with Deep Neural Networks for Object Recognition}.\hskip 1em plus 0.5em minus 0.4em\relax Springer International Publishing, Feb. 2012, p. 127–150. [Online]. Available: \url{http://dx.doi.org/10.1007/978-3-030-74478-6_6}
\BIBentrySTDinterwordspacing

\bibitem{Yurdem_2024}
\BIBentryALTinterwordspacing
B.~Yurdem, M.~Kuzlu, M.~K. Gullu, F.~O. Catak, and M.~Tabassum, ``Federated learning: Overview, strategies, applications, tools and future directions,'' \emph{Heliyon}, vol.~10, no.~19, p. e38137, Oct. 2024. [Online]. Available: \url{http://dx.doi.org/10.1016/j.heliyon.2024.e38137}
\BIBentrySTDinterwordspacing

\bibitem{Joshi_2022}
\BIBentryALTinterwordspacing
M.~Joshi, A.~Pal, and M.~Sankarasubbu, ``Federated learning for healthcare domain - pipeline, applications and challenges,'' \emph{ACM Transactions on Computing for Healthcare}, vol.~3, no.~4, p. 1–36, Oct. 2022. [Online]. Available: \url{http://dx.doi.org/10.1145/3533708}
\BIBentrySTDinterwordspacing

\bibitem{Sulaiman_2024}
\BIBentryALTinterwordspacing
R.~B. Sulaiman, \emph{A Proposed Semi-decentralized Approach of Federated Learning for Enhancing CCFD Using CS-SVM and K-Means Algorithms}.\hskip 1em plus 0.5em minus 0.4em\relax CRC Press, Feb. 2024, p. 105–122. [Online]. Available: \url{http://dx.doi.org/10.1201/9781032667478-9}
\BIBentrySTDinterwordspacing

\bibitem{Gupta_2023}
\BIBentryALTinterwordspacing
H.~Gupta, P.~Pareek, A.~Arora, M.~D. Kaur~Maini, R.~K. Bhukya, O.~P. Vyas, and A.~Puliafito, ``Fed-trace: An efficient model to traceback data poisoning attacks in federated learning,'' in \emph{2023 IEEE International Carnahan Conference on Security Technology (ICCST)}.\hskip 1em plus 0.5em minus 0.4em\relax IEEE, Oct. 2023, p. 1–6. [Online]. Available: \url{http://dx.doi.org/10.1109/iccst59048.2023.10474250}
\BIBentrySTDinterwordspacing

\bibitem{B__2023}
\BIBentryALTinterwordspacing
S.~B., Y.~R., M.~Dr.M.A.P., and G.~Dr.K., ``Federated learning and blockchain-enabled privacy-preserving healthcare 5.0 system: A comprehensive approach to fraud prevention and security in iomt,'' \emph{Journal of Internet Services and Information Security}, vol.~13, no.~4, p. 199–209, Dec. 2023. [Online]. Available: \url{http://dx.doi.org/10.58346/jisis.2023.i4.014}
\BIBentrySTDinterwordspacing

\bibitem{Ye_2024}
\BIBentryALTinterwordspacing
Y.~Ye, Y.~Chen, J.~Yang, M.~Ding, P.~Cheng, and H.~Zheng, ``Fedhelo: Hierarchical federated learning with loss-based-heterogeneity in wireless networks,'' \emph{IEEE Transactions on Network Science and Engineering}, vol.~11, no.~6, p. 6066–6079, Nov. 2024. [Online]. Available: \url{http://dx.doi.org/10.1109/tnse.2024.3447904}
\BIBentrySTDinterwordspacing

\bibitem{Godavarthi_2024}
\BIBentryALTinterwordspacing
S.~Godavarthi and D.~V.~R. G., ``Federated learning's dynamic defense against byzantine attacks: Integrating sift-wavelet and differential privacy for byzantine grade levels detection,'' \emph{International Journal of Computational and Experimental Science and Engineering}, vol.~10, no.~4, Oct. 2024. [Online]. Available: \url{http://dx.doi.org/10.22399/ijcesen.538}
\BIBentrySTDinterwordspacing

\bibitem{Qammar_2021}
\BIBentryALTinterwordspacing
A.~Qammar, J.~Ding, and H.~Ning, ``Federated learning attack surface: taxonomy, cyber defences, challenges, and future directions,'' \emph{Artificial Intelligence Review}, vol.~55, no.~5, p. 3569–3606, Nov. 2021. [Online]. Available: \url{http://dx.doi.org/10.1007/s10462-021-10098-w}
\BIBentrySTDinterwordspacing

\bibitem{Fotohi_2024}
\BIBentryALTinterwordspacing
R.~Fotohi, F.~Shams~Aliee, and B.~Farahani, ``Decentralized and robust privacy-preserving model using blockchain-enabled federated deep learning in intelligent enterprises,'' \emph{Applied Soft Computing}, vol. 161, p. 111764, Aug. 2024. [Online]. Available: \url{http://dx.doi.org/10.1016/j.asoc.2024.111764}
\BIBentrySTDinterwordspacing

\bibitem{Sarma_2023}
\BIBentryALTinterwordspacing
S.~Sarma, ``Zero-shot learning for computer vision applications,'' in \emph{Proceedings of the 31st ACM International Conference on Multimedia}, ser. MM '23.\hskip 1em plus 0.5em minus 0.4em\relax ACM, Oct. 2023, p. 9360–9364. [Online]. Available: \url{http://dx.doi.org/10.1145/3581783.3613435}
\BIBentrySTDinterwordspacing

\bibitem{Daneshfar_2024}
\BIBentryALTinterwordspacing
F.~Daneshfar, A.~Bartani, and P.~Lotfi, ``Image captioning by diffusion models: A survey,'' \emph{Engineering Applications of Artificial Intelligence}, vol. 138, p. 109288, Dec. 2024. [Online]. Available: \url{http://dx.doi.org/10.1016/j.engappai.2024.109288}
\BIBentrySTDinterwordspacing

\bibitem{Keita_2025}
\BIBentryALTinterwordspacing
M.~Keita, W.~Hamidouche, H.~Bougueffa~Eutamene, A.~Taleb‐Ahmed, D.~Camacho, and A.~Hadid, ``Bi‐<scp>lora</scp>: A vision‐language approach for synthetic image detection,'' \emph{Expert Systems}, vol.~42, no.~2, Jan. 2025. [Online]. Available: \url{http://dx.doi.org/10.1111/exsy.13829}
\BIBentrySTDinterwordspacing

\bibitem{Mao_2023}
\BIBentryALTinterwordspacing
Y.~Mao, W.~You, L.~Zhou, and Z.~Lu, ``Fixing domain bias for generalized deepfake detection,'' in \emph{2023 IEEE International Conference on Multimedia and Expo (ICME)}.\hskip 1em plus 0.5em minus 0.4em\relax IEEE, Jul. 2023, p. 2225–2230. [Online]. Available: \url{http://dx.doi.org/10.1109/icme55011.2023.00380}
\BIBentrySTDinterwordspacing

\bibitem{Huang_2022}
\BIBentryALTinterwordspacing
H.~Huang, N.~Sun, X.~Lin, and N.~Moustafa, ``Towards generalized deepfake detection with continual learning on limited new data,'' in \emph{2022 International Conference on Digital Image Computing: Techniques and Applications (DICTA)}.\hskip 1em plus 0.5em minus 0.4em\relax IEEE, Nov. 2022, p. 1–7. [Online]. Available: \url{http://dx.doi.org/10.1109/dicta56598.2022.10034569}
\BIBentrySTDinterwordspacing

\bibitem{Mancini_2020}
\BIBentryALTinterwordspacing
M.~Mancini, Z.~Akata, E.~Ricci, and B.~Caputo, \emph{Towards Recognizing Unseen Categories in Unseen Domains}.\hskip 1em plus 0.5em minus 0.4em\relax Springer International Publishing, 2020, p. 466–483. [Online]. Available: \url{http://dx.doi.org/10.1007/978-3-030-58592-1_28}
\BIBentrySTDinterwordspacing

\bibitem{Ziyadinov_2023}
\BIBentryALTinterwordspacing
V.~Ziyadinov and M.~Tereshonok, ``Low-pass image filtering to achieve adversarial robustness,'' \emph{Sensors}, vol.~23, no.~22, p. 9032, Nov. 2023. [Online]. Available: \url{http://dx.doi.org/10.3390/s23229032}
\BIBentrySTDinterwordspacing

\bibitem{Anbalagan_2022}
\BIBentryALTinterwordspacing
P.~Anbalagan, S.~Saravanan, and R.~Saminathan, ``Comprehensive and self‐contained introduction to deep reinforcement learning,'' p. 1–14, Feb. 2022. [Online]. Available: \url{http://dx.doi.org/10.1002/9781119821809.ch1}
\BIBentrySTDinterwordspacing

\bibitem{Gao_2019}
\BIBentryALTinterwordspacing
Y.~Gao, C.~Xu, D.~Wang, S.~Chen, D.~C. Ranasinghe, and S.~Nepal, ``Strip: a defence against trojan attacks on deep neural networks,'' in \emph{Proceedings of the 35th Annual Computer Security Applications Conference}, ser. ACSAC '19.\hskip 1em plus 0.5em minus 0.4em\relax ACM, Dec. 2019. [Online]. Available: \url{http://dx.doi.org/10.1145/3359789.3359790}
\BIBentrySTDinterwordspacing

\bibitem{2025}
\BIBentryALTinterwordspacing
\emph{International Journal of Intelligent Engineering and Systems}, vol.~18, no.~3, p. 730–745, Apr. 2025. [Online]. Available: \url{http://dx.doi.org/10.22266/ijies2025.0430.50}
\BIBentrySTDinterwordspacing

\bibitem{Gong_2020}
\BIBentryALTinterwordspacing
M.~Gong, Y.~Xie, K.~Pan, K.~Feng, and A.~Qin, ``A survey on differentially private machine learning [review article],'' \emph{IEEE Computational Intelligence Magazine}, vol.~15, no.~2, p. 49–64, May 2020. [Online]. Available: \url{http://dx.doi.org/10.1109/mci.2020.2976185}
\BIBentrySTDinterwordspacing

\bibitem{Miguelez_Tercero_2022}
\BIBentryALTinterwordspacing
R.~Miguelez-Tercero, A.~Jimenez-Ruiz, D.~Ruiz, G.~Fernandez-Escribano, and P.~Cuenca, ``Analysis of the capabilities of embedded systems in intraprediction video coding,'' \emph{IEEE Consumer Electronics Magazine}, vol.~11, no.~5, p. 25–40, Sep. 2022. [Online]. Available: \url{http://dx.doi.org/10.1109/mce.2021.3084817}
\BIBentrySTDinterwordspacing

\bibitem{Coccomini_2022}
\BIBentryALTinterwordspacing
D.~A. Coccomini, R.~Caldelli, F.~Falchi, C.~Gennaro, and G.~Amato, ``Cross-forgery analysis of vision transformers and cnns for deepfake image detection,'' in \emph{Proceedings of the 1st International Workshop on Multimedia AI against Disinformation}, ser. ICMR '22.\hskip 1em plus 0.5em minus 0.4em\relax ACM, Jun. 2022, p. 52–58. [Online]. Available: \url{http://dx.doi.org/10.1145/3512732.3533582}
\BIBentrySTDinterwordspacing

\end{thebibliography}

\vskip -2\baselineskip plus -1fil

\begin{IEEEbiography}[{\includegraphics[width=1in,height=1.25in,clip,keepaspectratio]{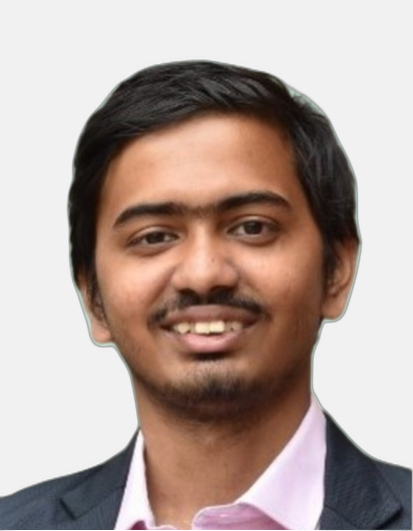}}]{Ayan Sar} \textit{Student Member, IEEE} is currently pursuing PhD from School of Computer Science, UPES Dehradun. He has a Bachelor of Technology in Computer Science at the School of Computer Science, UPES Dehradun, with a specialization in Big Data, Artificial Intelligence, and Machine Learning. His research interests include computer vision, geoinformatics, cybersecurity, and applied AI solutions.

Ayan has authored and co-authored several research papers published in prestigious venues such as IEEE and Springer Nature, focusing on AI applications, computer vision, geospatial intelligence, and computational security. His academic rigor is complemented by hands-on industry experience, including an internship at Microsoft India R\&D Pvt. Ltd., where he contributed to optimizing database queries and developing spam detection models for Microsoft 365 Copilot.

He is a National Record Holder in the India Book of Records (Academic Category) and a Gold Medalist in the International Mathematics Olympiad, underscoring his dedication to academic excellence and innovation. He has authored first-author publications in \textit{FnT Signal Processing}, \textit{IEEE Access}, \textit{Discover Food}, \textit{Engineering Applications of Artificial Intelligence}, \textit{NeuroComputing}, and co-authored many other peer-reviewed high-quality journals, along with Core B and A conferences.
\end{IEEEbiography}

\vskip -2\baselineskip plus -1fil

\begin{IEEEbiography}[{\includegraphics[width=1in,height=1.25in,clip,keepaspectratio]{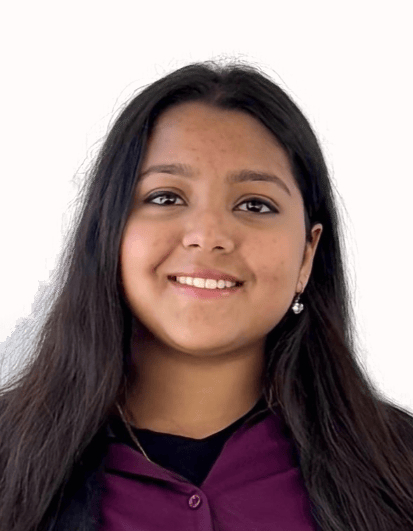}}]{Sampurna Roy} \textit{Student Member, IEEE} is currently pursuing a Bachelor of Technology in Computer Science and Engineering at the University of Petroleum and Energy Studies (UPES), Dehradun, specializing in Artificial Intelligence, Machine Learning, Cybersecurity, and UI/UX Design. She has a strong foundation in deep learning algorithms and natural language processing, with a keen interest in integrating AI-driven technologies and computer vision into practical, real-world applications.

Her notable projects include Comic Craft, a creative storytelling platform featuring automated character generation and narrative flow based on user input, and Emotion Analysis through Voice, a system designed to detect mental health patterns through vocal data. She also has experience in frontend development using Node.js and React, complemented by her growing expertise in UI/UX design for building intuitive and visually engaging user interfaces.

Sampurna has gained hands-on experience through internships at Quality AI and PortWol, where she worked on developing machine learning models and enhancing user interface designs. She has authored first-author publications in \textit{FnT Computer Graphics and Vision}, \textit{IEEE Transactions on AgriFood Electronics}, \textit{NeuroComputing} and many other peer-reviewed high-quality journals along with Core B and A conferences.
\end{IEEEbiography}

\vskip -2\baselineskip plus -1fil

\begin{IEEEbiography}[{\includegraphics[width=1in,height=1.25in, clip,keepaspectratio]{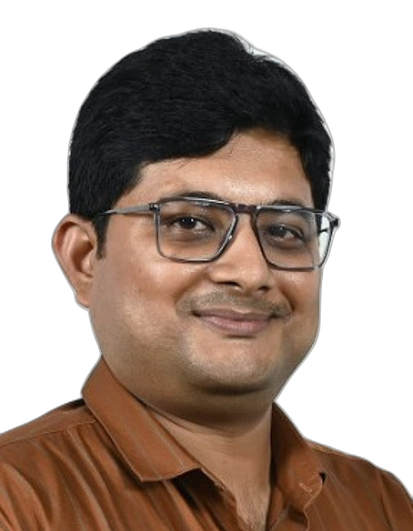}}]{Tanupriya Choudhury} \textit{Senior Member, IEEE} received his B.Tech. in Computer Science and Engineering from West Bengal University of Technology, Kolkata (2004–2008), and his M.Tech. in Computer Science and Engineering from Dr. M.G.R. University, Chennai (2008–2010). He earned his Ph.D. from Jagannath University, Jaipur, in 2016.

With over 15 years of experience in teaching and research, he is currently a Professor at the University of Petroleum and Energy Studies (UPES), Dehradun, India, and a Visiting Professor at Daffodil International University, Bangladesh. He has previously served at several prestigious institutions, including Symbiosis International Deemed University, Graphic Era Hill University (as Research Professor), and Amity University, where he was Assistant Professor (Grade-III) and International Department Head. He also held academic and industry roles at Dronacharya College of Engineering, Lingaya’s University, BBDIT, and Syscon Solutions Pvt. Ltd., among others.

His research interests include AIML, Deep Learning, Human Computing, Soft Computing, Cloud Computing, and Data Mining. A prolific innovator, he has filed 25 patents and received 16 software copyrights from the Ministry of Education (formerly MHRD), India. He has actively contributed to numerous national and international conferences as a speaker and participant.

Dr. Choudhury holds lifetime membership with IETA (International Engineering \& Technology Association) and is affiliated with several professional bodies, including IEEE, IET (UK), and other technical societies. He serves as a Technical Advisor to multiple corporate organizations such as Deetya Soft Pvt. Ltd., Noida, IVRGURU, and Mydigital360. He is also the Honorary Secretary of the Indian Engineering Teachers' Association (IETA-India) and holds Senior Advisory roles in the INDO-UK Confederation of Science, Technology and Research Ltd., London, UK, and the International Association of Professional and Fellow Engineers, Delaware, USA.
\end{IEEEbiography}

\vskip -2\baselineskip plus -1fil

\begin{IEEEbiography}[{\includegraphics[width=1in,height=1.25in, clip,keepaspectratio]{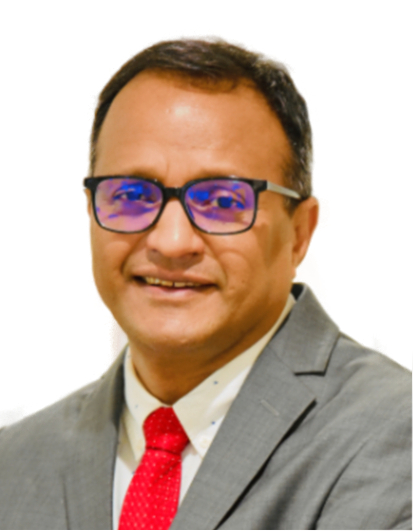}}]{Ajith Abraham} \textit{Senior Member, IEEE} received the B.Tech. degree in Electrical and Electronics Engineering from the University of Calicut in 1990, the M.S. degree from Nanyang Technological University, Singapore, in 1998, and the Ph.D. degree in Computer Science from Monash University, Melbourne, Australia, in 2001.

He is currently serving as the Vice Chancellor of Bennett University, India. Prior to this, he held several notable academic leadership positions, including Dean of the Faculty of Computing and Mathematical Sciences at FLAME University, Pune, and Founding Director of the Machine Intelligence Research Laboratories (MIR Labs), USA—an international, not-for-profit scientific research network fostering innovation and excellence in AI by bridging academia and industry.

Dr. Abraham has held professorial roles worldwide, including positions as a Professor of Artificial Intelligence at Innopolis University, Russia, and as the Yayasan Tun Ismail Mohamed Ali Professorial Chair of Artificial Intelligence at UCSI University, Malaysia.

He is an internationally renowned researcher with a multi-disciplinary focus, having authored or co-authored over 1500 research publications, including more than 100 books across various domains of computer science. His work has been widely recognized, with one of his books translated into Japanese, and several articles translated into Russian and Chinese. His scholarly impact is reflected in over 57,000 citations and an H-index exceeding 115 (Google Scholar).

Dr. Abraham has delivered over 250 plenary talks and tutorials across 20+ countries. From 2008 to 2021, he chaired the IEEE Systems, Man, and Cybernetics Society’s Technical Committee on Soft Computing, comprising over 200 members. He served as the Editor-in-Chief of Engineering Applications of Artificial Intelligence (EAAI) from 2016 to 2021 and continues to serve on the editorial boards of over 15 Thomson ISI-indexed journals. He also held the position of IEEE Computer Society Distinguished Lecturer for Europe from 2011 to 2013.
\end{IEEEbiography}

\vfill

\end{document}